\documentclass[12pt,twoside]{memoir}

\usepackage[T1]{fontenc}
\usepackage[utf8]{inputenc}
\usepackage{amsmath,amssymb,amsfonts}
\usepackage{amsthm} 
\usepackage{graphicx}
\usepackage{physics}
\usepackage{mathtools}
\usepackage[hidelinks]{hyperref}
\usepackage{cleveref}
\usepackage{geometry}
\usepackage{slashed}
\usepackage{xcolor}
\usepackage{tikz}
\usepackage{tikz-cd}
\usetikzlibrary {arrows.meta,positioning}
\usetikzlibrary{shadows}
\usetikzlibrary{calc,intersections,fadings}
\usetikzlibrary{shadows.blur}
\usepackage{tikz-3dplot}

\usepackage[absolute]{textpos}
\usepackage{tabularx}
\usepackage{ragged2e}

\usepackage{accents}
\newcommand{\flattilde}{\raisebox{0.0ex}{\scalebox{3.6}[0.7]{$\sim$}}}
\newcommand{\tconfcar}{\accentset{\flattilde}{\mathfrak{confcarr}}}

\setsecnumdepth{subsection}
\settocdepth{subsection}

\epigraphfontsize{\small\itshape}
\newenvironment{titlepage}{\cleardoublepage\thispagestyle{empty}}{}

\newcommand{\titleDISSde}[1]{{\large Dissertation} 
\\[3ex]
{\LARGE \textbf{#1}
\\[3.0ex]}
{\large
ausgeführt zum Zwecke der Erlangung des akademischen Grads
\\[1ex]
Doktor der technischen Wissenschaften
\\[1ex]
eingereicht an der TU Wien, Fakultät für Physik
}}

\newcommand{\thesistitle}{Carroll symmetries in field theory and gravity}
\newcommand{\titelvorgestellt}{Dipl.-Ing.}
\newcommand{\authorname}{Florian Ecker}
\newcommand{\titelnachgestellt}{, BSc}
\newcommand{\MatrNr}{01525800}
\newcommand{\supervisor}{Assoc. Prof. Dr.techn. \textbf{Daniel Grumiller}\newline
Institut für theoretische Physik \newline
Technische Universität Wien \newline
Wiedner Hauptstraße 8-10/136, 1040 Wien
}
\newcommand{\gutachterA}{Asst. Prof. Dr. \textbf{Laura Donnay}\newline
Scuola Internazionale Superiore di Studi Avanzati (SISSA) \newline
Via Bonomea 265, Triest, Italien
}
\newcommand{\gutachterB}{Dr. \textbf{Jan Rosseel}\newline
Rudjer Bo$\check{\textrm{s}}$kovi\'c Institut \newline
Bijeni$\check{\textrm{c}}$ka 54, HR-10000 Zagreb, Kroatien
}

\def\ruplus{\rotatebox[origin=c]{-90}{$\uplus$}}

\newcommand{\eq}[2]{\begin{equation} #1 \label{#2} \end{equation}}

\newtheorem{definition}{Definition}[section]
\newtheorem{theorem}{Theorem}[section]
\newtheorem{lemma}{Lemma}[section]
\newtheorem{example}{Example}[section]

\newcommand{\XH}{X_{\textrm{\tiny H}}}
\newcommand{\XP}{X_{\textrm{\tiny P}}}
\newcommand{\pot}{w}
\newcommand{\Xphi}{C}
\newcommand{\reldil}{X}
\newcommand{\rad}{\mathsf{r}} 

\definecolor{tabutter}{rgb}{0.98824, 0.91373, 0.30980}		
\definecolor{taorange}{rgb}{0.98824, 0.68627, 0.24314}

\begin{document}

\begin{titlepage}
\begin{textblock*}{5cm}(147mm,10mm)
\includegraphics[width=4cm]{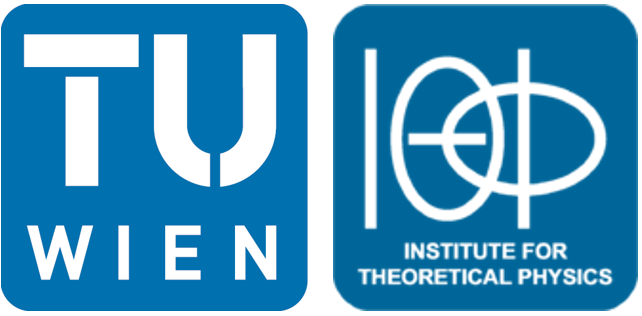}
\end{textblock*}

\vspace*{10mm}

\begin{center}
\titleDISSde{\thesistitle}
\\[1.5ex]
\rule{\linewidth}{0.2mm}
\\[10ex]
{\Large
\titelvorgestellt{} \textbf{\authorname}\titelnachgestellt}
\\[3ex]
Matr.Nr.: \MatrNr

\end{center}

\vfill

{\large

\begin{tabularx}{\linewidth-\parindent}{l>{\small\RaggedRight}X@{}} 
Betreuung: & \supervisor \\[2ex]
Begutachtung: & \gutachterA \\[2ex]
Begutachtung: & \gutachterB \\[2ex]
\end{tabularx}

\vfill

Wien, am 17.12.2025 \hfill \line(1,0){165}
}
\end{titlepage}

\frontmatter

\begin{abstract}
The past decade has seen a growing interest in Carroll symmetries, driven not only by developments in holography but also by applications in areas such as fluid mechanics and condensed matter physics.
They emerge as a contraction of Poincaré symmetries in the limit of vanishing speed of light and thus determine the structure of a particular non-Lorentzian type of kinematics. 
This thesis explores several facets of these symmetries through their applications to field theories and gravity.

The geometric description of curved Carroll manifolds is developed from a Cartan-geometric viewpoint, reviewed at the outset. On these backgrounds, we study various field theories, including scalar and vector Carroll swiftons. Imposing causality and locality, we derive a universal sector of the commutators between Carroll stress-energy tensor components valid for any Carroll quantum field theory. In two dimensions, we confirm the connection to holography by showing that a Carroll boost anomaly gives rise to additional Schwinger-like terms in these brackets, sourcing the familiar central extensions of the asymptotic symmetries of three-dimensional asymptotically flat Einstein gravity.

Afterwards, we come to theories of Carroll gravity which, as we argue, provide a valuable playground for understanding quantum gravity in a specific scaling limit which we refer to as the tantum gravity limit. At first, we review Carroll gravity in general dimensions and subsequently restrict to two spacetime dimensions where we introduce Carroll dilaton gravity. We define Carroll black holes as massive vacuum solutions to these theories that admit well-defined thermodynamic properties but have a Carroll extremal surface instead of an event horizon. After investigating several models and their solutions we finally add quantum matter to these backgrounds and study how the thermodynamic properties of a Carroll black hole reflect in its vacuum states. For the Carroll--Schwarzschild black hole we find a non-vanishing asymptotic energy density. We refer to this phenomenon as the Carroll--Hawking effect.

\end{abstract}
          
\cleardoublepage
\begin{center}
    {\bfseries Deutsche Kurzfassung}
\end{center}
Das vergangene Jahrzehnt zeigt ein stetig wachsendes Interesse an Carroll-Symmetrien, das nicht nur durch Entwicklungen in der Holografie, sondern auch durch Anwendungen in Bereichen wie der Fluidmechanik und der Festkörperphysik vorangetrieben wird. Die Carroll-Algebra l\"asst sich als Kontraktion der Poincaré-Algebra im Grenzfall verschwindender Lichtgeschwindigkeit ableiten und definiert eine charakteristische nicht-Lorentzsche Kinematik. Die vorliegende Arbeit untersucht verschiedene Aspekte dieser Symmetrien und ihre Anwendungen in Feldtheorien und Gravitationsmodellen.

Zunächst wird aus der Sichtweise der Cartan-Geometrie eine geometrische Beschreibung gekrümmter Carroll-Mannigfaltigkeiten entwickelt. Diese werden anschließend als festgehaltene Hintergrundgeometrien verwendet, um verschiedene Feldtheorien zu analysieren, darunter skalare und vektorielle Carroll-Swiftons. Unter den Annahmen von Kausalität und Lokalität wird ein universeller Satz von Kommutatoren zwischen den Komponenten des Carrollschen Energie-Impuls-Tensors abgeleitet, der für beliebige Carroll-Quantenfeldtheorien gilt. In zwei Raumzeitdimensionen wird zudem gezeigt, dass eine Carroll-Boost-Anomalie zus\"atzliche Schwinger-Terme erzeugt. Diese reproduzieren wiederum die bekannten zentralen Erweiterungen der asymptotischen Symmetriealgebra von dreidimensionaler asymptotisch flacher Einstein-Gravitation. 

Im zweiten Teil der Arbeit werden Theorien der Carroll-Gravitation untersucht, die ein vielversprechendes theoretisches Labor für das Verständnis eines bestimmten Skalierungsgrenzfalls der Quantengravitation darstellen, den wir als Tantengravitation bezeichnen. Nach einer Beschreibung von auf Carroll-Symmetrien basierenden Gravitationstheorien in allgemeinen Dimensionen wird der zweidimensionale Fall anhand von Carroll-Dilaton-Gravitation vertieft betrachtet. Es werden Carrollsche Schwarze L\"ocher als massive Vakuuml\"osungen identifiziert, die wohldefinierte thermodynamische Eigenschaften besitzen, jedoch anstelle eines Ereignishorizonts eine sogenannte Carroll-Extremalfl\"ache aufweisen. Daraufhin werden verschiedene Carroll-Dilaton-Gravitationsmodelle und deren L\"osungen detailliert analysiert. Schließlich wird das Verhalten von quantenmechanischen Materiefeldern auf fixen Geometrien untersucht. Unter bestimmten Regularit\"atsbedingungen zeigt sich dabei f\"ur die Carroll--Schwarzschild-L\"osung eine nicht verschwindende asymptotische Energiedichte. Wir bezeichnen dieses Ph\"anomen als Carroll--Hawking-Effekt.

\clearpage

\thispagestyle{empty}
\vspace*{\fill}

\begin{center}
    The research this thesis is based on was supported by \\
    the Austrian Science Fund (FWF) with the research\\
    projects P 32581, P 33789,  P 36619 and P 34562. \\[0.7\baselineskip]
    The present document was typeset in \LaTeX \\
    using the \texttt{memoir} class. 
    \\[0.7\baselineskip]
    $[$\texttt{arXiv} version$]$
\end{center}

\vspace*{2.0cm}   
\cleartorecto

\chapter*{Acknowledgments}
\addcontentsline{toc}{chapter}{Acknowledgments}

The past four years have been an incredibly enriching and rewarding time for me, both scientifically and personally. This was made possible above all by my PhD advisor, Daniel Grumiller, to whom I want to express my deepest gratitude. He was the one who first sparked my fascination for theoretical high-energy physics during my Master's studies through his lectures on black holes and holography, and I am grateful that this initial spark eventually led to the opportunity to work under his supervision. Throughout my PhD, he struck exactly the right balance between offering guidance when I needed it and giving me the freedom to explore ideas on my own. I always felt supported and encouraged. A considerable part of this experience was also shaped by the warm and supportive atmosphere in his research group. The regular meetings, open discussions, and the fact that there was always someone willing to help made it a truly fulfilling place to work at. In this regard I would also like to warmly thank all the current and former members of his research group that I had the pleasure to share office or collaborate with over the years, Ankit Aggarwal, Laura Donnay, Adrien Fiorucci, Iva Lovrekovic, Javier Matulich, Lucho Montecchio, Romain Ruzziconi, Patricio Salgado--Rebolledo, Mohanna Shams and Raphaela Wutte. Thanks to all of you! 

Second, I would like to thank my girlfriend Pia Kollermann for her support during the rough times, for sharing my musings about physics during the good times, and for being such a strong and kind person in general. You make me a better person every single day. 

My sincere thanks also go out to the brilliant people I had the privilege to collaborate with, who enriched this scientific experience by a large factor: Sudipta Dutta, Christian Ecker, Jelle Hartong, Marc Henneaux, Lucas Hörl, Tobias Jechtl, Bob McNees, Alfredo Pérez, Stefan Prohazka, Jakob Salzer, Ricardo Troncoso, Carlos Valcárcel, and Dima Vassilevich. This list naturally extends to all the very enjoyable encounters with various people from the high-energy physics community at conferences, workshops, and research visits. I am also grateful to Anton Rebhan for making an extension of my PhD possible, as well as to Laura Donnay and Jan Rosseel for accepting to be my external examiners and taking the time to read this manuscript.

Finally, I would like to express my gratitude to my family and friends for their unconditional support, also during the final stages of writing this thesis. Thanks to the members of $\Psi$ — especially Srinath Bulusu, David Globosits, Matthias Kettner, Anja Kositz, Karim El Syaad and Matthias Weinberger — for all the chats and discussions on different areas of physics, to Paul Hotzy and Florian Hechenberger for memorable bicycle adventures, and to Maximilian Ofner and Florian Lindenbauer for joining forces to revive the Pizza seminar. Thank you also to all the people I did not mention explicitly but who nonetheless had an impact on successfully finishing my PhD!

\clearpage
\cleartorecto

\tableofcontents*

\cleartoverso
\thispagestyle{empty}
\vspace*{\fill}
\begin{center}
    \includegraphics[width=0.7\textwidth]{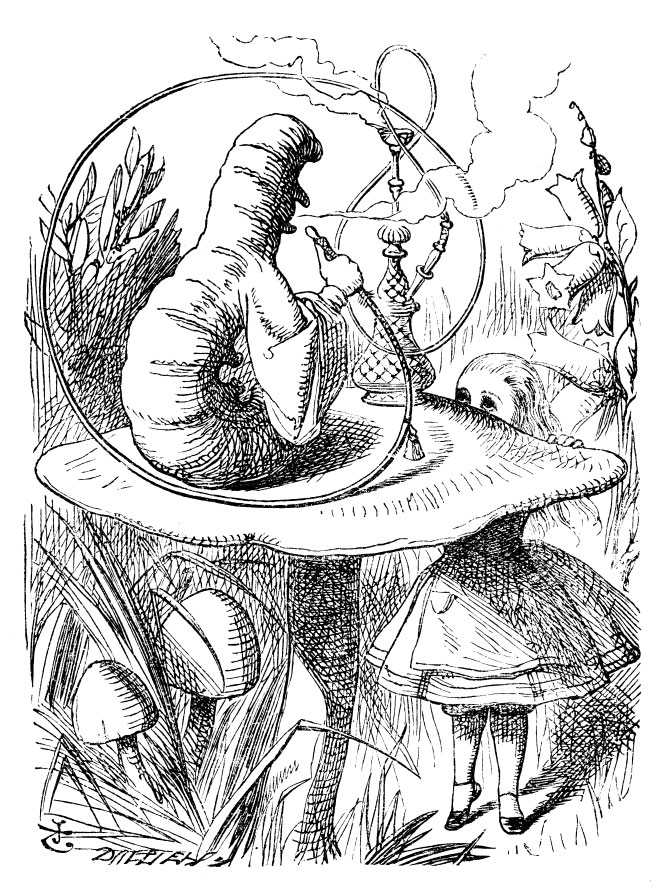}
\end{center}
\vspace*{\fill}

\mainmatter

\chapter{Introduction}
\epigraphfontsize{\small\itshape}
\epigraph{Manche meinen \\
lechts und rinks \\
kann man nicht velwechsern.\\
Werch ein Illtum!}{--- \textup{Ernst Jandl}\phantom{aaaaaa}}
\section{Black holes and a first glimpse of Carroll symmetry}
One of the biggest challenges in theoretical high-energy physics is the description of black holes. Our understanding of these objects has evolved over several decades by now dating back to the first such solution found by Schwarzschild just two weeks after Einstein published his general theory of relativity in 1915. The term black hole was only coined a few decades later by Wheeler who besides Bekenstein, Penrose and Hawking was one of the driving forces in the 60ies and 70ies to investigate the impact of the presence of a horizon on the spacetime around the black hole. A milestone in these investigations was the realization that black holes obey laws just like thermodynamic systems. In particular, a black hole with mass $M$ may be assigned an entropy \cite{Bekenstein:1973ur} as well as a temperature \cite{Hawking:1974rv}
\begin{align}
    S_{\textrm{BH}}=\frac{A\,k_B \,c^3}{4\,\hbar \,G_N} && T= \frac{\hbar\,c^3}{8\pi\,G_N\, k_B\,M} 
\end{align}
which together with the total energy satisfy the four laws of black hole thermodynamics \cite{Bardeen:1973gs}. In the expression above, $A$ stands for the area of the black hole horizon. Along with this realization, Hawking discovered his eponymous effect, which predicts that a black hole radiates as if it were a black body of temperature $T$. Subsequent developments made it evident that the emission of a strictly thermal spectrum by an evaporating black hole is fundamentally incompatible with the requirement of unitarity in any quantum field theory describing this process. Consider an initial pure quantum state undergoing gravitational collapse to form a black hole. After complete evaporation, the resulting Hawking radiation is thermal \cite{Hawking:1976ra} and is therefore described by a mixed density matrix. The overall process would thus map a pure state to a mixed state, in apparent violation of unitary time evolution. A widely accepted explanation for this ``information loss problem'' is that the emitted state is not in fact thermal but only exponentially close to being thermal. It is just the semi-classical analysis of Hawking that makes these small deviations from thermality inaccessible and once quantum gravitational effects are included the difference could be spotted in principle \cite{Almheiri:2020cfm}. 

This assertion is substantiated by lessons coming from the so-called holographic principle. Originating in the mid 90ies with the seminal works of 't Hooft \cite{'tHooft:1993gx} and Susskind \cite{Susskind:1995vu} it dictates a duality between quantum gravity in $d+1$ spacetime dimensions and a quantum field theory without gravity in one spacetime dimension less. As of today it counts as one of the most successful approaches to understanding quantum gravity. The currently most controlled implementation is the AdS/CFT correspondence proposed by Maldacena \cite{Maldacena:1997re} which describes a duality between quantum gravity of asymptotically anti de Sitter (AdS) spacetimes and conformal field theories (CFT). One important piece of evidence in this correspondence is symmetry: Given an asymptotic AdS structure one may compute the diffeomorphisms that leave it invariant, i.e. the asymptotic symmetries. For the AdS case this turns out to be just the conformal group in one dimension lower (see, e.g., \cite{Fiorucci:2021pha} for a review) which is reinterpreted as the global symmetry group of a holographically dual theory. 
Since the holographic principle predicts that any process on the gravitational side should have a corresponding description on the CFT side, and since the CFT is a unitary quantum field theory, it follows that the above process of black hole collapse and subsequent evaporation has to be a unitary process.  

The setup of the AdS/CFT correspondence, however, has the problem of being not very physical since our universe is not an asymptotically AdS spacetime. Instead, on scales still smaller than the cosmological scale, one may say to a good approximation that the asymptotics are described by flat Minkowski spacetime, e.g., for gravitational waves sent out by merging black holes that are detected by a gravitational wave detector on earth. An asymptotic symmetry analysis in this case yields the BMS group named after Bondi, van der Burg, Metzner and Sachs \cite{Bondi:1962,Sachs:1962}. In contrast to the case in AdS this is an infinite-dimensional group\footnote{Except in $2+1$ spacetime dimensions where the asymptotic symmetries of asymptotically $AdS$ spacetimes are typically also infinite-dimensional \cite{Brown:1986nw}.} which consists of the Poincaré group together with an infinite set of so-called supertranslations. These symmetries were shown to play several roles at once: Not only are they the candidate for the global symmetries of a holographically dual boundary theory but they also lead to Ward identities between S-matrix elements in the bulk which where identified \cite{Strominger:2013jfa,He:2014laa} as Weinberg's soft theorems \cite{Weinberg:1965nx} (see also \cite{Strominger:2017zoo} and references therein). These theorems state that in any scattering process between energetic (``hard'') particles in asymptotically flat spacetimes there will be additional low-energy (``soft'') particles emitted which in the limit of vanishing energy allow to factorize the corresponding scattering amplitude into a hard part multiplied by a universal soft factor. This factor is fixed entirely by the momenta of the hard particles and implies that soft gravitons are not emitted independently: their emission is correlated with the hard scattering data. 

In other words, the presence of BMS symmetries enforces correlations between soft and hard quanta. Viewing the collapse and evaporation of a black hole as a very complicated scattering process it was suggested \cite{Carney:2017jut,Strominger:2017aeh} that Hawking's computation only deals with the hard part and effectively traces over the soft quanta which is why a mixed state is obtained. According to this explanation the information loss problem thus has its origin in being agnostic about the soft sector in a scattering process. 

Coming back to the thermodynamic properties of black holes, a not completely unrelated question concerns the origin of the black hole entropy, i.e., whether there is some statistical mechanics description of black holes that it is assigned to. As stated by the no-hair theorem \cite{Chrusciel:2012jk} classical black holes are very simple objects in that they are just described by a mass, a spin as well as an electric charge. However, if their thermodynamic properties are more than a mere analogy, there should be a very large number of additional microstates associated to each such black hole. A step in this direction was made by Hawking, Perry and Strominger with the soft hair proposal \cite{Hawking:2016msc}. It asserts that there exists a large number of degrees of freedom associated with the black hole horizon that carry zero energy but whose configurations may nevertheless be distinguished by supertranslation charges. Thus, these soft degrees of freedom enlarge the classical phase space of black hole solutions and might form part of the microscopic degeneracy underlying the Bekenstein--Hawking entropy formula. Supporting evidence for this picture comes from the identification of near-horizon symmetries in $2+1$ \cite{Afshar:2016wfy,Afshar:2016kjj} as well as $3+1$ dimensions \cite{Donnay:2015abr,Donnay:2016ejv,Donnay:2019jiz}. In both cases these symmetries are related to BMS-like algebras in the respective spacetime dimension. 

As these examples make clear, the BMS group plays a central role in describing various aspects of gravity. The decisive discovery that connects it to the overall theme of this thesis was made by Duval, Gibbons and Horvathy \cite{Duval2014a,Duval:2014lpa} who showed that the BMS group is isomorphic to the conformal Carroll group. The non-conformal version of the latter was already studied in the 60ies by Lévy--LeBlond \cite{LevyLeblond1965} and Sen Gupta \cite{SenGupta1966OnAA} but remained largely unexplored for decades. Emerging as the $c \to 0$ limit of the Poincaré group, the Carroll group is associated to spacetime kinematics with restricted mobility \cite{deBoer:2021jej} -- a consequence of the relativistic light cones ``closing up'' in this limit. However, it also turns out to govern the intrinsic geometry of null hypersurfaces. Its relevance to null infinity and black hole horizons therefore follows directly, since both provide canonical examples of such null structures. The new perspective on holography of asymptotically flat spacetimes initiated an in part independent research field on Carroll physics that by now also extends to other areas in physics. This applicability of Carroll symmetries outside general relativity stems from the observation that whenever a system possesses an effective finite propagation speed for its excitations, one can identify analogues of null hypersurfaces relative to that effective causal structure. In such settings, Carrollian symmetries emerge naturally. We shall give a subjective list of some developments in the next section and refer to the recent reviews \cite{Bergshoeff:2022eog,Bagchi:2025vri,Ciambelli:2025unn,Nguyen:2025zhg,Ruzziconi:2026bix} for more details.

An overall remaining question in these approaches is how a quantum gravitational black hole should actually be defined in the first place. The classical definition cannot work any longer since it uses the notion of an event horizon in classical geometry, i.e. the existence of a region in spacetime which does not allow an escape towards future infinity \cite{Grumiller:2022qhx}. Such a description fails if black holes are really evaporating. To make some mileage in this direction it is desirable to have a simpler bulk model that on one hand reproduces the thermodynamic properties of black holes and at the same time does not rely that heavily on the existence of an event horizon. As it turns out and will be made more precise throughout this thesis gravitational theories built on Carroll symmetries provide such a setting \cite{Ecker:2023uwm}.

\section{The many faces of Carroll symmetry}
The richness of Carroll symmetries reflects in the many different areas of physics that it was discovered to play a role in. Here we shall provide a non-exhaustive list of interesting examples but refer to the recent reviews \cite{Bergshoeff:2022eog,Bagchi:2025vri,Ciambelli:2025unn,Nguyen:2025zhg,Ruzziconi:2026bix} for a more complete summary and more references.   

\paragraph{Holography}
As already mentioned in the previous section, the arguably most prominent context for Carroll physics is flat-space holography, where Carroll geometry provides the natural boundary structure at null infinity $\mathcal{I}$. The ingredients for what the actual geometric data on $\mathcal{I}$ in $3+1$ dimensions is have already been laid out by Ashtekar in the 80ies \cite{Ashtekar:1981sf,Ashtekar:2014zsa} but are now understood through a Carrollian looking glass. Nevertheless, this approach initially turned out to be very fruitful in $2+1$ dimensional gravity, where many interesting checks of the correspondence have been performed. This success can, in part, be attributed to the technical simplifications resulting from the absence of gravitational radiation in that dimension, which makes the theory topological \cite{Witten:1988hc}. A few examples are a match between the entropies for flat space cosmologies computed by a Cardy-like formula in a Carroll CFT as well as in the bulk \cite{Barnich:2012xq,Bagchi:2012xr}, a computation of entanglement entropy in a Ryu--Takayanagi-inspired way \cite{Li:2010dr,Bagchi:2014iea}, computation of correlation functions in the dual theory \cite{Bagchi:2009ca,Detournay:2014fva,Bagchi:2015wna,Bagchi:2017cpu} or the derivation of an effective action of the dual theory \cite{Barnich:2012rz,Barnich:2013yka,Merbis:2019wgk}. More recent work has also explored a coupling of additional bulk matter fields to introduce propagating degrees of freedom \cite{Bosma:2023sxn,Cotler:2024cia}. 

The situation in $3+1$-dimensional asymptotically flat Einstein gravity is more intricate due to the presence of gravitational radiation and the resulting richer boundary structure. A perspective suggested in \cite{Donnay:2022aba,Donnay:2022wvx} is to regard the Carroll CFT as being sourced by external fields representing the radiation reaching null infinity, such that the conservation equations for BMS charges become flux-balance laws \cite{Barnich:2011mi}. Subsequent developments include the construction of the corresponding Carroll boundary geometry and stress-tensor complex \cite{Hartong:2025jpp} and derivations of the flux-balance laws in a limit from AdS \cite{Campoleoni:2023fug} or an intrinsically Carroll way \cite{Fiorucci:2025twa}. Establishing entries of the holographic dictionary for correlation functions in four dimensions remains an active area of research \cite{Bagchi:2022emh,Bagchi:2023fbj,Mason:2023mti,Ruzziconi:2024kzo,Cotler:2025npu}.

Finally, Carroll symmetries are not only related to describing null infinity but also play a role at spacelike \cite{Gibbons:2019zfs} and timelike infinity of asymptotically flat spacetimes \cite{Figueroa-OFarrill:2021sxz}.  

\paragraph{Carroll field theories}
Motivated in part by holographic applications, it is interesting to consider Carroll (quantum) field theories on fixed background geometries. There are two main philosophies that are pursued. One starts with a Lorentzian theory and takes the $c\to 0$ limit in a predefined way, and the other directly constructs (conformal) Carroll invariant theories from scratch. While the two approaches agree in many cases, there is evidence that not all properties of Carroll field theories may be accessible via a limit \cite{Figueroa-OFarrill:2022mcy,Cotler:2024xhb}, see also the discussion in chapter \ref{ch:conclusion}. On the classical level, the zoo of theories reaches from Carroll versions of scalar fields \cite{Bagchi:2019xfx,Henneaux:2021yzg,deBoer:2021jej,Ecker:2024czx}, spinors \cite{Bagchi:2022eui,Bergshoeff:2023vfd,Grumiller:2025rtm}, electromagnetism \cite{Duval:2014uoa,Henneaux:2021yzg} to higher-form gauge fields \cite{Henneaux:2021yzg}. Also, a Carroll version of the Liouville action \cite{Barnich:2012rz}, as well as general ``electromagnetic'' duality invariant vector fields \cite{Bunster:2012hm} have been considered. 

On the quantum side, the most controlled setting arises when the theory enjoys additional conformal symmetry (see \cite{Bagchi:2025vri} and references therein), since, e.g., in $1+1$ dimensions, this symmetry is powerful enough to determine correlation functions directly. Studies of the general case have also been initiated in \cite{Banerjee:2023jpi,deBoer:2023fnj,Cotler:2024xhb}. Especially in \cite{Cotler:2024xhb} it became clear that Carroll quantum field theories are quite different from most Lorentzian/Euclidean counterparts: Starting, for example, with a lattice-regularized action for interacting Carroll scalar fields, it turns out that the theory obtained in the continuum limit depends on features of the lattice, i.e., it exhibits UV/IR mixing. 

\paragraph{Carroll gravity}
The applications mentioned above use Carroll symmetries as global spacetime symmetries, i.e., at the same level as special relativity uses Poincar\'e symmetries. Given the success of general relativity as a theory of gravity, which renders spacetime symmetries local, it is tempting to also make Carroll symmetries local \cite{Hartong:2015xda,Bekaert:2015xua,Bergshoeff:2017btm} and consider theories of Carroll gravity. Here, one may again begin by considering an appropriate limit of Einstein gravity, which can be implemented in at least two distinct ways \cite{Henneaux:2021yzg,Hansen:2021fxi}. One of them, often called the ``electric'' limit may be used to describe spacetimes close to singularities \cite{Henneaux:1979vn,Oling:2024vmq}, while the other, the ``magnetic'' limit, exhibits features more similar to general relativity, such as massive solutions \cite{Bergshoeff:2016soe,Hansen:2021fxi,deBoer:2023fnj,Ecker:2023uwm}. It will be one of the protagonists in this thesis, since it provides a setting for studying massive vacuum solutions in a theory of gravity that exhibit well-defined thermodynamic properties, yet do not have an event horizon. These Carroll black holes \cite{Ecker:2023uwm} are a step in the direction of finding a quantum gravitational definition of black holes outlined in the previous section. A special class of theories in $1+1$ dimensions are Carroll dilaton gravities \cite{Grumiller:2020elf,Gomis:2020wxp}, which we shall be mainly concerned with in this regard. 

Coupling Carroll gravity theories to Carroll matter fields is also an active subject \cite{Bergshoeff:2017btm,Aggarwal:2024yxy,Bergshoeff:2024ilz,Afshar:2025imp}. This also provides a link to condensed matter applications (see also the discussion below), where one is typically interested in the response functions of a system coupled to a prescribed background. A convenient way to compute some of these quantities is to covariantly couple an associated field theory to an arbitrary Carroll background and then vary the action with respect to that background.

An interesting viewpoint on curved Carroll geometries also comes from holography: Gravitational radiation passing through $\mathcal{I}$ may be equivalently viewed from the boundary perspective as determining a connection on that conformal Carroll manifold \cite{Ashtekar:2014zsa}. The curvature of this connection is related to the news tensor, which plays a central role in the description of radiative spacetimes. Thus, studying curved conformal Carroll geometries provides holographic insight into different radiative solutions of general relativity \cite{Herfray:2021qmp,Fiorucci:2025twa}.

\paragraph{Carroll hydrodynamics}  
Classical observers outside a black hole should be able to describe all relevant physics using a model that does not rely on information from the interior. This motivates an effective description in which the event horizon is replaced by a dynamical membrane, and it was shown that, as a consequence of Einstein’s equations, this membrane must behave like a fluid \cite{Thorne:1986iy}. Given the intimate connection between null hypersurfaces and Carroll symmetries, it is natural to identify this membrane fluid with a Carroll fluid \cite{Donnay:2019jiz}, whose dynamics is governed by the conservation equations of a Carroll stress energy tensor. This membrane paradigm has since been refined, and the Carroll-fluid perspective has proven capable of capturing black hole perturbations through the dissipative properties of the effective fluid \cite{Redondo-Yuste:2022czg}.

A seemingly different arena where Carroll hydrodynamics appears is in the description of the early-time dynamics of heavy-ion collisions. Shortly after the collision, the quark–gluon plasma evolves in a regime that is effectively modelled by Bjorken flow or its conformal generalization, Gubser flow. It was recently shown that both of these hydrodynamic models are, in fact, Carroll hydrodynamics in disguise \cite{Bagchi:2023ysc,Bagchi:2023rwd}.

Finally, Carroll fluids have also been proposed as effective models for the inflationary epoch of the early universe \cite{deBoer:2021jej}.

\paragraph{Condensed matter systems}
Systems with a conserved dipole moment are known to exhibit excitations with restricted mobility, commonly referred to as fractons. This links naturally to Carroll symmetries, since the collapse of lightcones in the Carrollian limit provides a geometric realisation of such constrained dynamics. This connection has been made explicit in \cite{Bidussi:2021nmp,Marsot:2022imf,Figueroa-OFarrill:2023vbj,Figueroa-OFarrill:2023qty}.

A related and equally rich setting arises in condensed matter systems with flat bands \cite{Bagchi:2022eui}. In these systems, the excitations obey a dispersion relation that is independent of momentum—precisely the Fourier transformed manifestation of collapsed lightcones. Examples include multilayer graphene, certain superconducting materials and effective models of shallow water waves, see \cite{Bagchi:2025vri} for more details and references. 

Finally, it was also realized that Carroll conformal field theories describe the region of phase separation in Tomonaga--Luttinger liquids \cite{Biswas:2025dte}.

\section{Outline of this thesis}
The present manuscript discusses various geometric and field theoretic applications based on Carroll symmetries. Special focus is laid on energy momentum tensor brackets in generic Carroll quantum field theories as well as Carroll gravity and some of its solutions, Carroll black holes. The basis is provided by the research carried out in the following papers: 
\begin{itemize}
     \setlength{\itemsep}{0.75em}
    \item F.~Ecker, D.~Grumiller, J.~Hartong, A.~Pérez, S.~Prohazka and R.~Troncoso,
``Carroll black holes,'' \emph{SciPost Phys.}\ \textbf{15} (2023), no. 6, 245,
\href{https://arxiv.org/abs/2308.10947}{arXiv:2308.10947}. 
\item F.~Ecker, D.~Grumiller, M.~Henneaux, and P.~Salgado--Rebolledo, ``Carroll swiftons,'' \emph{Phys. Rev. D}\ \textbf{110} (2024), no.4, L041901, \href{https://arxiv.org/abs/2403.00544}{arXiv:2403.00544}.
\item A.~Aggarwal, F.~Ecker, D.~Grumiller, and D.~Vassilevich, ``Carroll--Hawking effect,''  \emph{Phys. Rev. D}\ \textbf{110} (2024), no.4, L041506, \href{https://arxiv.org/abs/2403.00073}{arXiv:2403.00073}. 
\item F.~Ecker, A.~Fiorucci, and D.~Grumiller, ``Tantum gravity,'' \emph{Phys. Rev. D}\ \textbf{111} (2025), no. 2, L021901, \href{https://arxiv.org/abs/2501.00095}{arXiv:2501.00095}. 
\item S.~Dutta, F.~Ecker, L.~Montecchio, S.~Prohazka, and J.~Salzer, ``Energy momentum tensor brackets in Carroll field theories,'' \emph{in preparation}.
\end{itemize}
Other related works that did not directly make it into this thesis are:
\begin{itemize}
    \item F.~Ecker, D.~Grumiller, C.~Valc\'arcel, and D.~Vassilevich, ``Equivalences between 2D dilaton gravities, their asymptotic symmetries, and their holographic duals,'' \emph{JHEP}\ \textbf{06} (2023) 151, \href{https://arxiv.org/abs/2304.08523}{arXiv:2304.08523}.
    \item F.~Ecker, D.~Grumiller, and P.~Salgado--Rebolledo, ``Postcarrollian gravity,'' in \emph{24th Hellenic School and Workshops on Elementary Particle Physics and Gravity.} 4, 2025, \href{https://arxiv.org/abs/2504.16162}{arXiv:2504.16162}.
\end{itemize}

Let us outline the content of the following chapters in a bit more detail. In chapter \ref{ch:Carr_geom_and_symm}, several mathematical tools are introduced that allow defining what we actually mean by a Carroll spacetime. After a derivation of the Carroll algebra as a contraction of the Poincaré algebra in section \ref{sec:lim_poincare}, we delve into a systematic approach that takes as input a specific kinematical Lie algebra and delivers a way to describe general curved geometries based on this algebra. This framework, known as Cartan geometry, shall be introduced in section \ref{sec:gauging_kin_liealgebras}. In section \ref{sec:gauging_procedure}, it is applied to Carroll symmetries, where properties of Carroll metric variables and connections are discussed in more detail. Finally, we give a brief summary of various conformal extensions of the Carroll algebra that are relevant for holography in section \ref{sec:conformal_extensions}. 

Chapter \ref{ch:field_th_backgrounds} is about several aspects of Carroll field theories on fixed backgrounds. In section \ref{sec:cov_form}, we start out with a description for the various response functions that a field theory may exhibit when coupled covariantly to a Carroll background. This allows us to derive classical results such as the Carroll analogue of the stress tensor conservation and trace identities, which may in turn be used to show conservation of symmetry currents. While subsection \ref{eq:sec311} deals only with a coupling to the metric variables, the second subsection \ref{sec:coupl_conn} provides a quick look into the more general case where the field theory also couples to the degrees of freedom in the connection. Section \ref{sec:Quantum} proceeds with an application of these results for general Carroll quantum field theories: Based on locality and covariance, we derive a universal part in the commutators of Carroll energy momentum tensors. In subsection \ref{sec:Carroll_boost_anomaly}, the possibility of a Carroll boost anomaly is also discussed in $1+1$ dimensions, which connects with known results in holography. The last section \ref{sec:examples} deals with chosen examples for classical Carroll field theories and possible couplings to curved backgrounds. These consist of (non)-minimally coupled scalar fields, Maxwell fields and various versions of Carroll swiftons. 

We switch gears in chapter \ref{chap:carroll_gravity} and turn to gravity. A particular motivation for us is the exploration of definitions for black holes in a quantum setting. As we argue in section \ref{sec:bronstein}, a specific scaling limit of quantum gravity, tantum gravity, allows a classical approximation in terms of Carroll gravity. We then summarize some in part already existing results on Carroll gravitational theories for any dimension in section \ref{sec:GRforctozero}. Section \ref{sec:2d_dil} introduces our ``weapon of choice'' in the Carroll gravitational context, two-dimensional Carroll dilaton gravity. We study two ways for arriving at these theories, either from a $c\to 0$ limit of Lorentzian dilaton gravity or from spherically reducing higher-dimensional magnetic Carroll gravity. Ultimately, it is shown that both of these operations commute. General two-dimensional Carroll dilaton gravity theories are an infinite family of models, for which the spherical reduction-inspired ones are just a subset. A practical formulation of them is in terms of Poisson--Sigma models, which we introduce in subsection \ref{sec:PSMs}.

Chapter \ref{ch:CBHs} is about solutions to two-dimensional Carroll dilaton gravity, in particular about what we mean by Carroll black holes. We start out with deriving the most general solution for any two-dimensional Carroll dilaton gravity theory in section \ref{sec:5.1}, describing some special features like singularities, a Carroll version of the Birkhoff theorem and some global aspects of the corresponding spacetimes. In section \ref{sec:3}, we argue that some solutions admit thermal properties encoded in the geometry.  Energy and entropy may be defined as (boundary) Noether charges and a notion of temperature follows from demanding absence of conical defects. The ultimate consistency of these definitions is provided by a first law being satisfied by them, as shown in subsection \ref{sec:3.3}. For our definition of Carroll black holes, we need another ingredient besides thermality, which is the existence of a Carroll extremal surface. This notion is introduced in section \ref{sec:4}, before the final definition of Carroll black holes follows in subsection \ref{sec:4.4}. We provide further evidence for the internal consistency of the tantum gravity limit by evaluating the partition function in a saddle-point approximation in section \ref{sec:tant_partitionfunction}. This correctly reproduces all thermodynamic results. Finally, in sections \ref{sec:5}, \ref{sec:CS_4d} and \ref{sec:8}, we provide various examples for Carroll black holes in different settings. This includes intrinsically 2d models, 4d models as well as gravitational models coupled to additional fields such as Carroll Maxwell fields. 

The final chapter \ref{ch:Hawking} deals with Carroll quantum fields on Carroll black hole backgrounds in two spacetime dimensions. In particular, the goal here is to show that the thermal properties of the geometry also reflect in the behaviour of matter fields that couple to it. This is in analogy to the Hawking effect in general relativity. In section \ref{eq:rev_Hawking}, we review some results from the Lorentzian case before we come to the Carroll version in sections \ref{sec:Hawking_limit} and \ref{sec:Conf_anomaly_Carr}. The former takes the viewpoint of a $c\to 0$ limit, while the latter provides an intrinsic viewpoint for deriving the Carroll--Hawking effect.  

We finish off with some outlook and further perspectives in chapter \ref{ch:conclusion}.

\chapter{Carroll geometry and symmetries}
\label{ch:Carr_geom_and_symm}
\epigraphfontsize{\small\itshape}
\epigraph{``Begin at the beginning,'' the King said gravely, ``and go on till you
come to the end: then stop.''}{--- \textup{Lewis Carroll}, Alice in Wonderland}
In this chapter, we collect and review some tools for our planned journey. Specifically, we shall at first define the Carroll algebra as a contraction of the Poincaré algebra and show that a similar definition works in the case with non-zero cosmological constant. Next, we come to geometric realizations of these symmetries in terms of what is often called the gauging procedure. This allows us to systematically construct curved Carroll manifolds, first as so-called homogeneous spaces and then as more general Cartan geometries. A few definitions in connection with fiber bundles such as principal bundle connections or solder forms are necessary for this, which we introduce in subsection \ref{sec:fiber_bundles}. Finally, we mention some extensions of the Carroll algebra to various versions of conformal symmetries, and summarize some aspects about Weyl covariance of Carroll geometries.

\section{Carroll algebra as contraction of the Poincaré algebra}\label{sec:lim_poincare}
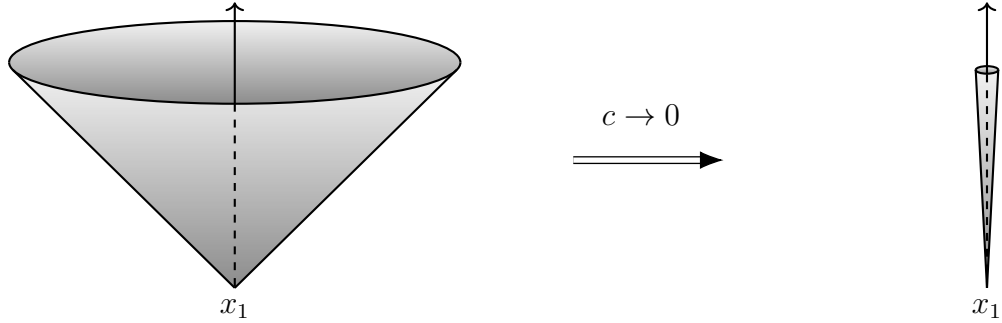
\begin{figure}
\begin{tikzpicture}[scale=1]

  \def\h{3}         
  \def\r{3}         
  \def\ellr{3}      
  \def\ellz{0.55}    

  \shade[bottom color=gray!90, top color=gray!10]
    (0,0) -- (-\r+0.05, \h-.1) arc[start angle=190, end angle=350, x radius=\ellr, y radius=\ellz] -- cycle;
  \shade[bottom color=gray!90, top color=gray!10]
    (-\r, \h) arc[start angle=180, end angle=360, x radius=\ellr, y radius=\ellz] (\r,\h) arc[start angle=0, end angle=180, x radius=\ellr, y radius=\ellz] (-\r, \h);

  \draw[thick] (0,0) -- (-\r+0.05, \h-.1);
  \draw[thick] (0,0) -- (\r-0.05, \h-.1);

  \draw[thick] (\r,\h) arc[start angle=0, end angle=180, x radius=\ellr, y radius=\ellz];
  \draw[thick] (-\r,\h) arc[start angle=180, end angle=360, x radius=\ellr, y radius=\ellz];

  \draw[dashed,thick] (0,\h-0.55) -- (0,0) node[below]{$x_1$};
  \draw[->, thick] (0,\h-0.55) -- (0,\h+0.8);


  \shade[bottom color=gray!90, top color=gray!10]
    (10,0) -- (-\r+0.05+12.8, \h-.1) arc[start angle=190, end angle=350, x radius=0.15, y radius=.05] -- cycle;
  \shade[bottom color=gray!90, top color=gray!10]
    (-\r+12.85, \h-.1) arc[start angle=180, end angle=360, x radius=.15, y radius=.05] (\r+7.15,\h-.1) arc[start angle=0, end angle=180, x radius=.15, y radius=.05] (-\r+12.85, \h-.1);

  \draw[thick] (10,0) -- (-\r+0.05+12.8, \h-.1);
  \draw[thick] (10,0) -- (\r-0.05+7.2, \h-.1);

  \draw[thick] (\r-0.05+7.2, \h-.1) arc[start angle=0, end angle=180, x radius=.15, y radius=.05];
  \draw[thick] (-\r+0.05+12.8, \h-.1) arc[start angle=180, end angle=360, x radius=.15, y radius=.05];

  \draw[dashed,thick] (10,\h-0.16) -- (10,0) node[below]{$x_1$};
  \draw[->, thick] (10,\h-0.16) -- (10,\h+0.8);
  
  \draw [line width=.5pt, double distance=2pt,
             arrows = {-Latex[length=0pt 3 0]}] (4.5,1.7) -- (6.5,1.7);
  \node[] at (5.4,2.3) {$c\to 0$};
\end{tikzpicture}
\label{fig:LC}
\caption{The lightcones of Lorentzian spacetime close in the Carroll limit $c\to 0$. }
\end{figure}

We start by considering Minkowski space, which forms the background for the theory of special relativity. It is defined by the data $(\mathbb{R}^{d+1},\eta )$, where the Minkowski metric 
\begin{align}\label{eq:minkowskimetric}
    \eta = -c^2\dd t^2 +\dd \vec{x}\,{}^2=\eta _{\mu \nu }\dd x^\mu \dd x^\nu && \mu,\nu =0,...,d
\end{align}
provides a means to measure the spacetime distance between two events at $x_1^\mu=(t_1,\vec{x}_1)$ and $x_2^\mu=(t_2,\vec{x}_2)$ by 
\begin{align}
    \Delta (x_1-x_2)=-c^2(t_1-t_2)^2+(\vec{x}_1-\vec{x}_2)^2=\eta _{\mu \nu }x_1^\mu x_2^\nu ~.
\end{align}
The Poincaré group $\text{ISO}(d,1)$ is defined as the group of transformations that leave this notion of distance invariant. Its elements are given by $(L ,a)$, where $L$ is an element of the Lorentz group $\text{SO}(d,1)$ satisfying $L^\mu {}_\alpha L^\nu {}_\beta \eta _{\mu \nu }=\eta _{\alpha \beta }$ and $a$ parametrizes a spacetime translation. They act on the coordinates as
\begin{align}
    x'{}^\mu =L^\mu {}_\nu x^\nu +a^\mu 
\end{align}
and from the composition of two such transformations one obtains that
\begin{align}\label{eq:Poincare_product}
    (L_1,a_1)\cdot (L_2,a_2)=(L_1L_2,a_1+L_1a_2)
\end{align}
showing the semi-direct product structure, $\text{ISO}(d,1)=\text{SO}(d,1)\ltimes \mathbb{R}^{d+1}$. This group lies at the heart of defining a notion of causality in Minkowski spacetime: While there is no invariant meaning to two events happening simultaneously, one can instead make statements about one event being able to influence another event. Sending out a signal at a specific spacetime point $x_1^\mu $, the finiteness of the speed of light $c$ dictates that only events that lie within or on the lightcone of $x_1^\mu $ can receive it, while events outside do not have access to the information. The Minkowski metric is a geometric way to express precisely this behaviour. If the observer of the signal is at the point $x_2^\mu $, one can distinguish three cases depending on the sign of $\Delta (x_1-x_2)$,
\begin{align}
    \Delta (x_1-x_2)\begin{cases}
        <0 \quad \text{timelike separation}\\
        =0 \quad \text{lightlike separation}\\
        >0 \quad \text{spacelike separation}
    \end{cases} 
\end{align}
and since the Poincaré group does not change $\Delta $, the nature of the separation is an invariant notion. For deriving the algebra of generators, we expand a Poincaré transformation around the identity element as
\begin{align}
    (L ,a)\sim (1+\lambda ,\epsilon )=1+\frac12 \lambda _{\mu \nu }J^{\mu \nu }+\epsilon _\mu P^\mu  ~,
\end{align}
where $\lambda _{\mu \nu }=-\lambda _{\nu \mu }$ and $\epsilon _\mu $ are the parameters for Lorentz transformations and translations and $J^{\mu \nu }$, $P^\mu $ are the respective generators. By the group product \eqref{eq:Poincare_product} we know that 
\begin{align}
     (\bar{L },\bar{a})\cdot (1+\lambda ,\epsilon )\cdot  (\bar{L },\bar{a})^{-1}=(1+\bar{L} \lambda \bar{L }^{-1},\bar{L }\epsilon -\bar{L}\lambda \bar{L}^{-1}\bar{a}) ~,
\end{align}
and from expanding $(\bar{L },\bar{a})$ as well we obtain the Poincaré algebra $\mathfrak{iso}(d,1)$ by comparing the linear terms,
\begin{align}\label{eq:poincare_compact}
    [J^{\mu \nu },J^{\alpha \beta }]&= \eta ^{\mu \beta }J^{\nu \alpha }-\eta ^{\nu \beta }J^{\mu \alpha }-\eta ^{\mu \alpha }J^{\nu \beta }+\eta ^{\nu \alpha }J^{\mu \beta }\\
    [P^\mu ,J^{\nu \alpha }]&= \eta ^{\mu \nu }P^\alpha -\eta ^{\mu \alpha }P^\nu \\
    [P^\mu ,P^\nu ]&=0
\end{align}
For taking the limit to the Carroll algbra, it is convenient to first decompose the Poincaré generators into Lorentz boosts $B_i:=J_{0i}$, temporal translations $H:=P_0$, spatial translations $P_i$ and angular rotations $J_{ij}$ in which case we get 
\begin{align}
\begin{split}\label{eq:poincare_algebra}
    [J_{ij},J_{kl}]&=\delta _{il}J_{jk}-\delta _{jl}J_{ik}-\delta _{ik}J_{jl}+\delta _{jk}J_{il}\\
    [P_i,J_{jk}]&=\delta _{ij}P_k-\delta _{ik}P_j\\
    [B_i,J_{jk}]&=\delta _{ij}B_k-\delta _{ik}B_j\\
    [B_i,P_j]&=\delta _{ij}H\\
    [B_i,B_j]&=c^2J_{ij}\\
    [B_i,H]&=c^2P_i
\end{split}
\end{align}
where we left out all the brackets that vanish. Written in this way, it is staightforward to take the contraction $c\to 0$ which leaves us with the Carroll algebra $\mathfrak{carr}(d+1)=\mathfrak{iso}(d)\;\ruplus \;\mathbb{R}^{d+1}$
\begin{equation}\label{eq:Carroll_algebra}
  \addtolength{\fboxsep}{7pt}
   \boxed{
\begin{aligned}
    [J_{ij},J_{kl}]&=\delta _{il}J_{jk}-\delta _{jl}J_{ik}-\delta _{ik}J_{jl}+\delta _{jk}J_{il}\\
    [P_i,J_{jk}]&=\delta _{ij}P_k-\delta _{ik}P_j\\
    [B_i,J_{jk}]&=\delta _{ij}B_k-\delta _{ik}B_j\\
    [B_i,P_j]&=\delta _{ij}H
\end{aligned}
   }
\end{equation} 
It was discovered by Lévy--LeBlond \cite{LevyLeblond1965} and Sen Gupta \cite{SenGupta1966OnAA} in a study of the possible kinematics of space and time. Already on this algebraic level one observes that the generator of time translations $H$ lies in the center of the algebra, making energy a Carroll invariant observable. Moreover, two Carroll boosts generated by $B_i$ now commute, as opposed to their Lorentzian counterparts. As shown in figure \ref{fig:LC}, the $c\to 0$ limit effectively closes the lightcones such that massive bodies can only move along lines of constant $\vec{x}$. 

This impossibility of spatial motion inspired Lévy--LeBlond to choose the name ``Carroll'', after the author of \textit{Alice's Adventures in Wonderland} \cite{carroll_alice}. There, at the Red Queen's race the red queen says to Alice 

\begin{quote}
\textit{"A slow sort of country!" said the Queen. "Now, here, you see, it takes all the running you can do, to keep in the same place. If you want to get somewhere else, you must run at least twice as fast as that!"}
\end{quote}
 
Various extensions and deformations of the Carroll algebra have been studied in recent years. Some examples are conformal extensions and their deformations \cite{Duval:2014lpa,Safari:2019zmc,Afshar:2024llh} or supersymmetric extensions \cite{Barnich:2015sca,Lodato:2016alv,Bagchi:2022owq}. We shall come back to the conformal versions in subsection \ref{sec:conformal_extensions}. For now, we generalize only slightly and allow a deformation by a cosmological constant.

\subsection{Adding a cosmological constant}
\label{sec:Ads_carroll_algebra}
It is straightforward to generalize the discussion to the case of a non-zero cosmological constant $\Lambda $. In that case, the algebra at $c\neq 0$ describes the isometries of Anti-de Sitter (AdS) spacetime for $\Lambda <0$ or de Sitter (dS) spacetime for $\Lambda >0$. These are both constant curvature solutions to general relativity, which are special in that they carry the maximal amount of symmetries, analogous to Minkowski spacetime for $\Lambda =0$. For further discussion of these spacetimes' properties see, e.g., \cite{Hawking:1973}. Depending on the sign of $\Lambda$, the Killing vector fields form the Anti-de
Sitter algebra $\mathfrak{so}(d,2)$ or the de Sitter algebra $\mathfrak{so}(d+1,1)$. Both of them are given by a deformation of the Poincaré algebra, 
\begin{align}\label{eq:ads_compact}
    [J^{\mu \nu },J^{\alpha \beta }]&= \eta ^{\mu \beta }J^{\nu \alpha }-\eta ^{\nu \beta }J^{\mu \alpha }-\eta ^{\mu \alpha }J^{\nu \beta }+\eta ^{\nu \alpha }J^{\mu \beta }\\
    [P^\mu ,J^{\nu \alpha }]&= \eta ^{\mu \nu }P^\alpha -\eta ^{\mu \alpha }P^\nu \\
    [P^\mu ,P^\nu ]&=-\frac{\sigma }{\ell ^2} J^{\mu \nu }
\end{align}
where $\sigma =\pm 1$ with the upper/lower sign corresponding to dS/AdS. $\ell $ is the characteristic length scale that these spacetimes are constructed with, and is related to the cosmological constant by 
\begin{align}\label{eq:cosm_const}
    \Lambda =\sigma \frac{d(d-1)}{2\ell ^2}~.
\end{align}
One can see that the translations do not commute anymore. Taking the $c\to 0$ limit in the same way as above and keeping $\Lambda $ fixed leads to the (A)dS Carroll algebra
\begin{equation}\label{eq:Carroll_ADS_algebra}
  \addtolength{\fboxsep}{7pt}
   \boxed{
\begin{aligned}
    [J_{ij},J_{kl}]&=\delta _{il}J_{jk}-\delta _{jl}J_{ik}-\delta _{ik}J_{jl}+\delta _{jk}J_{il} & &\\
    [P_i,J_{jk}]&=\delta _{ij}P_k-\delta _{ik}P_j & &\\
    [B_i,J_{jk}]&=\delta _{ij}B_k-\delta _{ik}B_j & [H,P_i]&=-\frac{\sigma }{\ell ^2} B_i\\
    [B_i,P_j]&=\delta _{ij}H & [P_i,P_j]&=-\frac{\sigma }{\ell ^2} J_{ij}
\end{aligned}
   }
\end{equation} 
As a side remark we note that for $\sigma =-1$ this algebra is isomorphic to the Poincaré algebra \eqref{eq:poincare_algebra} upon mapping $1/\ell^2 \mapsto c^2$, $B_i\mapsto -P_i$ and $P_i\mapsto B_i$. However, as we shall make more precise in the next section, when it comes to building a gravitational theory based on this algebra one additionally has to specify a subalgebra that generates the local tangent space transformations. This is where the two descriptions differ, since the subalgebra $\{J_{ij},B_i\}$ in the Poincaré case corresponds to the subalgebra $\{J_{ij},P_i\}$ in the AdS-Carroll case. In generic dimensions the gauge transformations associated to $P_i$ typically do not leave the action invariant.

\section{Minkowski spacetime in the limit \texorpdfstring{$c\to 0$}{ctozero}}
\label{sec:2.2}
So far, we have been looking only at the algebraic level of the $c\to 0$ limit. Let us now see what the implications for the geometry of Minkowski spacetime are. On the level of the metric we obtain
\begin{align}\label{eq:flat_carrVF}
    \lim _{c\to 0}\eta =\dd \vec{x}^2=h_{\mu \nu }\dd x^\mu \dd x^\nu
\end{align}
which is a degenerate positive-semidefinite metric called the \textit{Carroll metric} $h_{\mu \nu }$. Likewise, the limit of the inverse Minkowski metric, rescaled by a factor of $c^2$ yields
\begin{align}\label{eq:flat_carrmet}
    \lim _{c\to 0}c^2\eta ^{-1}=-\partial _t\otimes \partial _t =:-v^\mu v^\nu \partial _\mu \partial _\nu ~.
\end{align}
Since $\lim _{c\to 0}c^2\eta _{\mu \nu }\eta ^{\nu \alpha }=0$ we find that the \textit{Carroll vector field} $v^\mu $ spans the kernel of $h_{\mu \nu }$,
\begin{align}\label{eq:Carroll_constraints1}
    v^\mu h_{\mu \nu }=0 ~.
\end{align}
These two objects $(h_{\mu \nu },v^\mu )$ form the Carroll metric data and play the analogue role of the Lorentzian metric. Indeed, they still have the same number of independent components as the Minkowski metric: $h_{\mu \nu }$ has $(d+1)(d+2)/2$ components, $v^\mu$ has $(d+1)$ components, but there are also $(d+1)$ constraints \eqref{eq:Carroll_constraints1}, which gives $(d+1)(d+2)/2$ independent components. The important difference at this stage is that, because of its degeneracy, the Carroll metric cannot be used to raise or lower indices anymore. 

In consistency with taking the limit on the level of the Poincaré algebra in section \ref{sec:lim_poincare}, the Carroll symmetries \eqref{eq:Carroll_algebra} now emerge as spacetime symmetries that leave the structure $(v^\mu ,h_{\mu \nu })$ given in \eqref{eq:flat_carrVF}, \eqref{eq:flat_carrmet} invariant, i.e.
\begin{align}\label{eq:Killing_equations}
    \mathcal{L}_\xi v^\mu &=0 & \mathcal{L}_\xi h_{\mu \nu }&=0
\end{align}
where $\xi ^\mu $ is any Carroll Killing vector field corresponding to one of the algebra generators. Explicitly, we have
\begin{subequations}
\begin{align}
    \xi _{\mathrm{\scriptscriptstyle H}} &=\partial _t & \xi _{\mathrm{\scriptscriptstyle B_i}} &= -x_i \partial _t\\
    \xi _{\mathrm{\scriptscriptstyle P_i}} & =\partial _i & \xi _{\mathrm{\scriptscriptstyle J_{ij}}} &=x_i\partial _j-x_j \partial _i
\end{align}
\end{subequations}
which can be checked to satisfy \eqref{eq:Carroll_algebra} under the Lie bracket\footnote{In fact, just from solving \eqref{eq:Killing_equations} one obtains a much larger set of symmetries: Any vector field $\xi =f(\vec{x})\partial _t$ also solves this equation, for which $\xi _{\mathrm{\scriptscriptstyle H}}$ and $\xi _{\mathrm{\scriptscriptstyle B_i}}$ are just special cases. We shall discuss such infinite extensions in section \ref{sec:conformal_extensions}.}. 

For writing down covariant Lagrangians later on, we would like to have a volume form which we cannot build yet just from the data given.
Moreover, the normalization of $v^\mu $ is in fact undefined since there is no field that can be used to contract it into something non-zero. For this reason, following \cite{Henneaux:1979vn,Henneaux:2021yzg} we introduce an additional one-form $\tau _\mu $ with the property
\begin{align}
    v^\mu \tau _\mu =-1 
\end{align}
called the \textit{Carroll clock form}. Clearly, this does not determine $\tau _\mu $ uniquely since any shift $\tau _\mu \to \tau _\mu -\lambda _\mu $ with $\lambda _\mu v^\mu =0$ leaves this equation invariant. Once such a $\tau _\mu $ is chosen, we can use it to define a projection of an arbitrary vector field onto the subspaces parallel and normal to $v^\mu $, also referred to as the vertical and horizontal subspaces. To see this, we solve for a unique projective inverse metric $h^{\mu \nu }$ by demanding 
\begin{align}
    \delta ^\mu _\nu &=-v^\mu \tau _\nu +h^{\mu \alpha }h_{\alpha \nu } & h^{\mu \nu }\tau _\nu &=0 ~.
\end{align}
Any vector field $Y^\mu $ can now be split into its horizontal and vertical components by
\begin{align}\label{eq:verthor_proj}
    Y_{\scriptscriptstyle \text{V}}^\mu &=-v^\mu \tau _\nu Y^\nu & Y_{\scriptscriptstyle \text{H}}^\mu &=h^{\mu}_\nu Y^\nu &  Y_{\scriptscriptstyle \text{V}}^\mu +Y_{\scriptscriptstyle \text{H}}^\mu &=Y^\mu 
\end{align}
which gives $\tau _\mu $ the interpretation of an Ehresmann connection \cite{Ciambelli:2019lap}. Here, we defined the spatial projector $h^{\mu}_\nu :=h^{\mu \alpha }h_{\alpha \nu }$. It is important to keep in mind that this split inherits the non-uniqueness of $\tau _\mu $. Any transformation 
\begin{align}\label{eq:Carroll_boost}
    \tau _\mu &\to \tau _\mu -\lambda _\mu & h^{\mu \nu }&\to h^{\mu \nu }-2h^{\rho (\mu }\lambda _\rho v^{\nu )} +h^{\rho \sigma }\lambda _\rho \lambda _\sigma v^\mu v^\nu 
\end{align}
with $\lambda _\mu v^\mu =0$ provides an equally fine choice. However, some of theses choices lead to special properties of the geometry. One such example is the possibility to integrate the horizontal subspaces into a smooth submanifold, which provides a foliation of the whole spacetime by ``spacelike'' slices. By Frobenius' theorem the necessary and sufficient condition for this is that locally the horizontal subspaces close under the Lie bracket, i.e.
\begin{align}
    [Y_{\scriptscriptstyle \text{H}},Z_{\scriptscriptstyle \text{H}}]^\mu _{\scriptscriptstyle \text{V}}=-v^\mu \tau _\nu [Y_{\scriptscriptstyle \text{H}},Z_{\scriptscriptstyle \text{H}}]^\nu=0 
\end{align}
for any two vector fields $Y^\mu $, $Z^\mu $. By inserting \eqref{eq:verthor_proj}, it is straightforward to show that this corresponds to the condition
\begin{align}\label{eq:Frobenius}
    h_\alpha ^\mu h_\beta ^\nu (\partial _\mu \tau _\nu -\partial _\nu \tau _\mu )=0 \qquad \Leftrightarrow \qquad \dd \tau =a\wedge \tau 
\end{align}
for some one-form 
\begin{align}\label{eq:def_acceleration}
    a=\mathcal{L}_v\tau 
\end{align}
that is typically called the \textit{Carroll acceleration} \cite{Ciambelli:2018xat,Vilatte:2024jjr}. Since we have $d$ free parameters in $\lambda _\mu $, we can always choose $\tau =\tau _t(t,\vec{x})\,\dd t$, i.e., all the spacial legs are set to zero. This completely fixes the ambiguity and clearly solves \eqref{eq:Frobenius}. 

For arbitrary $\tau $ we may define a volume form
\begin{align}
  \mathfrak{e}=\sqrt{\det (\tau _\mu \tau _\nu +h_{\mu \nu })}
\end{align}
and show that it is invariant under \eqref{eq:Carroll_boost}. Indeed, for infinitesimal $\lambda _\mu $ we get
\begin{align}
    \delta _\lambda \mathfrak{e}=\mathfrak{e}\Big(-v^\mu \delta _\lambda \tau _\mu +\frac12 h^{\mu \nu }\delta _\lambda h_{\mu \nu }\Big)=\mathfrak{e}v^\mu \lambda _\mu =0 ~.
\end{align}
So far this discussion was purely kinematical but we shall see in the following sections that the ambiguity is realized as a gauge symmetry associated to local Carroll boosts for the case of Carroll gravity. Indeed, any sensible theory of Carroll gravity should not depend on the arbitrariness in the choice of $\tau _\mu $. 

Also, while we restricted to Carroll spacetimes obtained from the limit of Minkowski spacetime in this section, the generalization to the variables describing arbitrary curved spacetimes is immediate once the fields are written in covariant form. We shall come back to this once we introduced Carroll compatible connections in section \ref{sec:gauging_procedure}.

\section{Gauging kinematical Lie algebras}
\label{sec:gauging_kin_liealgebras}
In the previous subsection, we have seen how taking the limit $c\to 0$ affects the geometry of Minkowski spacetime and already wrote down the metric variables used for describing flat Carroll spacetimes. We shall put this on a more geometric footing now, which does not involve taking a limit but rather directly starts with the Carroll group, or one of its cousins with $\Lambda \neq 0$, and allows constructing geometries from it. The geometric framework we use here are so-called Klein geometries. The spacetimes obtained in this way typically carry a large amount of symmetries, and so we shall need to relax this concept further to be able to describe general curved spacetimes. This leads us to Cartan geometries which are a natural generalization of Klein geometries. They will not only prove to be a practical tool for organizing geometric variables, but also allow an elegant formulation of gravitational theories based on Carroll symmetries, as we shall see in the chapter \ref{chap:carroll_gravity}. The intuitive picture of how to arrive at them is summarized in the diagram \cite{Sharpe1997}
\[
\begin{tikzcd}[row sep=2.2cm, column sep=2.2cm]
  \substack{\text{\normalsize Euclidean} \\[.2em] \text{\normalsize geometry}} \arrow[r, "\substack{\text{allow} \\ \text{curvature}}"] \arrow[d,swap, "\substack{\text{generalize}\\ \text{symmetry group}}"]  & \substack{\text{\normalsize Riemannian} \\[.2em] \text{\normalsize geometry}}  \arrow[d,"\substack{\text{generalize tangent}\\ \text{space geometry}}"]\\
   \text{Klein geometry}  \arrow[r,swap, "\substack{\text{allow} \\ \text{curvature}}"] & \text{Cartan geometry}
\end{tikzcd}
\]
where it becomes evident that Riemannian geometry is just a special case where the tangent space geometry of the curved manifold is the Euclidean one. For Carroll spacetimes, it is precisely this geometry that will be replaced by a flat Carroll geometry in a sense that we shall make precise. Some useful references for this section are \cite{Sharpe1997} for a formal approach, \cite{Wise:2006sm,Wise:2009fu} for a pedagogic approach and \cite{nakahara2003geometry} for a mixture of both.

\subsection{Klein geometry}\label{sec:klein_geometry}
Consider a Lie group $G$ that acts smoothly on a manifold $\mathcal{M}$ from the left, i.e., for every $g\in G$ we get a diffeomorphism $\phi _g: \mathcal{M}\to \mathcal{M}$ that sends a point $p\in \mathcal{M}$ to $\phi _g(p)=:g\cdot p$ such that 
\begin{itemize}
    \item $\phi _g\circ \phi _{g'}=\phi _{gg'}$ with $g,g'\in G$
    \item $\phi _e(p)=p$ for any $p\in \mathcal{M}$, where $e$ is the identity element of $G$.
\end{itemize}
Additionally, this action is required to be transitive, meaning that for any two points $p$, $q\in \mathcal{M}$ there exists some $g\in G$ such that $g\cdot p=q$. 

For some fixed point $o\in \mathcal{M}$ one can determine the stabilizer subgroup $H=\{g\in G\;|\; g\cdot o=o\}$ consisting of all the group elements that leave the point unchanged\footnote{Note that in general the stabilizer depends on the point chosen. However, it can be shown that the stabilizers for any two points $o$, $o'$ with $o=go'$ the stabilizers are isomorphic, $H_{o'}=g^{-1}H_og$.}. The manifold $\mathcal{M}$ is then locally isomorphic to the space of left cosets $G/H$ built by the equivalence relation $g\sim hg$ and is called a \textit{homogeneous space}. This isomorphism $\chi:\mathcal{M}\to G/H$ is compatible with the group action both on $\mathcal{M}$ and $G/H$, i.e. $\chi (\phi _g(p))=g\chi (p)$. 
\begin{example}\label{ex:hom_example}
Consider the action of $G=SO(3)$ on the sphere $S^2$ by rotations, which is clearly transitive. Fixing the northpole $o$ yields a stabilizer $H=SO(2)$ consisting of rotations around the $z$-axis. We may construct the map $\chi : S^2\to SO(3)/SO(2)$ in the following way: Choose some $p\in S^2$ and find an element $g\in SO(3)$ such that $p=g\cdot o$. This element is only defined up to a previous rotation by an element in the stabilizer and so we get the equivalence relation $g\sim gh$ for $h\in SO(2)$. The map thus sends every $p$ to a unique coset in $SO(3)$, $p\mapsto [g]$. One may also show that it is smooth rendering $S^2$ into a homogeneous space of $SO(3)$.  
\end{example}

Reversing the order of introducing these notions, one can also start with a pair of Lie groups $(G,H)$ with $H\subset G$ being a connected subgroup and just construct the manifold $\mathcal{M}$ as a coset space. This defines a \textit{Klein geometry}. Depending on the specific properties of $(G,H)$ as well as the associated Lie algebras $(\mathfrak{g},\mathfrak{h})$ it is possible to characterize these geometries further, see \cite{Sharpe1997} for more details. The pair of Lie algebras $(\mathfrak{g},\mathfrak{h})$ is called a \textit{Klein pair} and often provides a more practical description. 

Since we are only interested in manifolds $\mathcal{M}$ that admit a spacetime interpretation we need to additionally require these Lie algebras to have a specific structure, related to the kinematical properties of a spacetime. 

\begin{definition}[Kinematical Lie algebra]\label{def:kin_lie}
    A $(d+1)$-dimensional kinematical Lie algebra $\mathfrak{g}$ is a real Lie algebra with a subalgebra $\mathfrak{so}(d)$ that allows a decomposition of $\mathfrak{g}=\mathfrak{so}(d)\oplus 2V\oplus S$ in terms of irreducible $\mathfrak{so}(d)$-representations. Here, $V$ is the $d$-dimensional vector representation and $S$ the scalar representation. A Lie group $G$ associated to $\mathfrak{g}$ is called a kinematical Lie group.
\end{definition}

We shall call the typical basis elements of such kinematical Lie algebras $\{ J_{ab},B_a,P_a,H \}$ with $J_{ab}$ denoting rotations, $B_a$ denoting boosts, $P_a$ spatial translations and $H$ temporal translations. One should keep in mind that $H$ here denotes both the subgroup and the algebra element but the distinction should be clear from the context in which it is mentioned. From the definition above it follows already that a subset of the brackets takes the form
\begin{align}
\begin{split}
    [J_{ab},J_{cd}]&=\delta _{ad}J_{bc}-\delta _{bd}J_{ac}-\delta _{ac}J_{bd}+\delta _{bc}J_{ad}\\
     [B_a,J_{bc}]&=\delta _{ab}B_c-\delta _{ac}B_b\\
    [P_a,J_{bc}]&=\delta _{ab}P_c-\delta _{ac}P_b\\
    [H,J_{ab}]&=0
\end{split}
\end{align}
with some additional non-zero brackets being possible as long as the Jacobi identities remain satisfied. Clearly, both the Poincaré algebra \eqref{eq:poincare_algebra} and the Carroll algebra \eqref{eq:Carroll_algebra} provide examples for such kinematical Lie algebras, differing from each other in the additional non-zero brackets. A classification of kinematical Lie algebras already goes back to works by Lévy--LeBlond, Bacry and Nuyts \cite{Bacry:1968zf,Bacry:1986pm}. Equipped with this definition we now make the connection to geometry.

\begin{definition}[Homogeneous kinematical spacetime]\label{eq:def_hom_spac}
A homogeneous kinematical spacetime is a Klein geometry $\mathcal{M}=G/H$ with $G$ being a kinematical Lie group and $H$ being a closed subgroup thereof. $H$ forms the stabilizer for the group action on $\mathcal{M}$ and is generated by $\mathfrak{h}=\mathfrak{so}(d)\oplus V$, giving the spacetime local spatial isotropy with $V$ acquiring the interpretation of boosts.
\end{definition}

A consequence of this definition is that the dimension of spacetime is always $\dim \mathcal{M}=\dim G-\dim H=d+1$. 
On the level of Klein pairs, the subalgebra $\mathfrak{h}$ must be chosen in such a way that the associated subgroup $H$ is closed, making the Klein pair \textit{geometrically realizable}. For further details of this constraint we refer to \cite{Figueroa-OFarrill:2018ilb} where all these connected homogeneous spacetimes built from kinematical Lie algebras were classified. 

A final property we need to define is the notion of a \textit{reductive} Klein pair. This means that one can find a complement $\mathfrak{m}$ of $\mathfrak{h}$ in $\mathfrak{g}$ such that
\begin{align}
    \mathfrak{g}=\mathfrak{h}\oplus \mathfrak{m}
\end{align}
seen as representation spaces for the adjoint action of $H$. For a connected $H$, this is equivalent to $[\mathfrak{h},\mathfrak{m}]\subset \mathfrak{m}$. Here, we shall only deal with reductive and geometrically realizable Klein pairs, although notable exceptions to reductivity exist in the literature \cite{Figueroa-OFarrill:2022mcy}, e.g., the Klein pair that generates the lightcone in Minkowski space.

A consequence of definition \ref{eq:def_hom_spac} is that one can obtain very different spacetimes from the same kinematical algebra, depending on the choice of subalgebra $\mathfrak{h}$. The archetypical example is the Klein pair $(\mathfrak{iso}(d,1),\mathfrak{so}(d,1))$ of the Poincaré algebra and its Lorentz subalgebra. The homogeneous spacetime associated to it is just Minkowski space in $d+1$ dimensions. Another homogeneous spacetime associated to the Poincaré group is $(\mathfrak{iso}(d,1),\mathfrak{iso}(d-1,1))$, replacing the Lorentz algebra with the $d$-dimensional Poincaré algebra. This spacetime appears as a blowup Spi$^{d+1}$ of spacelike infinity of asymptotically flat spacetimes \cite{Figueroa-OFarrill:2021sxz} and does not support a non-degenerate metric.  

As these two examples already indicate, different homogeneous spaces can often be distinguished—and in practice efficiently classified—by examining the invariant tensor fields that can be consistently defined on them. A general method for their construction is to look at linear isotropy representations \cite{Bergshoeff:2022eog}. The basic idea is the following: Fix again some origin $o\in \mathcal{M}$ whose stabilizer we denote by $H\subset G$. The elements $h\in H$ then induce linear maps $\lambda (h):T_o \mathcal{M} \to T_o \mathcal{M}$. The collection of these maps furnishes the isotropy representation of $H$. One can show (\cite{Besse1987}, p.282) that there is a one-to-one correspondence between $H$-invariant tensors in $T_o\mathcal{M}$ and $G$-invariant tensor fields on $\mathcal{M}$. In other words, every tensor at $o$ that is fixed by the isotropy representation extends uniquely to a $G$-invariant tensor field, and conversely, every $G$-invariant tensor field restricts to an $H$-invariant tensor at $o$. Since in the reductive case $T_o\mathcal{M}\cong \mathfrak{m}$ (for a proof see \cite{Sharpe1997}, p.163) this means that one has to find $H$-invariant tensors on $\mathfrak{m}$ and automatically knows that they can be extended to all of $\mathcal{M}$. 

Let us look at two examples to see how the structure from section \ref{sec:2.2} arises. In each case the algebra $\mathfrak{m}$ is spanned by $\{H,P_a\}$ with dual components $\{\pi ^0,\pi ^a\}\in \mathfrak{m}^\ast $ defined via the non-zero pairings $\pi ^0(H)=1$, $\pi ^b(P_a)=\delta ^b_a$. We can work on an infinitesimal level and look at the action of $\mathfrak{h}$ on $\mathfrak{m}$ and $\mathfrak{m}^\ast $. It is given by $\text{ad}_{\mathbf{y}}(\mathbf{x})=[\mathbf{y},\mathbf{x}]$ and $\text{ad}^\ast _{\mathbf{y}}(\alpha )=-\alpha \circ \text{ad}_{\mathbf{y}}$ for $\mathbf{y}\in \mathfrak{h}$, $\mathbf{x}\in \mathfrak{m}$ and $\alpha \in \mathfrak{m}^\ast $. 

\begin{example}\label{ex:poincare_tensors}
Consider the Klein pair $(\mathfrak{g},\mathfrak{h})=(\mathfrak{iso}(d,1),\mathfrak{so}(d,1))$ that models Minkowski space. The action of $\mathfrak{h}=\mathfrak{so}(d,1)$ on $\mathfrak{m}$ and $\mathfrak{m}^\ast $ can be worked out from the Poincaré algebra \eqref{eq:poincare_algebra},
    \begin{subequations}\label{eq:H_action_poincare}
    \begin{align}
        \text{ad}_{J_{ab}} H&=0 & \text{ad}^\ast _{J_{ab}} \pi ^0&=0 \\
        \text{ad}_{B_{a}} H&=c^2P_a & \text{ad}^\ast _{B_a} \pi ^0&=-\delta _{ab}\pi ^b\\
        \text{ad}_{J_{ab}} P_c&=2\delta _{c[b}P_{a]} & \text{ad}^\ast _{J_{ab}} \pi ^c&=-2\delta ^c_{[a}\delta _{b]d}\pi ^d \\
        \text{ad}_{B_{a}} P_b&=\delta _{ab}H & \text{ad}^\ast _{B_a} \pi ^b&=-c^2\delta _a^b\pi^0
    \end{align}
      \end{subequations}
By direct computation, one can show that the following two tensors are indeed invariant:
\begin{align}
    \eta &=-c^2\pi ^0\otimes \pi ^0 + \delta _{ab}\pi ^a\otimes \pi ^b \quad \in \quad \mathfrak{m}^\ast \odot \mathfrak{m}^\ast \\
    \epsilon &=\pi ^0 \wedge \pi ^{a_1} \wedge ... \wedge \pi^{a_d}\,\epsilon _{a_1...a_d} \quad \in \quad \bigwedge ^{d+1}\mathfrak{m}^\ast ~. 
\end{align}
As the names already suggest, $\eta $ just corresponds to the Minkowski metric while $\epsilon $ gives an invariant volume form on $\mathcal{M}$. 
\end{example}

\begin{example}\label{ex:carrooll_tensors}
Taking the limit $c\to 0$ of the previous example one arrives at the Klein pair $(\mathfrak{carr}(d+1),\mathfrak{iso}(d))$, which describes a flat Carroll manifold. The action of $\mathfrak{h}=\mathfrak{iso}(d)$ on $\mathfrak{m}$ and $\mathfrak{m}^\ast $ can be either worked out using the brackets of the Carroll algebra \eqref{eq:Carroll_algebra} or just by taking the $c\to 0$ limit of \eqref{eq:H_action_poincare}. In this case, one finds invariant tensors
\begin{align}
    v&=H \quad \in \quad \mathfrak{m}\\[.5em]
    h&= \delta _{ab}\pi ^a\otimes \pi ^b \quad \in \quad \mathfrak{m}^\ast \odot \mathfrak{m}^\ast\\
    \epsilon &=\pi ^0 \wedge \pi ^{a_1} \wedge ... \wedge \pi^{a_d}\,\epsilon _{a_1...a_d} \quad \in \quad \bigwedge ^{d+1}\mathfrak{m}^\ast 
\end{align}
where $\epsilon $ is defined as before. There is no possibility to define a non-degenerate invariant metric on $\mathcal{M}$ anymore. Instead, we get the degenerate Carroll metric $h$ we saw arise in section \ref{sec:2.2} together with an invariant vector $v$.
\end{example}
For a broader range of examples, as well as systematic classification results, we refer the reader to \cite{Figueroa-OFarrill:2018ilb} and the references cited therein. The geometries that arise from the Klein construction are, by design, very closely tied to the symmetries of the underlying homogeneous spaces.
This makes their construction on one hand very clear but on the other hand quite rigid in the sense that the obtained spacetimes are highly symmetric manifolds often admitting only a limited class of geometric deformations. While such maximally symmetric backgrounds serve as important prototypes, our ultimate aim is to describe dynamical geometries, where the fields and structures are allowed to vary from point to point in a nontrivial way. To achieve this, one must go beyond the strictly homogeneous setting provided by the Klein paradigm and seek a suitable generalization that can faithfully encode local geometric data. This naturally leads to the framework of Cartan geometries, which will be the focus of subsection \ref{sec:cartan_geometry}. Before, we however need to slightly adapt our language.

\subsection{Fiber bundles and connections}\label{sec:fiber_bundles}
We proceed by defining several notions about fiber bundles that will become important for Cartan geometries. Suppose we again have a smooth $(d+1)$-dimensional manifold $\mathcal{M}$ as our base manifold. 
\begin{definition}[Vector bundle]\label{def:vector_bundle}
    A real vector bundle $E\to \mathcal{M}$ of rank $k$ over a manifold $\mathcal{M}$ is a smooth manifold $E$, called the total space, with a projection map $\pi_E:E\to \mathcal{M}$ satisfying the following conditions:
    \begin{itemize}
        \item For every point $p\in \mathcal{M}$ the fiber $E_p:=\pi ^{-1}_E(p)$ is a real vector space of dimension $k$
        \item Locally, $E$ looks like a product: For every point in $\mathcal{M}$, there is a neighbourhood $\mathcal{U}$ with a map $\varphi _{\mathcal{U}}:\pi ^{-1}_E(\mathcal{U})\to \mathcal{U}\times \mathbb{R}^k$ called a local trivialization of $E$ over $\mathcal{U}$.
        \item If $(\mathcal{U},\varphi _{\mathcal{U}})$, $(\mathcal{V},\varphi _{\mathcal{V}})$ are two local trivializations with non-empty intersection $\mathcal{U}\cap \mathcal{V}$ then the composition
        \begin{align}
            \varphi _{\mathcal{U}}\circ \varphi _{\mathcal{V}}^{-1}: (\mathcal{U}\cap \mathcal{V})\times \mathbb{R}^k \to (\mathcal{U}\cap \mathcal{V})\times \mathbb{R}^k
        \end{align} 
        maps $(p,v)\mapsto (p,t_{\mathcal{U}\mathcal{V}}\,v)$ with a smooth GL$(k,\mathbb{R})$ valued transition function $t_{\mathcal{U}\mathcal{V}}: \mathcal{U}\cap \mathcal{V}\to GL(k,\mathbb{R})$. GL$(k,\mathbb{R})$ is called the structure group of the vector bundle.
    \end{itemize}
\end{definition}
The classic example is the tangent bundle $T\mathcal{M}$, i.e., the disjoint union of all the tangent spaces to $\mathcal{M}$,
\begin{align}
    T\mathcal{M}=\dot{\bigcup}_{p\in \mathcal{M}}T_p\mathcal{M} ~.
\end{align}
Here, $k=d+1$ and the local trivializations can be induced by coordinate patches on the base manifold: Given a chart $x^\mu : \mathcal{U}\to \mathbb{R}^{d+1}$ we get a basis $\partial /\partial x^\mu \vert _p$ for the tangent space $T_p\mathcal{M}$ at $p\in \mathcal{U}$. Then we define $\varphi _{\mathcal{U}}(v)=(p,v^\mu )$, where $v^\mu $ are the components of the tangent vector $v\in \pi ^{-1}_E(p)$ in that basis. Clearly, this is not a unique choice: We could have picked any other basis $\{E_0,..,E_d\}_p$, from here on referred to as a \textit{frame}, and decompose the vector at each $p\in \mathcal{U}$ like $v=v^AE_A$. The trivialization would then be defined like $\phi _{\mathcal{U}}(v)=(p,v^A)$. If the two bases are related like $E_A=E_A^\mu \partial _\mu $ with some GL$(d+1,\mathbb{R})$ matrix $E_A^\mu $, it follows that the transition functions from the first to the second choice of basis are $(E^{-1})^A_\nu =:e_\nu ^A$. 

A \textit{section} of a vector bundle is a smooth map $\sigma :\mathcal{M}\to E$ such that $\pi _E \circ \sigma =id$. Intuitively, to each base point $p\in \mathcal{M}$ it associates a single vector $\sigma (p)$ in the fiber $\pi ^{-1}_E(p)$. For example, sections of $T\mathcal{M}$ are just tangent vector fields. The space of sections of a given bundle $E$ is denoted by $\Gamma (E)$.

If there is further structure provided on the manifold, it is possible to choose the trivializations such that the transition functions only take values in a subgroup of GL$(k,\mathbb{R})$. One example is a Riemannian metric, which allows choosing orthonormal frames at every point $p\in \mathcal{M}$ and reduces the structure group to O$(d+1)$. For the case of a Carroll structure in the form of a vector field together with a degenerate metric, the structure group can be reduced to ISO$(d)$, the homogeneous Carroll group. 

There is a more general underlying structure to each vector bundle, which focuses on precisely this group theoretic aspect.

\begin{definition}[Principal bundle]
    A principal $G$-bundle $P\to \mathcal{M}$ is a smooth manifold $P$ with a projection map $\pi :P\to \mathcal{M}$ and a smooth right action $P\times G\to P$ that maps $(u,g)\mapsto u\cdot g$ satisfying the following properties:
    \begin{itemize}
        \item The fiber $\pi ^{-1}(p)$ has the structure of the Lie group $G$. In particular, the right action preserves the fiber, $\pi (u\cdot g)=\pi (u)$.
        \item Locally, $P$ looks like a product: For every point in $\mathcal{M}$ there is a neighbourhood $\mathcal{U}$ with a map $\varphi _{\mathcal{U}}:\pi ^{-1}(\mathcal{U})\to \mathcal{U}\times G$ called a local trivialization. Each such map should be $G$-equivariant, i.e. for $\varphi _{\mathcal{U}}(u)=(\pi (u),h)$ we should have $\varphi _{\mathcal{U}}(u\cdot g)=(\pi (u),hg)$ for some $g\in G$.
        \item If $(\mathcal{U},\varphi _{\mathcal{U}})$, $(\mathcal{V},\varphi _{\mathcal{V}})$ are two local trivializations with non-empty intersection $\mathcal{U}\cap \mathcal{V}$ then the composition
        \begin{align}
            \varphi _{\mathcal{U}}\circ \varphi _{\mathcal{V}}^{-1}: (\mathcal{U}\cap \mathcal{V})\times G \to (\mathcal{U}\cap \mathcal{V})\times G
        \end{align} 
        maps $(\pi (u),h)\mapsto (\pi (u),t_{\mathcal{U}\mathcal{V}}\,h)$ with a smooth G-valued transition function $t_{\mathcal{U}\mathcal{V}}: \mathcal{U}\cap \mathcal{V}\to G$. 
    \end{itemize}
\end{definition}
Similarly to the case of vector bundles, one can define sections of principal bundles, in general, however, only over some neighbourhoods $\mathcal{U}\subset \mathcal{M}$. The existence of a global section is a diagnostic for the principal bundle being trivial, i.e., $P=\mathcal{M}\times G$ globally \cite{nakahara2003geometry}. 

Let us provide two examples. The frame bundle $F\mathcal{M}$ of the manifold $\mathcal{M}$ is a principal GL$(d+1,\mathbb{R})$-bundle which consists of the set of frames at every point. A local trivialization can be again induced by a coordinate patch on the base manifold: Given a frame $E_A$ at a point $p\in \mathcal{U}$, we can decompose it like $E_A=E_A^\mu \partial _\mu $ and define a trivialization $\varphi _{\mathcal{U}}(E_A)=(p,E_A^\mu )$ where $E_A^\mu $ is seen as an element of GL$(d+1,\mathbb{R})$. In general, one can induce a local trivialization $\varphi _{\mathcal{U}}$ of a principal bundle by any choice of local section over $\mathcal{U}$. The coordinate frame $\partial _\mu $ is just one such example. The right action of GL$(d+1,\mathbb{R})$ on $F\mathcal{M}$ relates one frame at $p$ to another, $E_B\to E_Ag^A{}_B$. 

Another example is a Klein geometry: Given a Lie group $G$ together with a closed subgroup $H$, we construct the principal $H$-bundle $G\to G/H$, where $G/H$ is the space of left cosets. The projection map can just be defined as $g\to [g]$, which makes it clear that the typical fibers are isomorphic to $H$. A local trivialization is given by picking a local section $\sigma ([g])$ and defining $\varphi_\mathcal{U}:G\to \mathcal{U}\times H$ by $\varphi _{\mathcal{U}}(g)=([g],(\sigma ([g]))^{-1}g)=([g],a)$. In the last equality we have used that there exists a unique $a\in H$ such that $\sigma ([g])=ga$.

We can connect the two types of bundles by introducing the notion of an \textit{associated vector bundle}. The construction works as follows: Given a principal $G$-bundle $P\to \mathcal{M}$ and a vector space $V$ that $G$ acts on via some representation $\rho :G\to GL(V)$ we define a vector bundle $(P\times _\rho V)\to \mathcal{M}$ with projection map $\pi _E$. The total space is constructed by identifying two elements of $P\times V$ as 
\begin{align}
    (u,v)\sim (u\cdot g,\rho (g^{-1})v) ~,
\end{align} 
while the projection is defined as $\pi _E([u,v])=\pi (u)$, which is consistent because of the property $\pi (u\cdot g)=\pi (u)$. For the case of the frame bundle, picking $V=\mathbb{R}^{d+1}$ together with the fundamental matrix representation of GL$(d+1,\mathbb{R})$ we can construct the associated vector bundle $(F\mathcal{M}\times _\rho \mathbb{R}^{d+1})\to \mathcal{M}$. One can show that the result is isomorphic to the tangent bundle $T\mathcal{M}$ by identifying points in the total space by $[E_A,v^A]\sim [E_Ag^A{}_B,(g^{-1}){}^B{}_Av^A]\mapsto v^AE_A$. This is just an abstract way of encoding the typical inverse transformation of the components of a vector when the basis is transformed. 

Another example of an associated vector bundle is to pick the adjoint representation $\rho :G\to \text{GL}(\mathfrak{g})$ on its Lie algebra. This is defined by $\text{Ad}_g(\mathbf{x})=\frac{\dd }{\dd t}(ge^{t\mathbf{x}}g^{-1})\vert _{t=0}$ for $\mathbf{x}\in \mathfrak{g}$. The adjoint vector bundle is then $\text{ad}P:= (P\times _\rho \mathfrak{g})$.

In general, one can think of an associated bundle with total space $(P\times _\rho V)/G$ as a vector bundle over $\mathcal{M}$ with typical fiber $V$. The transition functions of the associated vector bundle are just given by $\rho (t_{\mathcal{U}\mathcal{V}})$ where $t_{\mathcal{U}\mathcal{V}}$ are those of $P$.

\bigskip

\begin{figure}
\centering
\begin{tikzpicture}[scale=1.7]

\shade[
    right color=gray!90,
    middle color=gray!35,
    left color=gray!15,
    draw=black!70,
    opacity=0.8
]
  (0,0)                             
      .. controls (3,-0.5) and (3,0.5) .. (5,0)   
      .. controls (3.8,-0.2) and (3.5,-0.5) .. (3,-1)  
      .. controls (1,-0.5) and (1,-1.5) .. (-2,-1)   
      .. controls (-1,-0.8) and (-0.2,-0.5) .. (0,0) 
    -- cycle;   

\node[above right] at (-1.4,-1.1) {$\mathcal{M}$};

\draw[black] (0.2,-0.7) -- (-0.5,1);
\draw[black] (1.1,-0.7) -- (1.1,1);
\draw[black] (2.2,-0.5) -- (2.5,1);

\draw[black] (0.64,-1.75) -- (0.36,-1.09);
\draw[black] (1.1,-1.7) -- (1.1,-0.96);
\draw[black] (1.95,-1.7) -- (2.13,-0.86);

\fill (0.2,-0.7) circle (0.6pt);
\fill (1.1,-0.7) circle (0.6pt);
\fill (2.2,-0.5) circle (0.6pt);
\node[right] at (0.2,-0.7) {$p$};
\node[above] at (-0.5,1) {$\pi ^{-1}(p)$};
\node[right] at (2.2,-0.5) {$p'$};
\node[above] at (2.5,1) {$\pi ^{-1}(p')$};

\draw[->, >={Stealth}] (-0.21,0.3) -- (0.155,0.88) node[midway, above] {};
\draw[black,dashed] (-0.33,0.58) -- (0.15,0.88);
\node[above right] at (0.12,0.85) {$X$};
\fill (-0.21,0.3) circle (0.6pt);
\node[left] at (-0.17,0.2) {$u$};

\draw[->, >={Stealth}] (-0.21,0.3) -- (0.26,0.61) node[midway, below] {};
\node[above right] at (-0.1,0.145) {$X_{\scriptscriptstyle \text{H}}$};
\draw[black,dashed] (0.25,0.61) -- (0.15,0.88);
\draw[->, >={Stealth}] (-0.21,0.3) -- (-0.33,0.58) node[midway, left] {$X_{\scriptscriptstyle \text{V}}$};

\fill (2.32,0.1) circle (0.6pt);
\node[right] at (2.32,0.1) {$u'$};
\fill (2.46,0.8) circle (0.6pt);
\node[right] at (2.46,0.8) {$u'\cdot g$};
\draw[-{Stealth}]
  (2.28,0.15)
    .. controls (2.2,0.3) and (2.2,0.6)
  .. (2.43,0.78);
\node[left] at (2.22,0.45) {$g$};

\end{tikzpicture}
\caption{The choice of a principal bundle connection allows to decompose each vector $X$ at some point $u\in \pi ^{-1}(p)$ into its horizontal and vertical components. We also drew an example for the right action of $G$ on a fiber $\pi ^{-1}(p')$.}
\label{fig:bundle1}
\end{figure}
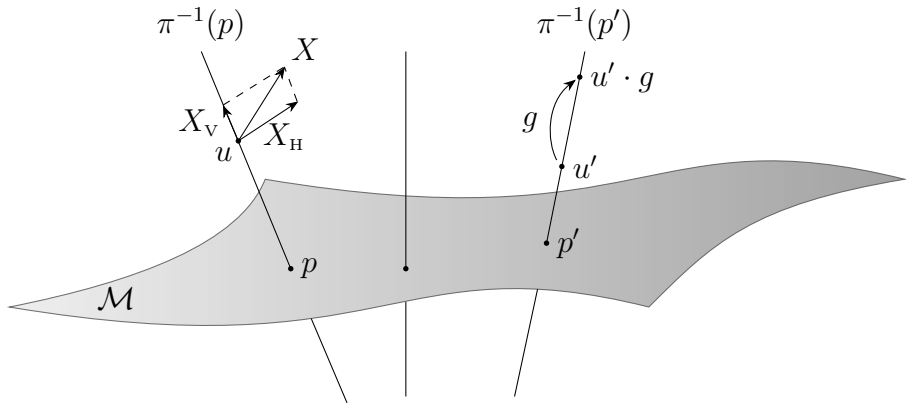

We now come to connections, a central class of objects on fiber bundles since they allow to define a notion of parallel transport and, as a further step, lead to the definition of curvature. We first introduce a connection on a principal fiber bundle and will afterwards see how this induces a connection on associated vector bundles such as the tangent bundle. Given a principal $G$-bundle $P\to \mathcal{M}$ with projection map $\pi $, there exists a natural map between the Lie algebra $\mathfrak{g}$ and the tangent space $T_uP$ of $P$ at the point $u$. We define 
\begin{align}\label{eq:vert_VF}
    X^\sharp f(u):=\frac{\dd }{\dd t}f(u\cdot e^{t\mathbf{x}})\big \vert _{t=0} \qquad X^\sharp \in T_uP \qquad \mathbf{x}\in \mathfrak{g}
\end{align}
for some function $f:P\to \mathbb{R}$. Since one can show that $\pi _\ast (X^\sharp)=0$ the collection of these vectors define a subspace $V_uP\subset T_uP$, called the \textit{vertical subspace}. Moreover, because of $[X^\sharp,Y^\sharp]=[\mathbf{x},\mathbf{y}]^\sharp$ the map $\mathbf{x}\to X^\sharp$ is a Lie algebra homomorphism. A connection on a principal bundle provides us with a projection onto this subspace given an arbitrary vector $X\in T_uP$. It is defined by
\begin{definition}[Connection on a principal bundle]\label{def:connection}
    A connection on a principal $G$-bundle $P\to \mathcal{M}$ is a $\mathfrak{g}$-valued one-form $\omega $ on $P$ satisfying the following properties:
    \begin{itemize}
        \item $\omega (X^\sharp )=\mathbf{x}$ with $\mathbf{x}\in \mathfrak{g}$.
        \item $R_g^\ast \omega =\text{Ad}_{g^{-1}}\circ \omega $.
        \item $\omega $ allows a unique decomposition of any vector $X=X_{\scriptscriptstyle \text{V}}+X_{\scriptscriptstyle \text{H}}$ with $\omega (X_{\scriptscriptstyle \text{H}})=0$. $X_{\scriptscriptstyle \text{H}}$ is an element of $H_uP\subset T_uP$ called the horizontal subspace, see also figure \ref{fig:bundle1}. 
    \end{itemize} 
\end{definition}
In the second property, we denoted the right action of $G$ on $P$ by $R_gu=u\cdot g$. It implies that the assignment of the horizontal subspace is consistent with the right group action on the fibers, i.e. $H_{ug}P=R_g{}_\ast H_uP$. In particular, the push-forward of a horizontal vector along the group action should again be a horizontal vector. One usually does not work with the abstract object $\omega $ on the total space $P$ but rather uses a local section $\sigma :\mathcal{U}\to P$ to define a $\mathfrak{g}$-valued one-form on $\mathcal{M}$,
\begin{align}
    \omega _{\mathcal{U}}=\sigma ^\ast \omega \quad \in \quad \Omega ^1(\mathcal{M})\otimes \mathfrak{g} ~.
\end{align}
This is the description we shall use in the later chapters. The definition is clearly not unique, since one could have equally well chosen a different section $\sigma ':\mathcal{U}\to P$. Because of the properties of the principal bundle we know that there exists a unique $G$-valued transition function $g(p):\mathcal{U}\to G$ such that $\sigma '(p)=\sigma (p)\cdot g(p)$ for each $p\in \mathcal{U}$. Defining also an associated $\omega '_{\mathcal{U}}=\sigma '{}^\ast \omega $, one can show that the two definitions of the connection one-form on the base manifold are related by
\begin{align}\label{eq:conn_trafo}
    \omega _{\mathcal{U}}'=g^{-1}\big(\dd +\omega _{\mathcal{U}}\big )g 
\end{align}
which is the typical transformation behaviour of a connection such as the Yang--Mills field under a gauge transformation. Therefore, in this picture ``choosing a gauge'' means to choose a local section of the underlying principal bundle by which one relates the globally defined object $\omega $ to a local description $\omega _{\mathcal{U}}$ on $\mathcal{M}$. This scheme will continue for all the quantities that are defined in the following: Once a description in terms of local tensorial objects is chosen, a certain gauge dependence enters the game. The used section should be the same for all quantities, making the transformation act on all fields simultaneously. It should be emphasized that the local and global descriptions are equivalent, in the sense that the full $\omega $ on the total space can always be reconstructed once a collection of local $\omega _{\mathcal{U}}$ covering all of $\mathcal{M}$ is given \cite{nakahara2003geometry}. We usually drop the subscript when we express $\omega _{\mathcal{U}}$ in terms of its components in a chart over $\mathcal{U}$, i.e., we just write $\omega _\mu $ in that case. 

The chosen principal bundle connection $\omega $ induces an exterior covariant derivative on $P$. Its action on a p-form $\alpha \in \Omega ^p(P)$ is
\begin{align}\label{eq:cov_der_P}
D\alpha (X_1,...X_{p+1}):=\dd _P\alpha (X_1{}_{\scriptscriptstyle \text{H}},...,X_{p+1}{}_{\scriptscriptstyle \text{H}}) 
\end{align}
where the connection enters in only taking the horizontal components of the vector fields $X_1,...,X_{p+1}$ and $\dd _P$ is just the exterior derivative on $P$. It is used to define the \textit{curvature two-form} of the connection,
\begin{align}
    \Omega =D\omega \quad \in \quad \Omega ^2(P)\otimes \mathfrak{g} 
\end{align} 
which can also be written as
\begin{align}
    \Omega =\dd _P\omega +\frac12 [\omega \wedge \omega ] ~.
\end{align}
Here, $[\cdot \wedge \cdot ]$ denotes the Lie bracket on the algebra part and the wedge product on the form part. From taking another derivative one learns about the Bianchi identity
\begin{align}\label{eq:Bianchi1}
    D\Omega =0
\end{align}
showing that the curvatures are not all independent. There is an elegant geometric interpretation of the curvature: Given two horizontal vector fields $X_{\scriptscriptstyle \text{H}}$, $Y_{\scriptscriptstyle \text{H}}$ and contracting them into $\Omega $ yields
\begin{align}
    \Omega (X_{\scriptscriptstyle \text{H}},Y_{\scriptscriptstyle \text{H}})=\dd _P\omega (X_{\scriptscriptstyle \text{H}},Y_{\scriptscriptstyle \text{H}})=-\omega ([X_{\scriptscriptstyle \text{H}},Y_{\scriptscriptstyle \text{H}}])
\end{align}
which is the vertical component of the commutator of two horizontal vector fields. Therefore, the curvature can be seen as an obstruction to being able to integrate the horizontal subspaces into a smooth submanifold of $P$, since vector fields on the latter are not in involution precisely when the curvature is non-zero.\footnote{This situation should be compared to the geometry of a Carroll manifold, where the field $\tau $ plays the role of the connection in equation \eqref{eq:Frobenius}. However, $\tau $ is in general not a principal connection, since this would require $\mathcal{L}_v \tau =0$ by the infinitesimal version of the second property in definition \ref{def:connection}. The Carroll acceleration $a$ can thus be seen as the obstruction for $\tau $ to be principal. \cite{Bekaert:2015xua}} Just as for the connection one-form, it is more practical to describe the curvature in terms of a local $\mathfrak{g}$-valued two-form on $\mathcal{U}\subset \mathcal{M}$ by choosing a section $\sigma $ of $P$ and defining
\begin{align}\label{eq:local_curvature}
    \Omega _{\mathcal{U}}:=\sigma ^\ast \Omega =\dd \omega _{\mathcal{U}}+\frac12 [\omega _{\mathcal{U}}\wedge \omega _{\mathcal{U}}] \quad \in \quad \Omega ^2(\mathcal{U})\otimes \mathfrak{g} ~.
\end{align}
Then, under a gauge transformation that affects $\omega _{\mathcal{U}}$ like in \eqref{eq:conn_trafo} one finds
\begin{align}
    \Omega '_{\mathcal{U}}=g^{-1}\, \Omega _{\mathcal{U}}\, g ~.
\end{align}

A connection on a principal bundle induces a covariant derivative on each associated vector bundle, also called a Koszul connection. This defines a way to take directional derivatives $\mathcal{D} _v s$ of a section $s\in \Gamma (P\times _\rho V)$ along a tangent vector $v$ of the base manifold, thereby producing another section. It satisfies the properties
\begin{equation}
\text{
\begin{minipage}{0.8\linewidth}
    \begin{itemize}
        \item $\mathcal{D} _{fv}s=f\mathcal{D} _vs $
        \item $\mathcal{D} _v(fs)=sv(f)+f\mathcal{D} _vs$
        \item $\mathcal{D} _v(as+bt)=a\mathcal{D} _vs +b\mathcal{D} _vt$
        \item $\mathcal{D} _{av+bw}s=a\mathcal{D} _vs+b\mathcal{D} _ws$
    \end{itemize}
    \end{minipage}
}
\label{eq:itemlist}
\end{equation}
    where $f\in \mathcal{C}^\infty (\mathcal{M})$ can be any smooth function on the base manifold, $a$ and $b$ are constants and $t$ is another section of the vector bundle. The above definitions also imply that $\mathcal{D}_v$ reduces to an ordinary directional derivative when acting on scalar functions. For constructing $\mathcal{D}$ we need the following fact \cite{Figueroa-OFarrill:2020gpr}:
\begin{lemma}
    Given a principal $G$-bundle and an associated vector bundle $(P\times _\rho V)$ there is a one-to-one correspondence between sections of $(P\times _\rho V)$ and $G$-equivariant, $V$-valued functions on $P$,
    \begin{align}\label{eq:module_isomorphism}
    \Gamma (P\times _\rho V)\cong \mathcal{C}^\infty _{G}(P,V) 
\end{align}
where the property of $G$-equivariance is defined by $\phi (u\cdot g)=\rho (g^{-1})\phi (u)$. 
\end{lemma}
The $G$-equivariance implies a corresponding change in the image of the function $\phi (u)$ when $G$ acts on the point $u\in P$. Because of this isomorphism, we can relate any notion of covariant derivative on one side to a notion of covariant derivative on the other side. But we already have a derivative \eqref{eq:cov_der_P} defined on $P$, so all we have to do is relate it to the associated bundle. Suppose that $\phi_s:P\to V$ is the unique function related to the section $s$ via \eqref{eq:module_isomorphism}. For some vector field $X$ on $P$, we get from \eqref{eq:cov_der_P}
\begin{align}\label{eq:cov_der2}
    D\phi _s(X)=\dd _P\phi _s(X_{\scriptscriptstyle \text{H}})=\dd _P\phi _s(X)-\dd _P\phi _s(X_{\scriptscriptstyle \text{V}})=\dd _P\phi _s(X)+\rho _\ast (\omega (X))\phi _s
\end{align}  
where in the second equality we used the unique split of the vector field $X$ into its horizontal and vertical components. In the last equality we used the fact that, because of \eqref{eq:vert_VF}, there exists a unique Lie algebra element associated to $X_{\scriptscriptstyle \text{V}}$ which can be written as $\omega (X)$ by the definition of the connection. Additionally, $G$-equivariance implies 
\begin{align}
    \dd _P\phi _s(X_{\scriptscriptstyle \text{V}})=X_{\scriptscriptstyle \text{V}}\phi _s=\frac{\dd }{\dd t}\phi _s\big(u\cdot e^{\omega (X)t}\big)\Big \vert _{t=0}=\frac{\dd }{\dd t}\rho \big(e^{-\omega (X)t}\big)\phi _s(u)\Big \vert _{t=0}=:-\rho _\ast\big(\omega (X)\big)\phi_s
\end{align}
with $\rho _\ast$ being the map $\rho _\ast: \mathfrak{g}\to\mathfrak{gl}(V)$ induced by the representation $\rho $. Since \eqref{eq:cov_der2} holds for any vector field on $P$, it is really an equation between $V$-valued one-forms on $P$. As such we can again use our usual tool of picking a local section $\sigma :\mathcal{U}\to P$ to pull it back to a $V$-valued one-form on the base manifold. Since the function $\sigma ^\ast \phi _s:\mathcal{M}\to V$ is just the representative of the original section $s$ restricted to some local region $\mathcal{U}\subset \mathcal{M}$, we can just choose the same letter, but have to keep in mind that its definition still depends on $\sigma $. Contracting in a tangent vector field $v$ we define 
\begin{align}
    \mathcal{D} _vs=\dd s (v)+\rho _\ast (\omega _{\mathcal{U}}(v))s \qquad v\in \Gamma (T\mathcal{M}), s\in \Gamma (P\times _\rho V) ~.
\end{align} 
One can show that this satisfies all the properties in \eqref{eq:itemlist}. In addition, under a change of gauge (the trivializing section) $\sigma '=\sigma \cdot g$ we find $s'=\rho (g^{-1})s$ and
\begin{align}
    \mathcal{D} _v s'=\rho (g^{-1})\mathcal{D} _vs
\end{align}
which is the typical behaviour of a covariant derivative. 

Let us look at the example of an SU$(N)$ bundle and its adjoint bundle ad$P:=(P\times _{\text{Ad}}\mathfrak{g})$. We write a local section of ad$P$ as $s=s^Ab_A$ with a Lie algebra basis $\{b_A\}_{A=1,...,\text{dim}(\mathfrak{g})}$, where $s^A$ is a collection of local functions on $\mathcal{M}$ and $[b_A,b_B]=f_{AB}{}^Cb_C$. For the case of the adjoint bundle, we have $\rho _\ast (\omega _{\mathcal{U}})=\text{ad}_{\omega _{\mathcal{U}}}=[\omega _{\mathcal{U}},\cdot ]$. The covariant derivative along a vector field $v\in \Gamma (T\mathcal{M})$ now reads explicitly 
\begin{align}
    \mathcal{D} _vs^A=v^\mu \big(\partial _\mu s^A +[\omega _{\mu},s]^A\big)=v^\mu \big(\partial _\mu s^A +f^A{}_{BC}\,\omega _{\mu}^B s^C\big)
\end{align}
where we also chose a local chart $x^\mu $ on the base manifold. Under a change of gauge $s^A\to (g^{-1})^A{}_Bs^B$ the connection coefficients transform as  
\begin{align}
    f^A{}_{CB}\,\omega ^C_{\mu}\to (g^{-1})^A{}_C\big(\delta ^C_D\,\partial _\mu +f^C{}_{ED}\,\omega ^E_{\mu}\big)g{}^D{}_B
\end{align}
which is just the right transformation such that $\mathcal{D} _vs^A \to (g^{-1})^A{}_B\mathcal{D} _vs^B$. With $g^A{}_B\approx \delta ^A_B+f^A{}_{CB}\lambda ^C$ the infinitesimal transformation\footnote{One can also define an adjoint version of $\omega ^A_\mu$ by $\omega ^B{}_{\mu C}=f^B{}_{AC}\omega ^A_\mu$ which transforms as $\delta _\lambda \omega ^A{}_{\mu B}=\partial _\mu \lambda ^A{}_B+[\omega _\mu,\lambda ]^A{}_B$ where $\lambda ^A{}_B$ is defined analogously and the bracket is just the commutator between matrices.} reads $\delta _\lambda \omega _\mu ^A=\mathcal{D} _\mu \lambda ^A$. The curvature two-form \eqref{eq:local_curvature} also induces a curvature two-form for this associated connection typically denoted by $\Omega ^A{}_{B}:=(\rho _\ast (\Omega _{\mathcal{U}}))^A{}_B=f^A{}_{CB} \Omega _{\mathcal{U}}^C$. The Bianchi identity \eqref{eq:Bianchi1} translates in this case to
\begin{align}
    \dd \Omega^A{}_B+f^A{}_{DC}\,\omega ^D_\mathcal{U}\wedge \Omega^C{}_B-\Omega^A{}_C\wedge \omega ^D _\mathcal{U}f^C{}_{DB} ~.
\end{align}
We need one more ingredient before we turn to the definition of Cartan geometries.
\begin{definition}[Solder form]
     A $V$-valued one-form $\theta $ on a principal $G$-bundle $P\to \mathcal{M}$ is called a solder form if 
     \begin{itemize}
        \item $V$ is a linear representation space of $G$ with representation $\rho :G\to \text{GL}(V)$.
        \item $\theta (X^\sharp )=0$ with $X^\sharp \in \Gamma (VP)$, i.e. $\theta $ is horizontal.
        \item $R_g^\ast \theta =\rho (g^{-1})\circ \theta $
        \item $\theta $ provides an isomorphism $T\mathcal{M}\cong (P\times _\rho V)$ and therefore $\dim V=\dim \mathcal{M}$.
     \end{itemize} 
\end{definition}
The idea of a solder form is thus to identify the vector space $V$ with the tangent space to $\mathcal{M}$ for each point of the base manifold. While, in general, this is additional data one may provide for a principal bundle, there are some special cases where a canonical choice exists. Take for instance the frame bundle $F\mathcal{M}$ together with the representation space $\mathbb{R}^{d+1}$. Define for any vector field $X\in \Gamma (TF\mathcal{M})$ 
\begin{align}
    \theta (X)=\chi ^{-1}_u(\pi _\ast (X)) ~,
\end{align} 
where the map $\chi _u:\mathbb{R}^{d+1}\to T_u\mathcal{M}$ just takes the components of a vector field and linearly combines them with the basis vectors of the frame at $u\in F\mathcal{M}$, i.e., $v^A\mapsto v^AE_A$. Its inverse $\chi _u^{-1}$ is given by the co-frame $e^A$ defined by $e^A(E_B)=\delta ^A_B$. 

Once a solder form is defined, one can construct the \textit{torsion two-form} $\Theta $ on $P$, 
\begin{align}
    \Theta =D\theta =\dd _P\theta +\rho _\ast (\omega )\wedge \theta \qquad \in \qquad \Omega ^2(P)\otimes V 
\end{align}
which also satisfies a Bianchi identity, as a result from taking the second exterior covariant derivative on $P$,
\begin{align}\label{eq:Bianchi2}
    D\Theta =\rho _\ast (\Omega )\wedge \theta ~.
\end{align}
Pulling $\Theta $ back to $\mathcal{M}$ with a section $\sigma $, its local version on the base manifold is given by 
\begin{align}\label{eq:local_torsion}
    T=\sigma ^\ast \Theta =\dd \theta _\mathcal{U}+\rho _\ast (\omega _{\mathcal{U}})\wedge \theta _\mathcal{U} \qquad \in \qquad \Omega ^2(\mathcal{M})\otimes V~.
\end{align}
Under a gauge transformation $\sigma '=\sigma \cdot g$ this transforms tensorially, i.e., $T'=\rho (g^{-1})T$.
To again look at a familiar example, for the case of the frame bundle we have $\theta _{\mathcal{U}}=\theta _\mu \dd x^\mu=e^A_\mu \dd x^\mu$ and $T^A_{\mu \nu }=2\partial _{[\mu } e_{\nu ]}^A+2\omega ^A{}_{[\mu B}e_{\nu ]}^B$, where $\omega ^A{}_{\mu B}$ are the components of a $\mathfrak{gl}(d+1,\mathbb{R})$-valued one-form. The operational meaning of $\theta _{\mathcal{U}}$ comes out especially simple here: It just gives the components of a vector field in the frame $E_A$. Since we can use the solder form to locally identify sections of $P\times _\rho V$ with sections of $T\mathcal{M}$, a Koszul connection on the former bundle automatically induces an affine connection $\nabla$ on the latter bundle, which is the usual object one works with in, e.g., general relativity. In a local chart it is given in terms of connection coefficients $\Gamma ^\alpha {}_{\mu \nu }$ that encode how the basis changes as one moves from one point to an infinitesimally close neighbouring point. For a coordinate basis $\{\partial _\alpha \}$ they are defined by
\begin{align}
    \nabla _\mu \partial _\nu =\Gamma ^\alpha {}_{\mu \nu }\partial _\alpha 
\end{align}
which automatically induces the action on the co-basis $\nabla _\mu \dd x^\nu =-\Gamma ^\nu {}_{\mu \alpha }\dd x^\alpha $ by using that $\nabla _\mu $ reduces to an ordinary partial derivative once acting on scalar functions. In the following, we shall use the common notation $\nabla _\mu v^\nu $ by which we mean the components of the covariant derivative of a vector field $v$ (instead of the covariant derivative acting on the components, which would just reduce to a partial derivative). The connection $\mathcal{D}$ determines $\nabla $ by requiring that taking the derivative should commute with performing the map between $\Gamma (T\mathcal{M})$ and $\Gamma (P\times _\rho V)$,
\[
\begin{tikzcd}
  & \theta _\nu \arrow[r, "\mathcal{D}"] & \partial _\mu \theta _\nu +\rho _\ast (\omega _\mu )\theta _\nu  \arrow[dd,leftrightarrow,""]\\
\partial _\nu  \arrow[ur, "\theta "] \arrow[dr, "\nabla "'] & &  \\
  & \Gamma ^\alpha {}_{\mu \nu }\partial _\alpha \arrow[r, "\theta "'] & \Gamma ^\alpha {}_{\mu \nu }\theta _\alpha 
\end{tikzcd}
\]
which is known as the \textit{vielbein postulate} for the case of the frame bundle but actually holds also in more general situations. Choosing a basis for $V$ such that $\rho _\ast (\omega _\mu )\sim \omega ^A{}_{\mu B}$ and equating the right sides of the diagram yields
\begin{align}\label{eq:vielbein_postulate}
    \partial _\mu \theta _\nu ^A +\omega ^A{}_{\mu B}\,\theta _\nu^B -\Gamma ^\alpha {}_{\mu \nu }\theta _\alpha^A =0 
\end{align}
which determines the coefficients $\Gamma ^\alpha {}_{\mu \nu }$ uniquely. If the connection is defined in this way, it automatically satisfies a key property: It is adapted to any tensor field induced by a $G$--invariant tensor on $V$ using the map provided by the solder form. To see this, we may first define an inverse to $\theta ^A_\mu $ by $\theta ^\nu _B\theta ^B_\mu =\delta ^\nu _\mu $ and find from contracting with \eqref{eq:vielbein_postulate} that 
\begin{align}\label{eq:inv_vielbein_postulate}
    \partial _\mu \theta _A^\alpha -\omega ^B{}_{A\mu }\,\theta _B^\alpha +\Gamma ^\alpha {}_{\mu \beta }\theta ^\beta _A=0 ~.
\end{align}
Suppose we have a $G$-invariant tensor $\Psi \in \otimes ^r _s V$. Infinitesimally, this means for any $\mathbf{x}\in \mathfrak{g}$
\begin{align}
    \mathbf{x}^{A_1}{}_C\Psi ^{C...A_r}_{B_1...B_s}+\dots +\mathbf{x}^{A_r}{}_C\Psi ^{A_1...C}_{B_1...B_s}-\mathbf{x}^{C}{}_{B_1}\Psi ^{A_1...A_r}_{C...B_s}-\dots-\mathbf{x}^{C}{}_{B_s}\Psi ^{A_1...A_r}_{B_1...C}=0
\end{align}
where $\mathbf{x}^A{}_B \in \mathfrak{gl}(V)$ is again shorthand for the representation of $\mathbf{x}$ on $V$. In particular, we may choose $\mathbf{x}^A{}_B=\omega ^A{}_{\mu B}$. Then, to any such $\Psi $ we associate a spacetime tensor
\begin{align}
    \Psi ^{\mu _1...\mu _r}_{\nu _1...\nu_s}=\theta ^{\mu _1}_{A_1}\dots\theta ^{\mu _r}_{A_r}\theta ^{B_1}_{\nu _1}\dots \theta ^{B_s}_{\nu _s}\;\Psi ^{A_1...A_r}_{B_1...B_s} 
\end{align}
that can be checked to satisfy 
\begin{align}
    \partial _\alpha \Psi ^{\mu _1...\mu _r}_{\nu _1...\nu_s} +\Gamma ^{\mu _1}{}_{\alpha \beta }\Psi ^{\beta...\mu _r}_{\nu _1...\nu_s}+\dots &+\Gamma ^{\mu _r}{}_{\alpha \beta }\Psi ^{\mu _1...\beta}_{\nu _1...\nu_s}\\
    &-\Gamma ^{\beta}{}_{\alpha \nu _1 }\Psi ^{\mu _1...\mu _r}_{\beta ...\nu_s}-\dots -\Gamma ^{\beta}{}_{\alpha \nu _s }\Psi ^{\mu _1...\mu _r}_{\nu _1 ...\beta}=0 \nonumber
\end{align}
as a consequence of \eqref{eq:vielbein_postulate} and \eqref{eq:inv_vielbein_postulate}. This is just the covariant derivative associated to the affine connection acting on $\Psi $ such that we established the mentioned relation
\begin{align}
    \big(\rho (g)\Psi \big)^{A_1...A_r}_{B_1...B_s}=\Psi ^{A_1...A_r}_{B_1...B_s} \quad \forall g\in G \quad \Rightarrow \quad \nabla _\alpha \Psi ^{\mu _1...\mu _r}_{\nu _1...\nu_s}=0 ~. 
\end{align}
To again have an example, take $V=\mathbb{R}^d$ in the fundamental representation of an $SO(d)$ bundle over a $d$-dimensional manifold. An invariant tensor is $\delta _{AB}\in V^\ast \odot V^\ast $ which, given a solder form, maps to the Riemannian metric $g_{\mu \nu }=\theta _\mu ^A\theta _\nu ^B\delta _{AB}$. Any principal bundle connection then induces an affine connection that satisfies $\nabla _\mu g_{\alpha \beta }=0$. One may specify further requirements like vanishing torsion, which makes it possible to uniquely express the connection in terms of the metric, leading to the Levi--Cività connection.

Finally, let us see how to express curvature and torsion \eqref{eq:local_torsion} of $\mathcal{D}$ and $\theta $ in terms of the more familiar curvature and torsion tensors of the affine connection. By a choice of basis for $V$ we may write $T \sim \frac12 T^A_{\mu \nu }\dd x^\mu \wedge \dd x^\nu $ and again use the map provided by the solder form at a point $p\in \mathcal{M}$ to establish
\begin{align}
    T_{\mu \nu }^A=2\Gamma ^\alpha {}_{[\mu \nu ]}\theta _\alpha ^A=:T^\alpha _{\mu \nu }\theta ^A_\alpha ~,
\end{align}
where $T^\alpha _{\mu \nu }$ is the torsion tensor associated to $\nabla $. Meanwhile, the curvature two-form, represented locally by $\Omega \sim \frac12 \Omega ^A{}_{B\mu \nu }\dd x^\mu \wedge \dd x^\nu $, may be rewritten as
\begin{align}
    \Omega ^A{}_{B\mu \nu }=\theta ^A_\alpha \theta _B^\beta R^\alpha {}_{\beta \mu \nu } 
\end{align} 
where $R^\alpha {}_{\beta \mu \nu }$ is just the Riemann curvature tensor
\begin{align}\label{eq:riemann}
    R^\alpha {}_{\beta \mu \nu }=\partial _\mu \Gamma ^\alpha {}_{\nu \beta }+\Gamma ^\sigma {}_{\nu \beta }\Gamma ^\alpha {}_{\mu \sigma }-(\mu \leftrightarrow \nu ) ~.
\end{align}
Torsion and curvature tensor are related by the identity 
\begin{align}
    [\nabla _\mu ,\nabla _\nu ]X^\alpha =R^\alpha {}_{\beta \mu \nu }X^\beta -T^\beta _{\mu \nu }\nabla _\beta X^\alpha
\end{align}
which holds for any vector field $X^\alpha $. The Bianchi identities \eqref{eq:Bianchi1} and \eqref{eq:Bianchi2} can now be translated to 
\begin{align}
    \nabla _{[\rho }R^\alpha {}_{\beta |\mu \nu ]}&=-R^\alpha {}_{\beta \sigma [\rho }T^\sigma _{\mu \nu] }  \\
    R^\alpha {}_{[\rho \mu \nu ]}&=\nabla _{[\rho }T^\alpha _{\mu \nu ]}-T^\sigma _{[\rho \mu }T^\alpha _{\nu ]\sigma }\label{eq:bianchi_riemm}
\end{align}
where we used that $R^\alpha {}_{\beta (\mu \nu )}=0$.

\subsection{Cartan geometry}\label{sec:cartan_geometry}
We are now ready to assemble the building blocks from the last two subsections. This leads back to works by Cartan \cite{Cartan1923,Cartan1937}, however the mathematical language he used was different from the by now more common way of a formulation in terms of principal fiber bundles \cite{Sharpe1997,Wise:2006sm,Figueroa-OFarrill:2020gpr}. Here we shall choose this more modern approach. The idea is to construct a geometry that is locally built out of a Klein pair $(\mathfrak{g},\mathfrak{h})$ in such a way that the algebraic information about the symmetries in $H$ is encoded in the tangent spaces. The manifold $\mathcal{M}$ itself, however, does not have to be a coset space anymore. This is reflected in the fact that the $H$-invariant tensors discussed in subsection \ref{sec:klein_geometry} now cannot be extended to invariant tensor fields on all of $\mathcal{M}$, since, in general, there is no $G$-action on $\mathcal{M}$ anymore. Instead, they are only defined on each tangent space. A well-known example is the spacetime of general relativity which still has the structure of Minkowski space imprinted on its tangent spaces but in general does not possess any Killing vector fields. As we shall see, the deviation from the Klein geometry is measured by Cartan curvature which consists of the curvature of a principal bundle connection adapted to the chosen Klein pair as well as the torsion of an adapted solder form. This subsection is meant to provide the basic kinematical variables that may be used for constructing theories of Carroll gravity. 
\begin{definition}[Cartan geometry \cite{Sharpe1997}]
    A Cartan geometry $(P,A)$ on $\mathcal{M}$ modelled on a Klein pair $(\mathfrak{g},\mathfrak{h})$ with group $H$ consists of the following data:
    \begin{itemize}
        \item A smooth manifold $\mathcal{M}$
        \item A principal $H$-bundle $P\to \mathcal{M}$
        \item A $\mathfrak{g}$-valued one-form $A$ on $P$ called the Cartan connection satisfying the following conditions:
        \begin{itemize}
            \item[i)] at each $u\in P$ the map $A_u:T_uP\to \mathfrak{g}$ is an isomorphism
            \item[ii)] $R_h^\ast A=\text{Ad}_{h^{-1}}\circ A$ for all $h\in H$
            \item[iii)] $A(X^\sharp )=\mathbf x $ for all $\mathbf x \in \mathfrak{h}$
        \end{itemize}
    \end{itemize}
\end{definition}
In the reductive case that we restrict to, one can always decompose $\mathfrak{g}=\mathfrak{h}\oplus \mathfrak{m}$ such that $\text{ad}_\mathfrak{h}:\mathfrak{m}\to \mathfrak{m}$. The Cartan connection $A$ then splits into two parts,
\begin{align}
    A=\omega +\theta  
\end{align}
where $\omega $ takes values in $\mathfrak{h}$ and $\theta $ takes values in $\mathfrak{m}$. We define the Cartan curvature two-form by
\begin{align}
    F=\dd A+\frac12 [A\wedge A]=F(\mathfrak{m})+F(\mathfrak{h})
\end{align}
which also splits into two parts. One can immediately check that $\omega $ satisfies all the requirements for a principal bundle connection on $P$ while $\theta $ is a solder form that attaches the associated bundle $(P\times _{\text{Ad}_H}\mathfrak{m})$ to the tangent bundle. The reason this works is that $\mathfrak{m}$ has exactly the right number of dimensions: Since $\dim P=\dim \mathcal{M}+\dim H $ it follows from property $i)$ above that $\dim \mathfrak{m}=\dim \mathcal{M}$. It is also horizontal by property $iii)$ which implies $\theta (X^\sharp )=0$. The Cartan curvature $F$ also satisfies a Bianchi identity
\begin{align}
    \dd F+[A\wedge F]=0
\end{align}
which, once split into its components along $\mathfrak{h}$ and $\mathfrak{m}$, implies the Bianchi identities \eqref{eq:Bianchi1}, \eqref{eq:Bianchi2} for curvature and torsion. With $\rho _\ast $ being the adjoint representation of $\mathfrak{h}$ they read explicitly
\begin{align}
    \dd \Omega +[\omega \wedge \Omega ]=0 && \dd \Theta +[\omega \wedge \Theta ]=[\Omega \wedge \theta ]~.
\end{align} 
One can see that from the perspective of the Klein pair, the ``gauge transformations'' of this model take values in the subgroup $H$, which for kinematical spacetimes consists of the homogeneous part like the Lorentz group in the case of the pair $(\mathfrak{iso}(d,1),\mathfrak{so}(d,1))$. Of course, as long as we do not specify a dynamical theory these are just kinematical redundancies of the description in terms of local quantities on the base manifold. The true interpretation as gauge symmetries only enters once we write down an action. Nevertheless, we shall assume from now on that we chose some trivializing section $\sigma :\mathcal{M}\supset \mathcal{U}\to P$ such that $A$ is given in terms of a $\mathfrak{g}$-valued one-form on $\mathcal{U}$. We can then express its components in a chart over $\mathcal{U}$ as $A=A_\mu \dd x^\mu $. Following the discussion in subsection \ref{sec:fiber_bundles} a change of the trivializing section $\sigma \mapsto \sigma \cdot h$ now infinitesimally induces a change in the local description of the Cartan connection that acts as
\begin{align}\label{eq:inf_trafo_H}
    \delta _\lambda A=\dd \lambda +[A ,\lambda ] && \delta _\lambda F=[F,\lambda ]
\end{align} 
where $h\approx 1+\lambda $ and $\lambda $ is some $\mathfrak{h}$-valued function on $\mathcal{M}$. These are just the generalization of local Lorentz transformations to any other choice of $\mathfrak{h}$. 

For theories of gravity, one might then wonder whether the action of diffeomorphisms on $\mathcal{M}$ can be packaged into this framework. Once all the fields are expressed in terms of local quantities on $\mathcal{M}$ their action is just given by the Lie derivative along some vector field $\xi $, i.e., $\delta _\xi A_\mu =\mathcal{L}_\xi A_\mu $. Let us consider a more general transformation \cite{Hartong:2015zia,Hartong:2015xda} than \eqref{eq:inf_trafo_H} given by
\begin{align}\label{eq:general_diffplusgauge_trafo}
    \bar{\delta }_{(\lambda ,\xi)} A=\dd \Sigma +[A,\Sigma ] + i_\xi F && \Sigma =\lambda +A_\mu \xi ^\mu 
\end{align}
where the parameter $\Sigma $ now also includes contributions in $\mathfrak{m}$. For $\xi =0$ this just reduces to \eqref{eq:inf_trafo_H}. It can be shown that
\begin{align}
    \bar{\delta }_{(\lambda ,\xi)}A =\delta _\lambda A +\mathcal{L}_\xi A ~.
\end{align}
Therefore, one can always relate a combined version of $\mathfrak{m}$- and $\mathfrak{h}$-transformations to spacetime diffeomorphisms. It might happen for specific theories that going on-shell sets some of the curvature components to zero, in which case the $\delta $ and $\bar{\delta }$ transformations coincide for the respective components of the Cartan connection. This happens, e.g., in the Palatini formulation of Einstein gravity where $F(\mathfrak{m})=0$ on-shell, corresponding to vanishing torsion. Another example is provided by BF-theories (see section \ref{sec:BF_theories}), where even all the curvatures vanish on-shell. It should be mentioned, however, that the invariance under $\bar{\delta}$ transformations does, in general, not imply invariance under $\mathfrak{m}$-transformations alone. 

Before we look at some examples let us state in which sense Cartan geometries generalize homogeneous spacetimes. This is the essence of the 
\begin{theorem}[Fundamental theorem of Cartan geometry (\cite{Sharpe1997}, p.212)]\label{th:fund_th}
    Let $(P,A)$ be a Cartan geometry on $\mathcal{M}$ modelled on a Klein pair $(\mathfrak{g},\mathfrak{h})$. The curvature $F=\dd A+\frac12 [A\wedge A]$ vanishes if and only if the underlying principal $H$-bundle is locally isomorphic to $G\to G/H$ with $A$ given by the Maurer--Cartan form on $G$. 
\end{theorem}
Therefore, in the case of vanishing Cartan curvature one indeed recovers a Klein geometry associated to the pair $(\mathfrak{g},\mathfrak{h})$, at least locally. For more details we refer to \cite{Sharpe1997}.

\subsection{Example: Gauging the (A)dS algebra}
\label{sec:example_ads}
Let us look at the example of the Poincaré algebra with non-zero cosmological constant to familiarize ourselves with the formalism. We choose the Klein pair as $(\mathfrak{g},\mathfrak{h})=(\mathfrak{so}(d,2),\mathfrak{so}(d,1))$ such that the Cartan gometry is given by an $SO(d,1)$ bundle together with a Cartan connection 
\begin{align}
    A=\frac12 \omega ^{AB}J_{AB}+e^AP_A
\end{align} 
where we denoted the (A)dS generators like in \eqref{eq:ads_compact} by $J_{AB}$ spanning $\mathfrak{h}$ and $P_A$ spanning $\mathfrak{m}$. We also changed the indices from spacetime indices to internal indices. They can be pulled with the internal Minkowski metric $\eta _{AB}$, an $\mathfrak{h}$-invariant tensor on $\mathfrak{m}$ that will induce a Lorentzian metric on $\mathcal{M}$.
Note that $\omega ^{AB}$ is already in the adjoint representation of $\mathfrak{h}$ on $\mathfrak{m}$, i.e., for any vector $v=v^AP_A$ it acts as $\rho _\ast (\omega )v = \omega ^A{}_B v^B P_A$. Like introduced in subsection \ref{sec:fiber_bundles}, we denote its associated curvature two-form by $\Omega^A{}_B:=(\rho _\ast (\Omega ))^A{}_B$. Infinitesimal $\mathfrak{h}$-transformations $\lambda =\frac12 \lambda ^{AB}J_{AB}$ act on the Cartan connection components as 
\begin{align}
    \delta _\lambda \omega ^{AB}&=\dd \lambda ^{AB}+ \lambda ^A{}_C\omega ^{CB}-\lambda ^B{}_C\omega ^{AC}\\
    \delta _\lambda e^A&=-\lambda ^A{}_B e^B 
\end{align}
and we can work out the components of the Cartan curvature two-forms to be
\begin{align}
    F^{AB}(J)&=\dd \omega ^{AB}+\omega ^A{}_C\wedge \omega ^{CB}-\frac{\sigma }{\ell ^2} e^A\wedge e^B \\
    F^A(P)&=\dd e^A +\omega ^A{}_B\wedge e^B ~.
\end{align}
They relate to curvature and torsion of $\omega ^{AB}$ and $e^A$ as
\begin{align}
    F^{AB}(J)=\Omega^{AB}-\frac{\sigma }{\ell ^2} e^A\wedge e^B && F^A(P)=T^A ~.
\end{align}
Therefore, the $\mathfrak{h}$-component of the Cartan curvature gives a different result than the curvature of the $\mathfrak{h}$-component of the Cartan connection precisely when the translations $P_A$ do not commute. The case $\Lambda =0$ is special in this regard. We can now use the solder form provided by $e^A$ to induce a connection on the tangent bundle. Solving the vielbein postulate
\begin{align}
    \partial _\mu e^A_\nu +\omega ^A{}_{\mu B}e^B_\nu -\Gamma ^\alpha _{\mu \nu }e_\alpha ^A=0 \quad \Rightarrow \quad \Gamma ^\alpha {}_{\mu \nu }=E^\alpha _A\big(\delta ^A_B \partial _\mu +\omega ^A{}_{\mu B}\big )e^B_\nu 
\end{align}
the covariant derivative $\nabla $ is automatically adapted to the spacetime metric $g_{\mu \nu }=\eta _{AB}e^A_\mu e^B_\nu $,
\begin{align}
    \nabla _\mu g_{\alpha \beta }=0 ~.
\end{align}
Often, one requires some additional properties from this connection. One example is the Einstein--Palatini action, where the equations of motion from varying with respect to $\omega ^{AB}$ yield a constraint of vanishing torsion, $T^A=0$. In that case, one can solve the equation 
\begin{align}
    \dd e^A+\omega ^A{}_B\wedge e^B=0
\end{align}
for the connection to find
\begin{align}
    \omega ^{AB}_\mu =2E^{\nu [A}\partial _{[\mu }e_{\nu ]}^{B]}-e_{\mu C}E^{\nu A}E^{\rho B}\partial _{[\nu }e_{\rho ]}^C ~.
\end{align}
When inserted into the vielbein postulate, one arrives at the Levi--Cività connection 
\begin{align}\label{eq:christoffels}
    \Gamma ^\alpha {}_{\mu \nu }=\frac12 g^{\alpha \rho }\big(\partial_\mu g_{\nu \rho }+\partial _\nu g_{\mu \rho }-\partial _\rho g_{\mu \nu }\big)
\end{align}
which is the unique torsionfree and metric-compatible connection on a Lorentzian manifold. It is manifestly invariant under $\mathfrak{h}$-transformations. For the special case $F^{AB}(J)=0$ the geometry is locally isomorphic to the Klein geometry given by (A)dS spacetime and the Riemann tensor associated to $\nabla $ is
\begin{align}
    R_{\mu \nu \alpha \beta }=\frac{\sigma }{\ell ^2} (g_{\mu \alpha }g_{\nu \beta }-g_{\mu \beta }g_{\nu \alpha }) && \Rightarrow && R=\Lambda  \frac{2(d+1)}{d-1}
\end{align}
where we used \eqref{eq:cosm_const} and defined the Ricci scalar $R:=R^\mu {}_{\nu \mu \beta }g^{\nu \beta }$.

\section{Geometric anatomy of Carroll spacetimes}\label{sec:gauging_procedure}
Let us now come to our main application of the Cartan formalism introduced in the previous subsection and set up the geometric variables used for describing general curved Carroll spacetimes. We shall directly start with the case for $\Lambda \neq 0$, since one can always revert to the flat case by setting the cosmological constant to zero. Aiming to connect with the variables introduced in subsection \ref{sec:2.2}, we pick a Cartan connection
\begin{align}
    A=\tau H +e^aP_a+\omega ^aB_a+\frac12 \omega ^{ab}J_{ab}
\end{align}
which is a one-form with values in the algebra \eqref{eq:Carroll_ADS_algebra}. Note, that we again use abstract algebra indices instead of spacetime indices, since these generators do not have a spacetime interpretation for general manifolds. The Klein pair we choose is the (A)dS Carroll algebra \eqref{eq:Carroll_ADS_algebra} together with the homogeneous Carroll subalgebra $\mathfrak{h}=\mathfrak{iso}(d)$ with basis $\{B_a,J_{ab}\}$. Its complement is denoted by $\mathfrak{m}=\mathbb{R}^{d+1}$ and has a basis $\{H,P_a\}$. Comparing with section \ref{sec:2.2}, we think of $\tau $ as the clock one-form and of $e^a$ as the spatial coframe, both taken together form the components of the solder form pulled back to the base manifold,
\begin{align}
    \theta =\tau H+e^aP_a=\theta^AP_A ~.
\end{align}
Capital Latin indices again label the basis elements of $\mathfrak{m}$ as $P_A=(H,P_a)$. According to the analysis in example \ref{ex:carrooll_tensors}, two $\mathfrak{h}$-invariant tensors on $\mathfrak{m}$ are  given by a degenerate metric $h _{AB}$ as well as a vector $v^A$ with components 
\begin{align}
  h _{AB}=  \begin{pmatrix}
        0 & 0 \\
        0 & \delta _{ab}
    \end{pmatrix} && v^A=(-1,0_a)^T
\end{align}
where we chose a conventional minus sign in $v$. Clearly, we now cannot use this metric to identify $\mathfrak{m}^\ast $ with $\mathfrak{m}$ as a whole, but we can still pull indices of purely spatial objects. To derive the transformation properties of the various fields we parametrize $\mathfrak{h}$-transformations by 
\begin{align}
    \lambda =\lambda ^aB_a+\frac12 \lambda ^{ab}J_{ab} && \lambda ^A{}_B=(\text{ad}_\lambda )^A{}_B=\begin{pmatrix}
        0 & \lambda _b \\
        0 & \lambda ^a{}_b
    \end{pmatrix}
\end{align}
and additionally write the connection as well as its curvature in the adjoint $\mathfrak{h}$-representation,
\begin{align}\label{eq:ome_in_adjoint}
    \omega ^A{}_B=
    \begin{pmatrix}
        0 & \omega _b \\
        0 & \omega ^a{}_b
    \end{pmatrix} && \Omega ^A{}_B= \begin{pmatrix}
        0 & \Omega _b \\
        0 & \Omega ^a{}_b
    \end{pmatrix} ~.
\end{align}
The action on the various gauge fields is then 
\begin{align}
     \delta _\lambda \omega ^A{}_B=\dd \lambda ^A{}_B+\omega ^A{}_C\lambda ^C{}_B-\lambda ^A{}_C\omega ^C{}_B && \delta _\lambda \theta ^A=-\lambda ^A{}_B\theta ^B 
\end{align}
which reads in components
\begin{subequations}\label{eq:conn_trafo_components}
\begin{align}
    \delta _\lambda \omega _\mu ^a&=\partial _\mu \lambda ^a+\omega _\mu ^{ab}\lambda _b -\lambda ^a{}_b\omega ^b_\mu \\
    \delta _\lambda \omega _\mu ^{ab}&=\partial _\mu \lambda ^{ab}+2\omega _\mu ^{c[a}\lambda ^{b]}{}_c \\
    \delta _\lambda e^a_\mu &=-\lambda ^a{}_b e_\mu ^b\\
    \delta _\lambda \tau _\mu &=-e_\mu ^a\lambda _a ~.
\end{align}
\end{subequations}
In the last line one can already recognize the transformation of $\tau _\mu $ matching with the ambiguity \eqref{eq:Carroll_boost} if one identifies $\lambda _\mu =e^a_\mu \lambda _a$. As described at the end of subsection \ref{sec:fiber_bundles}, the solder form $\theta $ relates the invariant tensors on $\mathfrak{m}$ to tensor fields on $\mathcal{M}$, 
\begin{align}
    h_{\mu \nu }=\theta ^A_\mu \theta ^B_\nu h_{AB} && v^\mu =\theta ^\mu _Av^A
\end{align}
where the inverse solder form is defined by 
\begin{align}
    \theta _A^\mu =(-v^\mu ,e_a^\mu ) && \theta ^A_\mu \theta _A^\nu =\delta _\mu ^\nu ~.
\end{align}
This corresponds to the dual components of the frame being defined by the relations 
\begin{align}
    v^\mu \tau _\mu =-1 && e^\mu _ae^b_\mu =\delta ^b_a && v^\mu e_\mu ^a=0 && \tau _\mu e^\mu _a=0
\end{align}
in accordance with subsection \ref{sec:2.2}. The inverse solder form then transforms as
\begin{align}\label{eq:Carr_frametrafo}
    \delta _\lambda \theta _A^\mu =\lambda ^B{}_A\theta _B^\mu ~.
\end{align}
Now that we set up the field content together with its transformation behaviour, we are ready to compute the various notions of curvature, analogously to the example \ref{sec:example_ads}. The Cartan curvature is $\mathfrak{g}$-valued and thus expands as 
\begin{align}\label{eq:ads_cartan_connection}
    F=F(H)H+F^a(P)P_a+F^a(B)B_a+\frac12 F^{ab}J_{ab}
\end{align}
with 
\begin{subequations}
\begin{align}
    F(H)&=\dd \tau +\omega ^a\wedge e_a\\
    F^a(P)&=\dd e^a+\omega ^a{}_b\wedge e^b\\
    F^a(B)&=\dd \omega ^a+\omega ^a{}_b\wedge \omega ^b-\frac{\sigma }{\ell ^2}\tau \wedge e^a\\
    F^{ab}(J)&=\dd \omega ^{ab}+\omega ^a{}_c\wedge \omega ^{cb}-\frac{\sigma }{\ell^2 }e^a\wedge e^b ~.
\end{align}
\end{subequations}
It should be noted that there is again a difference to the curvatures of $\omega ^A{}_B$,
\begin{align}
    F(H)=T && F^a(P)=T^a && F^a(B)=\Omega ^a-\frac{\sigma }{\ell ^2}\tau \wedge e^a && F^{ab}(J)=\Omega^{ab}-\frac{\sigma }{\ell ^2}e^a\wedge e^b 
\end{align}
which only vanishes for $\Lambda =0$. Under $\mathfrak{h}$-transformations the curvatures transform as dictated by \eqref{eq:inf_trafo_H} 
\begin{subequations}
\begin{align}
    \delta _\lambda \Omega^a&=\Omega^{ab}\lambda _b-\lambda ^a{}_b\Omega^b\\
    \delta _\lambda \Omega^{ab}&=\Omega^{ac}\lambda _c{}^b-\lambda ^{a}{}_c\Omega^{cb}\\
    \delta _\lambda T^a&=-T^b\lambda ^a{}_b \label{eq:torsion1}\\
    \delta _\lambda T&=-T^a\lambda _a \label{eq:torsion2}
\end{align}
\end{subequations}
or, equivalently with $T^A=(T,T^a)$, 
\begin{align}
    \delta _\lambda \Omega ^A{}_B&=\Omega ^A{}_C\lambda ^C{}_B-\lambda ^A{}_C\Omega ^C{}_B \\
    \delta _\lambda T^A&=-\lambda ^A{}_BT^B ~.
\end{align}
 Finally, the Bianchi identities \eqref{eq:Bianchi1}, \eqref{eq:Bianchi2} read
\begin{subequations}
\begin{align}
    \dd \Omega^a+\omega ^a{}_b\wedge \Omega^b-\Omega^a{}_b\wedge \omega ^b&=0\\
    \dd \Omega^{ab}+\omega ^a{}_c\wedge \Omega^{cb}-\Omega^{ac}\wedge \omega _c{}^b&=0\\
    \dd T+\omega ^a\wedge T_a-\Omega^a\wedge e_a&=0\\
    \dd T^a+\omega ^a{}_b\wedge T^b-\Omega^a{}_b\wedge e^b&=0 ~.
\end{align}
\end{subequations}

\subsection{Intrinsic torsion}\label{sec:intrin_torsion}
Often, further constraints are imposed on the geometric data that describe a curved Carroll spacetime. This might be done by hand or by the equations of motion resulting from an action principle and typically has the aim of making some gauge field components depend on others. The prime example is to impose some relations on the torsion components which is realized dynamically in the Einstein--Cartan formulation of general relativity. In that case, varying the Einstein--Cartan action with respect to the spin connection requires the torsion to vanish and allows to uniquely solve the spin connection in terms of the frame variables (see, e.g., \cite{Freedman:2012zz}). The action then reduces to the usual Einstein--Hilbert action. 

It might, however, happen that imposing vanishing torsion is a stronger condition than needed for making some connection components depend on others. For the Carroll geometries just discussed we have 
\begin{align}
    T=\dd \tau +\omega ^a\wedge e_a && T^a=\dd e^a+\omega ^a{}_b\wedge e^b 
\end{align}
and one finds that the part 
\begin{align}\label{eq:intrinsic_torsion}
T^{(a}(v,e^{b)})=2v^\mu e^{\nu (a}\partial _{[\mu }e_{\nu ]}^{b)}=-e^{\mu a}e^{\nu b}K_{\mu \nu }=:-K^{ab}
\end{align}
does not depend on $\omega ^a$ or $\omega ^{a}{}_b$. Therefore, this part of the torsion is a property of the metric data itself and imposing its vanishing would be too strong a constraint if one was just interested in integrating out the connection components $\omega ^a $ and $\omega ^{ab}$. In the last equality of \eqref{eq:intrinsic_torsion} we introduced the extrinsic curvature 
\begin{align}
    K_{\mu \nu }=-\frac12 \mathcal{L}_v h_{\mu \nu }
\end{align} 
which is an $\mathfrak{h}$-invariant and connection independent quantity. The part of the torsion that is controlled by $K_{\mu \nu }$ is referred to as \textit{intrinsic torsion} and does not have a Lorentzian analogue. It provides a means to further classify Carroll geometries into four categories \cite{Figueroa-OFarrill:2020gpr}:
\begin{itemize}
    \item[i)]$K_{\mu \nu }=0$ ($v$ is $h$ - Killing)
    \item[ii)]$h^{\mu \nu }K_{\mu \nu }=0 \quad \Leftrightarrow \quad \mathcal{L}_v\, \mathfrak{e}=0$ ($v$ is volume form preserving)  
    \item[iii)]$K_{\mu \nu }=fh_{\mu \nu }$ , $f\in \mathcal{C}^\infty (\mathcal{M})$ ($v$ is $h$ - conformal Killing)
    \item[iv)] general $K_{\mu \nu }$.
\end{itemize}
Referring to the above article for further details on this classification, we now want to see how much of the spin connection can be fixed without constraining the intrinsic part of the torsion. In other words, we choose to reduce the torsion to a ``minimal'' amount that cannot be changed further by restricting the connection. This amounts to solving the constraints 
\begin{align}\label{eq:torsion_constraints}
    T&=0 & T^{[a}(v,e^{b]})&=0 & T^a(e^b,e^c)&=0 
\end{align} 
and leaving the components \eqref{eq:intrinsic_torsion} untouched. The general solution is \cite{Bergshoeff:2017btm}
\begin{align}\label{eq:om1}
    \omega ^a_\mu &=\tau _\mu v^\nu e^{a\sigma }\partial _{[\nu }\tau _{\sigma ]}-e^{\nu a}\partial _{[\mu }\tau _{\nu ]} +C^{ab}e_{\mu b}\\
     \omega _\mu ^{ab}&=2e^{\rho [a}\partial _{[\mu }e_{\rho ]}^{b]}-e_{\mu c}e^{\rho a}e^{\nu b}\partial _{[\rho }e_{\nu ]}^c \label{eq:om2}
\end{align}
where $C^{ab}$ is an arbitrary transverse symmetric tensor corresponding to the components $\omega ^{(a}_\mu e^{b)\mu }$. This is in clear contrast to general relativity, where all the components can be determined uniquely.

There is one point to keep in mind when setting only the connection-dependent part of the torsion to zero. Looking again at \eqref{eq:torsion_constraints} together with the transformation behaviour of the torsion components \eqref{eq:torsion1}, \eqref{eq:torsion2} one can see that these constraints are not Carroll invariant. This already follows from $\delta _\lambda T\neq 0$ when evaluated on the constraints and is also reflected in the transformation behaviour of the solution \eqref{eq:om1}, \eqref{eq:om2} 
\begin{align}
    \delta _\lambda \omega _\mu ^a&=\partial _\mu \lambda ^a+\omega _\mu ^{ab}\lambda _b -\lambda ^a{}_b\omega ^b_\mu + \tau _\mu \delta ^{ab}K_{bc}\lambda ^c \\
    \delta _\lambda \omega _\mu ^{ab}&=\partial _\mu \lambda ^{ab}+2\omega _\mu ^{c[a}\lambda ^{b]}{}_c -2\lambda ^{[a}\delta ^{b]c}K_{cd}e_\mu ^d ~.
\end{align}
Here, we used that the transformation behaviour of $C^{ab}$ is dictated by the relation $C^{ab}=\omega ^{(a}_\mu e^{b)\mu }$ and reads
\begin{align}\label{eq:C_tafo_ab}
    \delta _\lambda C^{ab}=-\lambda ^a{}_cC^{cb}-\lambda ^b{}_cC^{ca}+\partial _\mu \lambda ^{(a}e^{b)\mu }+2v^\mu \lambda ^{(a}e^{b)\nu }\partial _{[\mu }\tau _{\nu ]}-2e^{\mu (a}\partial _{[\mu }e_{\nu ]}^{b)}e^{\nu}_c\lambda ^c ~.
\end{align}
This differs from \eqref{eq:conn_trafo_components} by $K$- and $\lambda ^a$-dependent terms and therefore makes this truncation seem inconsistent with Carroll boost invariance. However, we shall see in section \ref{sec:mag_carr} that magnetic Carroll gravity features a constraint that precisely sets the intrinsic torsion to zero. The additional terms in the transformation behaviour of the connection are then trivial in the sense that they vanish on-shell \cite{Henneaux:1992}. In that scenario the field $C^{ab}$ acts as a Lagrange multiplier. 

\subsection{Affine connection}\label{sec:aff_conn_Carroll}
Following the construction of subsection \ref{sec:fiber_bundles} we introduce a connection on the tangent bundle by imposing the vielbein postulate 
\begin{align}
    \partial _\mu \tau _\nu +\omega ^a_\mu e_{a\nu }-\Gamma ^\alpha {}_{\mu \nu }\tau _\alpha &=0\\
    \partial _\mu e^a_\mu +\omega ^{ab}_{\mu }e_{b\nu }-\Gamma ^\alpha {}_{\mu \nu }e^a_\alpha &=0
\end{align}
which can be solved for $\Gamma ^\alpha {}_{\mu \nu }$ by
\begin{align}
    \Gamma ^\alpha {}_{\mu \nu }=-v^\alpha \big(\partial _\mu \tau _\nu +\omega ^a_\mu e_{\nu a}\big)+e^\alpha _a\big(\delta ^a_b \partial _\mu +\omega ^a_{\mu }{}_b\big)e^b_\nu ~.
\end{align}
This leads to an associated covariant derivative that is automatically compatible with the Carroll structure,
\begin{align}\label{eq:Carr_metricity}
    \nabla _\mu h_{\alpha \beta }=0 && \nabla _\mu v^\nu =0 
\end{align}
but in general has non-vanishing torsion. The associated Riemann curvature tensor \eqref{eq:riemann} can be written as
\begin{align}
    R^\alpha {}_{\beta \mu \nu }=-v^\alpha e_{b\beta }\,\Omega ^b_{\mu \nu }+e^\alpha _ae^b_\beta \,\Omega ^a{}_{b\mu \nu }
\end{align}
and satisfies a number of identities as a consequence of the compatibility conditions \eqref{eq:Carr_metricity},
\begin{align}
    R^\alpha {}_{\beta \mu \nu }v^\beta &=0 \\
    h_{\alpha (\rho }R^\alpha {}_{\beta )\mu \nu }&=0 \\
    R^\alpha {}_{\alpha \mu \nu }&=0 ~.
\end{align}
The first of these also implies that the Ricci tensor $R_{\mu \nu }=R^\beta {}_{\mu \beta \nu }$ obeys
\begin{align}
    R_{\mu \nu }v^\mu =0 
\end{align}
but in general is not symmetric if the torsion is non-zero as a consequence of the Bianchi identity \eqref{eq:bianchi_riemm}. Since $\Gamma ^\alpha {}_{\mu \nu }$ is manifestly $\mathfrak{h}$-invariant at this point, we conclude that also $R^\alpha {}_{\beta \mu \nu }$ as well as $R_{\mu \nu }$ are. However, one needs to keep in mind that these quantities still depend on the connection coefficients $\omega ^a$ and $\omega ^{ab}$. Finally, to write down gravitational actions we need a curvature scalar. Its construction is not obvious since, e.g., a term like $e ^{a\mu }e _a^\nu R_{\mu \nu }$ cannot be $\mathfrak{h}$-invariant
as one can check by using \eqref{eq:Carr_frametrafo}. Instead, one finds $\delta _\lambda (e^{a\mu }e_a^\nu R_{\mu \nu })=-e^{\mu a}v^\nu \lambda _a R_{\mu \nu } \neq 0$, which only vanishes if torsion is restricted to zero. We shall come back to this below. There are, however, several other $\mathfrak{h}$-invariant terms that can be built according to the classification in \cite{Figueroa-OFarrill:2022mcy}. The most important one for this work is
\begin{align}\label{eq:curv_scalar}
    \mathcal{R}=-2v^\mu e_b^\nu \Omega ^b_{\mu \nu }+e^\mu _a e^{b\nu }\Omega ^a{}_b {}_{\mu \nu } = h^{\mu \nu }\big(R_{\mu \nu }-v^\rho \tau_\sigma R^\sigma {}_{\mu \rho \nu }\big)
\end{align}
which is also obtained from a $c\to 0$ limit of the Einstein--Cartan action \cite{Bergshoeff:2017btm,Campoleoni:2022ebj}. 

Let us now restrict to the minimal torsion case from subsection \ref{sec:intrin_torsion}. Inserting \eqref{eq:om1}, \eqref{eq:om2} one finds after some manipulations
\begin{align}\label{eq:gen_affine_connection}
    \Gamma ^\alpha {}_{\mu \nu }=&-v^\alpha \partial _{(\mu }\tau _{\nu )}-v^\alpha \tau _{(\mu }\mathcal{L}_v \tau _{\nu )}-v^\alpha C_{\mu \nu }\\
    &+\frac{1}{2}h^{\rho \alpha }\big(\partial _\mu h_{\rho \nu }+\partial _\nu h_{\rho \mu }-\partial _\rho h_{\mu \nu }\big)-h^{\alpha \rho }K_{\rho \mu }\tau _\nu \nonumber
\end{align}
whose torsion indeed only depends on $K_{\mu \nu}$,
\begin{align}
    T^\alpha _{\mu \nu }=2\Gamma ^\alpha {}_{[\mu \nu ]}=-2h^{\alpha \rho }K_{\rho [\mu }\tau _{\nu]} ~.
\end{align}
Here, we also defined $C_{\mu \nu }=e_{a\mu }e_{b\nu }C^{ab}$. Carroll boost invariance of $\Gamma ^\alpha {}_{\mu \nu }$ is broken as a consequence of the constraints \eqref{eq:torsion_constraints}. One finds instead
\begin{align}\label{eq:Gamma_Boostnoninvariance}
    \delta _\lambda \Gamma ^\alpha {}_{\mu \nu }=-\tau _\mu K_{\nu \sigma }\lambda ^\sigma v^\alpha -2h^{\alpha \sigma }K_{\mu [\nu }\lambda _{\sigma ]} 
\end{align}
and the transformation of $C_{\mu \nu }$ under $\mathfrak{h}$ follows from \eqref{eq:C_tafo_ab},
\begin{align}\label{eq:c_minimal_trafo}
    \delta _\lambda C_{\mu \nu }&=\partial _{(\mu }\lambda _{\nu )}+\lambda _{(\mu }a_{\nu )}+v^\sigma \big(\tau _{(\mu }\partial _\sigma \lambda _{\nu )}-\tau _{(\mu }\partial _{\nu )}\lambda _\sigma\big)\\
    &\quad -\frac12 \lambda ^\sigma \big(\partial _\mu h_{\nu \sigma }+\partial _\nu h_{\mu \sigma }-\partial _\sigma h_{\mu \nu }\big)+2\lambda ^\sigma K_{\sigma (\mu }\tau _{\nu )} \nonumber ~.
\end{align}
Since $C_{\mu \nu }v^\mu =0$ one can see that Carroll boost invariance of $\Gamma ^\alpha {}_{\mu \nu }$ cannot be repaired by altering the transformation behaviour of $C_{\mu \nu }$ since this would require adding non-transverse terms. In general, a Carroll compatible affine connection with its torsion only given by the intrinsic part cannot be made $\mathfrak{h}$-invariant when just realized on the fields $\tau _\mu $, $h_{\mu \nu }$ and $C_{\mu \nu }$. Only once the intrinsic torsion is set to zero $\mathfrak{h}$-invariance is recovered.\footnote{When realizing the $\mathfrak{h}$ transformations on a different set of fields this is not true any more \cite{Hartong:2015xda}.} Nevertheless, one can show that the curvature scalar \eqref{eq:curv_scalar} evaluated for this connection is invariant up to a total derivative, 
\begin{align}\label{eq:trafomathcalR}
    \delta _\lambda \mathcal{R}[\tau _\mu ,h_{\mu \nu },C_{\mu \nu }]=-\frac{2}{\mathfrak{e}}\partial _\mu (\mathfrak{e}\lambda ^\mu K)
\end{align}  
which will be the fundamental building block for magnetic Carroll gravity \eqref{sec:mag_carr}. For showing this it is important to use that total divergences of $\nabla $ associated to \eqref{eq:gen_affine_connection} read
\begin{align}
    \mathfrak{e}\nabla _\mu X^\mu =\partial _\mu (\mathfrak{e}X^\mu )-\mathfrak{e}K\tau _\mu X^\mu 
\end{align}
where $\partial _\mu \log\mathfrak{e} =\Gamma ^\nu {}_{\nu \mu }+K\tau _\mu $ and $K:=h^{\alpha \beta }K_{\alpha \beta }$. The Ricci tensor is still not symmetric, by contracting the Bianchi identity \eqref{eq:bianchi_riemm} one finds
\begin{align}
    R_{[\mu \nu ]}=\nabla _{[\mu }(\tau _{\nu ]}K)+\nabla _\rho (\tau _{[\mu }K^\rho {}_{\nu ]}) ~.
\end{align}

\subsection{Maximally symmetric spacetimes}
\label{sec:symmetries}
The case $F=0$ in \eqref{eq:ads_cartan_connection} corresponds to Cartan geometries that are locally isometric to the corresponding Klein geometries, i.e., (A)dS Carroll spacetimes in this case. Written in the geometric language of the tangent bundle introduced in the previous subsection, where curvature and torsion are given in terms of the affine connection $\Gamma ^\alpha {}_{\mu \nu }$ together with the Carroll metric variables this leads to the equations
\begin{align}\label{eq:Riem:AdSC3}
    R^\alpha {}_{\beta \mu \nu }=\frac{\sigma }{\ell ^2}\big(\delta ^\alpha _\mu h_{\nu \beta }-\delta ^\alpha _\nu h_{\mu \beta }\big) \qquad \Rightarrow \qquad h^{\mu \nu }R_{\mu \nu }=\Lambda  \frac{2(d+1)}{d-1} 
\end{align}
as well as
\begin{align}\label{eq:Torsion:AdSC3}
    T^\alpha _{\mu \nu }=2\Gamma ^\alpha {}_{[\mu \nu ]}=0 ~.
\end{align}
Here, we again used the identification \eqref{eq:cosm_const} to rewrite the curvature scalar in terms of $\Lambda $. Clearly, there can be no intrinsic torsion in this case, implying that $h^{\mu \nu }R_{\mu \nu }$ is a Carroll boost invariant quantity and that $R_{[\mu \nu ]}=0$. The general solution of \eqref{eq:Torsion:AdSC3} is just the corresponding restriction of \eqref{eq:gen_affine_connection}
\begin{align}
    \Gamma ^\alpha {}_{\mu \nu }=&-v^\alpha \partial _{(\mu }\tau _{\nu )}-v^\alpha \tau _{(\mu }\mathcal{L}_v \tau _{\nu )}-v^\alpha C_{\mu \nu }+\frac{1}{2}h^{\rho \alpha }\big(\partial _\mu h_{\rho \nu }+\partial _\nu h_{\rho \mu }-\partial _\rho h_{\mu \nu }\big)
\end{align}  
where the ambiguity associated to $C^{ab}$ in \eqref{eq:om1} is still present. However, since $K_{\mu \nu }=0$ it follows that the affine connection is boost-invariant, cf. \eqref{eq:Gamma_Boostnoninvariance}. The compatible transformation behaviour of $C_{\mu \nu }$ follows from \eqref{eq:c_minimal_trafo},
\begin{align}
    \delta _\lambda C_{\mu \nu }&=\partial _{(\mu }\lambda _{\nu )} +\lambda_{(\mu} a_{\nu)} + v^\alpha \big( \tau_{(\nu} \partial_\alpha \lambda_{\mu)}- \tau_{(\nu} \partial_{\mu)} \lambda_{\alpha} \big)\\
    \nonumber
    &\quad -\frac{1}{2}h^{\lambda \sigma }\lambda _\sigma \big(\partial _\mu h_{\nu \lambda }+\partial _\nu h_{\mu \lambda }-\partial _\lambda h_{\mu \nu }\big) ~.
\end{align}
This shows that a curved Carroll spacetime with a torsion free compatible connection is not just described by the Carroll metric variables but also by the tensor $C_{\mu \nu }$ that enters the connection. For obtaining the locally homogeneous spacetime we still need to solve the equations \eqref{eq:Riem:AdSC3} that typically restrict $C_{\mu \nu }$ further along with the metric variables. 

The spacetime symmetries of the resulting geometries manifest themselves in a Carroll analogue of the Killing equations. Since part of the affine connection plays an independent role in the Carroll case they are given by 
\begin{align}\label{eq:Carr_Killing}
    \mathcal{L}_\xi v^\mu =0 && \mathcal{L}_\xi h_{\mu \nu }=0 && (\mathcal{L}_\xi \nabla )_\mu =0
\end{align}
where the last condition expresses the invariance of the affine connection \cite{Morand:2018tke,Bekaert:2015xua} and reads explicitly\footnote{This condition can be shown to be equivalent to the concept of Cartan symmetries defined by $\mathcal{L}_\xi A+\delta _{\lambda(\xi )} A=0$. It states that the Cartan connection is left invariant up to a local $\mathfrak{h}$-transformation.} 
\begin{align}\label{eq:aff_Killing_cond}
    \xi ^\alpha \partial _\alpha \Gamma ^\sigma {}_{\mu \nu }+\Gamma ^\sigma {}_{\alpha \nu }\partial _\mu \xi ^\alpha +\Gamma ^\sigma {}_{\mu \alpha }\partial _\nu \xi ^\alpha -\Gamma ^\alpha {}_{\mu \nu }\partial _\alpha \xi ^\sigma +\partial _\mu \partial _\nu \xi ^\sigma =0 ~. 
\end{align}
Solving these equations one finds Carroll Killing fields that close into $\mathfrak{g}$ under the standard Lie bracket for vector fields, which in the present case is just the (A)dS Carroll algebra \eqref{eq:Carroll_ADS_algebra}. Sometimes the last condition is dropped, in which case one obtains an infinite-dimensional lift of this algebra. Then, any vector of the form $\xi ^\mu =f(x^\nu )v^\mu $ with $f$ being an arbitrary function satisfying $v^\alpha \partial _\alpha f=0$ solves the first two equations. Using the terminology of \cite{Afshar:2024llh} the corresponding algebra is referred to as $\mathfrak{isocarr}_\Lambda (d+1)$ and given by 
\begin{subequations}\label{eq:isocarrlamb}
\begin{gather}
        [P_a,J_{bc}]=2\delta _{a[b}P_{c]} \qquad \qquad [P_a,P_b]=-\frac{\sigma }{\ell ^2}J_{ab} \qquad \qquad [J_{ab},J_{cd}]=4\delta _{[a[c}J_{d]b]}\\
    [P_a,M_f]=M_{f_a} \qquad \qquad [M_f,J_{ab}]=M_{f_{ab}} 
\end{gather}
\end{subequations}
where the functions $f_a$, $f_{ab}$ again are annihilated by $v^\alpha \partial _\alpha $ but otherwise depend on the specific coordinates chosen to compute the Carroll Killing vectors.

Let us illustrate this with an example in $d=2$ and choose $\sigma =-1$. Consider a one-parameter family of solutions to Eqs. \eqref{eq:Riem:AdSC3} and \eqref{eq:Torsion:AdSC3} given by
\begin{subequations}\label{eq:cads3_Data}
\begin{align}
    v=-\frac{1}{\rho (r)}\partial _t && \tau =\rho(r)\dd t && \dd s^2=\frac{\dd r^2}{\rho^2(r)}+r^2\dd \phi ^2 \\
    \Gamma ^t{}_{tr}=\frac{r}{\ell ^2\rho^2(r)} && \Gamma ^t{}_{rr}=\frac{\gamma }{\rho^{3}(r)} && \Gamma ^t{}_{\phi \phi }=\frac{\gamma \,r^2}{\rho(r)} \\
    \Gamma ^r{}_{rr}=-\frac{r}{\ell ^2\rho^2(r)} && \Gamma ^r{}_{\phi \phi }=-r\rho^2(r) && 
    \Gamma ^\phi {}_{r\phi }=\frac{1}{r}
\end{align}
\end{subequations}
where $\rho^2(r)=1+\frac{r^2}{\ell ^2}$ and $\gamma \in \mathbb{R}$. By definition, these geometries locally describe AdSC$_3$ spacetime but include different choices of connections parametrized by $\gamma $. The case $\gamma =0$ corresponds to the naive $c\to 0$ limit of AdS$_3$ spacetime in static coordinates. We solve the first two equations in \eqref{eq:Carr_Killing} and obtain the vector fields
\begin{subequations}
\begin{align}
    M_f &=f(r,\phi )\partial _t & P_1&=\frac{rt}{\ell ^2\rho(r)}\cos \phi \,\partial _t- \rho(r)\cos \phi \, \partial _r +\frac{\rho(r)}{r}\sin \phi \, \partial _\phi \\
    J&=\partial _\phi & P_2&=\frac{rt}{\ell^2 \rho(r)}\sin \phi \,\partial _t- \rho(r)\sin \phi \, \partial _r -\frac{\rho(r)}{r}\cos \phi \, \partial _\phi
\end{align}
\end{subequations}
where $f(r,\phi )$ is an arbitrary function. They close into the three-dimensional version of \eqref{eq:isocarrlamb} where $M_f$ represents a whole tower of supertranslations.

Solving the last condition \eqref{eq:aff_Killing_cond} restricts this tower to the independent solutions
\begin{align}
    H=\partial _t && B_1=\frac{r}{\ell f(r)}\cos \phi \,\partial _t && B_2=\frac{r}{\ell f(r)}\sin\phi \,\partial _t
\end{align}
and makes the Carroll Killing fields form the AdS Carroll algebra \eqref{eq:Carroll_ADS_algebra} in $d=2$ with $J_{ab}=J\epsilon _{ab}$. In this Carroll boost frame the tensor $C_{\mu \nu }$ is given by $C_{\mu \nu }=\gamma h_{\mu \nu }$ but any boost of the form 
\begin{align}
    \lambda _t=0 && \lambda _r =\delta \frac{r}{\ell ^2\, \rho^2(r)} && \lambda _\phi =0
\end{align}
with $\delta \in \mathbb{R}$ can be used to change $C_{\mu \nu }\to C_{\mu \nu }+\delta h_{\mu \nu }$. In particular, there is a boost frame where $C_{\mu \nu }=0$ and all the fields in \eqref{eq:cads3_Data} are unchanged except 
\begin{align}
    \tau =\rho(r)\dd t +\gamma \frac{r}{\ell ^2\rho^2(r)}\dd r ~.
\end{align}

\section{Conformal extensions of the Carroll algebra}\label{sec:conformal_extensions}
Before we switch gears and tackle Carroll field theories, we discuss conformal extensions of the Carroll algebra since after all these are the relevant symmetries in a holographic context. The reason is that the null boundary of asymptotically flat spacetimes only carries an equivalence class of Carroll metrics and vector fields \cite{Ashtekar:2014zsa} with two members $(h_{\mu \nu },v^\mu )$, $(h'_{\mu \nu },v{}'^\mu )$ being identified by
\begin{align}\label{eq:conf_BMS_def}
    (h'_{\mu \nu },v{}'^\mu )\sim (\eta^2 h_{\mu \nu },\eta ^{-z}v^\mu ) \qquad \qquad z=1~.
\end{align}
Here, $\eta $ is an arbitrary function of the coordinates and $z$ is called the dynamical or anisotropy exponent. While $z=1$ in special cases it may be an arbitrary real number in generic situations. The diffeomorphisms that preserve this structure are generated by vector fields $\xi $ with 
\begin{align}\label{eq:CKeqs_z}
    \mathcal{L}_\xi h_{\mu \nu }=2\rho h_{\mu \nu } \qquad \qquad \mathcal{L}_\xi v^\mu =-z\rho v^\mu 
\end{align}
for some function $\rho (\xi )$. They form an infinite lift of the conformal Carroll group which is just a different name for the BMS group, the group of asymptotic symmetries at null infinity \cite{Duval2014a,Bondi:1962,Sachs:1962}. While one may also consider further generalizations of \eqref{eq:conf_BMS_def} which lead to an even larger symmetry group (for a summary see, e.g., appendix E of \cite{Chandrasekaran:2021vyu}), we shall only stick to this simplest case here. However, we keep the dimension of the Carroll manifold $d+1$ general for most of the discussion and only look at special cases in the end. 

Following \cite{Afshar:2024llh}, it is possible to extend the Carroll algebra \eqref{eq:Carroll_algebra} to a slightly bigger finite-dimensional algebra called $\mathfrak{confcarr}_z(d+1)$ by adding generators $\{D,K\}$ representing $z$-dependent dilatations and temporal special conformal transformations. The bracket relations read
\begin{equation}\label{eq:confcarr_z}
\begin{aligned}
    [P_i,B_j]&=-\delta _{ij}H & [D,H]&=-zH \\
    [P_i,J_{jk}]&=2\delta _{i[j}P_{k]} & [D,P_i]&=-P_i \\
    [B_i,J_{jk}]&=2\delta _{i[j}B_{k]} & [D,B_i]&=(1-z)B_i\\
    [J_{ij},J_{kl}]&=4\delta _{[i[k}J_{l]j]} & [D,K]&=(2-z)K\\
     & & [K,P_i]&=2B_i 
\end{aligned}
\end{equation}
where the left column is the Carroll algebra and the right column are the additional brackets. For generic dimensions, there exists an infinite lift of this algebra called $\tconfcar_z(d+1)$
\begin{equation}\label{eq:zconfcarrtilde}
    \begin{aligned}
        [P_i,J_{jk}]&=2\delta _{i[j}P_{k]} & & \\
    [J_{ij},J_{kl}]&=4\delta _{[i[k}J_{l]j]} & [M_f,J_{jk}]&=-M_{x_j\partial _kf-x_k\partial _jf} \\
    [D,P_i]&=-P_i & [P_i,M_f]&=M_{\partial _if} \\
    [D,K_i]&=K_i & [D,M_f]&=M_{(x_i \partial_i -z)f} \\
    [K_i,J_{jk}]&=2\delta _{i[j}K_{k]} & [K_i,M_f]&=M_{(2x_ix_j\partial_j-2zx_i-x_jx_j\partial _i)f} \\
    [K_i,P_j]&=-2J_{ij}-2\delta _{ij}D & & 
    \end{aligned}
\end{equation}
where the various generators have a spacetime realization as
\begin{gather}\label{generatorrep}
    P_i=\partial _i \qquad \qquad J_{ij}=x_i\partial _j-x_j\partial _i \qquad \qquad M_f=f(x_i)\partial _t \\[.4em]
     D=zt\partial _t+x_i\partial _i \qquad \qquad K_i=2x_i(zt\partial _t+x_j\partial _j)-x_jx_j\partial _i \label{eq:D_SCT}
\end{gather}
and we defined the supertranslation generator $M_f$. As special cases it contains the generators for $\{H,B_i,K\}=\{M_1,M_{-x_i},M_{x_ix_i}\}$. This algebra can also be directly obtained by solving \eqref{eq:CKeqs_z} for $h_{\mu \nu }=\delta _{ij}\delta ^i_\mu \delta ^j _\nu $ and $v=-\partial _t$ and is thus the $z$-dependent generalization of the BMS algebra in any dimension\footnote{The option of general $z$ was also considered in \cite{Duval2014a} but called the level $N$. They are related by $z=2/N$.}. Since the left column in \eqref{eq:zconfcarrtilde} is just the conformal algebra of Euclidean space $\mathbb{R}^d$, we can also write
\begin{align}\label{eq:def:conf_carr}
   \tconfcar_z(d+1) \cong \mathfrak{conf}(\mathbb{R}^d)\,\ruplus _z \,C^\infty (\mathbb{R}^d) 
\end{align} 
which is the definition provided in \cite{Afshar:2024llh}. While further extensions are still possible we shall stop here and only mention that once the $K_i$ have been added a truncation to a finite dimensional algebra is only possible for $z\in \mathbb{N}$. The case $z=1$ is the most prominent here since its finite dimensional truncation is the standard (finite-dimensional) conformal Carroll algebra $\mathfrak{cca}(d+1)\cong\mathfrak{iso}(d+1,1)$. This algebra is isomorphic to the Poincaré algebra in one dimension higher. Its bracket relations are given by \eqref{eq:confcarr_z} at $z=1$ together with
\begin{align}\nonumber
    [D,K_i]&=K_i & [K_i,H]&=2B_i \\
    [K_i,P_j]&=-2J_{ij}-2\delta _{ij}D & [K_i,B_j]&=\delta _{ij}K\\
    \nonumber
    [K_i,J_{jk}]&=2\delta _{i[j}K_{k]} &  &
\end{align}
\subsubsection*{Restricting to $d=2$}
Let us now focus on the case $d=2$. Since in this case we have by definition
\begin{align}\label{eq:confcarrz2d}
    \tconfcar_z(2+1)\cong (\text{Witt}\oplus \text{Witt})\, \ruplus _z\, C^\infty (\mathbb{R}^2)
\end{align}
the algebra is now even larger as the local conformal transformations of two-dimensional Euclidean space are infinite-dimensional themselves and given by a direct sum of two Witt algebras. In a holographic context, these transformations are referred to as superrotations and play a decisive role in the formulation of celestial holography \cite{Barnich:2009se,Strominger:2017zoo}. To bring the algebra in a recognizable form it is convenient to dualize the rotation generator by defining $J=\epsilon ^{ij}J_{ij}$. Then, switching to complex coordinates for the spatial directions $w=x_1+ix_2$, $\Bar{w}=x_1-ix_2$ the algebra \eqref{eq:confcarrz2d} reads
\begin{align}
    [L_n,L_m]&=(n-m)L_{n+m} & [L_n,M_{(r,s)}]&=\Big(\frac{z}{2}(n+1)-r\Big)M_{(r+n,s)}\label{eq:extBMS1}\\
    [\Bar{L}_n,\Bar{L}_m]&=(n-m)\Bar{L}_{n+m} & [\Bar{L}_n,M_{(r,s)}]&=\Big(\frac{z}{2}(n+1)-s\Big)M_{(r,s+n)}\label{eq:extBMS2}
\end{align}
where $n,m,r,s\in \mathbb{Z}$ and
\begin{subequations}\label{eq:CKVs_d2}
\begin{align}
    &M_{(r,s)}=w^r\Bar{w}^s\partial _t \\
    &L_n=-\frac{z}{2}(n+1)w^nt\partial _t-w^{n+1}\partial _w\\
    &\Bar{L}_n=-\frac{z}{2}(n+1)\Bar{w}^nt\partial _t-\Bar{w}^{n+1}\partial _{\Bar{w}} ~.
\end{align}
\end{subequations}
The generators $\{P_i,J,D,H,B_i,K,K_i\}$ are packaged into these for the values $n,m,r,s=-1,0,1$ (for details see section 3.2 of \cite{Afshar:2024llh}). However, it can be shown that they only form a subalgebra for $z=1$. The resulting algebra is referred to as the extended BMS algebra and admits a truncation to the standard $\mathfrak{bms}_4\cong \mathfrak{so}(3,1)\, \ruplus \, C^\infty (\mathbb{R}^2)$ by restricting the superrotations $L_n$, $\Bar{L}_n$ to $n=-1,0,1$. Another special case is $z=0$ where it was found that \eqref{eq:extBMS1}, \eqref{eq:extBMS2} form near horizon symmetries of four-dimensional black holes \cite{Donnay:2015abr,Donnay:2016ejv,Grumiller:2019fmp}.
We finally mention that the Carroll conformal Killing vectors \eqref{eq:CKVs_d2} can also be written in the concise form
\begin{align}
    \xi =\Big[f+\frac{zt}{2}(\partial _wY^w+\partial _{\Bar{w}}Y^{\Bar{w}})\Big]\partial _t +Y^w\partial _w+Y^{\Bar{w}}\partial _{\Bar{w}}
\end{align}
where $f=f(w,\Bar{w})$, $Y^w=Y^w(w)$ and $Y^{\Bar{w}}=Y^{\Bar{w}}(\Bar{w})$. This makes it especially simple to check \eqref{eq:CKeqs_z} resulting in the function on the left hand side being given by $\rho =\frac{1}{2}(\partial _wY^w+\partial _{\Bar{w}}Y^{\Bar{w}})$.

\subsubsection*{Restricting to $d=1$}
At $d=1$ one finds from the definition \eqref{eq:def:conf_carr}
\begin{align}\label{eq:confcarrz1d}
    \tconfcar_z(1+1)\cong \text{Witt}\,\, \ruplus _z\, C^\infty (\mathbb{R})
\end{align}
which is generated by the Carroll conformal Killing vectors
\begin{align}
    \xi =\Big[f(x)+tz\partial _xY(x)\Big]\partial _t+Y(x)\partial _x ~.
\end{align}
They satisfy \eqref{eq:CKeqs_z} with $\rho =\partial _xY$. Here, $f$ again generates the tower of supertranslations and $Y$ is associated to the two-dimensional version of superrotations. They can be put again into a more recognizable form by defining
\begin{align}
    L_n&=-zt(n+1)x^{n}\partial _t-x^{n+1}\partial _x  & M_n&=x^{n+z}\partial _t
\end{align}
which gives the Lie brackets
\begin{align}
[L_n,L_m]&=(n-m)L_{n+m} \\
[L_n,M_m]&=\big(zn-m\big)M_{n+m}\\
[M_n,M_m]&=0 ~.    
\end{align}
Some special cases of this algebra are $z=1$, which yields the superrotation extended $\mathfrak{bms}_3$ algebra \cite{Barnich:2006av}. Up to central extensions these are the asymptotic symmetries of asymptotically flat three-dimensional Einstein gravity. The case $z=0$ corresponds to the non-centrally extended version of the warped conformal algebra where $M_n$ are the modes of a $U(1)$ current (see, e.g. \cite{Detournay:2012pc}). Finally, $z=-1$ is called the $\mathfrak{bms}_2$ algebra. It features in studies of asymptotically flat spacetimes in two-dimensional dilaton gravity \cite{Afshar:2019axx,Afshar:2021qvi}.

\section{Weyl-transformations}\label{sec:Weyl}
As a final point, we mention various Weyl transformation properties of Carroll geometries since they shall play an important role for various instances of Carroll field theories. Given an arbitrary local function $\rho (x)$, we define infinitesimal Weyl transformations of the metric variables as
\begin{align}
    \delta _\rho \tau _\mu &= z\rho \tau _\mu & \delta _\rho h_{\mu \nu }&=2\rho h_{\mu \nu } \\
    \delta _\rho v^\mu &=-z\rho v^\mu & \delta _\rho h^{\mu \nu }&=-2\rho h^{\mu \nu } 
\end{align}
where $z\in \mathbb{R}$ is the anisotropy exponent. The various composite fields therefore transform as
\begin{align}
    \delta _\rho K_{\mu \nu }&=(2-z)\rho K_{\mu \nu }-h_{\mu \nu }v^\alpha \partial _\alpha \rho \\
    \delta _\rho K&=-z\rho K-dv^\alpha \partial _\alpha \rho \\
    \delta _\rho \,a_\mu &=zh_{\mu \alpha }h^{\alpha \nu }\partial _\nu \rho ~.
\end{align}
They can be used to define the object 
\begin{align}
    W_\mu =\frac{K}{d}\tau _\mu +\frac{1}{z}a_\mu  
\end{align}
which transforms like a $U(1)$-connection, $\delta _\rho W_\mu =\partial _\mu \rho $. 

\chapter{Field theories on Carroll backgrounds}
\label{ch:field_th_backgrounds}
\setlength{\epigraphwidth}{0.6\textwidth}
\epigraphfontsize{\small\itshape}
\epigraph{``We can't stop here, this is bat country!''}{--- \textup{Hunter S. Thompson}, Fear and Loathing in Las Vegas}

Now that we defined Carroll geometries, we may consider them as backgrounds for field theories. As we shall see, depending on how the fields couple to the geometry one arrives at different conclusions about the physical content these theories are able to capture. This chapter is structured as follows: In section \ref{sec:cov_form} we derive the stress-energy tensor as a response to varying the background metric- and connection data. In both cases, we compute the classical Ward identities associated to the background gauge symmetries. Once a symmetric background is chosen, these allow us to construct charges and conservation laws associated to these symmetries. For deriving the brackets between these charges we need brackets between the stress-energy tensor components which are derived for a general Carroll quantum field theory in section \ref{sec:Quantum}. Finally, we discuss some examples in section \ref{sec:examples}. The material of this chapter in part is based on \cite{Dutta:dummy} and \cite{Ecker:2024czx}.

\section{General structure and covariant formulation}\label{sec:cov_form}
Consider a local action functional $I_m[\phi ,h_{\mu \nu },\tau _\mu ]$ for some field $\phi $ (not necessarily a scalar field) that couples to the Carroll metric data $h_{\mu \nu }$, $\tau _\mu $ in a covariant way. The general variation of the action then takes the form
\begin{align}\label{var}
    \delta I_m[\phi ,h_{\mu \nu },\tau _{\mu }]=\int \dd ^{d+1}x\, \mathfrak{e}\Big[\frac12 T^{\mu \nu }\delta h_{\mu \nu }+T^\mu \delta \tau _\mu +(\text{EOM})_\phi \,\delta \phi \Big] 
\end{align}
where the last part are just the equations of motion for the matter fields. We take the background geometry here in the language of the second order formulation of section \ref{sec:aff_conn_Carroll}, where the affine connection is chosen to be metric compatible and has only intrinsic torsion. As we shall see in chapter \ref{chap:carroll_gravity}, these are typical off-shell geometries of magnetic Carroll gravity. While they also contain independent data $C_{\mu \nu }$ in the connection that the field theory could couple to, we assume here that this is not the case. This assumption is valid for a large part of the Carroll field theories discussed in the literature \cite{Henneaux:2021yzg,Bergshoeff:2023vfd,Ecker:2024czx}, but there do exist counterexamples \cite{Baiguera:2022lsw}, see also section \ref{sec:nonminsca}.  

The independent background metric fields are chosen as the Carroll clock form $\tau _\mu $ and Carroll metric $h_{\mu \nu }$ which uniquely determine the fields $v^\mu $, $h^{\mu \nu }$ with upper indices by (see also section \ref{sec:2.2})
\begin{align}
    v^\mu \tau _\mu &=-1 & v^\mu h_{\mu \nu }&=0 & \tau _\mu h^{\mu \nu }&=0 & h^{\mu \rho }h_{\rho \nu }&=\delta ^\mu _\nu +v^\mu \tau _\nu ~.
\end{align}
We also use these expressions to determine the variations of the latter fields by
\begin{align}
    \delta v^\mu &=v^\mu v^\nu \delta \tau _\nu -h^{\mu (\nu }v^{\alpha )}\delta h_{\nu \alpha } & \delta h^{\mu \nu }&=2v^{(\mu }h^{\nu )\alpha }\delta \tau _\alpha -h^{\rho (\mu }h^{\nu )\alpha }\delta h_{\rho \alpha } ~.
\end{align}
Except for the symmetry in the indices of $\delta h_{\alpha \beta }$, the variations $\delta \tau _\mu $ and $\delta h_{\mu \nu }$ are completely unconstrained which is needed for deriving the response functions unambiguously. Had we instead taken the Carroll boost invariant variations $\delta v^\mu $ and $\delta h_{\mu \nu }$ to define the response functions we would not be able to access the same information by computing the background variation, since the relation $v^\mu h_{\mu \nu }=0$ makes these variations dependent, see also the discussion in \cite{Baiguera:2022lsw}.

\subsection{Spacetime symmetries and conservation laws}\label{eq:sec311}
\subsubsection{Non-conformal case}
For the coupling to be consistent with the background gauge symmetries, the response functions $T^\mu $ and $T^{\mu \nu }$ need to satisfy further requirements when the equations of motion for the dynamical fields hold. These (classical) Ward identities are determined by demanding 
\begin{align}
    \delta _\xi I_m &= 0 & \delta _\lambda I_m &= 0 
\end{align}
on-shell and up to boundary terms leading to the Carroll boost Ward identity
\begin{align}
    T^\mu h_{\mu \nu }=0
\end{align}
as well as the diffeomorphism Ward identity 
\begin{align}\label{eq:diff_0.0}
    \partial_{\mu} \big( \mathfrak{e} \, T^\mu \tau _\alpha +\mathfrak{e}T^{\mu \rho }h_{\rho \alpha } \big) = \mathfrak{e}T^\mu \, \partial_\alpha \tau_\mu + \frac{1}{2} \mathfrak{e}T^{\mu \nu} \partial_\alpha h_{\mu \nu} ~.
\end{align}
By taking a second variation of the action one may as well determine the boost transformation behaviour of the response functions 
\begin{align}
    \delta _\lambda T^\mu &=0 & \delta _\lambda T^{\mu \nu }&=2\lambda ^{(\mu }T^{\nu )} 
\end{align}
which motivates the boost invariant definition 
\begin{align}\label{eq:Carroll_EMT}
    \mathbf{T}^\mu {}_\nu :=T^\mu \tau _\nu +T^{\mu \rho }h_{\rho \nu }
\end{align}
called the \textit{Carroll stress-energy tensor}. With this definition, the diffeomorphism Ward identity \eqref{eq:diff_0.0} and Carroll boost Ward identity can be written in the covariant form 
\eq{
\boxed{
\phantom{\Bigg(}
   \nabla _\mu \mathbf{T}^\mu {}_\nu =2\mathbf{T}^\mu {}_\alpha h^{\alpha \rho }K_{\rho [\mu }\tau _{\nu ]}-K\tau _\mu \mathbf{T}^\mu {}_\nu \qquad \qquad v^\nu h_{\mu \rho }\mathbf{T}^\mu {}_\nu =0  ~.
\phantom{\Bigg)}
    }
}{eq:ward_identities_covariant}
where we used the Carroll connection \eqref{eq:gen_affine_connection} associated to the chosen background which satisfies $\partial _\mu \log \mathfrak{e}=\Gamma ^\alpha {}_{\alpha \mu }+\tau _\mu K$. One can see that on torsion free backgrounds, where $K_{\mu \nu }=0$, the diffeomorphism Ward identity reduces to the usual covariant conservation equation. 

In the same manner as for Riemannian geometries, any symmetry of the background visible to the field theory leads to an on-shell conserved current density defined by
\begin{align}\label{eq:def_Carr_current}
    j^\mu =\mathfrak{e}\mathbf{T}^\mu {}_\nu \xi ^\nu 
\end{align}
where the vector field $\xi ^\mu $ is the infinitesimal generator of that symmetry. Since the field theory is only assumed to couple to the Carroll metric data $\xi ^\mu $ only needs to solve the Carroll Killing equations 
\begin{align}\label{eq:ahhhh}
    \mathcal{L}_\xi v^\mu &=0 & \mathcal{L}_\xi h_{\mu \nu }&=0
\end{align} 
and not the affine Killing condition \eqref{eq:aff_Killing_cond}. Then, according to the discussion in section \ref{sec:symmetries} one obtains infinitely many symmetries and thus infinitely many conservation laws. Indeed, using the diffeomorphism Ward identity one shows that 
\begin{align}\label{eq:curr_cons}
    \partial _\mu j^\mu &= \mathfrak{e}T^\mu \mathcal{L}_\xi \tau _\mu +\frac{\mathfrak{e}}{2}T^{\mu \rho }\mathcal{L}_\xi h_{\mu \rho } \\
    &=\mathfrak{e}T^\alpha \tau _\alpha \tau _\mu \mathcal{L}_\xi v^\mu+\frac{\mathfrak{e}}{2}T^{\mu \rho }\mathcal{L}_\xi h_{\mu \rho } 
\end{align}
where we used the Carroll boost Ward identity in the second line. Thus, on account of \eqref{eq:ahhhh} we obtain $\partial _\mu j^\mu =0$. Choosing a codimension$-1$ spatial hypersurface $\Sigma $ with normal one-form $n^\mu $, one can define a charge by 
\begin{align}\label{eq:codim1_charge}
    Q(\Sigma )=\int _\Sigma \dd ^d x \,n_\mu j^\mu \big \vert _\Sigma 
\end{align}
and show that for two such hypersurfaces the change in charge is given by the flux across the spatial boundary of the region enclosed by the surfaces,
\begin{align}
    Q(\Sigma ')-Q(\Sigma )=-\int _\mathcal{B}d^dy\, b_\mu j^\mu 
\end{align}
where $b_\mu $ is the outward-pointing normal one-form to $\mathcal{B}$. Therefore, if we interpret $n_\mu $ as giving a time direction one finds charge conservation if the flux falls off sufficiently fast. 

Let us look at this scenario for the simplest background, i.e., the flat Carroll manifold as obtained from taking the limit of Minkowski space (see section \ref{sec:2.2}). We have (renaming $t=x^0$)
\begin{align}\label{eq:flat_background}
    v&=-\partial _0 & h_{\mu \nu }&=\delta _{ij}\delta ^i_\mu \delta ^j _\nu & \tau &=\dd x^0 
\end{align}
and the Carroll Killing fields preserving this structure are given by
\begin{align}\label{eq:Carr:kv1}
    \xi _{\mathrm{\scriptscriptstyle M}_f} &=f(x^i)\partial _0 &
    \xi _{\mathrm{\scriptscriptstyle P_i}} & =\partial _i & \xi _{\mathrm{\scriptscriptstyle J_{ij}}} &=x_i\partial _j-x_j \partial _i ~,
\end{align}
closing into the infinite lift $\mathfrak{isocarr}(d+1)$ of the Carroll algebra given by setting $\sigma =0$ in \eqref{eq:isocarrlamb}. Picking an $x^0=\text{cst.}$ surface with $n=\dd x^0$ one arrives at the corresponding charges
\begin{align}\label{eq:carr_charges}
    M_f&=\int \dd ^dx\, \mathbf{T}^0{}_0f(x^i) & P_i&=\int \dd ^dx\, \mathbf{T}^0{}_i & J_{ij}&=\int \dd ^dx\, (x_i\mathbf{T}^0{}_j-x_j\mathbf{T}^0{}_i) ~.
\end{align}
They denote the supertranslation charge $M_f$, spatial momentum charge $P_i$ and angular momentum charge $J_{ij}$. The special cases $M_{f=-x_i}$ and $M_{f=1}$ are the charges for Carroll boosts $B_i$ and energy $H$, respectively. Since the Carroll boost Ward identity implies vanishing energy flux, $\mathbf{T}^i{}_0=0$, one finds that on a fixed time slice $\Sigma $
\begin{align}
    \frac{\dd }{\dd x^0} M_f &= 0 & \frac{\dd }{\dd x^0}P_i&=-\int _{\partial \Sigma }\dd \sigma _k \mathbf{T}^k{}_i & \frac{\dd }{\dd x^0}J_{ij}&=-2\int _{\partial \Sigma }\dd \sigma _k x_{[i}\mathbf{T}^k{}_{j]} 
\end{align}
where $\partial \Sigma =\Sigma \cap \mathcal{B}$. This shows that the total supertranslation charge, and in particular the energy, is conserved for any a given region of spacetime while linear and angular momentum charges can in principle be changed by flux through the boundary. This is the field theory avatar of saying that massive bodies are immobile in Carroll theories (see also section \ref{sec:lim_poincare}) - no energy flux is possible. We shall see in subsection \ref{sec:coupl_conn} that this statement can change once the theory couples to the connection of the background.  

\subsubsection{Scale invariant case}
There are examples, where the set of Carroll symmetries just discussed is extended by a $z$-dependent scale symmetry. The classical covariantly coupled action is then additionally invariant under constant $z$-dependent Weyl rescalings defined in section \ref{sec:Weyl}. One may construct an associated conserved current in the following way. First, covariance of the action implies that for an arbitrary, non-constant Weyl parameter we have
\begin{align}
    \delta _\rho I_m=\int \dd ^{d+1}x\, \mathfrak{e}\,  \mathcal K^\mu \partial _\mu \rho 
\end{align}
where $\mathcal K^\mu $ is some covariant expression that depends on the dynamical and background fields. Second, an on-shell Weyl-variation can always be written as 
\begin{align}
    \delta _\rho I_m=\int \dd ^{d+1}x\, \mathfrak{e}\big(zT^\mu \tau _\mu +T^{\mu \nu }h_{\mu \nu }\big)\rho 
\end{align}
where we assumed that $\rho $ falls of sufficiently fast such that boundary terms can be dropped. Putting these expressions together yields 
\eq{
\boxed{
\phantom{\Bigg(}
   zT^\mu \tau _\mu +T^{\mu \nu }h_{\mu \nu }=-\mathfrak{e}^{-1}\partial _\mu \big(\mathfrak{e} \mathcal K^\mu ) 
\phantom{\Bigg)}
    }
}{eq:scale_identity}
setting the generalized trace of the stress-energy tensor to a total derivative. For a generator of anisotropic scale transformations $\xi _{\textrm{D}}$ satisfying $\mathcal{L}_{\xi _{\textrm{D}}}h_{\mu \nu }=2h_{\mu \nu }$, $\mathcal{L}_{\xi _{\textrm{D}}}v^\mu =-zv^\mu $, one may then identify the associated on-shell conserved current density by 
\begin{align}
    j_D^\mu =\mathfrak{e}\mathbf{T}^\mu {}_\nu \xi ^\nu _{\textrm{D}}+\mathfrak{e}\mathcal K^\mu ~.
\end{align}
where conservation is shown by using the diffeomorphism Ward identity \eqref{eq:diff_0.0}. For a flat background, these scale transformations are just generated by 
\begin{align}
    \xi _{\textrm{D}}=zx^0\partial _0 +x^i\partial _i ~.
\end{align}

From this discussion it becomes evident that scale invariance in general does not imply a vanishing trace of the stress-energy tensor, even for the case $z=1$. It might, however, happen for some theories that the term $\mathcal{K}^\mu $ can be absorbed into the definition of the stress-energy tensor by an improvement
\begin{align}
    \mathbf{T}^\mu {}_\nu \to \mathbf{T}^\mu {}_\nu +\partial _\rho B^{\rho \mu }{}_\nu 
\end{align}
where the tensor $B^{\rho \mu }{}_\nu $ has to satisfy
\begin{align}
    B^{(\rho \mu )}{}_\nu =0 && \partial _\rho B^{\rho \mu }{}_\nu v^\nu h_{\mu \sigma }=0
\end{align}
to ensure stress-energy tensor conservation and local Carroll boost invariance. If this improvement renders the stress-energy tensor traceless the theory is not just scale invariant any more but rather conformally invariant, as explained in the next point. We have not pursued this direction further but it could be worth to explore and possibly show some results like \cite{Polchinski:1987dy,Callan:1970ze} for the Carroll case.

\subsubsection{Conformal case}
If the field theory is additionally conformally coupled to the background, it should be invariant under local Weyl rescalings of secetion \ref{sec:Weyl} with anisotropy exponent $z$. This makes the quantity $\mathcal{K}^\mu $ of the previous point vanish identically. Then, in addition to the Ward identities \eqref{eq:ward_identities_covariant} one obtains a generalized trace identity 
\eq{
\boxed{
\phantom{\Bigg(}
   zT^\mu \tau _\mu +T^{\mu \nu }h_{\mu \nu }=0 
\phantom{\Bigg)}
    }
}{eq:z-trace identity}
which reduces to $\mathbf{T}^\mu {}_\mu =0$ for the isotropic case $z=1$.
In this case, depending on the particular background, one may obtain an even larger class of symmetries and charges by solving the conformal Carroll Killing equations \eqref{eq:CKeqs_z} (see also section \ref{sec:conformal_extensions}). For the flat background \eqref{eq:flat_background}, the solutions are 
\begin{align}\label{eq:CCKVs}
    \xi =Y^i(x^j)\partial _i +\Big(f(x^i)+\frac{zx^0}{d}\partial _k Y^k\Big)\partial _0 && \partial _{(i}Y_{j)}=\frac{1}{d}\partial _k Y^k \delta _{ij}
\end{align}
i.e., $Y^i$ needs to be a conformal Killing vector of the Euclidean transverse plane. Besides the choices \eqref{eq:Carr:kv1}, these symmetries contain dilations and special conformal transformations represented by the vectors \eqref{eq:D_SCT}. One may again construct conserved currents associated to these charges by $j^\mu =\mathfrak{e}\mathbf{T}^\mu {}_\nu \xi ^\nu $. Their conservation can be readily checked by using \eqref{eq:curr_cons} together with the trace Ward identity \eqref{eq:z-trace identity}. 

Any theory that admits a covariant formulation for an arbitrary background such that Weyl and diffeomorphism invariance are manifest will automatically descend to a Carroll conformal field theory (CCFT) on a flat background. However, in general this is a stronger condition than necessary for a CCFT to exist on a flat background. A counterexample is the Carroll version of Liouville theory \cite{1210.0731} which is not Weyl invariant on a curved background but still descends to a true CCFT once the background is flat.

\subsection{Coupling to the connection}
\label{sec:coupl_conn}
One may consider a generalization of the couplings allowed in \eqref{var} and write
\begin{align}\label{var2}
    \delta I_m[\phi ,h_{\mu \nu },\tau _{\mu },C_{\mu \nu }]=\int \dd ^{d+1}x\, \mathfrak{e}\Big[\frac12 T^{\mu \nu }\delta h_{\mu \nu }+T^\mu \delta \tau _\mu +\frac12 \Psi ^{\mu \nu }\delta C_{\mu \nu }+(\text{EOM})_\phi \,\delta \phi \Big] 
\end{align}
where $C_{\mu \nu }$ is additional free data in the Carroll compatible connection that describes the background geometry. Since this is a transverse symmetric tensor there is no loss of generality in assuming the same properties for the corresponding response function $\Psi ^{\mu \nu }$. The field $C_{\mu \nu }$ transforms non-trivially both under diffeomorphisms and local Carroll boosts, which makes the Ward identities pick up some additional terms. For local Carroll boosts we use the transformation \eqref{eq:c_minimal_trafo} and find that the requirement $\delta _\lambda I_m=0$ leads to the Carroll boost Ward identity
\begin{align}\label{eq:boost_WI_general}
    \Big(T^\mu +\frac{1}{2}\big(\nabla _\nu -a_\nu \big)\Psi ^{\mu \nu }\Big)h_{\mu \rho }=0 
\end{align}
where we imposed the equations of motion for the matter field. Thus, due to the coupling to the connection the energy flux component acquires a non-zero value that is controlled by the response to the connection. Similarly, invariance under diffeomorphisms leads to
\begin{equation}\label{eq:diff_WI_general}
\frac{1}{\mathfrak{e}}\partial_{\mu} \big(  \mathfrak{e} \, T^\mu \tau _\alpha +\mathfrak{e}T^{\mu \rho }h_{\rho \alpha } + \mathfrak{e} \, \Psi^{\mu \nu} C_{\nu \alpha}\big) = T^\mu \, \partial_\alpha \tau_\mu + \frac{1}{2} T^{\mu \nu} \partial_\alpha h_{\mu \nu} + \frac{1}{2} \Psi^{\mu \nu} \partial_\alpha C_{\mu \nu} 
\end{equation}
which can also be rewritten covariantly and in terms of $\mathbf{T}^\mu {}_\alpha $ as
\begin{align}
    \nabla _\mu \big(\mathbf{T}^\mu {}_\nu +\Psi ^{\mu \alpha }C_{\alpha \nu }\big)=2\big(&\mathbf{T}^\mu {}_\alpha +\Psi ^{\mu \beta }C_{\beta \alpha }\big)h^{\alpha \rho }K_{\rho [\mu }\tau _{\nu ]}-K\tau _\mu \mathbf{T}^\mu {}_\nu \\
    &\qquad +\frac12 \Big(\Psi ^{\mu \alpha }\nabla _\nu C_{\mu \alpha }-(\nabla _\nu \tau _\mu )(\nabla _\rho -a_\rho )\Psi ^{\rho \mu }\Big) ~.\nonumber 
\end{align}
These generalized Ward identities may again be used to construct conservation laws once the theory is put on a symmetric background. By definition, a symmetry of the background should leave all the geometric data, including the independent part of the connection invariant. We saw the general condition written in terms of Carroll boost invariant quantities in \eqref{eq:Carr_Killing}. In the present case, it is more convenient to phrase the Carroll Killing conditions in terms of the background variables the theory is chosen to couple to. They read
\begin{align}\label{eq:alt_Carr_Killing}
    \delta_{\xi}\tau _\mu =0&& \delta_{\xi }h_{\mu \nu } =0&& \delta_{\xi}C_{\mu \nu }=0
\end{align} 
where the variations are given by $\delta _{\xi}=\mathcal{L}_\xi +\delta _{\lambda (\xi)}$ and include a local Carroll boost variation that depends on $\xi ^\mu$. This is necessary to take the gauge dependence of $\tau _\mu $ and $C_{\mu \nu }$ into account where $\delta _\lambda C_{\mu \nu }$ is given in \eqref{eq:c_minimal_trafo}. One can show that on the space of minimal torsion geometries these conditions are equivalent to \eqref{eq:Carr_Killing}. Consider a current density given by 
\begin{align}
    j^\mu =\mathfrak{e}\mathbf{T}^\mu {}_\nu \xi ^\nu +\mathfrak{e}\Psi ^{\mu \rho }C_{\rho \nu }\xi ^\nu +\frac{\mathfrak{e}}{2} \Psi ^{\mu \nu }\lambda _\nu (\xi ) ~.
\end{align}
Upon inserting the Ward identities \eqref{eq:boost_WI_general} and \eqref{eq:diff_WI_general} one can show that 
\begin{align}\label{eq:gen_cons}
    \partial _\mu j^\mu =\mathfrak{e}T^\mu \delta _\xi \tau _\mu +\frac{\mathfrak{e}}{2}T^{\mu \nu }\delta _\xi h_{\mu \nu }+\frac{\mathfrak{e}}{2}\Psi ^{\mu \nu }\delta _\xi C_{\mu \nu }
\end{align}
and thus obtains conservation once \eqref{eq:alt_Carr_Killing} is satisfied. In the maximally symmetric case, we know that the local solutions to these conditions form the Carroll algebra (or its $\Lambda$-deformation) under the Lie bracket. Therefore, we again obtain a conserved charge for each generator of the Carroll algebra. 

This framework of coupling a Carroll field theory to the connection has recently been used to perform a holographic derivation of the flux balance laws associated to asymptotically flat Einstein gravity in four dimensions \cite{Fiorucci:2025twa}. While the authors extended the available background symmetries further to also include local Weyl rescalings, the basic mechanism can already be spotted in this more simple scenario. First, we rewrite equation \eqref{eq:gen_cons} such that the left hand side is just built by the responses to the metric fields, 
\begin{align}\label{eq:further_eq}
    \partial _\mu(\mathfrak{e}\mathbf{T}^{\mu }{}_\nu \xi ^\nu )=\mathfrak{e}T^\mu \delta _\xi \tau _\mu +\frac{\mathfrak{e}}{2}T^{\mu \nu }\delta _\xi h_{\mu \nu }+\text{(flux)}
\end{align}
where we absorbed all terms that are proportional to $\Psi ^{\mu \nu }$ into a flux-term. In a holographic setting one then chooses boundary conditions that fix some geometric properties of the boundary manifold. If one demands that the Carroll metric data is fixed completely one arrives at $\delta _\xi \tau _\mu =0$, $\delta _\xi h_{\mu \nu }=0$, which yields the vector fields of the supertranslation-extended Carroll algebra\footnote{In the conformal case superrotations are also included.} for a flat background. Equation \eqref{eq:further_eq} then reduces to a relation of the type
\begin{align}
    \partial _\mu j^\mu = \text{(flux)}
\end{align}
which is a flux-balance law where the right hand side consists of the $\Psi ^{\mu \nu }$ dependent terms. In a holographic setting, these terms consist of the news tensor which describes the gravitational radiation passing through the boundary. This description still misses the conformal aspect of the symmetries to describe the actual situation for asymptotically flat spacetimes, but we wish to point out the difference to the philosophy we used so far: While the invariance of $C_{\mu \nu }$ was a requirement for having a true variational symmetry of a matter field coupled to this background this requirement is dropped in a holographic setting. This renders the conservation laws into controlled non-conservation laws. They capture the leakiness of asymptotically flat spacetimes (see, e.g., \cite{Fiorucci:2021pha} and refs therein). In other words, a Carroll theory holographically dual to asymptotically flat four-dimensional Einstein gravity should be sourced by the data corresponding to gravitational radiation \cite{Troessaert:2015nia,Donnay:2022aba,Donnay:2022wvx}.  

\section{Commutators and universal relations}\label{sec:Quantum}
We saw how symmetries of Carroll field theories represented by Carroll (conformal) Killing vectors lead to conservation laws and flux balance relations. As in quantum field theory, we expect these symmetries to be realized on the level of the codimension-1 charges \eqref{eq:codim1_charge} which, however, requires knowing the brackets between the various stress-energy tensor components that are involved. In this subsection, we shall present a way to obtain these brackets from the assumptions of covariance and locality, based on the upcoming work \cite{Dutta:dummy}. This reveals a certain universal structure appearing in the stress-energy tensor brackets which is intimately tied to the Carroll nature of the field theory. In the Lorentzian case, the result of a universal part in the stress-energy tensor commutators goes back to the seminal works by Dirac \cite{Dirac1962RelativisticQFT} and Schwinger \cite{Schwinger1963CRCL}. Here, we shall use the slightly more modern formulation by Boulware and Deser \cite{Deser:1967zzf}. 

Given a matter theory formulated in terms of a local action $I_m$ on a fixed Carroll background, we assume that its correlation functions can be extracted from a generating functional 
\begin{align}
    Z[\tau _\mu ,h_{\mu \nu }]=\int \mathcal{D}\phi \,e^{iI_m} =e^{iW}
\end{align}
where $\phi $ again should be seen as a dummy for any possible field content and $W$ is the generating functional of connected correlation functions denoted by $\langle ...\rangle _c$ in the following. Given some local operator $\mathcal{O}(x)$, we may write, e.g., a two-point function as 
\begin{align}
    \langle \mathcal{T}\mathcal{O}(x)\mathcal{O}(y)\rangle = \frac{1}{Z}\int \mathcal{D}\phi \, \mathcal{O}(x)\mathcal{O}(y)e^{iI_m}
\end{align}
where time-ordering $\mathcal{T}$ is defined in the usual way
\begin{align}
    \mathcal{T}(\mathcal{O}(x)\mathcal{O} (y))=\Theta (x^0-y^0)\mathcal{O}(x)\mathcal{O} (y)+\Theta (y^0-x^0)\mathcal{O} (y)\mathcal{O}(x) ~.
\end{align}
As in the previous subsection, we assume here that no coupling to the connection degrees of freedom $C_{\mu \nu }$ exists. Since we are mostly interested in the representation of spacetime symmetries, we will evaluate expressions on a symmetric Carroll geometry. In the present case, this will just be the flat Carroll spacetime \eqref{eq:flat_background} but in principle one could also choose some other symmetric manifold such as a homogeneous spacetime. In any case, this evaluation always has to be performed at the very end of each calculation. 
For $\mathcal{O}$ given by stress-energy tensor components the correlation functions are generated by a variation of the background,
\begin{align}
    \langle T^\mu (x)\rangle &= \frac{1}{\mathfrak{e}}\frac{\delta W}{\delta \tau _\mu (x)} & \langle T^{\mu \nu }(x)\rangle &=\frac{2}{\mathfrak{e}}\frac{\delta W}{\delta h_{\mu \nu } (x)} 
\end{align}
and likewise, for obtaining a two-point function one computes the second variation,
\begin{align}
\frac{1}{\mathfrak{e}}\frac{\delta }{\delta \tau _\mu (y)}\frac{1}{\mathfrak{e}}\frac{\delta }{\delta \tau _\nu (x)}W&=i\langle \mathcal{T}T^\mu (x)T^\nu (y)\rangle -i\langle T^\mu (x)\rangle \langle T^\nu (y)\rangle+\frac{1}{\mathfrak{e}(y)}\Big\langle \frac{\delta T^\mu (x)}{\delta \tau _\nu (y)}\Big\rangle\\
&=i\langle \mathcal{T}T^\mu (x)T^\nu (y)\rangle _c +\frac{1}{\mathfrak{e}(y)}\Big\langle \frac{\delta T^\mu (x)}{\delta \tau _\nu (y)}\Big\rangle ~.\label{eq:TT_conn2pt}
\end{align}
This expression splits into a bilocal piece given by the connected two-point function as well as a local piece that schematically can be written as
\begin{align}
    \frac{1}{\mathfrak{e}(y)}\Big\langle \frac{\delta T^\mu (x)}{\delta \tau _\nu (y)}\Big\rangle = \mathcal{D}^{\mu \nu }_x\delta ^{d+1}(x-y)
\end{align}
and includes only contact terms. $\mathcal{D}^{\mu \nu }_x$ is some derivative operator with $c$-number coefficients. The origin of this term lies in the variation of the measure as well as the variation of the stress-energy tensor component with respect to the background. As we shall see, it is precisely terms of this form that contain universal information about the Carroll nature of these field theories. One may work out analogous expressions for the other second variations but since we only want to set up the formalism here we shall leave it at this one.  

In the Lorentzian case, the generating functional $W$ is typically ill-defined and needs some renormalization procedure. This comes with the usual ambiguities of adding further counterterms to $W$ that do not change the finiteness properties but may change the structure of the contact terms appearing on the right hand side of \eqref{eq:TT_conn2pt}. For obtaining consistent equal-time commutators between stress-energy tensor components one has to partially fix these terms which in the Lorentzian case goes back to the work by Boulware and Deser \cite{Deser:1967zzf}. The important point is that this always has to be possible in a relativistic quantum field theory as otherwise there would be no consistent way of realizing the classical symmetries on the Hilbert space of the quantum theory. We shall see that something analogous can be done for Carroll quantum field theories that admit a path integral formulation as put forward in this section. 

It may happen that some classical symmetries are violated in the quantum theory, i.e., the theory exhibits an anomaly. This is a consequence of the path integral measure being not invariant under the respective symmetry \cite{Fujikawa:1979ay} and there being no possibility of adding counterterms to remedy this. We shall see in an explicit example how such anomalous contributions are tied to central extensions appearing in the representation of the symmetries.

\subsection{Ward identities revisited}
Requiring that the generating functional is consistent with diffeomorphism invariance of the action $I_m[\phi ,\tau _\mu ,h_{\mu \nu }]$, we can infer some differential relations between one-point functions. This is easy to see by noting that a change of variables $\phi \to \phi '$ leaves $Z$ invariant. Choosing $\phi '=\phi +\delta _\xi \phi $, where $\delta _\xi \phi =\mathcal{L}_\xi \phi $ denotes the action of a diffeomorphism we obtain infinitesimally
\begin{align}
    0=\int \mathcal{D}\phi \, \Big(i\int \dd ^{d+1}x\, \frac{\delta I_m}{\delta \phi (x)}\delta _\xi \phi (x)\Big)\,e^{iI_m}
\end{align}
where it was assumed that the path integral measure is invariant under this change of variables, i.e., $\mathcal{D}\phi =\mathcal{D}\phi '$. Since the matter action is covariantly coupled, we can write
\begin{align}
    \int \dd ^{d+1}x\, \frac{\delta I_m}{\delta \phi (x)}\delta _\xi \phi (x)&=-\int \dd ^{d+1}x\, \frac{\delta I_m}{\delta \tau _\mu (x)}\delta _\xi \tau_\mu (x)-\int \dd ^{d+1}x\, \frac{\delta I_m}{\delta h_{\mu \nu } (x)}\delta _\xi h_{\mu \nu } (x)\\
    &=-\int \dd ^{d+1}x\,\mathfrak{e} \Big(T^\mu (x)\delta _\xi \tau _\mu (x)+\frac12 T^{\mu \nu }(x)\delta _\xi h_{\mu \nu }(x)\Big)\label{eq:secondlineaha}
\end{align}
where we used the definition of the stress-energy tensor components \eqref{var}. Assuming that $\xi $ is chosen such that no boundary terms contribute, we can partially integrate and obtain the diffeomorphism Ward identity 
\begin{align}\label{eq:quant_WI_diff}
    \partial_{\mu} \big( \mathfrak{e} \, \langle T^\mu \rangle\tau _\alpha +\mathfrak{e}\langle T^{\mu \rho }\rangle h_{\rho \alpha } \big) = \mathfrak{e}\langle T^\mu \rangle\, \partial_\alpha \tau_\mu + \frac{1}{2} \mathfrak{e}\langle T^{\mu \nu} \rangle\partial_\alpha h_{\mu \nu} ~.
\end{align}
At this point, this holds for an arbitrary background configuration. From \eqref{eq:secondlineaha} it also becomes clear that this statement is equivalent to $W$ being invariant under a diffeomorphism variation of the background. 

One may proceed in a similar fashion for the Ward identity associated to local Carroll boosts. However, we take into account the option of an anomaly for these transformations. Upon performing an infinitesimal change of variables $\phi '= \phi + \delta _\lambda \phi $ one obtains 
\begin{align}
    0=\int \mathcal{D}\phi \, \Big(-i\int \dd ^{d+1}x \, \mathfrak{e}\, \mathcal{A}^\mu \lambda _\mu +i\int \dd ^{d+1}x\, \frac{\delta I_m}{\delta \phi (x)}\delta _\lambda \phi (x)\Big)\,e^{iI_m}
\end{align} 
where $\mathcal{A}^\mu $ is some local expression that captures the contribution from the anomaly, i.e., 
\begin{align}
    \mathcal{D}\phi '=\mathcal{D}\phi \,\exp \Big({-i\int \dd ^{d+1}x\,\mathfrak{e}\,\mathcal{A}^\mu \lambda _\mu }\Big) ~.
\end{align}
From this we obtain the anomalous Carroll boost Ward identity
\begin{align}\label{eq:boost_an_WI}
    \langle T^\mu (x)\rangle h_{\mu \nu }= \mathcal{A}^\mu (x) h_{\mu \nu } ~.
\end{align}
It is a fair question to ask in which sense such an anomaly is physical since given our presentation of local Carroll boosts so far it is associated to a gauge symmetry of the background. Indeed, when such theories are coupled to Carroll gravity this leads to an inconsistency. However, in a holographic context where the dual boundary theory does not include gravity and Carroll boosts act as global symmetries it is not inconsistent per se to have anomalous Carroll boosts. In fact, this situation is even expected \cite{Campoleoni:2022wmf,Baulieu:2025itt,Hartong:2025jpp}.

\subsection{Equal-time commutators of a Carroll stress-energy tensor}
The commutator structure of a local quantum field theory is intimately tied to the contact terms appearing in two-point functions through the time ordering prescription. This can be observed explicitly by taking a background variation of the diffeomorphism Ward identity \eqref{eq:quant_WI_diff}. Similarly to \eqref{eq:TT_conn2pt}, this produces contact terms along a two-point function. The variations with respect to $\tau _\nu (y)$ and $h_{\nu \rho }(y)$ read
\begin{align}
   & i\partial _\mu \big\langle\mathcal{T}\big(\mathfrak{e}\mathbf{T}^\mu {}_\alpha (x)\big)\mathfrak{e}T^\nu (y) \big\rangle -i \big\langle\mathcal{T}\partial _\mu\big(\mathfrak{e}\mathbf{T}^\mu {}_\alpha (x)\big)\mathfrak{e}T^\nu (y) \big\rangle =\\
   &\qquad \qquad \mathfrak{e}\langle T^\nu (x)\rangle \partial _\alpha \delta ^{d+1}(x-y)-\partial _\mu \Big\langle \frac{\delta (\mathfrak{e}\mathbf{T}^\mu {}_\alpha (x))}{\delta \tau _\nu (y)}\Big\rangle + (\text{curv.}) \nonumber 
\end{align}
\begin{align}
   & i\partial _\mu \big\langle\mathcal{T}\big(\mathfrak{e}\mathbf{T}^\mu {}_\alpha (x)\big)\mathfrak{e}T^{\nu \rho } (y) \big\rangle -i \big\langle\mathcal{T}\partial _\mu\big(\mathfrak{e}\mathbf{T}^\mu {}_\alpha (x)\big)\mathfrak{e}T^{\nu \rho } (y) \big\rangle =\\
   &\qquad \qquad \mathfrak{e}\langle T^{\nu \rho } (x)\rangle \partial _\alpha \delta ^{d+1}(x-y)-2\partial _\mu \Big\langle \frac{\delta (\mathfrak{e}\mathbf{T}^\mu {}_\alpha (x))}{\delta h _{\nu \rho } (y)}\Big\rangle + (\text{curv.}) \nonumber
\end{align}
where we used that the diffeomorphism Ward identity holds as an operator equation and absorbed all terms that vanish once the background is chosen to be flat on the right hand sides. The left hand sides are rewritten in terms of commutators by letting the derivatives act on the time ordering. For the first equation this yields, e.g.,    
\begin{align}
    i\partial _\mu \big\langle\mathcal{T}\big(\mathfrak{e}\mathbf{T}^\mu {}_\alpha (x)\big)\mathfrak{e}T^\nu (y) \big\rangle -i \big\langle\mathcal{T}\partial _\mu\big(\mathfrak{e}&\mathbf{T}^\mu {}_\alpha (x)\big)\mathfrak{e}T^\nu (y) \big\rangle=\\[.5em]
    &i\delta _\mu ^0\big\langle[\mathfrak{e}\mathbf{T}^\mu {}_\alpha (x),\mathfrak{e}T^\nu (y)] \big\rangle \delta (x^0-y^0) \nonumber
\end{align}
with an analogous expression for the second equality. Evaluating on the flat background \eqref{eq:flat_background}, we therefore obtain commutators between the stress-energy tensor components
\begin{align}
    i\delta _\mu ^0\big\langle[\mathbf{T}^\mu {}_\alpha (x),T^\nu (y)] \big\rangle \delta (x^0-y^0)&=\langle T^\nu (x)\rangle \partial _\alpha \delta ^{d+1}(x-y)-\partial _\mu \Big\langle \frac{\delta (\mathfrak{e}\mathbf{T}^\mu {}_\alpha (x))}{\delta \tau _\nu (y)}\Big\rangle \Big \vert _{\text{flat }} \label{eq:uhu1}\\
    i\delta _\mu ^0\big\langle[\mathbf{T}^\mu {}_\alpha (x),T^{\nu \rho } (y)] \big\rangle \delta (x^0-y^0)&=\langle T^{\nu \rho } (x)\rangle \partial _\alpha \delta ^{d+1}(x-y)-2\partial _\mu \Big\langle \frac{\delta (\mathfrak{e}\mathbf{T}^\mu {}_\alpha (x))}{\delta h _{\nu \rho } (y)}\Big\rangle \Big \vert _{\text{flat }}\label{eq:uhu2}
\end{align}
with the left hand sides supported only on equal time slices. To isolate the corresponding pieces on the right hand sides we shall eventually integrate over a small interval $x^0\in [y^0-\epsilon,y^0+\epsilon]$ and take the limit $\epsilon \to 0$. This effectively picks out only the contact contributions at equal times which form the commutator and also removes total time derivatives of contact terms since, e.g.,
\begin{align}
    \int _{y^0-\epsilon}^{y^0+\epsilon}\dd x^0\, \partial _0(\mathcal{O}(x)\delta ^{d+1}(x-y))=(\mathcal{O}(x)\delta ^{d+1}(x-y))\big \vert  _{y^0-\epsilon}^{y^0+\epsilon}=0 ~.
\end{align}
For proceeding, it is convenient to combine the two equations \eqref{eq:uhu1} and \eqref{eq:uhu2} into a single one,
\begin{align}
    i\delta _\mu ^0\big\langle[\mathbf{T}^\mu {}_\alpha (x),\mathbf{T}^\nu {}_\beta (y)] \big\rangle &\delta (x^0-y^0)=\langle \mathbf{T}^\nu {}_\beta(x)\rangle \partial _\alpha \delta ^{d+1}(x-y) \label{eq:almostdone}\\
    &-\partial _\mu \Big\langle \frac{\delta (\mathfrak{e}\mathbf{T}^\mu {}_\alpha (x))}{\delta \tau _\nu (y)}\Big\rangle \Big \vert _{\text{flat }} \delta _\beta ^0-2\partial _\mu \Big\langle \frac{\delta (\mathfrak{e}\mathbf{T}^\mu {}_\alpha (x))}{\delta h _{\nu \rho } (y)}\Big\rangle \Big \vert _{\text{flat }} \delta _\rho ^i\delta _\beta ^j \delta _{ij} \nonumber ~.
\end{align}
The second line of this expression is still somewhat impractical for evaluating the commutators. One may simplify it further by noting that the variations of the boost Ward identity \eqref{eq:boost_an_WI} with respect to the background geometry is
\begin{align}
    \Big\langle \frac{\delta T^\mu (x)}{\delta \tau _\alpha (y)}\Big\rangle h_{\mu \nu }&=\frac{\delta \mathcal{A}^\mu (x)}{\delta \tau _\alpha (y)}h_{\mu \nu }\label{eq:zoo1}\\
    \Big\langle \frac{\delta T^\mu (x)}{\delta h _{\alpha \beta}(y)}\Big\rangle h_{\mu \nu }&=\frac{\delta \mathcal{A}^\mu (x)}{\delta h _{\alpha \beta} (y)}h_{\mu \nu }+\big(\mathcal{A}^\mu (x) -\langle T^\mu (x)\rangle \big)\delta _{(\mu }^\alpha \delta _{\nu )}^\beta \delta ^{d+1}(x-y)\label{eq:zoo2}
\end{align}
where we dropped all non-contact contributions by means of the time integration prescription. The anomaly $\mathcal{A}$ typically vanishes when evaluated on a flat background which we shall assume here as well. However, its variations in general do not vanish. We may insert equations \eqref{eq:zoo1} and \eqref{eq:zoo2} in \eqref{eq:almostdone} to show that the brackets between the stress-energy tensor components that are relevant for the charge algebra are
\begin{align}
     i\big\langle [\mathbf{T}^0{}_0 (x)&,\mathbf{T}^0{}_0 (y)]\big\rangle=-\delta ^i_\mu \delta _\nu ^0\, \partial _i \frac{\delta \mathcal{A}^\mu (x)}{\delta \tau _\nu (y)} \Big \vert _{\textrm{flat}}\nonumber\\
    i\big\langle[\mathbf{T}^0{}_0 (x)&,\mathbf{T}^0{}_i (y)]\big\rangle= \big(\langle\mathbf{T}^0 {}_0 (y)\rangle \delta^k_i -\langle\mathbf{T}^k {}_i (x)\rangle\big) \partial _k \delta ^{d}(x-y) -2\delta _{\rho i}\delta _\nu ^0\delta _\mu ^j \partial _j\frac{\delta \mathcal{A}^\mu (x)}{\delta h_{\nu \rho }(y)}\Big \vert _{\textrm{flat}}\label{SET}\\
    i\big\langle[\mathbf{T}^0{}_i (x)&,\mathbf{T}^0{}_j (y)]\big\rangle=   \langle\mathbf{T}^0 {}_j (x)\rangle \partial_i  \delta ^{d}(x-y) -\delta _{ij}\partial _k\big(\delta ^{km}\langle\mathbf{T}^0{}_m(x)\rangle\delta ^{d}(x-y)\big) \nonumber\\[.5em]
    &\hspace{2cm}+\partial _k \langle\mathcal{S}^k {}_i {}^{0 }{}_j(x,y)\rangle \nonumber
\end{align}
where all the expressions are evaluated at $x^0=y^0$ through the above-mentioned integration prescription. The operators
\begin{align}
    \mathcal{S}^k {}_i {}^{0 }{}_j(x,y)=-2\frac{\delta T^{\mu \sigma }(x)}{\delta h_{\nu \rho }(y)}\delta _\mu ^k\delta _\nu ^0\delta _{\rho i}\delta _{\sigma j}
\end{align}
denote a part that is not uniquely fixed by symmetry and depends on the theory at hand. However, it still contains a universal part that is common to all theories. To see this we invoke antisymmetry of the last bracket which leads to the condition 
\begin{align}
    -\langle \mathbf{T}^0{}_j(x)\rangle &\partial _i^y\delta ^{d}(x-y)-\delta _{ij}\partial _k\big(\delta ^{km}\langle\mathbf{T}^0{}_m(x)\rangle\delta ^{d}(x-y)\big) +\partial _k \langle\mathcal{S}^k {}_i {}^{0 }{}_j(x,y)\rangle \label{eq:cons_cond}\\[.3em]
    &\overset{!}{=}\langle \mathbf{T}^0{}_i(y)\rangle \partial _j\delta ^{d}(x-y)+\delta _{ij}\partial _k^y\big(\delta ^{km}\langle\mathbf{T}^0{}_m(y)\rangle\delta ^{d}(x-y)\big) -\partial _k^y \langle\mathcal{S}^k {}_j {}^{0 }{}_i(y,x)\rangle \nonumber ~.
\end{align}
This shows that the operator $\mathcal{S}^k {}_i {}^{0 }{}_k$ is tied to the consistency of the bracket itself and may not be chosen at will without destroying the symmetry of the brackets. We may isolate the part associated to symmetry by writing the condition \eqref{eq:cons_cond} as
\begin{align}
    \partial _k\Big(\langle\mathcal{S}^k {}_i {}^{0 }{}_k(x,y)\rangle&-\delta _{ij}\big(\delta ^{km}\langle\mathbf{T}^0{}_m(x)\rangle\delta ^{d}(x-y)\big)-\langle \mathbf{T}^0{}_i(y)\rangle \delta ^k_j\delta ^{d}(x-y)\Big) \\
    &\overset{!}{=}-\partial _k^y \Big(\langle\mathcal{S}^k {}_j {}^{0 }{}_i(y,x)\rangle -\delta _{ij}\big(\delta ^{km}\langle\mathbf{T}^0{}_m(y)\rangle\delta ^{d}(x-y)\big)-\langle \mathbf{T}^0{}_j(x)\rangle \delta ^k_i\delta ^{d}(x-y)\Big) \nonumber ~.
\end{align}
Thus, defining the operators
\begin{align}
    t^k {}_i {}^{0 }{}_j(x,y):=\mathcal{S}^k {}_i {}^{0 }{}_j(x,y)-\delta _{ij}\big(\delta ^{km}\mathbf{T}^0{}_m(x)\delta ^{d}(x-y)\big)- \mathbf{T}^0{}_i(y) \delta ^k_j\delta ^{d}(x-y)
\end{align}
the symmetry condition reduces to
\begin{align}
    \partial _k\langle t^k {}_i {}^{0 }{}_j(x,y)\rangle +\partial _k^y\langle t^k {}_j {}^{0 }{}_i(y,x) \rangle =0 
\end{align}
and the functions $\langle t^k {}_i {}^{0 }{}_j(x,y)\rangle $ parametrize the additional theory- and counterterm dependence of this last bracket. We arrive at the stress-energy tensor equal time brackets of a Carroll quantum field theory on a flat background
\begin{equation}\label{eq:SET_final}
  \addtolength{\fboxsep}{7pt}
   \boxed{
\begin{aligned}
    i [\mathbf{T}^0{}_0 (x)&,\mathbf{T}^0{}_0 (y)]=-\delta ^i_\mu \delta _\nu ^0\, \partial _i \frac{\delta \mathcal{A}^\mu (x)}{\delta \tau _\nu (y)} \Big \vert _{\textrm{flat}} \\
    i[\mathbf{T}^0{}_0 (x)&,\mathbf{T}^0{}_i (y)]= \big(\mathbf{T}^0 {}_0 (y) \delta^k_i -\mathbf{T}^k {}_i (x)\big) \partial _k \delta ^{d}(x-y) -2\delta _{\rho i}\delta _\nu ^0\delta _\mu ^j \partial _j\frac{\delta \mathcal{A}^\mu (x)}{\delta h_{\nu \rho }(y)}\Big \vert _{\textrm{flat}}\\
     i[\mathbf{T}^0{}_i (x)&,\mathbf{T}^0{}_j (y)]=   \big( \mathbf{T}^0 {}_j (x) \partial_i + \mathbf{T}^0 {}_i (y) \partial_j \big)   \delta ^{d}(x-y) + \partial _k t^k {}_i {}^{0 }{}_j(x,y)
\end{aligned}
   }
\end{equation} 
where the expectation values were left out since these identities hold for any state and are therefore elevated to operator identities. One can see that the Carroll boost anomaly may source non-trivial terms on the right hand sides of two of these brackets. In the two-dimensional example we shall see, however, that only the second of these survives. These brackets do not form an algebra as can be seen from the appearance of $\mathbf{T}^k{}_i$ on the right hand side. 

\subsection{Carroll algebra from stress-energy tensor brackets}
Let us use the just derived brackets to construct a representation of the Carroll algebra in terms of Hermitean operators. For this simple example, assume that no Carroll boost anomaly is present, i.e., $\mathcal{A}^\mu (x)=0$. We directly promote \eqref{eq:carr_charges} to operators and from evaluating \eqref{eq:SET_final} with suitable smearing we obtain the infinitely extended Carroll algebra $\mathfrak{isocarr}(d+1)$,
\begin{align}
    i[M_f,M_g]&=0 &  i[P_i,J_{jk}]&=2\delta _{i[j}P_{k]} \\
   i[P_i,M_f]&=M_{\partial _if} & i[M_j,J_{jk}]&=-M_{x_j\partial _kf-x_k\partial _jf} \\
   & & i[J_{ij},J_{kl}]&=4\delta _{[i[k}J_{l]j]} ~.
\end{align} 
The operators $\partial _k t^k {}_i {}^{0 }{}_j(x,y)$ thereby have to satisfy 
\begin{align}
    \int \dd ^dx\, x_{[i}\partial _mt^m{}_{j]}{}^0{}_{k}(x,y)&=0
\end{align}
from requiring the bracket $i[J_{ij},J_{kl}]$ to take the given form. 

\subsection{Carroll boost anomaly in 2d}
\label{sec:Carroll_boost_anomaly}
If the Carroll theory is viewed in a holographic context, one expects a Carroll boost anomaly to be present \cite{Hartong:2025jpp} that is given by a local expression in terms of background quantities,
\begin{align}
    \mathcal{A}^\mu=\frac{c_M}{12\pi }\,h^{\mu \sigma }v^\rho \partial _{[\rho }\big(a_{\sigma ]}+K\tau _{\sigma ]}\big) ~.
\end{align}
Here, $c_M$ is some real number that plays the role of a central charge. To see this, we evaluate \eqref{eq:SET_final} leading to
\begin{align}
    i [\mathbf{T}^0{}_0 (x)&,\mathbf{T}^0{}_0 (y)]=0 \\
    i[\mathbf{T}^0{}_0 (x)&,\mathbf{T}^0{}_1 (y)]= \big(\mathbf{T}^0 {}_0 (y)  -\mathbf{T}^1 {}_1 (x)\big) \partial  \delta (x-y) +\frac{c_M}{24\pi } \partial ^3\delta (x-y)\\
     i[\mathbf{T}^0{}_1 (x)&,\mathbf{T}^0{}_1 (y)]=   \big( \mathbf{T}^0 {}_1 (x)  + \mathbf{T}^0 {}_1 (y)  \big)\partial   \delta (x-y) + \partial  t^1 {}_1 {}^{0 }{}_1(x,y) 
\end{align}
where we denote spatial derivatives by $\partial _1=:\partial $. Consider a conformally coupled theory without a Weyl anomaly such that the stress-energy tensor additionally obeys the trace Ward identity \eqref{eq:z-trace identity}. We may construct the charges associated to symmetries of the chosen flat background which are given by the two-dimensional Carroll conformal Killing vectors, \eqref{eq:CCKVs} for $d=1$. We get 
\begin{align}
    Q_{(f,Y)}=\int dx\, \Big[Y(x)\mathbf{T}^0{}_1(x)+\big(zx^0\partial Y(x)+f(x)\big)\mathbf{T}^0{}_0(x)\Big]
\end{align}
for two free functions $Y(x)$ and $f(x)$, and it can be checked that the stress-energy tensor brackets imply
\begin{align}
    i[Q_{(f_1,0)},Q_{(f_2,0)}]&=0\\
    i[Q_{(0,Y_1)},Q_{(f_2,0)}]&=Q_{(Y_1\partial f_2-zf_2\partial Y_1,0)}-\frac{c_M}{24\pi}\int \dd x\, \partial ^3Y_1f_2 \\
    i[Q_{(0,Y_1)},Q_{(0,Y_2)}]&= Q_{(0,Y_1\partial Y_2-Y_2\partial Y_1)}+\int \dd x\dd y\, Y_1(x)Y_2(y)\partial t^1{}_1{}^0{}_1(x,y) ~.
\end{align}
To bring this into a more recognizable form from the perspective of the Carroll CFT literature, one compactifies the spatial domain to a circle and defines Fourier modes 
\begin{align}
    L_n:=Q_{(0,e^{-inx})} && M_n:=-Q_{(e^{-inx},0)}
\end{align}
such that 
\begin{equation}\label{eq:exteneded_bms3}
\begin{aligned}
    [L_n,L_m]&=(n-m)L_{n+m}\\
    [L_n,M_m]&=(zn-m)M_{n+m}+\frac{c_M}{12} n^3\delta _{n+m,0}\\
    [M_n,M_m]&=0 ~.
\end{aligned}
\end{equation}
Here, we set the model dependent terms $t^1{}_1{}^0{}_1(x,y)$ to zero for simplicity. For $z=1$, this is the centrally extended $\mathfrak{bms}_3$ algebra (see also section \ref{sec:conformal_extensions}) which was previously derived from a bulk perspective as the asymptotic symmetry algebra of three-dimensional asymptotically flat Einstein gravity \cite{Barnich:2006av} where $c_M=3/G_N$. The derivation above shows that the central extension appearing in the mixed commutator is intimately tied to the Carroll boost anomaly and is therefore a quantum effect.

While we set the model-dependent terms $t^1{}_1{}^0{}_1(x,y)$ to zero in this case, there are indeed situations where they contribute in a non-trivial way to the brackets. A notable example is the case where they are just given by the c-numbers
\begin{align}
    t^1{}_1{}^0{}_1=-\frac{c_L}{24\pi }\partial ^2\delta (x-y)
\end{align}
which leads to the typical Virasoro-like central term in the $[L_n,L_m]$ bracket,
\begin{align}
    [L_n,L_m]&=(n-m)L_{n+m}+\frac{c_L}{12} n^3\delta _{n+m,0}~.
\end{align} 
From a bulk perspective this central charge is non-zero for the asymptotic symmetries of topologically massive gravity \cite{Bagchi:2012yk} while in an intrinsically Carroll context one example is the Carroll--Ising model, see \cite{Grumiller:2025rtm} and references therein.

\section{Examples of Carroll field theories}
\label{sec:examples}
We proceed by presenting several examples for classical Carroll field theories. Since the literature on this subject is already quite extensive, we are only able to focus on a small cross section. In particular, we shall start with the simplest yet non-trivial case of Carroll scalar fields in subsection \ref{sec:Carr_sc}, followed by a discussion of Carroll Maxwell fields in subsection \ref{sec:Carr_Max}. The last subsection \ref{sec:Car_swift} discusses Carroll swiftons \cite{Ecker:2024czx}, which is a class of interacting models.

In a canonical formulation, each of these models provides us with the classical version of the stress-energy tensor commutators \eqref{eq:SET_final} in the previous section. They hold, provided the theory is put on a flat background, which one may check explicitly in each case.

\subsection{Minimally coupled scalar fields}\label{sec:Carr_sc}
Carroll scalar fields were first studied in \cite{Henneaux:2021yzg} and further developed in their conformal aspects in 2d \cite{Bagchi:2022eav} and higher dimensions \cite{Baiguera:2022lsw,deBoer:2021jej,Rivera-Betancour:2022lkc}. We shall also use them as a toy model for coupling to a Carroll black hole background \cite{Aggarwal:2024gfb} in chapter \ref{ch:Hawking}. 

In fact there are currently two inequivalent scalar theories known that are referred to as ``magnetic'' and ``electric''. This terminology is borrowed from the respective case in electrodynamics and refers to two different ways of taking the Carroll limit starting from a Lorentzian scalar field. In the following, it is shown how one arrives at these limits, for which we find it most instructive to proceed in a Hamiltonian formulation. Additionally, we shall put the theory on an arbitrary background first which makes it especially easy to read off the canonical expressions for energy and momentum density.

Starting from the Lorentzian scalar field
\begin{align}
    I_{\textrm{Lorentz}}[\phi ]=-\frac12 \int \dd ^{d+1}x\, \sqrt{G}\, \Big(G^{\mu \nu }\partial _\mu \phi \partial _\nu \phi +V(\phi) \Big)
\end{align}
 with some potential $V(\phi )$ we parametrize the Lorentzian metric $G_{\mu \nu }$ with ADM-like variables by
\begin{align}\label{eq:mod_ADM}
    G_{\mu \nu }&=\begin{pmatrix}
        N^iN_i-c^2N^2 & N_i \\
        N_i & g_{ij}
    \end{pmatrix}
     & G^{\mu \nu }&=\begin{pmatrix}
        -\frac{1}{c^2N^2} & \frac{N^i}{c^2N^2} \\
        \frac{N^i}{c^2N^2} & g^{ij}-\frac{N^iN^j}{c^2N^2} 
     \end{pmatrix}
\end{align}
where, in contrast to the usual ADM split, we absorbed a factor of $c$ into the shift vector $N^i$ for convenience. The volume element in this case becomes $\sqrt{G}=cN\sqrt{g}$ and $g^{ij}$ is defined as the inverse of $g_{ij}$. Performing a Legendre transformation, one finds the canonical momentum
\begin{align}
    \pi =\frac{\sqrt{g}}{Nc}\big(\dot{\phi }-N^i\partial _i \phi )
\end{align} 
and the scalar field action in Hamiltonian form for a general background 
\begin{align}\label{eq:scalar_action_lorentz}
    I_{\textrm{Lorentz}}[\phi, \pi ]=\int \dd t\, \dd ^d x\, \big(\pi \dot{ \phi } - N\mathcal{H}-N^i\mathcal{H}_i \big) ~.
\end{align}
Here, we denote $t$-derivatives with a dot and set $x^0=t$ as compared to the previous section. The energy and momentum densities read explicitly
\begin{align}
    \mathcal{H}=\frac{c}{2\sqrt{g}}\pi ^2+\frac{c\sqrt{g}}{2}\big(g^{ij}\partial _i\phi \partial _j \phi+V\big) \qquad \qquad \mathcal{H}_i=\pi \partial _i \phi
\end{align}
and the fields $\phi (x)$, $\pi(x)$ obey a canonical equal-time Poisson bracket $\{\phi (x),\pi (y)\}=\delta ^d(x-y)$. For taking a limit $c\to 0$ the action \eqref{eq:scalar_action_lorentz} is one viable starting point, albeit not a very interesting one: The energy density just vanishes in the limit. One may, however, perform a canonical transformation before taking the limit. If this canonical transformation depends on $c$, it transforms between inequivalent starting points as far as taking the limit is concerned: Two theories related by a canonical transformation before the limit do not have to be related by one afterwards. 

To give the two prominent examples, the electric action is obtained by performing the transformation $\phi \to \phi \sqrt{c}$, $\pi \to \pi /\sqrt{c}$ which in the limit yields 
\begin{align}\label{eq:elec_scalar}
    I_{E}[\phi ,\pi]=\int \dd t\,\dd ^dx\,\big(\pi \dot{\phi }-N\mathcal{H}_E-N^i\mathcal{H}_i \big)\approx \frac{1}{2}\int \dd ^{d+1}x\, \mathfrak{e}\big(v^\mu \partial _\mu \phi \big )^2
\end{align}
with the electric energy density
\begin{align}
    \mathcal{H}_E=\frac{\pi ^2}{2\sqrt{g}} ~.
\end{align}
In the second equality, we integrated out $\pi $ and translated the ADM variables back into Carroll metric data by
\begin{align}\label{eq:backtranslation}
    h_{\mu \nu }=\lim _{c\to 0}G_{\mu \nu }&& -v^\mu v^\nu =\lim _{c\to 0}c^2G^{\mu \nu } ~. 
\end{align}
We thus choose the Carroll ADM variables as
\begin{align}\label{eq:Carr_ADM_variables_dummy}
    v^\mu =\Big(-\frac{1}{N},\frac{N^i}{N}\Big) && \tau _\mu = (N,0) && h_{\mu \nu }=\begin{pmatrix}
        N^iN_i & N_i \\
        N_i & g_{ij}
    \end{pmatrix}
\end{align}
such that $h^{00}=h^{i0}=0$, $h^{ij}=g^{ij}$ and $\mathfrak{e}=N\sqrt{g}$. One can see that on a flat background where $v=-\partial _t$ this action only keeps the part of the Lorentzian free scalar that involves the temporal derivatives. This is why it is sometimes also referred to as the ``timelike'' action \cite{Baiguera:2022lsw,deBoer:2021jej,Rivera-Betancour:2022lkc}. 

Alternatively, doing the opposite transformation $\phi \to \phi /\sqrt{c}$, $\pi \to \pi \sqrt{c}$ yields the magnetic action 
\begin{align}
    I_{M}[\phi ,\pi]=\int \dd t\,\dd ^dx\,\big(\pi \dot{\phi }-N\mathcal{H}_M-N^i\mathcal{H}_i \big)
\end{align}
with energy density
\begin{align}
    \mathcal{H}_M=\frac{\sqrt{g}}{2}\big(g^{ij}\partial _i\phi \partial _j \phi+V\big) 
\end{align}
and where we assumed that the coupling constants in the potential $V$ scale in such a way that it remains finite in the limit. In this case one cannot integrate out the momentum anymore since it only features linearly in the action. It acts as a Lagrange multiplier setting the time dependence of $\phi $ to zero. One may bring this action into a covariant form by pulling out the density weight of $\pi $ and defining a scalar field $\chi$ by $-\sqrt{g}\chi =\pi $. Using the same ADM split as above one obtains 
\begin{align}\label{eq:mag_scalar}
    I_M[\chi ,\phi ]=\int \dd ^{d+1}x\,\mathfrak{e}\,\Big(\chi v^\mu \partial _\mu \phi -\frac12 h^{\mu \nu }\partial _\mu \phi \partial _\nu \phi -\frac12 V(\phi ) \Big ) 
\end{align}
which can be checked to be Carroll boost invariant under the infinitesimal version of the transformations \eqref{eq:Carroll_boost} together with 
\begin{align}
    \delta _\lambda \chi =-h^{\mu \nu }\lambda _\mu \partial _\nu \phi ~.
\end{align}
Thus, the magnetic action is a two-field realization of a Carroll field theory. The field $\chi $ is an essential part of the action and ensures Carroll boost invariance. 

\subsection{Non-minimally coupled scalar fields}\label{sec:nonminsca}
Once written in a second order formulation, one may explore a few further possibilities to couple the two types of scalar fields just discussed to a Carroll geometry. We shall restrict to the minimal-torsion geometries from section \ref{sec:aff_conn_Carroll} which are obtained from a Cartan perspective by ``integrating out'' the boost- and rotation connections. As we shall see in section \ref{sec:mag_carr} these are typical off-shell geometries of magnetic Carroll gravity. The independent background fields are $\tau _\mu $, $h_{\mu \nu }$ and $C_{\mu \nu }$ which are used to define a metric compatible affine connection \eqref{eq:gen_affine_connection}. Under the assumption of maintaining local Carroll boost invariance, the fields $\phi $ and $\chi$ of the magnetic action may for example be coupled as 
\begin{align}\label{eq:nonminmag}
    I_M^{\xi ,\eta }[\chi ,\phi ]=\int \dd ^{d+1}x\, \mathfrak{e}\, \Big(\chi (v^\mu \partial _\mu \phi +2\eta \xi K\phi ^{\eta -1})-\frac12 h^{\mu \nu }\partial _\mu \phi \partial _\nu \phi +\xi \phi ^\eta \mathcal{R}\Big)  && \xi ,\eta  \in \mathbb{R} ~.
\end{align}
Here, $\mathcal{R}[\tau _\mu ,h_{\mu \nu },C_{\mu \nu }]=h^{\mu \nu }\big(R_{\mu \nu }-v^\rho \tau_\sigma R^\sigma {}_{\mu \rho \nu }\big)$ is the Carroll curvature scalar associated to the connection \eqref{eq:gen_affine_connection}. Under local Carroll boosts it transforms as \eqref{eq:trafomathcalR}. Due to the coupling to $C_{\mu \nu }$ this theory falls outside the class of models we restricted to initially. Indeed, the corresponding response function as defined in section \ref{sec:coupl_conn} reads
\begin{align}
    \Psi ^{\mu \nu }=\frac{2}{\mathfrak{e}}\frac{\delta I_M^{\xi,\eta }}{\delta C_{\mu \nu }}=4\phi ^\eta K^{\mu \nu }+4 h^{\mu \nu }v^\alpha (\partial _\alpha +\tau _\alpha K)\phi^\eta ~.
\end{align}
Once coupled to magnetic Carroll gravity (see section \ref{sec:mag_carr}), this theory sources intrinsic torsion since the field $C_{\mu \nu }$ then becomes dynamical and introduces an equation of motion that couples $K^{\mu \nu }$ to $\Psi ^{\mu \nu }$. For $\eta =2$ a similar action was proposed in \cite{Baiguera:2022lsw} where the field $C_{\mu \nu }$ translates to an additional Lagrange multiplier $\chi _{\mu \nu }$ that is part of the dynamical field content. In the present case, however, this field is part of the background structure and thus should not be varied when computing equations of motion. An interesting open question is whether the action \eqref{eq:nonminmag} can be rendered Weyl-invariant for some choices of $\xi $ and $\eta $. 

As a final example we also propose a non-minimally coupled version of the electric scalar \eqref{eq:elec_scalar},
\begin{align}
    I_E^{\xi }[\phi ]=\frac12 \int \dd ^{d+1}x\, \mathfrak{e}\, \Big(v^\mu (\partial _\mu -\xi W_\mu ) \phi \Big )^2 && \xi \in \mathbb{R}
\end{align}
where 
\begin{align}
    W_\mu =\frac{K}{d}\tau _\mu +\frac{1}{z}a_\mu ~.
\end{align}
The action can be checked to be Carroll boost invariant and is additionally invariant under anisotropic Weyl transformations of subsection \ref{sec:Weyl} for $\xi = \frac{z-d}{2}$ with the scalar field transforming as $\delta _\rho \phi =\xi \rho \phi$. From its transformation property $\delta _\rho W_\mu =\partial _\mu \rho $ one observes that $W_\mu $ behaves like a Weyl connection. Such a field was previously introduced in the Carroll hydrodynamics context \cite{Ciambelli:2018xat}. In this case there is no coupling to the field $C_{\mu \nu }$. 

\subsection{Carroll--Maxwell fields}\label{sec:Carr_Max}
A Carroll limit of the Maxwell field was first investigated in \cite{Duval:2014uoa} and further developed in \cite{Henneaux:2021yzg}. Like in the case of the free scalar field, there are again at least two distinct contractions which project out either the magnetic or the electric components of the electromagnetic field, hence the name. Proceeding again on an arbitrary background parametrized by \eqref{eq:mod_ADM} we start from the Maxwell action
\begin{align}
    I_{\textrm{Lorentz}}[A]=-\frac{1}{4}\int \dd ^{d+1}x\sqrt{G}\,G^{\mu \alpha }G^{\nu \beta }F_{\mu \nu }F_{\alpha \beta }
\end{align}
with $F_{\mu \nu }=\partial _\mu A_\nu -\partial _\nu A_\mu $. Performing a Legendre transformation in the sector $(A_i,\pi ^i)$ leads to its Hamiltonian form\footnote{This is a practical shortcut to the fully-fledged canonical analysis that already solves the primary constraint $\pi ^t\approx 0$, see chapter 19 of \cite{Henneaux:1992}.} 
\begin{align}
    I_{\textrm{Lorentz}}[\pi ^i,A_i,A_t]=\int \dd t\, \dd ^dx\, \Big(\pi ^i \dot{A}_i-N\mathcal{H}-N^i\mathcal{H}_i+A_t\mathcal{G}\Big )
\end{align}
where $\mathcal{G}=\partial _i\pi ^i$ is the Gauss constraint and $A_t$ the corresponding Lagrange multiplier. Energy and momentum densities are given by 
\begin{align}
    \mathcal{H}=\frac{c}{2\sqrt{g}}\pi ^i\pi _i+\frac{c\sqrt{g}}{4}g^{ik}g^{jl}F_{ij}F_{kl} && \mathcal{H}_i=F_{ij}\pi ^j
\end{align}
with the canonical variables satisfying the Poisson bracket $\{A_i(x),\pi ^j(y)\}=\delta _i^j\delta ^d(x-y)$. 

The electric limit is obtained by first performing the canonical transformation $A_i\to \sqrt{c}A_i$, $\pi ^i\to \pi ^i/\sqrt{c}$ and then sending $c\to 0$ which leads to 
\begin{align}
    I_E[A_i,\pi ^i ,A_t]=\int \dd t\, \dd ^dx\, \Big(\pi ^i \dot{A}_i-N\mathcal{H}_E-N^i\mathcal{H}_i+A_t\mathcal{G}\Big )\approx \frac12 \int \dd ^{d+1}x \mathfrak{e}\, \big(v^\mu F_{\mu \nu }\big)^2 ~.
\end{align}
In the second equality we again integrated out the momenta $\pi ^i$ and used that $v^\mu F_{\mu \nu }$ can be squared in a Carroll boost invariant way using $h^{\mu \nu }$ because it is a transverse quantity. The translation between ADM and second order variables is again given by \eqref{eq:Carr_ADM_variables_dummy}. The energy density in this case reads 
\begin{align}
    \mathcal{H}_E=\frac{\pi ^i\pi _i }{2\sqrt{g}}=\frac{\sqrt{g}}{2} E^iE_i
\end{align}
which only involves the electric field. The translation into second order geometric variables works in a similar fashion as in the scalar case described around equation \eqref{eq:backtranslation}.

For the magnetic contraction we perform the converse transformation $A_i\to A_i/\sqrt{c}$, $\pi ^i\to \pi ^i\sqrt{c}$ and obtain the action 
\begin{align}
    I_M[A_i,\pi ^i,A_t]=\int \dd t\, \dd ^dx\, \Big(\pi ^i \dot{A}_i-N\mathcal{H}_M-N^i\mathcal{H}_i+A_t\mathcal{G}\Big )
\end{align}
with 
\begin{align}
    \mathcal{H}_M=\frac{\sqrt{g}}{4}g^{ik}g^{jl}F_{ij}F_{kl}  ~.
\end{align}
Just like for the magnetic scalar field we again cannot integrate out the canonical momenta which instead act as Lagrange multipliers. Together with the Gauss constraint they set the electric field to zero on-shell. In four spacetime dimensions we may write the energy density as $\mathcal{H}_M=\sqrt{g}B^iB_i/2$ which, as already mentioned, consists only of the magnetic part.

\subsection{Scalar Carroll swiftons}\label{sec:Car_swift}
Following \cite{Ecker:2024czx} we now come to a class of interacting models called Carroll swiftons. In a sense they are a generalization of the electric/magnetic scalar- and Maxwell models in that their actions are at least of second order in both temporal and spatial derivatives. We first focus on the scalar case and then come to the electromagnetic case in subsection \ref{eq:ele_swift}.

Consider a ``bi-scalar'' model that couples two scalar fields $\phi,\chi$ with canonically normalized kinetic terms and coupling constant $g$, thereby generalizing the model introduced in \cite{Baig:2023yaz} (see their eq. (2.1) as well as \cite{Kasikci:2023tvs}) to any Carroll background. Its action in covariant form 
\begin{align}\label{eq:cips01}
I_{\textrm{bi-scalar}} [\phi ,\chi ]= \frac12 \int\dd^{d+1}x \,\mathfrak{e}\,\left(\left(v^\mu \partial_\mu \phi \right)^2 + \left(v^\mu \partial_\mu \chi \right)^2 + \alpha \, B_\mu B^\mu\right)
\end{align}
contains the manifestly transverse covariant vector
\eq{
B_\nu \equiv v^\mu \left(\partial_\mu \phi \partial_\nu \chi - \partial_\mu \chi \partial_\nu \phi \right) \equiv 2 v^\mu \partial_{[\mu} \phi \partial_{\nu]} \chi\,
}{eq:Bdef}
and a coupling constant $\alpha $. Because $B_\nu$ involves simultaneously time and spatial derivatives, the model allows propagation off the Carroll lightcone (see below). Antisymmetry of the coefficient of $v^\mu$ in $B_\nu$ is crucial for transversality, which would not hold if $\phi=\chi$ since then $B_\nu$ would be identically zero. By contrast, two distinct scalar fields can ``mutualize'' their derivatives in a non-trivial way through the identically transverse vector $B_\nu$. Each scalar field is crucial for its mutualistic partner in this construction.  

The Hamiltonian for this model can be derived straightforwardly, noting that the above Lagrangian density can be rewritten in terms of a doublet $\phi^A\equiv(\phi,\chi)$ as
\begin{equation}
\mathcal L = \frac{N\sqrt{g}}{2 } H_{AB} (v^\mu \partial _\mu {\phi}^A)(v^\nu \partial _\nu {\phi}^B) \quad \,\;\; H_{AB} = \begin{pmatrix} 1+  \alpha (\partial \chi)^2& - \alpha\partial\phi\cdot\partial\chi \\ - \alpha \partial\phi\cdot\partial\chi & 1+  \alpha (\partial \phi)^2\end{pmatrix}
\end{equation}
where we used the notation $\partial\phi^A\cdot\partial\phi^B \equiv g^{ij} \partial_i \phi^A \partial_j \phi^B$ and the same ADM split as in the previous subsections \eqref{eq:Carr_ADM_variables_dummy}. The inverse matrix $H^{AB}$ is $H^{AB}=\frac{1}{D}\left(\delta^{AB}+\alpha (\partial\phi^A)\cdot(\partial\phi^B)\right)$ where $D$ is the determinant 
\begin{align}
    D=1+\alpha (\partial\phi)^2+\alpha (\partial\chi)^2+\alpha^2\left( (\partial\phi)^2(\partial \chi)^2-(\partial\phi\cdot\partial\chi)^2\right) ~,
\end{align}
which obeys $D\geq1$ for $\alpha \geq0$, implying that the field space metric $H_{AB}$ has Euclidean signature. The Hamiltonian is $N \mathcal H + N^i \mathcal H_i$ where the energy and momentum densities are  
\begin{align}
\mathcal H = \frac{1}{2\sqrt{g}}\, H^{AB}\, \pi_A \pi_B && \mathcal H_i = \pi_A \partial_i \phi^A ~.
\end{align}
We note that $\mathcal{H}$ is bilinear in the conjugate momenta $\pi_A$. Since the quadratic form $H^{AB}\,\pi_A\pi_B$ is positive definite, the energy density is bounded from below by zero.

The equations of motion following from the mutualistic scalar action \eqref{eq:cips01}
\begin{subequations}
\label{eq:cips3}
\begin{align}
\partial_\mu\Big(\mathfrak{e}\,\big(v^\mu v^\nu\partial_\nu\chi-\alpha h^{\alpha\nu}B_\nu v^\mu\partial_\alpha\phi+\alpha h^{\mu\nu}B_\nu v^\alpha\partial_\alpha\phi\big)\Big)&=0\\
\partial_\mu\Big(\mathfrak{e}\,\big(v^\mu v^\nu\partial_\nu\phi+\alpha h^{\alpha\nu}B_\nu v^\mu\partial_\alpha\chi-\alpha h^{\mu\nu}B_\nu v^\alpha\partial_\alpha\chi\big)\Big)&=0
\end{align}
\end{subequations}
are coupled non-linear partial differential equations, and we have not tried to devise a general method to solve them. 

On a flat background the scalar swifton model is not just Carroll invariant but also exhibits an anisotropic scaling symmetry (see section \ref{eq:sec311}). This is readily seen by computing an infinitesimal Weyl transformation 
\begin{align}
    \delta _\rho I_{\textrm{bi-scalar}}=\frac12 \int \dd ^{d+1}x\, \mathfrak{e}\Big[\big((v^\mu \partial _\mu \phi )^2&+(v^\mu \partial _\mu \chi )^2\big)(d-z+2\Delta )\rho  \\
    &+\alpha B^\mu B_\mu (d-z-2+4\Delta)\rho +2 \mathcal
    K^\mu \partial _\mu \rho  \Big] \nonumber
\end{align} 
where we used the transformations of section \ref{sec:Weyl} together with 
\begin{align}
    \delta _\rho \phi =\Delta \rho \phi && \delta _\rho \chi =\Delta \rho \chi 
\end{align}
for $\Delta \in \mathbb{R}$. The vector $\mathcal{K}^\mu $ is given by 
\begin{align}
    \mathcal K^\mu =v^\mu \Delta \big(\phi v^\alpha \partial _\alpha \phi +\chi v^\alpha \partial _\alpha \chi +&\alpha \phi B^\alpha \partial _\alpha \chi -\alpha \chi B^\alpha \partial _\alpha \phi \big)\\
    &+\alpha \Delta B^\mu \big(\chi v^\alpha \partial _\alpha \phi -\phi v^\alpha \partial _\alpha \chi \big) ~.
\end{align}
Thus, for the choice
\begin{align}
    \Delta =1 && z=d+2
\end{align}
this action is left invariant by constant Weyl rescalings. According to the discussion in section \ref{eq:sec311}, the term $\mathcal{K}^\mu \partial _\mu \rho $ prevents the stress-energy tensor from satisfying the generalized traceless condition and poses an obstruction for the scale symmetry to lift to a conformal symmetry. Indeed, we find on-shell
\begin{align}
    (d+2)T^\mu \tau _\mu + T^{\mu \rho }h_{\mu \rho }=-\frac{1}{\mathfrak{e}}\partial _\mu (\mathfrak{e}\mathcal{K}^\mu )
\end{align}
where the components $T^\mu $ and $T^{\mu \rho }$ are given after the definition \eqref{var} by 
\begin{align}
    T^\mu &=\frac12 v^\mu \big(\left(v^\nu \partial_\nu \phi \right)^2 + \left(v^\nu \partial_\nu \chi \right)^2 + \alpha \, B_\nu B^\nu\big) \\
    T^{\mu \rho }&=\frac12 h^{\mu \rho }\big(\left(v^\alpha \partial_\alpha \phi \right)^2 + \left(v^\alpha \partial_\alpha \chi \right)^2 + \alpha \, B_\nu B^\nu\big)-\alpha B^{\mu }B^\rho \\
    &\qquad -2v^{(\rho }h^{\mu )\beta }\partial _\beta \phi v^\alpha \partial _\alpha \phi -2v^{(\rho }h^{\mu )\beta }\partial _\beta \chi v^\alpha \partial _\alpha \chi +\alpha v^{(\rho }h^{\mu )\beta }B^\alpha (\partial _\alpha \phi \partial _\beta \chi -\partial _\alpha \chi \partial _\beta \phi ) ~. \nonumber 
\end{align}
The scale current is constructed by 
\begin{align}
    j^\mu =\mathfrak{e}\mathbf{T}^\mu {}_\nu \xi ^\nu _{\textrm{D}}+\mathfrak{e}\mathcal{K}^\mu 
\end{align}
and satisfies $\partial _\mu j^\mu =0$ on-shell provided $\xi _{\textrm{D}}$ generates a scale transformation for the given background., i.e., $\mathcal{L}_{\xi _{\textrm{D}}}\tau _\mu =(d+2)\tau _\mu $, $\mathcal{L}_{\xi _{\textrm{D}}}h_{\mu \nu } =2h _{\mu \nu }$.

\subsubsection{Taylor expanded Swiftons}

To gain insight into the solutions of the scalar swifton model, we use perturbative methods. A soluble case arises when one of the scalar fields, say $\chi=\chi_{\textrm{\tiny BG}}+{\mathcal{O}}(\epsilon^2)$, is a background field in addition to the geometric background (which we assume to be static) and the other one, $\phi=\epsilon\,\varphi$, is a small fluctuation on top ($\epsilon\ll1$). To leading order we obtain from \eqref{eq:cips3} the background solution
\eq{
\chi_{\textrm{\tiny BG}} = \chi_0(x^i) + \chi_1(x^i)\,t
}{eq:cips4}
where we used adapted coordinates such that $v=f(x^i)\,\partial_t$. For simplicity, we set $\chi_0=0$ and $\chi_1=1$.

Inserting this solution back into the mutualistic scalar action \eqref{eq:cips01} yields a quadratic action
\eq{
I^{\textrm{Taylor}}_{\textrm{Swifton}}[\varphi]=\frac12 \int\dd^{d+1}x\,\mathfrak{e}\,\Big(\epsilon^2\,\big(v^\mu\partial_\mu\varphi\big)^2 + \epsilon^2\,\alpha \,h^{\mu\nu}\,(\partial_\mu\varphi)(\partial_\nu\varphi)  + 1\Big)
}{eq:cips5}
for the fluctuations $\varphi$ on such a background. 

To get hyperbolic equations of motion, we need a negative coupling constant $\alpha $, which may seem at odds with positivity of energy density. However, even for negative $\alpha $, the energy density remains positive as long as the coupling constant obeys the inequality 
\begin{align}
    \alpha >-1/[(\partial\phi)^2+(\partial\chi)^2] ~.
\end{align}
In our perturbative context where both $(\partial\phi)^2$ and $(\partial\chi)^2$ are small the bound on the coupling constant is very weak.

Up to a cosmological constant term and conventions, the action \eqref{eq:cips5} coincides precisely with the one proposed in \cite{Ciambelli:2023xqk}, i.e., it combines electric and magnetic Carroll scalar field actions to a single electromagnetic one. However, as opposed to \cite{Ciambelli:2023xqk}, the full theory maintains Carroll boost invariance.

\subsubsection{Multi-scalar models}
The bi-scalar model \eqref{eq:cips01} can be generalized by the same ``mutualization trick'' to a multi-scalar theory. For instance, with three scalars we have the action [$\phi^A\equiv(\phi,\chi,\psi)$]
\eq{
I_{\textrm{tri-scalar}}[\phi ^A] = \frac12\int\dd^{d+1}x\,\mathfrak{e}\, \Bigg( \sum_{A= 1}^3\left(v^\mu \partial_\mu \phi^A \right)^2  + \alpha \, B_{\mu \nu}B^{\mu \nu}\Bigg)
}{eq:multi_scalar}
with the transverse tensor
\eq{B_{\mu \nu} \equiv v^\rho B_{\mu \nu \rho}\qquad \qquad  B_{\mu \nu \rho} \equiv \partial_{[\mu} \phi \partial_\nu \chi \partial_{\rho]} \psi\,.}
{eq:whatever_new}
The interaction term in \eqref{eq:multi_scalar} is of order six in the derivatives, but only quadratic in the time derivatives. 

Yet another generalization is by imposing internal $O(N)$ symmetry and working with the multiplett $\phi ^A$, $A=1,...,N$. After defining the transverse tensor
\begin{align}
    B^{AB}_\mu =v^\nu (\partial _\nu \phi ^A \partial _\mu \phi ^B-\partial _\nu \phi ^B \partial _\mu \phi ^A) 
\end{align} 
we can write an $O(N)$-vector model 
\begin{align}
    I_{O(N)}[\phi ^A]=\frac12 \int \dd ^{d+1}x\, \mathfrak{e}\, \Big((v^\mu \partial _\mu \phi ^A)^2+\frac{\alpha }{2}h^{\mu \nu }B^{AB}_\mu B^{AB}_\nu \Big)
\end{align}
which is still of order four in spatial derivatives and order two in time derivatives. Once written in this formulation one may perform a large-$N$ expansion and try to quantize the theory in that limit \cite{Cotler:2024xhb}. See also \cite{Jensen:2022ibn} for an example of this procedure in the context of fractonic systems.

\subsection{Electromagnetic Carroll Swiftons}
\label{eq:ele_swift}
A non-trivially interacting electromagnetic model can also be constructed using the same idea as in the scalar case. Consider 
\begin{equation}
C_{\mu \nu \rho} \equiv v^\sigma C_{\mu \nu \rho \sigma} \qquad\qquad  C_{\mu \nu \rho \sigma} \equiv F_{[\mu \nu} F_{\rho \sigma]}
\end{equation}
where $F_{\mu\nu}=\partial_\mu{A}_\nu-\partial_\nu{A}_\mu$ is the electromagnetic field.  This is a transverse three-form, $C_{\mu\nu\rho}\nu^\rho=0$. Hence, again its square $C_{\mu\nu\rho}C^{\mu\nu\rho}$ is well-defined. 

The electromagnetic Carroll swifton action
\eq{
\boxed{\phantom{\Big(}
I_{\textrm{\tiny EM--swifton}} =\frac12 \int\dd^{d+1}x\,\mathfrak{e}\,\Big(\big(v^\mu F_{\mu \nu} \big)^2 + \alpha \, C_{\mu \nu \rho}C^{\mu \nu \rho}\Big)
\phantom{\Big)}}
}{eq:EM}
yields the equations of motion
\eq{
\partial_\lambda\Big(\mathfrak{e}\big(v^\mu v^{[\sigma}h^{\lambda]\nu}+\alpha v^\nu C^{\sigma\lambda\mu}+\alpha v^{[\lambda}C^{\sigma]\mu\nu}\big)F_{\mu\nu}\Big)=0
}{eq:EMeom}
that are again coupled non-linear partial differential equations. 

The energy density is given by a similar expression as in the scalar case
\begin{align}
    \mathcal{H}=\frac{1}{2\sqrt{g}}H_{ij}\pi ^i\pi ^j 
\end{align}
where $H_{ij}$ is the inverse of the symmetric matrix
\begin{align}
    H^{ij}=g^{ij}+\frac{2\alpha }{3}\Big(g^{il}g^{mk}g^{nj}+\frac{1}{2}g^{ij}g^{lk}g^{mn}\Big)F_{lm}F_{kn} ~.
\end{align}
In four dimensions the determinant of the latter is given by $D=g^{-1}(1+2\alpha B^2/3)$ with $B^2=g^{ik}g^{jl}F_{kl}F_{ij}/2$ which makes $\mathcal{H}$ positive definite if $\alpha >-3/(2B^2)$, allowing in particular again negative values of $\alpha $ in a perturbative context.

For the special case of a four-dimensional flat Carroll background we linearize the equations of motion \eqref{eq:EMeom} with a constant electric background field plus fluctuations on top,
\begin{equation}
F_{ij}
\equiv \epsilon\,\mathcal B_{ij}\qquad\qquad F_{ti}= -\delta_i^x\,E-\epsilon\,\mathcal E_i \,.
\end{equation} 
The linearized equations of motion are solved by plane waves 
\eq{
\mathcal E_i = A_i e^{i(k_yy+k_zz-\omega t)}\qquad\mathcal{B}_{ij}=B\,\big(\delta_i^y\delta_j^z-\delta_i^z\delta_j^y\big)e^{i(k_yy+k_zz-\omega t)}
}{eq:planewave}
subject to the dispersion relation $\omega^2=c_{\textrm{\tiny{eff}}}^2\,(k_y^2+k_z^2)$ with the effective speed of light 
\begin{align}
    c_{\textrm{\tiny{eff}}}^2=-\frac{2\alpha }{3}\,E^2 ~,
\end{align}
the transversality condition $k_yA_y+k_zA_z=0=A_x$, and the normalization $B^2=A_y^2+A_z^2$. As in the bi-scalar model, we need negative $\alpha $ to have hyperbolic equations of motion with a real effective propagation speed for the swiftons.

Denoting $J_1=(v^\mu{F}_{\mu\nu})^2$ and $J_2=C_{\mu\nu\rho}C^{\mu\nu\rho}$, this model can be generalized to Lagrange densities of the form $\mathcal{L}=\mathfrak{e}\,f(J_1,J_2)$. It would be interesting to determine which choice of $f$ leads to the duality-invariant theory constructed in \cite{Bunster:2012hm}.

Finally, we mention that the electromagnetic model can be directly coupled to the bi-scalar action \eqref{eq:cips01}. Indeed, in terms of the complex field $\Phi=\phi+i\chi$ the bi-scalar action in \eqref{eq:cips01} takes the simple form
\begin{align}
    I_{\textrm{bi-scalar}}[\Phi ]=\frac12 \int \dd ^{d+1}x\,\mathfrak{e}\, \Big(|v^\mu \partial _\mu \Phi|^2+\alpha B^\mu B_\mu \Big)
\end{align}
where 
$B_\mu =(v^\mu \partial _\mu \Phi ^\ast  \partial _\nu \Phi - v^\mu \partial _\mu \Phi   \partial _\nu \Phi^\ast) /2i$ .
Thus, the global $U(1$) symmetry $\Phi \to e^{i\alpha }\Phi $ of the model can be made local by introducing standard minimal coupling $\partial_\mu\rightarrow \partial_\mu-iA_\mu$.
\chapter{Carroll gravity}
\label{chap:carroll_gravity}
\epigraphfontsize{\small\itshape}
\epigraph{``Jazz isn't dead. It just smells funny.''}{--- \textup{Frank Zappa}\phantom{aaaaaaaaaa}}
So far, we worked on fixed Carroll background geometries that provided the stage for investigating specific Carroll field theories. One of the strongest motivations for this was the holographic principle, which, applied to asymptotically flat spacetimes, predicts a duality between quantum gravity and some conformal Carroll field theory. While we already had a brief detour into the quantum world, defining Carroll quantum field theories from scratch turns out to be a thorny problem even for the simplest Carroll field theories like the electric or magnetic scalar field \cite{deBoer:2023fnj,Cotler:2024xhb}, and the last word on this matter might be far from being spoken. 

We now switch gears and come to gravitational theories built on Carroll symmetries. This takes us a step away from holographic applications but instead opens up a different angle for understanding quantum gravity in a simplified setting. In section \ref{sec:bronstein}, we shall propose a specific limit, coined the \textit{tantum gravity limit} which is understood to be directly taken in a putative fully-fledged theory of quantum gravity. This limit is defined such that it preserves specific semi-classical features like a sensible notion of thermodynamics and a finite Schwarzschild radius. The classical approximation of this limit is described by magnetic Carroll gravity, which we shall introduce in section \ref{sec:GRforctozero}. In section \ref{sec:2d_dil}, we then focus on a class of two-dimensional Carroll gravity theories which provide a useful arena for investigating Carroll black hole solutions together with their thermodynamics. The main original contributions to this chapter are \cite{Ecker:2023uwm,Ecker:2024czh}.

\section{Corners of the Bronstein cube}\label{sec:bronstein}
So far, quantum gravity is best understood in scaling limits where either Newton's constant vanishes (quantum field theory) or Planck's constant vanishes (Einstein gravity). As a further simplification, a large speed of light is often assumed (Galilean limit), which suppresses particle creation in quantum field theories and reduces Einstein gravity to Newton gravity. 

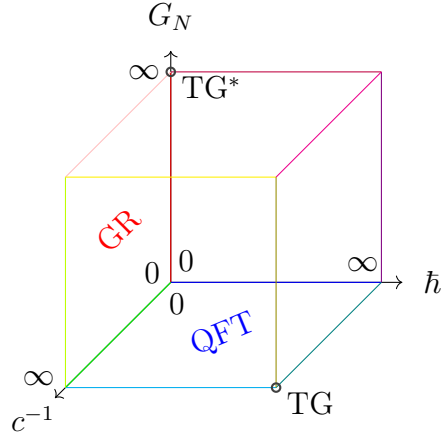
\begin{figure}[hbt]
\def\L{3.5}
\begin{center}
\begin{tikzpicture}[scale=0.8]
\draw[black,thin,->] (0,0) coordinate (orig) -- (0,1.1*\L);
\draw[black,thin,->] (orig) -- (1.1*\L,0);
\draw[black,thin,->] (orig) -- (-0.55*\L,-0.55*\L);
\draw[red,thin,-] (orig) -- (0,\L) coordinate (y);  
\draw[blue,thin,-] (orig) -- (\L,0) coordinate (x);
\draw[green,thin,-] (orig) -- (-0.5*\L,-0.5*\L) coordinate (z);
\draw[purple,thin,-] (y) -- (\L,\L) coordinate (xy);
\draw[violet,thin,-] (x) -- (xy);
\draw[cyan,thin,-] (z) -- (0.5*\L,-0.5*\L) coordinate (xz);
\draw[teal,thin,-] (x) -- (xz);
\draw[lime,thin,-] (z) -- (-0.5*\L,0.5*\L) coordinate (yz);
\draw[pink,thin,-] (y) -- (yz);
\draw[yellow,thin,-] (yz) -- (0.5*\L,0.5*\L) coordinate (xyz);
\draw[olive,thin,-] (xz) -- (xyz);
\draw[magenta,thin,-] (xy) -- (xyz);
\draw[darkgray,thick] (xz) circle(0.02*\L);
\draw[black] (xz) node[right,yshift=-0.5em] {TG};
\draw[darkgray,thick] (y) circle(0.02*\L);
\draw[black] (y) node[right,yshift=-0.5em] {TG$^\ast$};
\draw[red] (-0.25*\L,0.25*\L) node[rotate=45] {GR};
\draw[blue] (0.25*\L,-0.25*\L) node[rotate=22.5] {QFT};
\draw[black] (x) node[right,xshift=1.0em] {$\hbar$};
\draw[black] (y) node[above,yshift=1.0em] {$G_N$};
\draw[black] (z) node[left,yshift=-1.0em] {$c^{-1}$};
\draw[black] (orig) node[above,xshift=0.5em] {0};
\draw[black] (orig) node[below,xshift=0.2em] {0};
\draw[black] (orig) node[left,yshift=0.3em] {0};
\draw[black] (x) node[above,xshift=-0.6em] {$\infty$};
\draw[black] (y) node[left,xshift=0em] {$\infty$};
\draw[black] (z) node[left,yshift=0.3em] {$\infty$};
\end{tikzpicture}
\vspace*{-0.6truecm}
\end{center}
\caption{Bronstein cube with tantum gravity limit highlighted as TG and its antipodal as TG$^\ast$.}
\label{fig:1}
\end{figure}

These various limits can be captured by the Bronstein cube, depicted in Fig.~\ref{fig:1}. The three axes in the Bronstein cube are Newton's constant $G_N$, Planck's constant $\hbar$, and the inverse vacuum speed of light $c^{-1}$. Unlike traditional renderings, we have compactified all axes so that the faces, edges, and corners of the cube correspond to zero or infinite values of the corresponding coupling constants. Quantum gravity fills the volume of the cube, and each of the limiting cases corresponds to a face, a double-scaling limit to an edge, and a triple-scaling limit to a corner. This means that we have (at least) $6+12+8=26$ limiting theories of quantum gravity: the latter are listed in Table \ref{tab:1}, where the entries $0$, $1$, or $\infty$ mean that the corresponding quantity is zero, finite, or infinite.

The first nine limiting theories have common acronyms, \textit{i.e.}~GM stands for Galilean mechanics, SR for special relativity, QM for quantum mechanics, NG for Newtonian gravity, QFT for quantum field theory, GR for general relativity, GQG for Galilean quantum gravity, CM for Carroll mechanics and CQM for Carroll quantum mechanics. The limits 10-24 have no common names yet. The final two entries are tantum gravity (denoted as \textbf{TG}) and its dual limit (denoted as TG$^\ast$).

\begin{table}[hbt]
\centering
\begin{tabular}{c|ccccccccc}
& GM & SR & QM & NG & QFT & GR & GQG & CM & CQM \\\hline
$G_N$ & 0 & 0 & 0 & 1 & 0 & 1 & 1 & 0 & 0 \\
$\hbar$ & 0 & 0 & 1 & 0 & 1 & 0 & 1 & 0 & 1  \\
$c^{-1}$ & 0 & 1 & 0 & 0 & 1 & 1 & 0 & $\infty$ & $\infty$ \\\hline\hline
& 10 & 11 & 12 & 13 & 14 & 15 & 16 & 17 & 18 \\\hline
$G_N$ & $\infty$ & $\infty$ & $\infty$ & 1 & $\infty$ & 1 & 1 & $\infty$ & $\infty$ \\
$\hbar$ & $\infty$ & $\infty$ & 1 & $\infty$ & 1 & $\infty$ & 1 & $\infty$ & 1 \\
$c^{-1}$ & $\infty$ & 1 & $\infty$ & $\infty$ & 1 & 1 & $\infty$ & 0 & 0 \\\hline\hline
& 19 & 20 & 21 & 22 & 23 & 24 & TG$^\ast$ & \textbf{TG} & \\\hline
$G_N$ & 1 & 1 & $\infty$ & 0 &
0 & $\infty$ & $\infty$ & $\boldsymbol{0}$ & \\
$\hbar$ & $\infty$ & 0 & 0 & $\infty$ & $\infty$ & 0 & 0 & $\boldsymbol{\infty}$ & \\
$c^{-1}$ & 0 & $\infty$ & 1 & 1 & 
0 & $\infty$ & 0 & $\boldsymbol{\infty}$ & \\
\end{tabular}
\caption{26 limits of quantum gravity.}
\label{tab:1}
\end{table}

It can be rewarding to go through the 26 entries in Table \ref{tab:1} and specify how the limits are taken, which combinations of $G_N$, $\hbar$, and $c$ remain finite, which sector of quantum gravity this theory maintains, which physical effects can be described by the limiting theory, \textit{etc}. In this manner, one may find more than one way of taking such limits for the same entry in the Table, so there could be more than 26 limiting theories, though not of equal interest for applications. Extending the Bronstein cube by additional axes (cosmological constant $\Lambda$, Boltzmann constant $k_{\textrm{\tiny B}}$ \cite{Cohen-Tannoudji:2009hwn}, number of degrees of freedom $N$ \cite{Oriti:2018tym}, \textit{etc}.) can also be fruitful but is unnecessary for our purposes.

Here we focus on the highlighted entry in Table \ref{tab:1} or, equivalently, the corner in Fig.~\ref{fig:1} labeled ``TG,'' corresponding to infinite Planck's constant and vanishing Newton's constant and speed of light. We call this theory ``tantum gravity,'' where the Latin word ``tantum'' means ``that much!'' --- a fitting answer to the question ``quantum?''. 

However, it is not obvious that such a limit leads to a gravity theory. After all, we send Newton's constant to zero. To investigate this issue, we need to be more specific about these limits. We employ an inductive approach.

Our main motivation comes from black hole thermodynamics, which lies at the heart of many quantum gravity mysteries, including the information paradox, black hole microstates, and the holographic principle (see, \textit{e.g.}, \cite{'tHooft:1993gx,Susskind:1995vu,Maldacena:1997re,Gubser:1998bc,Witten:1998qj,Ryu:2006bv,Mathur:2009hf,Raju:2020smc,Almheiri:2020cfm} and references therein). Thus, we check under which conditions the Schwarzschild radius, entropy, temperature, and energy remain finite. 

Let us start with the latter. For a given black hole of mass $M$, its energy $E$ is given by $E=Mc^2$. Thus, we leave $c$ and $M$ finite or scale $M$ inversely to $c^2$. Written in terms of energy, the formula on Hawking's tombstone (with Boltzmann's constant set to one) 
\begin{equation}
    T = \frac{\hbar\,c^5}{8\pi\,G_N\,E} 
    \label{eq:tg10}
\end{equation}
shows that any single scaling limit in the Bronstein cube leads to infinite or zero Hawking temperature. Since we want to avoid this scenario, we deduce that our desired theory can only live at the edges or corners of the Bronstein cube and not at one of its faces. This eliminates, in particular, GR and QFT. 

At this stage, we still have many possibilities for double- and triple-scaling limits that maintain finite Hawking temperature (for instance, the Newton gravity limit $\hbar\to 0$, $c\to\infty$). Finiteness of the Bekenstein--Hawking entropy does not add any new conditions due to the Smarr formula $E=2TS$ or the first law $\delta E = T\,\delta S$, see for instance \cite{Bardeen:1973gs}.

Finiteness of the Schwarzschild radius $\rad_{\textrm{\tiny S}}=2G_N\,E/c^4$ imposes a second constraint, namely finiteness of $G_N/c^4$, implying that $\hbar\,c$ needs to remain finite to maintain finite Hawking temperature \eqref{eq:tg10}. This condition removes all the edges from the Bronstein cube and leaves us with the two marked points in Fig.~\ref{fig:1}, one of which is the tantum gravity limit $G_N,c\to 0$, $\hbar\to\infty$ and the other its dual version where all limits are taken oppositely. In the Bronstein cube, the latter corresponds to the antipodal point labeled as TG$^\ast$ in Fig.~\ref{fig:1}~\footnote{%
We keep finite the Planck length and energy in both limits but send the Planck time and mass to $\infty$ in tantum gravity and to $0$ in TG$^{\ast}$.
}.

We stress that the two points TG and TG$^\ast$ are the only points on the whole Bronstein cube (besides its quantum gravity interior) where all three thermodynamic quantities and the Schwarzschild radius are finite. So from the perspective of black hole thermodynamics, the tantum gravity limit is unique, up to dualization.

In summary, to obtain tantum gravity from quantum gravity, we have to take the Carroll limit $c\to 0$ and retain finite combinations
\begin{equation}
    G_M := G_N\, c^{-4}\qquad\qquad \kappa :=   \hbar\,c~ .
    \label{eq:tg1}
\end{equation}
The first one is the gravitational coupling constant in front of the Einstein--Hilbert action. So it is not automatically true that sending $G_N\to 0$ removes gravity --- it is still there if simultaneously the speed of light is sent to zero while keeping $G_M$ fixed, which is precisely what is done to reach the so-called magnetic Carroll gravity limit \cite{Hansen:2021fxi,Campoleoni:2022ebj}, hence the notation $G_M$. We shall describe this theory in more detail in the coming subsections. This elementary observation justifies calling our limit theory ``gravity.'' Similarly, the limit $\hbar\to\infty$ does not automatically imply a breakdown of the semi-classical approximation since we still have two combinations of the coupling constants \eqref{eq:tg1} that could be either large or small. We shall demonstrate this explicitly in section \ref{sec:tant_partitionfunction}. Thus, one should consider the points in the Bronstein cube labeled as TG and TG$^\ast$ as two-dimensional planes spanned by the coupling constants $G_M$ and $\kappa$.

An efficient way to parametrize the triple scaling limit leading to TG is to rescale 
\begin{equation}
c\to c\,\epsilon\qquad\qquad\hbar\to\hbar/\epsilon\qquad\qquad G_N\to G_N\,\epsilon^4
\label{eq:whynot}
\end{equation}
and then take the limit $\epsilon\to0$. This has the added advantage that the limiting parameter is dimensionless, and we shall use this procedure implicitly below. The parameter $\epsilon$ drops out in the TG coupling constants \eqref{eq:tg1}. 

By construction, the tantum gravity limit not only leads to a gravity theory, but the limiting theory also contains black-hole-like states with finite entropy, temperature, and energy, related to each other by the first law. At first glance, this is again surprising since we take a Carroll limit, $c\to 0$, for which the lightcone collapses \cite{Levy1965, SenGupta1966OnAA} and makes obsolete all notions of horizons. However, as we have seen above, one has to be careful with naive arguments based on singular limits. Indeed, there can be ``Carroll black holes,'' though one needs to extend the notion of black holes and avoid basing them on event horizons. Instead, they are defined by their thermal properties and the existence of a so-called Carroll extremal surface, which is reminiscent of the bifurcation sphere of the Schwarzschild black hole \cite{Ecker:2023uwm}. The details of these definitions shall be spelt out in chapter \ref{ch:CBHs}.

From a quantum gravity perspective, it is an asset that we are forced to go beyond the classical notion of a black hole in tantum gravity, since quantum black holes should not be defined in terms of event horizons either --- after all, quantum black holes evaporate \cite{Hawking:1974rv,Hawking:1975vcx}, so event horizons are artifacts of the classical approximation. This is one of several ways in which tantum gravity is closer to quantum gravity than classical gravity.

\section{General relativity in the limit \texorpdfstring{$c\to 0$}{ctozero}}\label{sec:GRforctozero}
Tantum gravity admits a semi-classical description in terms of gravitational theories based on Carroll symmetries. In this section, we shall give an overview on how to arrive at these by taking appropriate scaling limits of Einstein gravity. Just like for the matter theories discussed in the last chapter, there are two distinct ways to take the limit $c\to 0$. One corresponds to the scaling \eqref{eq:whynot} and is often termed ``magnetic'' after the corresponding limit in electrodynamics. The other ``electric'' limit also leads to a theory of Carroll gravity but does not maintain a well-defined notion of thermodynamics. In the literature one finds essentially two equivalent approaches for deriving these two limits. Either one starts with the Einstein--Hilbert action and does a manifestly covariant expansion for small $c$ \cite{Hansen:2021fxi}, or one starts with the Hamiltonian formulation of general relativity and takes the limit there \cite{Henneaux:2021yzg}. Since the first path is slightly more involved, we shall present the second one here. The action of general relativity reads \cite{Arnowitt:1962hi,Arnowitt:1960es}
\begin{align}
    I_{\text{\tiny ADM}}[g_{ij},\pi ^{ij},N,N^i]=\int \dd t\, \dd ^dx\; \big(\pi ^{ij}\dot{g}_{ij}-N\mathcal{H}-N^i\mathcal{H}_i\big)
\end{align} 
where
\begin{align}
    \mathcal{H}&=\frac{16\pi G_N }{c^2}G_{ijkl}\pi ^{ij}\pi ^{kl}-\frac{c^4\sqrt{g}}{16\pi G_N}(R^{(d)}[g]-2\Lambda ) \\
    \mathcal{H}_i&=-2\hat{\nabla }_j\pi ^j{}_i
\end{align}
are Hamiltonian constraint and the momentum constraints, respectively. $\hat{\nabla }$ denotes the Levi--Civitá connection associated to the $d$-dimensional metric $g_{ij}$ on a given $t=\text{cst.}$ hypersurface and $t$-derivatives are denoted by a dot. To emphasize the parallels with the simple matter models of the previous section we introduced the supermetric
\begin{align}
    G^{ijmn}=\sqrt{g}\big(g^{i(m}g^{n)j}-g^{ij}g^{mn}\big) && G_{ijmn}=\frac{1}{\sqrt{g}}\Big(g_{i(m}g_{n)j}-\frac{1}{d-1}g_{ij}g_{mn}\Big)
\end{align}
which satisfies $G_{ijkl}G^{ijmn}=\delta _{(k}^m\delta _{l)}^n$. It becomes evident that the Hamiltonian constraint consists of two parts, a term quadratic in $\pi ^{ij}$ reminiscent of a kinetic term, as well as a momentum independent term that behaves like a potential. The two different Carroll limits described above each suppress one of these terms. However, they both agree in the way they change the constraint algebra. Indeed, from the fundamental equal-time Poisson bracket
\begin{align}
    \{g_{ij}(x),\pi ^{mn}(y)\}=\delta _{(i}^m \delta _{j)}^n \delta ^d(x-y)
\end{align}
one deduces 
\begin{align}
    \{\mathcal{H}(x),\mathcal{H}(y)\}&=c^2\big(g^{ij}(x)\mathcal{H}_i(x)+g^{ij}(y)\mathcal{H}_i(y)\big)\partial _j \delta ^d(x-y) \\
    \{\mathcal{H}(x),\mathcal{H}_i(y)\}&=\mathcal{H}(y)\partial _i \delta ^d(x-y) \\
    \{\mathcal{H}_i(x),\mathcal{H}_j(y)\}&=\big(\mathcal{H}_i(y)\partial _j +\mathcal{H}_j(x)\partial _i \big)\delta ^d(x-y)
\end{align}  
such that in the limit $c\to 0$ one finds the algebra of first class constraints of Carroll gravity,
\begin{equation}\label{eq:Carroll_FCC_algebra}
  \addtolength{\fboxsep}{7pt}
   \boxed{
\begin{aligned}
    \{\mathcal{H}(x),\mathcal{H}(y)\}&=0\\
    \{\mathcal{H}(x),\mathcal{H}_i(y)\}&=\mathcal{H}(y)\partial _i \delta ^d(x-y) \\
    \{\mathcal{H}_i(x),\mathcal{H}_j(y)\}&=\big(\mathcal{H}_i(y)\partial _j +\mathcal{H}_j(x)\partial _i \big)\delta ^d(x-y)
\end{aligned}
   }
\end{equation} 
Since the scaling of the gravitational constant does not enter these brackets it follows that this result holds both for the electric and the magnetic theory. 

Before we turn to these two examples, let us remark that once such a theory of Carroll gravity is coupled to matter in a covariant way, all of these constraints acquire contributions from the matter field, i.e., we have $\mathcal{H}=\mathcal{H}_{\textrm{grav}}+\mathcal{H}_m$ and likewise for $\mathcal{H}_i$. The algebra of first class constraints \eqref{eq:Carroll_FCC_algebra} is, however, still satisfied. This is analogous to the Lorentzian case as discussed, e.g., in \cite{Kuchar1977GeometrodynamicsIV} but does not in general imply that the matter parts alone satisfy \eqref{eq:Carroll_FCC_algebra} since on curved backgrounds there might be mixed brackets $\{\mathcal{H}_{\textrm{grav}},\mathcal{H}_m\}$ which contribute in a nontrivial way. Typically, this happens when the matter field couples to the canonical gravitational momenta $\pi ^{ij}$.

\subsection{Electric Carroll gravity}
The electric limit of general relativity is obtained by keeping fixed the combination $G_E=c^{-2}G_N$. As mentioned previously, this is in conflict with the scaling that maintains finite energy, temperature, and Schwarzschild radius and therefore does not describe a valid semi-classical description of tantum gravity. The Hamiltonian action reduces to
\begin{align}
    I_E[g_{ij},\pi ^{ij},N,N^i]=\int \dd t \, \dd ^d x\; \Big(\pi ^{ij}\dot{g}_{ij}-N\mathcal{H}_E-N^i\mathcal{H}_i\Big)
\end{align}
where the momentum constraints are unchanged, and the Hamiltonian constraint reads
\begin{align}
    \mathcal{H}_E&=16\pi G_E \,G_{ijkl}\pi ^{ij}\pi ^{kl} ~.
\end{align}
In consistency with \eqref{eq:Carroll_FCC_algebra}, this can be checked to satisfy $\{\mathcal{H}_E,\mathcal{H}_E\}=0$. This action was in fact the first instance of a theory of Carroll gravity going back to the original work \cite{Henneaux:1979vn}. Because of its ultra-local dynamics, it was used to model spacetime close to a singularity in, e.g., a black hole or the big bang \cite{Oling:2024vmq}. It also has a covariant second order form, which is obtained by integrating out $\pi ^{ij}$. The corresponding solution is
\begin{align}
    \pi ^{ij}=\frac{1}{16\pi G_E}G^{ijmn}K_{mn} && K_{mn}=\frac{1}{2N}\big(\dot{g}_{mn}-2\hat{\nabla }_{(m}N_{n)}\big)
\end{align}
and upon reinsertion into the action one finds
\begin{align}
    I_E[g_{ij},N^i,N]=\frac{1}{16\pi G_E}\int \dd t\, \dd ^d x\,\sqrt{g}N\big(K^{ij}K_{ij}-K^2\big) ~.
\end{align}
Finally, we reorganize the ADM variables in terms of the Carroll metric fields as
\begin{align}\label{eq:Carr_ADM_variables}
    v^\mu =\Big(-\frac{1}{N},\frac{N^i}{N}\Big) && \tau _\mu = (N,0) && h_{\mu \nu }=\begin{pmatrix}
        N^iN_i & N_i \\
        N_i & g_{ij}
    \end{pmatrix}
\end{align}
such that $K_{\mu \nu }$ is just the extrinsic curvature of the Carroll structure, $K_{\mu \nu }=-\frac12 \mathcal{L}_v h_{\mu \nu }$. The action can now be written in covariant form \cite{Henneaux:1979vn,Henneaux:2021yzg}
\begin{equation}\label{eq:electric_action_2nd}
  \addtolength{\fboxsep}{7pt}
   \boxed{
\begin{aligned}
   I_{2^{\textrm{nd}}}[h_{\mu \nu },\tau _\mu ]=\frac{1}{16\pi G_E}\int \dd ^{d+1}x \,\mathfrak{e}\, \big(K^{\mu \nu }K_{\mu \nu }-K^2\big) 
\end{aligned}
   }
\end{equation} 
which is equivalently obtained from a direct $c\to 0$ limit of the Einstein--Hilbert action \cite{Hansen:2021fxi} or from a gauging perspective \cite{Hartong:2015xda}. It is functionally dependent on the data $h_{\mu \nu }$ and $\tau _\mu $ which are the independent variables that allow to fix $v^\mu $ and $h^{\mu \nu }$ uniquely by orthonormality and completeness relations. 

Clearly, \eqref{eq:electric_action_2nd} is manifestly invariant under local Carroll boosts. One may wonder why these gauge transformations did not appear as an additional set of first class constraints, besides the Hamiltonian and momentum constraints. Doing a Legendre transformation of \eqref{eq:electric_action_2nd} indeed reveals additional first class constraints $\pi ^i\approx 0$, where $\pi ^i$ are the canonical momenta associated to the components $\tau _i$. Since they generate precisely those transformations that shift $\tau _i$, one can see that the Hamiltonian action of electric Carroll gravity is just a gauge-fixed version of the full theory where the gauge condition $\tau _i=0$ chooses a specific Carroll boost-frame. At the same time, this ensures that the space transverse to $v^\mu $ is tangent to the $t=\text{cst.}$ surfaces. These surfaces then correspond to the integrated distribution generated by $\tau $ (see also \eqref{eq:Frobenius}).

Finally, we remark that there is presently no known way to describe electric Carroll gravity within the framework of the Cartan formalism. There does exist a first order formulation in terms of Cartan-like variables \cite{Pekar:2024ukc} which, however, needs to assume a different action of Carroll boosts than dictated by Cartan geometry. 

\subsection{Magnetic Carroll gravity}\label{sec:mag_carr}
Applying the scaling \eqref{eq:whynot} the $\epsilon\to 0$ limit leads to the Hamiltonian action of magnetic Carroll gravity 
\begin{align}\label{eq:mag_Hamil}
    I_{M}[g_{ij},\pi ^{ij},N,N^i]=\int \dd t \, \dd ^d x \, \Big(\pi ^{ij}\dot{g}_{ij}-N\mathcal{H}_M-N^i\mathcal{H}_i\Big)
\end{align}
where the Hamiltonian constraint reads
\begin{align}
    \mathcal{H}_M=-\frac{\sqrt{g}}{16\pi G_M}(R^{(d)}[g]-2\Lambda ) ~.
\end{align}
Since it does not depend on the canonical momenta, the vanishing of the Poisson bracket with itself is trivial. The canonical momenta in this case act as Lagrange multipliers for the constraint 
\begin{align}
    \dot{g}_{ij}-2\hat{\nabla }_{(i}N_{j)}=2NK_{ij}\approx 0
\end{align}
which sets the extrinsic curvature to zero. We now show that \eqref{eq:mag_Hamil} is equivalent to an action obtained from a Cartan perspective by gauging a Klein pair associated to the Carroll (A)dS algebra as in section \ref{sec:gauging_procedure}. Following \cite{Campoleoni:2022ebj}, the Cartan action reads
\begin{align}\label{eq:Carroll_Cartan_action}
    I_{1^{\mathrm{st}}}[\tau ,e^a,\omega ^a,\omega ^{ab}]=\frac{1}{16\pi G_M}\int \dd ^{d+1}x\, \mathfrak{e}\,\big(\mathcal{R}-2\Lambda \big)
\end{align} 
where $\mathcal{R}$ is the invariant curvature scalar defined in \eqref{eq:curv_scalar}.
One first needs to integrate out the connection components by their equations of motion. The latter are computed most easily in a formulation in terms of differential forms where 
\begin{align}\label{eq:blub}
    I_{1^{\mathrm{st}}}=\frac{1}{16\pi G_M (d-2)!}\int \Big(\frac{2}{d-1}\Omega^{a_1}\wedge e^{a_2}
    +\tau \wedge \Omega ^{a_1a_2} \Big)\wedge e^{a_3}\wedge \dots \wedge e^{a_d}\,\epsilon _{a_1...a_d}
\end{align}
with the orientation chosen such that $\dd ^{d+1}x\, \mathfrak{e} = \tau \wedge e^1\wedge \dots \wedge e^d$. 
Varying \eqref{eq:blub} with respect to the connections yields the equations
\begin{align}
    \delta \omega ^a&: & T^{[a_1}\wedge e^{a_2}\wedge \cdots \,\wedge  e^{a_{d-1}]}&=0\\
    \delta \omega ^{ab}&: & \Big(T\wedge e^{[a_1}-(d-2)T^{[a_1}\wedge \tau \Big)\wedge e^{a_2}\wedge \cdots \, \wedge e^{a_{d-2}]}&=0
\end{align}
and by expanding the various torsion components in terms of the coframe one can show that they imply $T^a=0=T$. We now follow the considerations of section \ref{sec:intrin_torsion} and solve only the part of these equations that depends on the connection, leaving the part associated to intrinsic torsion arbitrary. This makes $\omega ^a$ and $\omega ^{ab}$ dependent variables given in terms of the coframe and the arbitrary symmetric tensor $C^{ab}$ by \eqref{eq:om1}, \eqref{eq:om2}. Inserting this back into the curvature scalar and using \eqref{eq:curv_scalar} yields 
\begin{align}
    \mathcal{R}=\mathcal{R}\big \vert _{C_{\mu\nu}=0}+2\big(C^{\mu \nu }-h^{\mu \nu }C\big)K_{\mu \nu }-\frac{2}{\mathfrak{e}}\partial _\mu (\mathfrak{e}v^\mu C) 
\end{align}
where we pulled out the part that depends on $C_{\mu \nu }$ with $C=h^{\mu \nu }C_{\mu \nu }$, and the affine connection in the curvature is given by \eqref{eq:gen_affine_connection}. Thus, the action of magnetic Carroll gravity in a second order form reads up to boundary terms \cite{Bergshoeff:2017btm}
\begin{equation}\label{eq:second_order_mag_action}
  \addtolength{\fboxsep}{7pt}
   \boxed{
\begin{aligned}
    I&_{2^{\textrm{nd}}}[\tau _\mu ,\,h_{\mu \nu },C_{\mu \nu }]=\\[.5em]
    &\frac{1}{16\pi G_M}\int \dd ^{d+1}x\, \mathfrak{e}\, \Big(h^{\mu \nu }\big(R_{\mu \nu }-v^\rho \tau_\sigma R^\sigma {}_{\mu \rho \nu }\big)\big \vert _{C^{\mu \nu }=0} +2\big(C^{\mu \nu }-h^{\mu \nu }C\big)K_{\mu \nu }-2\Lambda \Big)
\end{aligned}
   }
\end{equation} 
 One can check that, again up to boundary terms, this action is invariant under local Carroll boosts and diffeomorphisms, with the former acting as 
\begin{align}
    \delta _\lambda I_{2^{\textrm{nd}}}=\frac{1}{8\pi G_M}\int \dd ^{d+1}x \, \partial _\mu \big(\mathfrak{e}v^\mu\partial _\lambda C-\mathfrak{e}\lambda ^\mu K \big)
\end{align}
where we used \eqref{eq:c_minimal_trafo}. The equivalence to the Hamiltonian form now proceeds similarly to the electric case. One fixes a boost frame as well as an ADM-like split for the geometry given by \eqref{eq:Carr_ADM_variables} such that
\begin{align}\label{eq:int_blab}
    I_{2^{\textrm{nd}}}[N,N^i,g_{ij},C_{ij}]=\frac{1}{16\pi G_M}\int \dd t\, \dd ^{d}x\, \sqrt{g}N\,\big(R^{(d)}[g]-2\Lambda +2(C^{ij}-g^{ij}C)K_{ij}\big) ~.
\end{align} 
Finally, after renaming 
\begin{align}\label{eq:renaming_C}
    \pi ^{ij}=\frac{\sqrt{g}}{16\pi G_M}(C^{ij}-g^{ij}C)
\end{align}
it is straightforward to show that \eqref{eq:int_blab} reduces to the Hamiltonian action \eqref{eq:mag_Hamil}, again up to a total derivative. For more details on this computation we refer to \cite{Campoleoni:2022ebj}.

\section{Two-dimensional Carroll dilaton gravity}\label{sec:2d_dil}
We now come to two-dimensional Carroll gravity theories, which shall be our main tool for describing Carroll black holes. Similarly to the Lorentzian case (see e.g. \cite{Grumiller:2002nm}), there is not just one of these theories but rather an infinite number of different models. Some of them indeed describe a subsector of magnetic Carroll gravity in $d+1$ dimensions obtained by imposing spherical symmetry. We shall come to this class of spherically reduced models and how to construct them in subsection \ref{sec:spher-reduct-magn}. Others have to be regarded as intrinsically two-dimensional but may also provide interesting examples for Carroll black holes. What they all have in common is that their dynamical field content does not only consist of the Carroll metric data but also of additional scalar fields such as the dilaton $X$ in order to make the actions non-trivial. The resulting theories are of topological type in the sense that they do not have any local propagating degrees of freedom.

\subsection{BF theories}\label{sec:BF_theories}
The simplest class of 2d Carroll dilaton gravity theories can be written as certain two-dimensional gauge theories called BF theories. Their field content consists of a Cartan connection $A$ (see section \ref{sec:cartan_geometry}) for a specific Klein pair $(\mathfrak{g},\mathfrak{h})$ as well as a scalar field $\mathcal{X}$ taking values in the dual of $\mathfrak{g}$ with an action given by
\begin{align}\label{eq:bf_action}
    I_{\mathrm{BF}}[A,\mathcal{X}]=\frac{k}{2\pi }\int \, \mathcal{X}(F) = \frac{k}{2\pi }\int \, X^IF_I ~.
\end{align}
Here, $F$ is the curvature two-form of $A$ and $k$ a coupling constant. For the second equality we chose a basis $\{f^I\}$ for $\mathfrak{g}$ with $[f^I,f^J]=c^{IJ}{}_Kf^K$ such that, e.g., $A=A_{I_\mu }f^I\dd x^\mu $. The action \eqref{eq:bf_action} is gauge invariant under $\mathfrak{g}$-valued transformations $\lambda =\lambda _If^I$ which act as
\begin{align}\label{eq:2d_gen_trafo}
    \delta _\lambda A=\dd \lambda +[A,\lambda ] && \delta _\lambda \mathcal{X}(\cdot )=-\mathcal{X}([\cdot ,\lambda]) 
\end{align}
and the equations of motion from varying with respect to $\mathcal{X}$ and $A$ are 
\begin{subequations}\label{eq:BF_EOM}
\begin{align}
    \delta X^I:& \qquad F_I=0\\
    \delta A_I:& \qquad \dd X^I+c^{IJ}{}_KX^KA_J=0 ~.
\end{align}
\end{subequations}
The first of these implies that all the solutions to these models are locally isomorphic to the homogeneous spacetime associated to the Klein pair (see Theorem \ref{th:fund_th}). Let us further specify these geometries by taking $\mathfrak{g}$ to be the 2d AdS Carroll algebra (see section \ref{sec:Ads_carroll_algebra}) with generators $\{B,H,P\}$ satisfying the non-zero brackets
\begin{align}
    [B,P]=H && [H,P]=\frac{1}{\ell ^2}B ~.
\end{align}
The subalgebra $\mathfrak{h}$ is the homogeneous Carroll algebra spanned by $B$. We write the components as
\begin{align}
    A_I=(\omega ,\tau ,e) && X^I=(X,\XH,\XP) && \lambda _I=(\lambda _{\textrm{\tiny B}},\lambda_{\textrm{\tiny H}}, \lambda_{\textrm{\tiny P}}) 
\end{align}
such that the action becomes
\begin{align}
    I_{\mathrm{BF}}=\frac{k}{2\pi }\int X\,\dd\omega+\XH\,\big(\dd\tau+\omega\wedge e\big)+\XP\,\dd e + \frac{X}{\ell ^2}\,\tau\wedge e  
\end{align}
which is referred to as the Carroll--Jackiw--Teitelboim (CJT) model \cite{Grumiller:2020elf,Gomis:2020wxp}.
The $1$-forms are spatial einbein $e$,
temporal einbein $\tau$ and Carroll boost connection $\omega$. The
scalar fields are dilaton $X$ and Lagrange multipliers for torsion
constraints $\XH$, $\XP$. This identification of variables gives the BF model the interpretation of a Carroll gravitational theory. Indeed, comparing with the cases from section \ref{sec:GRforctozero} it is of ``magnetic'' type as a consequence of the constraint generated by $\XP$ setting the intrinsic torsion to zero. What remains is to show that the gauge transformations can be interpreted as internal local Carroll boosts together with spacetime diffeomorphisms. For this we follow the discussion around \eqref{eq:general_diffplusgauge_trafo} and take $\lambda =\lambda _{\textrm{\tiny B}} B$. Then, with $\Sigma =\lambda +i_\xi A$, the action is left invariant by the transformations 
\begin{subequations}\label{eq_mod_BF_Trafo}
\begin{align}
    \bar{\delta }_{(\lambda ,\xi )}A_I&=\dd \Sigma _I+c^{JK}{}_I A_J\Sigma _K +i_\xi F_I \\
    \bar{\delta }_{(\lambda ,\xi )}X^I&=-c^{IJ}{}_KX^K\Sigma _J+i_\xi (\dd X^I+c^{IJ}{}_KX^KA_J)
\end{align}
\end{subequations}
as a consequence of 
\begin{align}
    \bar{\delta }_{(\lambda ,\xi )}A_I=\mathcal{L}_\xi A_I +\delta _\lambda A_I && \bar{\delta }_{(\lambda ,\xi )}X^I=\mathcal{L}_\xi X^I +\delta _\lambda X^I ~.
\end{align}
But, since the rightmost terms in \eqref{eq_mod_BF_Trafo} are just the equations of motion \eqref{eq:BF_EOM} themselves, one can see that on-shell there is no difference between the $\delta $ and $\bar{\delta }$ transformations. Thus, the parameters $\lambda_{\textrm{\tiny H}}$ and $\lambda_{\textrm{\tiny P}}$ directly generate spacetime diffeomorphisms for the choices $(\lambda_{\textrm{\tiny H}},\lambda_{\textrm{\tiny P}})= (i_\xi\tau ,i_\xi e)$. This should be put in contrast to the higher-dimensional Cartan actions where $\bar{\delta }$- and $\delta $-transformations typically do not differ by terms that are proportional to the equations of motion, see \eqref{eq:general_diffplusgauge_trafo}.

Clearly, having $F=0$ as an equation of motion is quite restrictive, and it is desirable to generalize these models. We shall come to this in the next subsection.

\subsection{First-order formulation}
\label{sec:2.1.1}
Generic 2d Carroll dilaton gravity was first constructed in \cite{Grumiller:2020elf} and includes BF-type theories as a special case. The 2d Carroll dilaton gravity bulk action is given by \cite{Grumiller:2020elf}
\begin{align}\label{eq:car24}
    I_{1^{\mathrm{st}}}[\omega,\,\tau,\,e,\,X,\,\XH,\,\XP]=\frac{k}{2\pi}\,\int _{\mathcal{M}}{\mathcal{L}}
\end{align}
on a 2d manifold $\mathcal{M}$ with coupling $k$ and the Lagrange-2-form
\begin{equation}
  \label{eq:car1}
\boxed{
\phantom{\Bigg(}
{\mathcal{L}} = X\,\dd\omega+\XH\,\big(\dd\tau+\omega\wedge e\big)+\XP\,\dd e + {\mathcal{V}}(X,\,\XH)\,\tau\wedge e
\phantom{\Bigg(}}
\end{equation}
where the potential ${\mathcal{V}}(X,\,\XH)$ is an arbitrary function. Whenever it is linear in either of its arguments the theory reduces to a BF-type model. The
composite 2-forms are curvature $\Omega =\dd\omega$, torsion
$T=\dd\tau+\omega\wedge e$, and intrinsic torsion $T_{\text{int}} =\dd e$
where the latter is defined as the part of the torsion independent of
the boost connection. Just as in the BF case, we summarily refer to the scalar fields as
$X^I=(X,\XH,\XP)$ and to the $1$-forms as $A_I=(\omega,\tau,e)$.

The Lagrange-2-form \eqref{eq:car1} [and hence also the action
\eqref{eq:car24}] is invariant under local Carroll boosts 
\begin{subequations}
\label{eq:car25}
\begin{align}
 \delta_{\lambda _{\textrm{\tiny B}}} X &= 0 &  \delta_{\lambda _{\textrm{\tiny B}}} \XH &= 0 &  \delta_{\lambda _{\textrm{\tiny B}}} \XP &= \XH\,\lambda _{\textrm{\tiny B}} \\
 \delta_{\lambda _{\textrm{\tiny B}}} \omega &= \dd\lambda _{\textrm{\tiny B}} &  \delta_{\lambda _{\textrm{\tiny B}}}\tau &= -e\,\lambda _{\textrm{\tiny B}} &  \delta_{\lambda _{\textrm{\tiny B}}} e &= 0 \, .
\end{align}
\end{subequations}
The transformations \eqref{eq:car25} show that the dilaton $X$, the
field $\XH$ and the spatial einbein $e$ are Carroll boost invariant,
while the 1-form $\omega$ is the Carroll boost connection. The
non-invariances of the temporal einbein $\tau$ and the scalar $\XP$
conspire such that the sum of the middle two terms in the
Lagrange-2-form is Carroll boost-invariant. To simplify the notation a bit from now on, we shall denote Carroll boosts just by $\delta _\lambda := \delta_{\lambda _{\textrm{\tiny B}}}$ and make it explicit everywhere where something else is implied. 

Moreover, the action \eqref{eq:car24} is invariant under two
additional gauge symmetries $\lambda_{\textrm{\tiny H}}$ and
$\lambda_{\textrm{\tiny P}}$ (we define
$\partial_X:=\partial/\partial X$ and
$\partial_{\textrm{\tiny H}}:=\partial/\partial\XH$)
\begin{subequations}
\label{eq:car26}
\begin{align}
 \delta_{\lambda_{\textrm{\tiny H}}} X            & = 0                                    & \delta_{\lambda_{\textrm{\tiny H}}} \XH  & = 0  & \delta_{\lambda_{\textrm{\tiny H}}} \XP & = {\mathcal{V}}\,\lambda_{\textrm{\tiny H}} \\
 \delta_{\lambda_{\textrm{\tiny H}}} \omega       & = -(\partial_X {\mathcal{V}})\,e\lambda_{\textrm{\tiny H}}   & \delta_{\lambda_{\textrm{\tiny H}}} \tau & = \dd\lambda_{\textrm{\tiny H}}-(\partial_{\textrm{\tiny H}} {\mathcal{V}})\,e\lambda_{\textrm{\tiny H}} & \delta_{\lambda_{\textrm{\tiny H}}} e   & = 0      
\end{align}
\end{subequations}
and 
\begin{subequations}
\label{eq:car26a}
\begin{align}
 \delta_{\lambda_{\textrm{\tiny P}}} X            & = -\XH\,\lambda_{\textrm{\tiny P}}                         & \delta_{\lambda_{\textrm{\tiny P}}} \XH  & = -{\mathcal{V}}\,\lambda_{\textrm{\tiny P}}                                              & \delta_{\lambda_{\textrm{\tiny P}}} \XP & = 0                   \\
 \delta_{\lambda_{\textrm{\tiny P}}} \omega       & = (\partial_X {\mathcal{V}})\,\tau\lambda_{\textrm{\tiny P}} & \delta_{\lambda_{\textrm{\tiny P}}} \tau & = \omega\,\lambda_{\textrm{\tiny P}} + (\partial_{\textrm{\tiny H}} {\mathcal{V}})\,\tau\lambda_{\textrm{\tiny P}}             & \delta_{\lambda_{\textrm{\tiny P}}} e   & = \dd\lambda_{\textrm{\tiny P}}
\end{align}
\end{subequations}
On-shell they generate diffeomorphisms along a vector field $\xi^\mu$ by virtue of the relations\footnote{Additionally, we need a compensating Carroll boost generated by $\lambda = \omega_{\mu}\,\xi^\mu$.} $\lambda_{\textrm{\tiny H}} = i_\xi \tau_{\mu}\,$ and $\lambda_{\textrm{\tiny P}} = i_\xi e_{\mu}$:
\begin{subequations}
\label{eq:car27}
 \begin{align}
  \delta_\xi X                   & \approx \xi^\mu\partial_\mu X          & \delta_\xi \omega_\mu   & \approx \xi^\nu\partial_\nu\omega_\mu + \omega_\nu\partial_\mu\xi^\nu \\
  \delta_\xi \XH                 & \approx \xi^\mu\partial_\mu \XH        & \delta_\xi \tau_\mu     & \approx \xi^\nu\partial_\nu\tau_\mu + \tau_\nu\partial_\mu\xi^\nu \\
  \delta_\xi \XP                 & \approx \xi^\mu\partial_\mu \XP        & \delta_\xi e_\mu        & \approx \xi^\nu\partial_\nu e_\mu + e_\nu\partial_\mu\xi^\nu 
 \end{align}
\end{subequations}
The Lie variations above follow from the gauge symmetries together
with the Carroll equations of motion displayed below in
\eqref{eq:eom} ($\approx$ denotes on-shell equivalence).

\subsubsection{Equations of motion}
\label{sec:2.2.1}

Varying the action \eqref{eq:car24} with respect to all fields yields the equations of motion.
\begin{subequations}
\label{eq:eom}
\begin{align}
&\delta X&\textrm{Carroll\;curvature:} &&& \Omega =\dd\omega = - \partial_X{\mathcal{V}}(X,\,\XH)\,\tau\wedge e \label{eq:1} \\
&\delta\XH&\textrm{Carroll\;torsion:} &&& T=\dd\tau + \omega\wedge e = -\partial_{\textrm{\tiny H}} {\mathcal{V}}(X,\,\XH)\,\tau\wedge e \label{eq:2}\\
&\delta\XP&\textrm{No\;intrinsic\;torsion:} &&& T_{\mathrm{int}} =\dd e  = 0 \label{eq:3}\\
 &\delta\omega&\textrm{Carroll\;metric:} &&& \dd X + \XH\,e = 0 \label{eq:4}\\
&\delta\tau&\textrm{Carroll\;Casimir:} &&& \dd \XH + {\mathcal{V}}(X,\,\XH)\,e = 0 \label{eq:5}\\
&\delta e&\textrm{Auxiliary\;field:} &&& \dd \XP -{\mathcal{V}}(X,\,\XH)\,\tau-\XH\,\omega = 0 \label{eq:6}
\end{align}
\end{subequations}
The first equation \eqref{eq:1} determines the Carroll curvature, which generally is non-zero but trivially vanishes whenever the potential is independent of the dilaton field. The second equation \eqref{eq:2} shows that on-shell Carroll torsion vanishes whenever the potential is independent of $\XH$. The third equation \eqref{eq:3} reveals that there is never intrinsic torsion, regardless of how the potential is chosen. The fourth equation~\eqref{eq:4} allows algebraically determining the spatial einbein (and hence the Carroll metric) in terms of the Carroll boost invariant scalars, $X$ and $\XH $. The fifth equation \eqref{eq:5} entails a conserved Casimir function, which we shall uncover below when discussing linear dilaton vacua. The final equation \eqref{eq:6} allows determining the auxiliary field $\XP$ in terms of the potential ${\mathcal{V}}(X,\,\XH)$ and the geometric data extracted from the other five equations of motion or, alternatively, if $\XP$ is gauge fixed suitably it provides an algebraic constraint relating $\tau$ and $\omega$.

It can be useful to map solutions of different models to each other if they can be related by suitable Weyl rescalings. Therefore, consider a dilaton-dependent Weyl rescaling, parametrized by $\alpha$, of the Carroll metric
\begin{equation}
  \label{eq:weyl1}
    e\to\tilde e = e\, e^{\alpha(X)}
\end{equation}
that leaves invariant the dilaton, $X\to \tilde X = X$. Such Weyl
rescalings are compatible with the absence of intrinsic torsion (this
would not be the case if the Weyl factor did depend on time).
Consistency with Carroll boosts demands that also $\tau$ scales in the
same way as $e$,
\eq{
\tau\to\tilde\tau = \tau\, e^{\alpha(X)}\,.
}{eq:weyl2}
The Carroll metric equation \eqref{eq:4} implies that $\XH$ transforms inversely to $e$,
\eq{
\XH\to\tilde\XH = \XH\, e^{-\alpha(X)}
}{eq:weyl3}
and consistency with Carroll boosts demands the same scaling for $\XP$.
\eq{
\XP\to\tilde\XP = \XP\, e^{-\alpha(X)}
}{eq:weyl4}
The Carroll Casimir equation \eqref{eq:5}
\eq{
\dd\XH+{\mathcal{V}}(X,\,\XH)\, e = 0 \qquad\leftrightarrow\qquad\dd\tilde\XH+\tilde {\mathcal{V}}(\tilde X,\,\tilde \XH)\,\tilde e = 0
}{eq:weyl5}
establishes the transformation behaviour of the potential
\eq{
\mathcal{V}(X,\,\XH)\to\tilde {\mathcal{V}}(\tilde X,\,\tilde \XH) = e^{-2\alpha(X)}\, \big({\mathcal{V}}(X,\,\XH) - \XH^2\,\partial_X\alpha\big) \, .
}{eq:weyl6}
The auxiliary field equation \eqref{eq:6} yields an inhomogeneous shift for $\omega$,
\eq{
\omega\to\tilde\omega = \omega + (\partial_X\alpha)(\XH\tau+\XP e)\,.
}{eq:weyl7}
The Carroll torsion and curvature equations (\ref{eq:1},\ref{eq:2}) are compatible with all the transformations above, replacing consistently everywhere quantities with their tilde counterparts.

Thus, dilaton-dependent Weyl rescalings \eqref{eq:weyl1} act pretty
much in the same way as in the Lorentzian case (see
\cite{Grumiller:2002nm}) and can be used to introduce or eliminate a
kinetic potential function $U(X)$ in potentials of the type
\eqref{eq:UV} below.

\subsection{Second-order formulation}
\label{sec:2.1.2}

To set the stage for other discussions, we translate the
first-order/PSM formulation to the second-order formulation.

While the functional dependence of $\mathcal{V}$ can, in principle, be arbitrary, we consider it to be of the form
\eq{
    \mathcal{V}(X,\,\XH)=-\frac{U(X)}{2}\XH^2+V(X)
}{eq:UV}
for some kinetic and potential function of the dilaton, $U(X)$ and $V(X)$, respectively. (This is also the most commonly used form of the potential in Lorentzian 2d dilaton gravity \cite{Grumiller:2002nm}.)

To get the second-order formulation, one needs to integrate out
$\omega$ and $\XH$ by their own equations of motion \eqref{eq:eom}.
For this we introduce the dual vectors $v^\mu $ and $e^\mu $ satisfying
\begin{align}
    v^\mu \tau _\mu =-1 \qquad e^\mu e_\mu =1 \qquad \delta ^\mu _\nu =-v^\mu \tau _\nu +e^\mu e_\nu  ~.
\end{align}
The $\omega$-equation \eqref{eq:4} can be solved algebraically for $\XH$ 
\begin{equation}
  \XH=-e^\mu \partial_\mu X ~,
\end{equation}  
and also leads to a constraint $v^\mu \partial _\mu X=0$. The $\XH$-equation \eqref{eq:2} is solved by splitting
\begin{align}\label{eq:spin_conn_split}
    \omega =\hat{\omega}+t+\Xphi \, e
\end{align}
with a part $\hat{\omega}$ satisfying $\dd \tau +\hat{\omega}\wedge e =0$, a torsion part $t$, and an arbitrary undetermined function $\Xphi $. The latter embodies the usual ambiguity that the part $e^\mu \omega _\mu $ of the Carroll spin connection is not entirely determined by the torsion constraints. In contrast to the higher dimensional setting, we shall see that it does not play the role of a Lagrange multiplier that sets the intrinsic torsion to zero but rather implements the constraint $v^\mu \partial _\mu X=0$. Explicitly, the different parts read
\begin{align}
    \hat{\omega}_\mu=-e^\nu \partial_{\mu}\tau_{\nu}+e^\nu \partial_{\nu}\tau_{\mu} :=-2e^\nu \partial_{[\mu}\tau_{\nu]} \qquad \qquad t=U(X)\, \XH\tau 
\end{align}
where the latter is determined by the $\XH$-equation. Plugging these
solutions into the first-order action \eqref{eq:car24} with
\eqref{eq:car1} yields \eq{ I_{1^{\mathrm{st}}}=\frac{k}{2\pi} \int
  _{\mathcal{M}}\Big( X\,\dd\hat{\omega}-\Xphi \dd X\wedge e+\XP\dd
  e+\Big(-\frac{U(X)}{2}\big(e^\mu \partial_\mu
  X\big)^2+V(X)\Big)\,\tau\wedge e\Big) \,. }{eq:lalapetz}  
In the second and third terms one can already see the structure of two constraints arise. As a last step, we express the curvature two-form in the first term in terms of an affine connection. Following section \ref{sec:fiber_bundles}, this is done by solving vielbein postulates for a given boost connection. We choose to only do this for the part $\hat{\omega }$,
\begin{subequations}\label{eq:vb_posts}
\begin{align}
    \partial _\mu \tau _\nu +\hat{\omega }_\mu e_\nu -\Gamma ^\alpha {}_{\mu \nu }\tau _\alpha &=0 & \partial _\mu v^\nu +\Gamma ^\nu {}_{\mu \alpha }v^\alpha &=0\\
    \partial _\mu e_\nu - \Gamma ^\alpha {}_{\mu \nu }e _\alpha &=0 & \partial _\mu e^\nu +\hat{\omega }_\mu v^\nu +\Gamma ^\nu {}_{\mu \alpha }e^\alpha &=0
\end{align}
\end{subequations}
such that the Riemann tensor associated to $\Gamma ^\alpha {}_{\mu \nu }$ 
is related to the connection
$\hat{\omega}$ by
$R^\lambda {}_{\sigma \mu \nu }=-2v^\lambda e_\sigma \partial _{[\mu }
\hat{\omega}_{\nu]}$. The solution for the affine connection reads
\begin{align}
    \Gamma ^\lambda{}_{\mu\nu} 
     =-v^\lambda \partial _{(\mu }\tau_{\nu )}-v^\lambda \tau_{(\mu }\mathcal{L}_v \tau_{\nu )}+e^\lambda \partial _\mu e_\nu 
\end{align}
which encodes the intrinsic torsion,
\begin{align}
    T^\lambda _{\mu \nu }=2e^\lambda \partial _{[\mu }e_{\nu ]}=2K e^\lambda \tau _{[\mu }e_{\nu ]}~. 
\end{align}
Here, we defined $K=-e^\mu \mathcal{L}_ve_\mu $ as the only component of the extrinsic curvature in 2d.
On the other hand, using the change of basis
$\dd x^\mu =-v^\mu \tau +e^\mu e$, we write
$\dd\hat{\omega }=\partial _{[\mu }\hat{\omega }_{\nu ]}\dd
x^\mu\wedge \dd x^\nu =2\partial _{[\mu }\hat{\omega }_{\nu ]}e^\mu
v^\nu\tau \wedge e$ implying
$\dd \hat{\omega }=\frac{R}{2} \, \tau \wedge e$ where we
defined the curvature scalar $R=e^\mu e^\nu (R^{\lambda }{}_{\mu \lambda \nu}-v^\rho \tau _\sigma R^\sigma {}_{\mu \rho \nu })$ analogous to \eqref{eq:curv_scalar} but with the arbitrary part associated to $\Xphi$ already pulled out. Finally, we
define the volume form
$\tau \wedge e=\tau _\mu e_\nu \dd x^\mu \wedge \dd x^\nu
=\varepsilon^{\mu \nu }\tau _\mu e_\nu \dd ^2x =\det (\tau ,e)\, \dd^2x$.

Inserting these results into the first-order action
\eqref{eq:lalapetz} yields the second-order Carroll dilaton gravity
action
\begin{align}\label{eq:smith}
    I_{2^{\mathrm{nd}}}[e,\tau ,\Xphi ,\XP ,X]=\frac{k}{4\pi}\int _{\mathcal{M}}\,\mathcal{L}_{2^{\mathrm{nd}}}
\end{align}
with
\eq{
\boxed{
\phantom{\Bigg(}
   \mathcal{L}_{2^{\mathrm{nd}}}=\dd ^2x\, \det (\tau ,e)\,\Big(X R+2\Xphi \,v^\mu \partial _\mu X+2\XP K-U(X)\big(e^\mu \partial _\mu X\big)^2+2V(X)\Big) ~.
\phantom{\Bigg)}
    }
}{eq:angelinajolie}
As we will show in subsection \ref{sec:spher-reduct-magn}, comparing with the literature
(e.g.~\cite{Bergshoeff:2017btm,Henneaux:2021yzg,Campoleoni:2022ebj})
allows identifying this action with the magnetic Carroll theory. It is Carroll invariant if the Lagrange multipliers transform
under Carroll boosts as
\eq{
    \delta _\lambda \Xphi =-U\lambda e^\mu \partial _\mu X+\nabla _\mu (e^\mu \lambda )\qquad\qquad
    \delta _\lambda \XP=-\lambda e^\mu \partial _\mu X 
}{eq:just1line}
where $\nabla $ is the connection associated with $\hat{\omega }$ via
the vielbein postulates. The left equality is compatible with the
transformation behaviour of $\omega$ in \eqref{eq:car26} by using
\eqref{eq:spin_conn_split}. The right equality agrees with the
corresponding on-shell transformation in the first-order formalism
\eqref{eq:car25}. One should keep in mind that the affine connection itself is not Carroll boost invariant in this case, similarly to \eqref{eq:gen_affine_connection}.

\subsection{Spherical reduction of magnetic Carroll gravity}
\label{sec:spher-reduct-magn}

For this, we start with the magnetic
limit of $(d+1)$-dimensional Einstein gravity \eqref{eq:Carroll_Cartan_action}, impose spherical symmetry,
and reduce the corresponding action by the angular variables. We
approach this in the second-order formulation, i.e., we work with the
spin connection expressed in terms of the vielbein variables by using
the torsion constraints. Nevertheless, we will not insert the
expressions explicitly to keep the notation cleaner. The
$(d+1)$-dimensional torsion components read
\eq{
    T^0=\dd \tau +\omega ^a \wedge e_a \qquad\qquad
    T^a=\dd e^a+\omega ^{ab}\wedge e_b
}{eq:edit1}
where $a=1,...,d$ and $\omega ^{ab}=-\omega ^{ba}$. By spherical
symmetry, we write the Carroll metric as
\begin{align}\label{eq:carr_spher_ansatz}
    h_{MN} \dd x^M \dd x^N &=h_{\mu \nu }(x^\sigma )\dd x^\mu \dd x^\nu +\Phi^2(x^\sigma)\,\dd \Omega^2_{S^{(D-2)}} 
\end{align}
where capital latin letters denote the higher dimensional spacetime indices. 
It is convenient to phrase things in terms of a choice of vielbein,
\eq{
    h=e\otimes e +\delta _{lm}e^l\otimes e^m
    =\Bar{e}\otimes \Bar{e}+\Phi ^2 \delta _{lm}\Bar{e}^l\otimes \Bar{e}^m 
}{eq:edit2}
where the internal indices take values $l,m=2,...,d$ and in the
second equality we defined a transverse vielbein $\Bar{e}^l$ normalized to
the unit sphere $S^{(d-1)}$. To keep the notation simple we set
$e^1\equiv e$, $\omega ^1\equiv \omega $. The relations between the
vielbeins are
\begin{align}
    \Bar{\tau }_M&=\tau _M& \Bar{e}_M&=e_M & \Bar{e}^l_M&=\Phi ^{-1}e^l_M \\
   \Bar{v}^M&=v^M & \Bar{e}^M&=e^M & \Bar{e}_l^M&=\Phi \,e_l^M
\end{align}
where $\tau_M v^M =-1$, $e_M e^M =1$, and $e_M^le_m^M =\delta^l_m$ define
the dual versions. We assume throughout that the coordinates are
chosen in such a way that only $\bar e^l_M$ depend on the internal
coordinates. The barred vielbeins are assumed to satisfy the
Carroll torsion constraints, which as a reminder, are not obtained
by setting all the torsion to zero but only those components that
allow the elimination of the spin connection from the first-order
action. We have
\eq{
    \dd \Bar{\tau }+\Bar{\omega }\wedge \Bar{e}=0 \qquad\qquad
    \dd \Bar{e}^l+\Bar{\omega }^{lm}\wedge \Bar{e}_m =0 
}{eq:edit3}
where the remaining torsion component $\dd \Bar{e}$ crucially is not
set to zero. The second of these equations corresponds to the
spherical part which is thus fixed to be torsion-free. Let us assume
    $\Bar{\omega} =\omega$ and   $\Bar{\omega}^{lm}=\omega ^{lm}$.
Furthermore, we impose vanishing torsion in the unbarred space by the
same principle. In particular, this means that the following
conditions are imposed
\eq{
    T^0(e_a,v)=T_a(e_b,e_c)=
    T^0(e_a,e_b)=T_{[a}(e_{b]},v)=0 ~.
}{eq:edit4}
This sets to zero all the higher-dimensional torsion except the components
$T_{(a}(e_{b)},v)$ which correspond to the intrinsic torsion (see section \ref{sec:intrin_torsion}). The conditions can
be solved for $\omega ^{1l}=-\omega ^{l1}$ and $\omega ^l$ by
\begin{align}\label{eq:sol_omega2}
    \omega ^l=\varphi \, \Bar{e}^l && \omega ^{l1}=(e^\mu \partial _\mu \Phi )\Bar{e}^l 
\end{align}
where $\varphi $ is an arbitrary function. This can be inserted to compute the decomposition of the higher-dimensional curvature scalar
\begin{align}\label{eq:scalarcurvhighdim}
    \mathcal{R}=-2v^M e_a^N \Omega ^a_{MN}+e_a^M e_b^N \Omega ^{ab}_{MN}
\end{align}
where
\eq{
    \Omega ^a=\dd \omega ^a+\omega ^{ab}\wedge \omega _b \qquad\qquad
    \Omega ^{ab}=\dd \omega ^{ab}+\omega ^{a}{}_c\wedge \omega ^{cb} ~.
}{eq:edit5}
Explicitly, the components are given by
\begin{align}
    \Omega^1&=\Bar{\Omega}^1=\dd \Bar{\omega} & \Omega^{1l}&=-\dd \,(e^\mu \partial _\mu \Phi )\wedge \Bar{e}^l=-\Omega^{l1}\\
    \Omega^l&=\dd \varphi \wedge \Bar{e}^l+(e^\mu \partial _\mu \Phi )\Bar{e}^l\wedge \Bar{\omega } & \Omega^{lm}&=\Bar{\Omega}^{lm}-(e^\mu \partial _\mu \Phi )\Bar{e}^l\wedge \Bar{e}^m 
\end{align}
leading to the scalar curvature taking the form
\begin{multline}
    \mathcal{R}=-4v^\mu e^\nu \partial _{[\mu }\Bar{\omega } _{\nu ]}+\frac{2(d-1)}{\Phi }\Big((e^\mu \partial _\mu \Phi )v^\nu \Bar{\omega }_\nu -v^\mu \partial _\mu \varphi -e^\mu \partial _\mu (e^\alpha \partial _\alpha \Phi )\Big)\\
    -(d-1)(d-2)\frac{(e^\mu \partial _\mu \Phi )^2}{\Phi ^2} +\frac{(d-1)(d-2)}{\Phi ^2}
\end{multline}
where we used $R_{S^{(d-1)}}=(d-1)(d-2)$. Plugging this result into the
$(d+1)$-dimensional action (see \cite{Campoleoni:2022ebj}), using the
field redefinition 
\begin{align}\label{eq:redef_dilaton}
    \Phi =\frac{d-1}{\lambda }X^{\frac{1}{d-1}} ,
\end{align}
and integrating over the angles yields
\begin{align}
    I_{2^\textrm{nd}}&=\frac{1}{16\pi G_M}\int \dd ^{d+1} x\det (\tau ,e^a)\, \mathcal{R}\\
    &=\frac{k}{4\pi }\int \dd ^2 x \det (\tau ,e)\Big(4X e^\mu v^\nu \partial _{[\mu }\Bar{\omega }_{\nu ]}-2\lambda \varphi X^{\frac{2-d}{1-d}}K+2\lambda \varphi \frac{d-2}{d-1}X^{\frac{1}{1-d}}v^\mu \partial _\mu X \nonumber \\
    &\hspace{3cm} +2v^\mu \Bar{\omega }_\mu e^\nu \partial _\nu X 
    -2e^\mu \partial _\mu (e^\alpha \partial _\alpha X)-U(e^\mu \partial _\mu X)^2+2V\Big)
\end{align}
with $k$, $U$, and $V$ defined as 
\begin{align}\label{eq:choice_UV}
    k&=\frac{\pi ^{d/2}}{2\Gamma (d/2)G_M}\Big(\frac{d-1}{\lambda }\Big)^{d-1} & U&=-\frac{d-2}{(d-1)X} & V&=\lambda ^2\frac{d-2}{2(d-1)}X^{\frac{d-3}{d-1}}~.
\end{align}
The curvature terms simplify by noticing that
\eq{
    -2e^\mu \partial _\mu (e^\alpha \partial _\alpha X)=-2X\nabla _\mu a^\mu \qquad\qquad
    2v^\mu \Bar{\omega }_\mu e^\nu \partial _\nu X=2X\nabla _\mu a^\mu
}{eq:edit6}
up to total derivatives. These two terms, therefore, cancel in the
action. The quantity $\Bar{\omega }$ is the 2d Carroll boost connection
evaluated on a solution of the torsion constraints, i.e.,
\begin{align}
    \Bar{\omega }=\hat{\omega }+\Xphi e 
\end{align}
with an undetermined component given by the arbitrary function
$\Xphi (x^\alpha )$. Together with the definition of the curvature
scalar $R=4e^\mu v^\nu \partial _{[\mu }\hat{\omega }_{\nu ]}$ the
action reads
\begin{align}
    I_{2^\textrm{nd}}&=\frac{k}{4\pi }\int \dd ^2x \det (\tau ,e)\Big(XR+2 \Xphi v^\mu \partial _\mu X+2\XP K -U(e^\mu \partial _\mu X)^2+2V\Big) 
\end{align}
where we discarded all boundary terms and redefined the Lagrange
multipliers $\Xphi $, $\XP $ as specific combinations of the free
functions $\Xphi $ and $\varphi $,
\begin{equation}
    \Xphi \to \Xphi - \varphi \lambda \frac{d-2}{d-1}X^{\frac{1}{1-d}} \qquad \qquad  \varphi \to -\frac{\XP }{\lambda }X^{\frac{2-d}{d-1}}~.
\end{equation}
This matches with the result \eqref{eq:angelinajolie} for the choice \eqref{eq:choice_UV} such that one can interpret the two-dimensional theory as an effective description of the spherically symmetric sector of the higher-dimensional theory. This is analogous to the Lorentzian case, see, e.g., \cite{Grumiller:2002nm,Ecker:2023uwm}.

We shall see in the next subsection that taking the magnetic limit after spherically reducing Einstein gravity also leads to the same theory. It follows that spherical reduction and taking the Carroll limit commute. A similar result was found already for the Galilean case \cite{VandenBleeken:2015rzu}.

\subsection{Magnetic Carroll dilaton gravity from ultra-relativistic expansion}\label{sec:mag_limdil}
Here, we show that the second order action of 2d Carroll dilaton gravity \eqref{eq:smith} can also be understood as coming from an ultrarelativistic expansion of 2d dilaton gravity. The technique we use here is analogous to the one used in \cite{Hansen:2021fxi} for general relativity. We start by considering Lorentzian 2d dilaton gravity given by the second-order action \cite{Grumiller:2002nm} 
\begin{align}\label{eq:rela_2d-action}
    I_{\mathrm{rel.}}[g_{\mu \nu },X]=\frac{k_{\mathrm{rel.}}}{4\pi }c^3\int\dd ^2x\, \sqrt{-g}\, \Big(X{\overset{\textrm{\tiny (LC)}}{R}}-U(X)(\partial X)^2+2V(X)\Big) 
\end{align}
where we allow for arbitrary potentials $U(\reldil )$ and $V(\reldil )$. The connection chosen here is just the Levi--Civitá connection associated to $g_{\mu \nu }$. Using the conventions of \cite{Hansen:2021fxi}, we
switch to pre-Carroll variables
\begin{align}\label{eq:URansatz}
    g_{\mu \nu }=-c^2T_\mu T_\nu +E_\mu E_\nu && g^{\mu \nu }=-\frac{1}{c^2}V^\mu V^\nu +E^\mu E^\nu 
\end{align}
where 
\begin{align}
    V^\mu T_\mu =-1 && E^\mu E_\mu =1 && E^\mu E_\nu -V^\mu T_\nu =\delta ^\mu _\nu 
\end{align}
and the other contractions are zero. The fields $T_\mu $, $V^\mu $, $E_\mu $, $E^\mu $ and $\reldil $ are assumed to be Taylor-expandable in $c^2$. In particular, we have to leading-order (LO) the Carroll fields
\eq{
    V^\mu =v^\mu +\mathcal{O}(c^2) \qquad\qquad E_\mu =e_\mu +\mathcal{O}(c^2) \qquad \qquad \reldil=X+\mathcal{O}(c^2)\,,
}{eq:expansions}
where we denoted the leading-order term in the dilaton expansion by
the same letter. As the subleading terms in the $X$-expansion do not
play a role in the following and both sides are invariant under
Carroll boosts this is just a convenient definition. The first goal is
to rewrite all quantities in the relativistic action
\eqref{eq:rela_2d-action} in terms of the variables on the left-hand side of
\eqref{eq:expansions}. The relativistic Levi--Civit\'a connection is
thereby organized in a specific way,
\eq{
    {\overset{\textrm{\tiny (LC)}}{\Gamma }}{}^\rho{}_{\mu \nu}=\frac{1}{2}g^{\rho \alpha }\left(\partial_\mu g_{\alpha \nu}+\partial _\nu g_{\alpha \mu}-\partial_\alpha g_{\mu \nu }\right)
    =\frac{1}{c^2}{\overset{\scriptscriptstyle (-2)}{C}}{}^\rho{}_{\mu \nu }+\Tilde{C}^\rho{}_{\mu \nu }+S^\rho{}_{\mu \nu }+c^2{\overset{\scriptscriptstyle (2)}{C}}{}^\rho{}_{\mu \nu}
}{eq:conn_exp}
where all orders in $c^2$ transform tensorially except the $c^0$ part. This part is further split into a connection $\Tilde{C}^\rho{}_{\mu \nu}$ satisfying
\eq{
    \overset{\scriptscriptstyle (\Tilde{C})}{\nabla }{}_\nu E_\mu =0 \qquad\qquad \overset{\scriptscriptstyle (\Tilde{C})}{\nabla }{}_\nu V^\mu =0
}{eq:whatever}
and a tensorial part $S^\rho{}_{\mu \nu}$. In this way, a true
Carroll connection is obtained at leading order in the limit
$c\to 0$ (for further details see \cite{Hansen:2021fxi}). While one
has to keep in mind that the pre-Carroll variables $E_\mu $ and
$V^\mu $ are still Lorentz-covariant, it is useful to treat them as if
they were Carroll in order to determine the split of the
$\mathcal{O}(c^0)$ part of the relativistic Levi--Civit\'a connection into
$\Tilde{C}^\rho{}_{\mu \nu}$ and $S^\rho{}_{\mu \nu}$. In other words,
we introduce a pre-Carroll connection\footnote{%
  In 2d, there is no rotational part of the connection and the boost
  part has only one component denoted by the one-form
  $\tilde{\Omega}$.} $\tilde{\Omega }_\mu $ on the frame bundle which
satisfies the Carroll vielbein postulates (see also \eqref{eq:vb_posts})
\begin{align}\label{vielbein_post1}
    \partial_\mu T_\nu-\Tilde{C}^\lambda{}_{\mu\nu}\,T_\lambda + \Tilde{\Omega }_\mu E_\nu &=0 & \partial_\mu E_\nu-\Tilde{C}^\lambda{}_{\mu\nu}\,E_\lambda &=0\\
    \partial_\mu V^\nu + \Tilde{C}^\nu{}_{\mu\lambda}\,V^\lambda &= 0 & \partial_\mu E^\nu + \Tilde{C}^\nu{}_{\mu\lambda}E^\lambda + V^\nu\Tilde{\Omega }_\mu &= 0 \label{vielbein_post2}
\end{align}
where the off-diagonal equations are compatible with \eqref{eq:whatever}. Solving for $\Tilde{C}^\rho{}_{\mu \nu}$ in terms of $\tilde{\Omega}_\mu$ and the vielbein leads to
\begin{align}
     \Tilde{C} ^\lambda{}_{\mu\nu} &= -V^\lambda\big(\partial_\mu T_\nu+\Tilde{\Omega }_\mu E_\nu\big)+\frac{1}{2}\,\Pi ^{\lambda\rho}\big(\partial_\mu \Pi _{\nu\rho}+\partial_\nu \Pi _{\mu\rho}-\partial_\rho \Pi _{\mu\nu}\big) -\Pi ^{\lambda \rho } T_\nu \mathcal{K}_{\mu \rho } \nonumber \\
    &=-V^\lambda\big(\partial_\mu T_\nu+\Tilde{\Omega }_\mu E_\nu\big)+E^\lambda \partial _\mu E_\nu \label{eq:conn_expr}
\end{align}
where $\Pi _{\mu \nu }=E_\mu E_\nu $ and $\mathcal{K}_{\mu \rho }$ is the extrinsic curvature. In 2d, the latter object only has one component,
\eq{
    \mathcal{K}_{\mu \nu }=\mathcal{K}E_\mu E_\nu =-\frac{1}{2}\mathcal{L}_V(E_\mu E_\nu ) \qquad\Rightarrow \qquad \mathcal{K}=2E^\mu V^\nu \partial _{[\mu }E_{\nu ]}~. 
}{eq:yetanother} 
As usual in a second-order formulation, $\tilde{\Omega }_\mu $ is not an independent dynamical variable but is determined in terms of the vielbein such that its torsion vanishes. However, in a \mbox{(pre-)} Carroll geometry, this can only be achieved to a certain degree as there are torsion components independent of $\tilde{\Omega }_\mu $, corresponding to intrinsic torsion. One, therefore, sets only those torsion components to zero that depend on the pre-Carroll connection. In 2d, this leads to only one equation,
\begin{equation}
    \dd T+\Tilde{\Omega }\wedge E=0
\end{equation}
solved by
\eq{
    \Tilde{\Omega } _\mu =-2E^\nu \partial _{[\mu }T_{\nu ]}+\gamma (x^\alpha )E_\mu ~.
}{eq:spin_conn}
Here, $\gamma$ is some arbitrary function representing the remaining freedom in the pre-Carroll connection \cite{Bergshoeff:2017btm,Campoleoni:2022ebj}. Inserting into \eqref{eq:conn_expr} yields
\eq{
     \Tilde{C} ^\lambda{}_{\mu\nu} 
     =-V^\lambda \partial _{(\mu }T_{\nu )}-V^\lambda T_{(\mu }\mathcal{L}_V T_{\nu )}+E^\lambda \partial _\mu E_\nu -\gamma V^\lambda E_\mu E_\nu 
}{eq:nolabel}
which is not torsion-free, but contains the expected intrinsic torsion $\Tilde{C}^\rho{}_{[\mu \nu]}=E^\rho \partial _{[\mu }E_{\nu ]}$ even in the limit $c\to 0$. While the arbitrary function $\gamma$ is set to zero in \cite{Hansen:2021fxi}, we keep it, for now, to see how it contributes at later stages of the expansion. We will see below that $\gamma$ does not play a role.

If we go back to \eqref{eq:conn_exp} and use the pre-Carroll variables we get at order $c^0$
\eq{
    {\overset{\textrm{\tiny (LC)}}{\Gamma }}{}^\lambda{}_{\mu \nu}\Big \vert _{c^0}=-V^\lambda \partial _{(\mu }T_{\nu )}-V^\lambda T_{(\mu }\mathcal{L}_V T_{\nu )}+E^\lambda \partial _\mu E_\nu +E^\lambda E^\alpha T_\nu \mathcal{K}_{\mu \alpha }=:\Tilde{C}^\lambda{}_{\mu \nu}+S^\lambda{}_{\mu \nu}\,.
}{eq:noidea}
Using \eqref{eq:nolabel}, the expansion of the Levi--Civit\'a connection \eqref{eq:conn_exp} has coefficients
\begin{align}
    {\overset{\scriptscriptstyle (-2)}{C}}{}^\rho{}_{\mu \nu}&=-V^\rho \mathcal{K}_{\mu \nu }\\
    \Tilde{C}^\rho{}_{\mu \nu}&=-V^\rho \partial _{(\mu }T_{\nu )}-V^\rho T_{(\mu }\mathcal{L}_V T_{\nu )}+E^\rho \partial _\mu E_\nu -\gamma V^\rho E_\mu E_\nu \\
    S^\rho{}_{\mu \nu}&=E^\rho E^\lambda T_\nu \mathcal{K}_{\mu \lambda }+\gamma V^\rho E_\mu E_\nu \\
     {\overset{\scriptscriptstyle (2)}{C}}{}^\rho{}_{\mu \nu }&=-E^\rho E^\alpha T_{(\mu }(\dd T)_{\nu )\alpha } ~.
\end{align}
The Riemann tensor associated with the Levi--Civit\'a connection is defined by 
\eq{
    {\overset{\textrm{\tiny (LC)}}{R}}{}^\lambda {}_{\mu \nu \sigma }=\partial _\nu {\overset{\textrm{\tiny (LC)}}{\Gamma }}{} ^{\lambda }{}_{\sigma \mu }+{\overset{\textrm{\tiny (LC)}}{\Gamma }}{} ^\lambda {}_{\nu \beta }{\overset{\textrm{\tiny (LC)}}{\Gamma }}{}^\beta {}_{\sigma \mu }-\big (\nu \leftrightarrow \sigma \big ) \,.
}{eq:noalign}
We work directly with the Ricci tensor in the following obtained by contracting first and third indices. It can be organized as
\eq{
    {\overset{\textrm{\tiny (LC)}}{R}}{}_{\mu \nu }=\frac{1}{c^4}\overset{\scriptscriptstyle (-4)}{R}{}_{\mu \nu }+\frac{1}{c^2}\overset{\scriptscriptstyle (-2)}{R}{}_{\mu \nu }+\overset{\scriptscriptstyle (0)}{R}{}_{\mu \nu }+c^2\overset{\scriptscriptstyle (2)}{R}{}_{\mu \nu }+c^4\overset{\scriptscriptstyle (4)}{R}{}_{\mu \nu }
}{eq:yadayada}
with the coefficients up to $\mathcal{O}(c^2)$ given by
\begin{subequations}
\begin{align}
    \overset{\scriptscriptstyle (-4)}{R}{}_{\mu \nu }&=0\\
    \overset{\scriptscriptstyle (-2)}{R}{}_{\mu \nu }&=\overset{\scriptscriptstyle (\Tilde{C})}{\nabla }{}_\lambda {\overset{\scriptscriptstyle (-2)}{C}}{}^\lambda {}_{\nu \mu }-2\tilde{C}^\alpha {}_{[\nu \lambda ]}{\overset{\scriptscriptstyle (-2)}{C}}{}^\lambda {}_{\alpha \mu }+S^\lambda {}_{\lambda \beta }{\overset{\scriptscriptstyle (-2)}{C}}{}^\beta {}_{\nu \mu }-S^\alpha {}_{\nu \lambda }{\overset{\scriptscriptstyle (-2)}{C}}{}^\lambda {}_{\alpha \mu } \\
    \overset{\scriptscriptstyle (0)}{R}{}_{\mu \nu }&=\overset{\scriptscriptstyle (\tilde{C})}{R}{}_{\mu \nu }+\overset{\scriptscriptstyle (\Tilde{C})}{\nabla }{}_\lambda S^\lambda {}_{\nu \mu }-\overset{\scriptscriptstyle (\Tilde{C})}{\nabla }{}_\nu S^\lambda {}_{\lambda \mu }-2\tilde{C}^\lambda {}_{[\nu \beta ]}S^\beta {}_{\lambda \mu }-{\overset{\scriptscriptstyle (-2)}{C}}{}^\lambda {}_{\nu \beta }{\overset{\scriptscriptstyle (2)}{C}}{}^\beta {}_{\lambda \mu }-{\overset{\scriptscriptstyle (2)}{C}}{}^\lambda {}_{\nu \beta }{\overset{\scriptscriptstyle (-2)}{C}}{}^\beta {}_{\lambda \mu }\\
     \overset{\scriptscriptstyle (2)}{R}{}_{\mu \nu }&=\overset{\scriptscriptstyle (\Tilde{C})}{\nabla }{}_\lambda {\overset{\scriptscriptstyle (2)}{C}}{}^\lambda {}_{\nu \mu }-2\tilde{C}^\alpha {}_{[\nu \lambda ]}{\overset{\scriptscriptstyle (2)}{C}}{}^\lambda {}_{\alpha \mu }-{\overset{\scriptscriptstyle (2)}{C}}{}^\lambda {}_{\nu \beta }S^\beta {}_{\lambda \mu }-{\overset{\scriptscriptstyle (2)}{C}}{}^\lambda {}_{\alpha \mu }S^\alpha {}_{\nu \lambda }~.
\end{align}
\end{subequations}
We use these expressions to compute the Ricci scalar expansion
\begin{align}
{\overset{\textrm{\tiny (LC)}}{R}}
    =-\frac{1}{c^4}V^\mu V^\nu \overset{\scriptscriptstyle (-2)}{R}{}_{\mu \nu }+\frac{1}{c^2}\Big(E^\mu E^\nu \overset{\scriptscriptstyle (-2)}{R}{}_{\mu \nu }-&V^\mu V^\nu \overset{\scriptscriptstyle (0)}{R}{}_{\mu \nu }\Big)\nonumber \\
    &-V^\mu V^\nu \overset{\scriptscriptstyle (2)}{R}{}_{\mu \nu }+E^\mu E^\nu \overset{\scriptscriptstyle (0)}{R}{}_{\mu \nu }+\mathcal{O}(c^2) 
    \label{eq:runningoutoflabels}
\end{align}
leading to
\begin{align}
    \overset{\scriptscriptstyle (-4)}{R}&=0\\
    \overset{\scriptscriptstyle (-2)}{R}&=-\overset{\scriptscriptstyle (\Tilde{C})}{\nabla }{}_\mu \left(V^\mu \mathcal{K}\right)+\mathcal{K}^2\\
    \overset{\scriptscriptstyle (0)}{R}
    &=E^\mu E^\nu \overset{\scriptscriptstyle (\Tilde{C})}{R}{}_{\mu \nu }-\overset{\scriptscriptstyle (\Tilde{C})}{\nabla }{}_\rho \Big (E^\rho  \overset{\scriptscriptstyle (\Tilde{C})}{\nabla }{}_\mu E^\mu \Big) +\overset{\scriptscriptstyle (\Tilde{C})}{\nabla }{}_\rho \big(V^\rho \gamma \big)-\mathcal{K}\gamma \label{eq:secdontolast}\\
    &=E^\mu E^\nu \overset{\scriptscriptstyle (\Tilde{C})}{R}{}_{\mu \nu }\Big \vert _{\gamma =0}-\overset{\scriptscriptstyle (\Tilde{C})}{\nabla }{}_\rho \Big (E^\rho  \overset{\scriptscriptstyle (\Tilde{C})}{\nabla }{}_\mu E^\mu \Big) ~.
\end{align}
In the last equality, we used that the last two terms in \eqref{eq:secdontolast} cancel with the $\gamma $-dependence in the first term such that all the $\gamma $-dependence in the last line is in the total derivative term. In the following, we use that the Carroll compatible derivative allows writing a total divergence as
\eq{\int \dd ^2 x\det (T,E)\,\overset{\scriptscriptstyle (\Tilde{C})}{\nabla }{}_\mu X^\mu =-\int \dd ^2x \det (T,E)\, \mathcal{K}T_\mu X^\mu 
}{eq:tot_div}
up to boundary terms \cite{Hansen:2021fxi}. We are ready to expand in $c^2$ and rewrite the relativistic dilaton gravity action \eqref{eq:rela_2d-action} in terms of the pre-Carroll variables. 
\eq{
    I_{\textrm{\tiny rel.}}=c^2 \overset{\scriptscriptstyle (2)}{I}+ c^4\overset{\scriptscriptstyle (4)}{I}+\mathcal{O}(c^6)
}{eq:actionexpanded}
The first two terms in this expansion are 
\begin{align}
     \overset{\scriptscriptstyle (2)}{I}&=\frac{k_{\textrm{\tiny rel}}}{4\pi} \int  \dd ^2x\,\det (T,E)\, \Big(\mathcal{K}V^\mu \partial _\mu \reldil +U(V^\mu \partial _\mu \reldil )^2\Big) \\
      \overset{\scriptscriptstyle (4)}{I}&=\frac{k_{\textrm{\tiny rel}}}{4\pi} \int  \dd ^2x\,\det (T,E)\,\Big(\reldil E^\mu E^\nu \overset{\scriptscriptstyle (\Tilde{C})}{R}{}_{\mu \nu }\Big \vert _{\gamma =0}-U(\reldil )(E^\mu \partial _\mu \reldil )^2+2V(\reldil ) \nonumber \\
    &\hspace{4cm}  -\reldil \overset{\scriptscriptstyle (\Tilde{C})}{\nabla }{}_\rho \big(E^\rho \overset{\scriptscriptstyle (\Tilde{C})}{\nabla }{}_\mu E^\mu \big)\Big) ~.
\end{align}
To extract the magnetic action, we rewrite $\overset{\scriptscriptstyle (2)}{I}$ as
\begin{align}
    \overset{\scriptscriptstyle (2)}{I}&=\frac{k_{\textrm{\tiny rel}}}{4\pi} \int  \dd ^2x\,\det (T,E)\, \Big(\frac{1}{2}\mathcal{K}V^\mu \partial _\mu \reldil+\frac{1}{2}\big(\mathcal{K}+2UV^\mu \partial _\mu \reldil \big)V^\mu \partial _\mu \reldil \Big)\nonumber\\
    &= \frac{k_{\textrm{\tiny rel}}}{4\pi} \int  \dd ^2x\,\det (T,E)\, \Big(-c^4\frac{2\mathcal{K}}{V^\mu \partial _\mu \reldil }\XP ^2+c^22\XP \mathcal{K}\nonumber\\
    &\hspace{4cm}-c^4\frac{2V^\mu \partial _\mu \reldil }{\mathcal{K}+2UV^\mu \partial _\mu \reldil }\Xphi ^2+c^22\Xphi V^\mu \partial _\mu \reldil \Big)
\end{align}
where on-shell evaluation of the auxiliary fields $\XP$ and $\Xphi $
reproduces the action in the first line. The introduction of these auxiliary
fields\footnote{This is sometimes also referred to as a Hubbard--Stratonovich transformation.} effectively redistributes the powers of $c$ such that
$\overset{\scriptscriptstyle (2)}{I}$ is converted into terms
contributing to $\overset{\scriptscriptstyle (4)}{I}$ and
$\overset{\scriptscriptstyle (6)}{I}$. Like in the higher dimensional case the magnetic theory is obtained
by taking the tantum gravity scaling \eqref{eq:whynot}, which in the present case amounts to defining
$k:=k_{\textrm{\tiny rel}}c^{4}$ and taking the limit $c\to 0$.
Effectively, this picks out the action
$\overset{\scriptscriptstyle (4)}{I}$ together with the auxiliary
field contributions from $\overset{\scriptscriptstyle (2)}{I}$ at
leading order. In this case the connection $\Tilde{C}$ reduces to the
Carroll compatible connection $\Gamma $ satisfying
$\nabla (e_\mu e_\nu )=0=\nabla v^\mu $, where
$\nabla  $ is the associated derivative. The dust settles and
we obtain the magnetic Carroll dilaton gravity action
\begin{align}
    I_{\textrm{\tiny mag}}^L[e,\tau ,\rho ,\XP ,X]=\frac{k}{4\pi}\int \,\mathcal{L}^L_{\textrm{\tiny mag}}
\end{align}
with
\begin{align}
     \mathcal{L}^L_{\textrm{\tiny mag}}=\dd ^2x\, \det (\tau ,e)\,\Big(X(e^\mu e^\nu R_{\mu \nu }\big \vert _{\gamma =0}-\nabla _\mu a^\mu )&+2\Xphi \, v^\mu \partial _\mu X+2\XP K\nonumber\\
     &-U(X)(e^\mu \partial _\mu X)^2+2V(X) \Big) 
     \label{eq:mag_carroll}
\end{align}
where it was assumed that the potential $V$ does not scale with $c$
and the leading term of $\mathcal{K}$ is denoted by $K$. The field
$\XP$ acts as a Lagrange multiplier setting the intrinsic torsion of
$\Gamma^\rho{}_{\mu \nu}$ given by the extrinsic curvature
$K$ to zero. On-shell this leaves us with a Carroll torsion-free
connection satisfying the requirements of \cite{Campoleoni:2022ebj}.
The spatial acceleration vector $a^\mu $ is defined in \eqref{eq:def_acceleration}. As divergence terms $\nabla _\mu X^\mu $ are independent of the ambiguous $\gamma $-dependent term in the Carroll compatible connection, we see that $\gamma $ does not enter this action. Therefore, we could have set it to zero from the beginning, without loss of generality, like in \cite{Hansen:2021fxi}. 
The curvature term can be further massaged by using the identity 
\begin{align}
    e^\mu e^\nu R_{\mu \nu }\big \vert _{\gamma =0}=-e^\mu e^\nu \nabla _\mu a_\nu -(e^\mu a_\mu )^2=-\nabla _\mu a^\mu 
\end{align}
which holds only in 2d. Using additionally the definition of the
Carroll curvature scalar
$R=2e^\mu e^\nu R_{\mu \nu }\big \vert _{\gamma =0}$ we find
that the Lagrange-2-form \eqref{eq:mag_carroll} matches with \eqref{eq:angelinajolie}.

\begin{figure}
\begin{center}
\tikzstyle{relblocksmall} = [rectangle, draw, fill=taorange, text width=8em, text centered, rounded corners, minimum height=4em, drop shadow]
\tikzstyle{relblockbig} = [rectangle, draw, fill=taorange, text width=12em, text centered, rounded corners, minimum height=4em, drop shadow]
\tikzstyle{relblockmid} = [rectangle, draw, fill=taorange, text width=10em, text centered, rounded corners, minimum height=4em, drop shadow]
\tikzstyle{carblocksmall} = [rectangle, draw, fill=tabutter, text width=8em, text centered, rounded corners, minimum height=4em, drop shadow]
\tikzstyle{carblockbig} = [rectangle, draw, fill=tabutter, text width=12em, text centered, rounded corners, minimum height=4em, drop shadow]
\tikzstyle{carblockmid} = [rectangle, draw, fill=tabutter, text width=10em, text centered, rounded corners, minimum height=4em, drop shadow]
\tikzstyle{line} = [draw, -Latex]
\hspace*{-0.6truecm}
\begin{tikzpicture}[node distance = 3cm,auto]
    \node [relblockbig] (grD) {General relativity in $d+1$ dimensions\\ $g_{MN}$};
    \node [relblockbig, below of=grD, node distance=4.25cm] (reldil) {2d dilaton gravity\\
    UV family, \eqref{eq:rela_2d-action} \\ $g_{\mu \nu }$, $X$};
    \node [carblockbig, right of=grD, node distance=8cm] (carD) {Magnetic Carroll gravity in $d+1$ dimensions \eqref{eq:second_order_mag_action} \\ $h_{MN}$, $\tau _M$, $C_{MN}$};
    \node [carblockbig, below of=carD, node distance=4.25cm] (cardil) {Magnetic Carroll 2d dilaton gravity \eqref{eq:angelinajolie} \\
    $h_{\mu \nu }$, $\tau _\mu $, $X$, $\XP$, $\Xphi $};
    \node [right of=cardil, node distance=3cm](aux2){};
    \path [line] (grD) -- (carD) node [midway, below] {\begingroup
            \renewcommand{\arraystretch}{0.5}
            \begin{tabular}{c}
                \tiny{$c\to 0$ }\\
                \tiny{\& magnetic limit }
            \end{tabular}
        \endgroup}
        node [midway, above] {\begingroup
            \renewcommand{\arraystretch}{0.5}
           \text{\tiny \cite{Bergshoeff:2017btm,Hansen:2021fxi,Campoleoni:2022ebj}}
        \endgroup};
    \path [line] (grD) -- (reldil) node [midway, left] {\begingroup
            \renewcommand{\arraystretch}{0.5}
            \begin{tabular}{c}
                \tiny{spher.\ red.: choose}\\
                \tiny{$U$,$V$ as in \eqref{eq:choice_UV}}
            \end{tabular}
        \endgroup};
    \path [line] (carD) -- (cardil) node [midway, right] {\begingroup
            \renewcommand{\arraystretch}{0.5}
            \begin{tabular}{c}
                \tiny{spher.\ red.: choose}\\
                \tiny{$U$,$V$ as in \eqref{eq:choice_UV}}
            \end{tabular}
        \endgroup};
    \path [line] (reldil) -- (cardil) node [midway, below] {\begingroup
            \renewcommand{\arraystretch}{0.5}
            \begin{tabular}{c}
                \tiny{$c\to 0$ }\\
                \tiny{\& magnetic limit }
            \end{tabular}
        \endgroup}
        node [midway, above] {\begingroup
            \renewcommand{\arraystretch}{0.5}
           \text{\tiny Sec. \ref{sec:mag_limdil}}
        \endgroup};
\end{tikzpicture}
\caption{Left (orange): Lorentzian theories. Right (yellow): Carroll theories.}
 \label{fig:CBH_flow}
\end{center}
\end{figure}
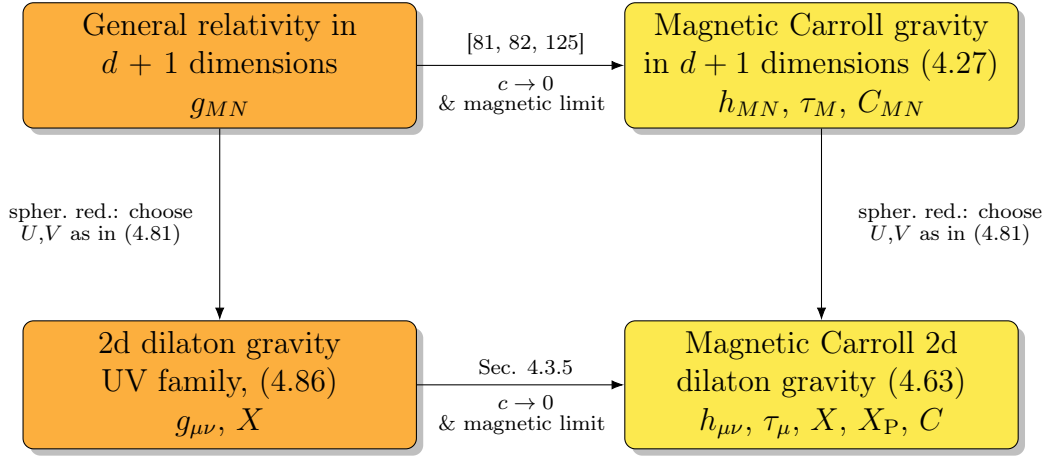

This completes our analysis of ultrarelativistic limits and spherical reductions, showing the commutativity of the diagram in figure \ref{fig:CBH_flow}. For solving the equations of motion of the two-dimensional theory it is most convenient to work in the first order formulation of subsection \ref{sec:2.1.1}. 

\subsection{Formulation as a Poisson--Sigma model}
\label{sec:PSMs}
There is another, gauge-theoretic, formulation of 2d Carroll dilaton gravity that lies at the heart of the original construction in \cite{Grumiller:2020elf}. Since some statements are phrased succinctly in this PSM formulation, we briefly summarize its main aspects.

PSMs are topological, non-linear gauge theories in 2d
\cite{Schaller:1994es, Ikeda:1994fh} and arise as the most general
consistent deformation of Abelian BF-theories \cite{Izawa:1999ib}. The
PSM bulk action~\cite{Grumiller:2020elf}
\begin{align}
 \label{eq:PSM}
\boxed{    I_{\textrm{\tiny PSM}}[A_I,X^I] = \frac{k}{2\pi}\,\int _{\mathcal{M}}\Big(X^I\dd A_I+\frac12\,P^{IJ}(X^K)\,A_I\wedge A_J\Big)
}  \end{align}
with the field content 
\begin{equation}\label{eq:PSMmap}
A_I=(\omega,\,\tau,\,e) \qquad \qquad X^I=(X,\,\XH,\,\XP)
\end{equation}
and the Poisson tensor
\eq{
P^{IJ} = \begin{pmatrix}
          0 & 0 & \XH \\
          0 & 0 & {\mathcal{V}}(X,\,\XH) \\
          -\XH & -{\mathcal{V}}(X,\,\XH) & 0
         \end{pmatrix}
}{eq:car28}
is equivalent to the first-order action \eqref{eq:car24}. The scalar fields $X^I$ are target space coordinates of a Poisson manifold, and for each of them, there is an associated gauge connection 1-form $A_I$. The non-linear Jacobi identities
\eq{
P^{IL}\partial_LP^{JK} + P^{JL}\partial_LP^{KI} + P^{KL}\partial_LP^{IJ}  = 0
}{eq:Jacobis}
hold, essentially because the potential depends only on Carroll boost-invariant combination of the target space coordinates $X^I$, viz., $X$ and $\XH$.

The action \eqref{eq:PSM} is invariant under the non-linear gauge symmetries
\eq{
\delta_\lambda X^J = \lambda_I\,P^{IJ} \qquad\qquad \delta_\lambda A_I = \dd\lambda_I + \big(\partial_I P^{JK}\big)\,A_J\lambda_K \,.
}{eq:gaugesymmetries}
It is easy to verify that \eqref{eq:gaugesymmetries} with the choices above is equivalent to \eqref{eq:car25}-\eqref{eq:car26a}. Also, for the special case of a linear Poisson tensor, $P^{JK}=c^{JK}{}_IX^I$ with structure constants $c^{IJ}{}_K$ one recovers the non-abelian BF theories of subsection \ref{sec:BF_theories}.

PSMs have no local physical degrees of freedom. So, all physical
excitations can be considered holographically as boundary states.
Since the Poisson tensor is anti-symmetric and hence must have an even
rank, a Poisson tensor associated with a 3-dimensional target space
like in \eqref{eq:car28} necessarily has a non-trivial kernel.
Physically, this kernel corresponds to a conserved Casimir that can be
interpreted as mass, as we shall see in section \ref{sec:3}.
\chapter{Carroll black holes}
\label{ch:CBHs}
\epigraphfontsize{\small\itshape}
\epigraph{``Aaaah, so a Kaffee is herrlich! \\
Jaaa, direkt unentbehrlich.''}{--- \textup{Wolfgang Ambros}, Augustin}
We are now ready to investigate the solution space to the theories of Carroll dilaton gravity introduced in the last chapter. Classically, the models may be solved in full generality and in closed form which is carried out in section \ref{sec:5.1}. 
Afterwards, we argue that some solutions can be interpreted as black hole-like states given two ingredients: The first is a sensible notion of thermodynamics being defined for these solutions, which we shall come to in section \ref{sec:3}. The other key entity is a Carroll extremal surface, a Carroll analogue of a Lorentzian extremal surface (see e.g.~\cite{Engelhardt:2014gca} for their classical and quantum definitions). Lorentzian extremal surfaces arise as bifurcation surfaces on Killing horizons and are thus part of eternal black hole geometries. Despite the absence of horizons in Carroll geometries, the analogue of a bifurcation surface still exists, and we declare it to be a defining property of Carroll black holes. We shall be more precise and detailed in section \ref{sec:4}, but for now, a schematic formula that summarizes our main definition is
\begin{equation}
  \text{Carroll black hole}\;= \;\text{Carroll extremal surface}\; +\; \text{Carroll thermal properties}\,.
\end{equation} 
Afterwards, in section \ref{sec:tant_partitionfunction} we turn to an evaluation of the tantum gravity partition function in the saddle-point approximation which provides another independent check for the thermodynamics.  
Finally, we discuss several examples for Carroll black holes in different models in sections \ref{sec:5}, \ref{sec:CS_4d} and \ref{sec:8}. This chapter to a large amount follows \cite{Ecker:2023uwm} as well as \cite{Ecker:2024czh}.

\section{Solutions of Carroll dilaton gravity}
\label{sec:5.1}

In order to derive all classical solutions, the first-order/PSM formulation is advantageous. In the end, we translate our solutions to the second-order formulation.

\subsection{Constant dilaton vacua}
\label{sec:5.1.1}

Constant dilaton vacua are defined to be solutions where $\XH$
vanishes everywhere. The equation that normally determines the Carroll
metric \eqref{eq:4} leads to constant dilaton (hence the name). The
Carroll Casimir equation~\eqref{eq:5} shows that this constant cannot
be anything but has to solve the non-differential equation
\eq{
{\mathcal{V}}(X,\,0) = 0\,.
}{eq:car100}
In particular, this means constant dilaton vacua need infinite finetuning of the dilaton field and may not even exist for some models (the simplest example is the Carroll CGHS model where ${\mathcal{V}}=\Lambda\neq 0$).

The Carroll curvature of all constant dilaton vacua is a constant times the volume-form,
\eq{
\Omega = -\partial_X {\mathcal{V}}(X,\,0)\,\tau\wedge e\,.
}{eq:car101}
Similar remarks apply to torsion. Finally, the auxiliary field equation \eqref{eq:6} implies that also $\XP$ is some (arbitrary) constant. 

Since constant dilaton vacua are non-generic, require infinite finetuning, and are not very rich in structure, we move on to the generic sector, the linear dilaton vacua.

\subsection{Linear dilaton vacua}
\label{sec:5.1.2}

Linear dilaton vacua are solutions where $\XH$ does not vanish
everywhere. Thus, let us start by assuming a patch where
$\XH\neq 0$.\footnote{ Below, many statements implicitly come with the
  qualifier ``assuming $\XH\neq 0$''.} This allows to solve the
Carroll metric equation \eqref{eq:4} as
\eq{
e = -\frac{\dd X}{\XH}\,.
}{eq:car4}
Inserting this result into the Carroll Casimir equation \eqref{eq:5} yields 
\eq{
\frac12\,\dd\, (\XH^2)-{\mathcal{V}}(X,\,\XH)\,\dd X = 0
}{eq:car5}
which allows expressing $\XH$ as function of the dilaton $X$ and of an
integration constant $M$. We refer to $M$ as Carroll mass or
Carroll Casimir. The latter nomenclature was chosen since in the
PSM formulation, the function $M(X,\,\XH)$ spans precisely the kernel
of the degenerate Poisson tensor \eqref{eq:car28}. I.e., if we went to
Casimir--Darboux coordinates the expression $M(X,\,\XH)$ would
correspond to the Casimir direction in the Poisson manifold.

To simplify the discussion, we assume, for now, ${\mathcal{V}}=V(X)$ and return to more general cases in the end. It is useful to define the integrated potential
\eq{
\pot(X):=\int^XV(y)\dd y
}{eq:car6}
in terms of which the conserved Carroll mass is given by
\eq{
M = \pot(X)-\frac12\,\XH^2\qquad\qquad \dd M = 0\,.
}{eq:car7}

The equation that establishes no intrinsic torsion, \eqref{eq:3}, is solved trivially,
\eq{
\dd e = 0 \qquad \Longrightarrow \qquad e = \dd r \, .
}{eq:car8}
We use $r$ as our Carroll radial coordinate without loss of generality.\footnote{%
This is true only if we disregard edge modes, which we do for the time being. Once a boundary is considered with specific boundary conditions imposed on the fields, there could be a loss of generality in assuming $e = \dd r$.} 
Expressing $\XH$ as a function of $X$ using the Carroll mass \eqref{eq:car7} and inserting it into \eqref{eq:car4} yields a simple differential equation for the dilaton
\eq{
\boxed{
\phantom{\Bigg(}
\frac{\dd X}{\mp\sqrt{2(\pot(X)-M)}} = \dd r
\phantom{\Bigg)}}
}{eq:car9}
where the signs refer to the two branches of the square-root function.  

To solve the remaining equations, it is convenient to fix the Carroll boosts such that $\XP=0$, which is always possible locally. The auxiliary equation \eqref{eq:6} simplifies to a constraint
\eq{
\omega = -\frac{V}{\XH}\,\tau
}{eq:car102}
that renders the remaining two equations, for Carroll curvature \eqref{eq:1} and torsion \eqref{eq:2}, identical to each other. 

Thus, there is only one more equation we need to solve, e.g., the Carroll torsion equation \eqref{eq:2}. By virtue of the constraint \eqref{eq:car102} it simplifies to
\eq{
\dd\tau + \big(\partial_X\ln\XH\big)\,\tau\wedge\dd X = 0
}{eq:car103}
solved by 
\eq{
\tau = -\XH\,\dd t\,.
}{eq:car104}
Here, we used the scaling ambiguity $t\to\alpha\tilde t$ with $\alpha\in\mathbb{R}^+$ to choose some time coordinate $t$ and fixed the residual Carroll boost invariance by assuming $\tau$ has no $\dd r$-component. The result \eqref{eq:car104} implies
\eq{
\omega = V(X)\,\dd t
}{eq:car105}
for the Carroll boost connection. As a consequence, the Carroll curvature of our solutions is given by
\eq{
\boxed{
\phantom{\Bigg(}
\Omega = -V'(X)\,\tau\wedge e =- V'(X)\,\dd t \wedge \dd X\,.
\phantom{\Bigg)}}
}{eq:car106}
In summary, in the chosen gauge, the solution reads
\begin{subequations}
 \label{eq:sol}
\begin{align}
    X &= \textrm{given\;by\;integrating\;(\ref{eq:car9})}& \omega &= V(X)\,\dd t\\
    \XH &= \pm\sqrt{2(\pot(X)-M)} & \tau &= - \XH\,\dd t\\
    \XP &= 0 & e &= \dd r \,. 
\end{align}
\end{subequations}
Translating our solution to the second-order formulation as described in section \ref{sec:2.1.2} yields the metric
\eq{
\boxed{
\phantom{\Bigg(}
\dd s^2 = h_{\mu\nu}\,\dd x^\mu\dd x^\nu = e_\mu e_\nu \,\dd x^\mu\dd x^\nu = \dd r^2
\phantom{\Bigg)}}
}{eq:car107}
and the timelike vector field
\eq{
\boxed{
\phantom{\Bigg(}
v=v^\mu\partial_\mu = \frac{1}{\XH}\,\partial_t=\pm\frac{1}{\sqrt{2(\pot(X)-M)}}\,\partial_t\,.
\phantom{\Bigg)}
}
}{eq:car108}
The dilaton field $X$ is still given by integrating \eqref{eq:car9}.

Finally, we come back to more general cases when the potential $\mathcal{V}$ does not only depend on the dilaton $X$ but additionally depends on the boost invariant scalar $\XH$. We discuss here the family of models \eqref{eq:UV} and refer by analogy to \cite{Grumiller:2021cwg} for further generalizations. 

We continue to fix the radial coordinate by $e=\dd r$, so the Carroll metric is still given by \eqref{eq:car107}. Exploiting the Weyl rescalings discussed after \eqref{eq:weyl1}, we find that we need a modified definition of the function $\pot$, namely
\eq{
\pot(X) := \int^X e^{Q(y)} V(y)\dd y\qquad\qquad e^{Q(X)} := e^{\int^X U(y)\dd y}\,.
}{eq:car111}
The additive integration constant implicit in the function $w(X)$ can be absorbed by shifting the mass $M$. The multiplicative integration constant implicit in $e^{Q(X)}$ can be chosen to give this expression the desired physical dimensions, discussed in subsection \ref{sec:dimensions} below. 

Relatedly, the Carroll mass is changed slightly
\eq{
M = \pot(X)-\frac{1}{2}\,\XH^2\,e^{Q(X)}
}{eq:car112}
which changes also the boost invariant scalar 
\eq{
\XH = \pm\sqrt{2e^{-Q(X)}(\pot(X)-M)} \, .
}{eq:car115}
The dilaton is obtained by integrating
\eq{
\frac{\dd X}{\mp\sqrt{2e^{-Q(X)}(\pot(X)-M)}} = \dd r\,.
}{eq:car114}
Fixing, again, Carroll boosts suitably recovers $\XP=0$ and
\eq{
\tau = -e^{Q(X)}\,\XH\,\dd t\,.
}{eq:car116}
The analogue of the constraint \eqref{eq:car102} implies
\eq{
\omega = e^{Q(X)}\,{\mathcal{V}}(X,\,\XH)\,\dd t\,.
}{eq:car117}
Finally, the timelike vector field is given by
\eq{
v=v^\mu\partial_\mu =\pm\frac{1}{\sqrt{2e^{Q(X)}(\pot(X)-M)}}\,\partial_t\,.
}{eq:car113}
Carroll curvature evaluates as
\eq{
\Omega =-\big(V'(X)-\tfrac12\XH^2U'(X)\big)\,\tau\wedge e = e^{Q(X)}\big(\tfrac12\XH^2U'(X)-V'(X)\big)\,\dd t\wedge\dd X\,.
}{eq:car118}

\subsection{Carroll Birkhoff theorem}
\label{sec:5.1.3}

In Lorentzian dilaton gravity, there is a generalized Birkhoff
theorem, in the sense that all solutions to all models have at least
one Killing vector, see e.g.~\cite{Grumiller:2002nm} and references
therein.

In the Carroll case, we see similar features: in the constant
dilaton sector, all solutions have constant curvature and constant
scalar fields, so there is a sense in which these configurations are
maximally symmetric. However, let us focus on the less trivial linear
dilaton vacua.

A minimal requirement to define a Carroll Killing vector is that all boost-invariant fields are invariant under Lie transport along it. This establishes the conditions
\eq{
{\mathcal{L}}_\xi h_{\mu\nu} = {\mathcal{L}}_\xi v^\mu = {\mathcal{L}}_\xi X = 0\,.
}{eq:car109}
The last condition implies $\xi^r=0$, the second condition yields $\partial_t\xi^t=0$, and the first condition imposes no further restriction. Thus, any vector field 
\eq{
\xi^\mu\,\partial_\mu=f(r)\,\partial_t
}{eq:car110}
preserves the boost-invariant structure. Analogous to the arguments in subsection \ref{sec:symmetries} one may additionally require invariance of the Cartan variables up to local tangent space transformations which are just Carroll boosts in this case. Invariance of the connection $\omega $ leads to a second order ODE for $f(r)$,
\begin{align}
    \mathcal{L}_\xi \omega - \dd \lambda =0 \qquad \Rightarrow \qquad \tau _t \partial _r ^2f+\partial _rf (\omega _t+\partial _r \tau _t)=0
\end{align}
where the Carroll boost parameter is given by $\lambda =-\tau _t \partial _r f$ upon using \eqref{eq:car25}. Further, for the solutions in the previous subsection we automatically have $\mathcal{L}_\xi \XH =0 $ but get another condition from $\XP $, 
\begin{align}
    \mathcal{L}_\xi \XP - \lambda \XH =0 \qquad \Rightarrow \qquad \partial _r f =0 
\end{align}
This just leaves us with $f=\text{cst.}$ which is the generator of ordinary time translations. The remaining condition $\mathcal{L}_\xi \tau +\lambda e =0$ is automatically satisfied. This Carroll Killing vector exists for all solutions of all 2d Carroll dilaton gravity models. We refer to this statement as ``Carroll Birkhoff theorem''. 

\subsection{Singularities of Carroll manifolds}
\label{sec:5.1.4}
It is not the purpose of this work to provide a comprehensive
discussion of Carroll singularities. Nevertheless, we need to
confront three types of singularities since they naturally and rather
generically occur in 2d Carroll dilaton gravity.
\begin{enumerate}
\item Carroll coordinate singularities
\item Carroll curvature singularities
\item Carroll structure singularities
\end{enumerate}
The first type of coordinate singularity can arise much in the same
way as in general relativity. The prototypical example is
Schwarzschild-gauge, which in our context is obtained by using the
radial coordinate
\eq{
\dd\rho = e^{Q(X)}\,\dd X
}{eq:car120}
in terms of which the Carroll metric reads
\eq{
h_{\mu\nu}\,\dd x^\mu\dd x^\nu = \frac{\dd\rho^2}{2e^{Q(X)}(\pot(X)-M)}\,.
}{eq:car121}
There is a (coordinate-)singularity at zeros of $\pot(X)-M$. We shall use this singular coordinate system when comparing with Schwarzschild--Tangher\-lini solutions in section \ref{sec:carr_schwarzsch}.

Concerning curvature singularities, we merely note that they can (and
quite often do) arise, typically in the strong-coupling region where
the dilaton tends to zero. Perhaps there is nothing more to this type
of singularity than the observation that curvature can become infinite
if gravity is infinitely strong.

The third type is a singularity reminiscent of what happens at the
bifurcation sphere of the Schwarzschild black hole (see, e.g.,
\cite{Grumiller:2022qhx}), in the following sense. The attentive
reader will have realized already that there is a remaining issue in
our classification of solutions: first, we assumed $\XH=0$ everywhere
(constant dilaton vacua) and then we assumed $\XH\neq 0$ everywhere
(linear dilaton vacua) but what if $\XH$ vanishes or diverges at
isolated loci? That this can happen is evident from the explicit
solution \eqref{eq:sol} for linear dilaton vacua: The equation $\XH=0$
can have solutions for certain values of the radial coordinate $r$,
depending on the choice of the potential and the value of the mass
$M$. Thus, we have potentially singular points in the interior or the
boundary of linear dilaton vacua. The difference to the Lorentzian
case is that there the singularity analogous to the one at $\XH=0$ is
merely a coordinate singularity of Eddington--Finkelstein patches and
can be removed by going into, say, Kruskal coordinates, see
e.g.~\cite{Klosch:1995qv}. By contrast, in the Carroll case, there
is no coordinate system where both the metric and the vector field are
non-zero and finite at $\XH=0$.

The physical interpretation of these potential singularities is
simple: the timelike vector field either blows up ($\XH\to 0$) or
collapses to zero ($|\XH|\to\infty$). We call this a singularity in
the Carroll structure. Note that curvature \eqref{eq:car106} may
remain finite at such loci.

In section \ref{sec:4}, where we define Carroll extremal surfaces, we employ these intriguing loci where $\XH$ vanishes, independently of the behaviour of curvature.
\eq{
\XH=0\qquad\leftrightarrow\qquad e^{-Q(X)}(\pot(X)-M)=0
}{eq:car122}
We move on to the global properties of Carroll thermal manifolds next, where such loci also play a decisive role.

\subsection{Global aspects of Carroll thermal solutions}
\label{sec:thermal_sols}

Before we provide further motivation and details, let us define a Carroll thermal (C-thermal) manifold:

\begin{definition}[C-thermal manifolds]\phantom{()}\\
    C-thermal manifolds (in 2d) are smooth manifolds $\mathcal{M}$ carrying a Carroll structure up to isolated points, with a boundary $\partial \mathcal{M}$ diffeomorphic to $S^1$.
\end{definition}

Thus, we are relaxing the original definition of a Carroll manifold
\cite{Duval:2014uva}, which disallowed these isolated singularities in
the Carroll structure. For a more mathematically rigorous treatment of such situations we refer to \cite{Bruce:2025irk}. As we do not investigate any form of matter coupled to this theory yet, the word ``thermal'' cannot be related to its full physical meaning. We shall see, however, in section \ref{sec:3} that it is still
natural to use this terminology because of a corresponding term
appearing in the first law. Moreover, it is shown in chapter \ref{ch:Hawking} that once matter is introduced on such backgrounds it is in fact possible to have true thermal behaviour of its states. To amalgamate these conflicting indicators
in favour and against a Carroll version of temperature, we still refer to
these geometries as C-thermal.  

In analogy to the Euclidean case, we compactify the orbits of a
Carroll Killing vector field $\xi $ associated with time
translations. Taking $\xi =\partial _t$ we identify\footnote{%
  As in the Lorentzian case, we analytically continue to imaginary
  time and change to complexified frame variables and spin connection,
  see section \ref{sec:CS_4d} for an explicit example. Unlike the
  Lorentzian case, this has no consequence on the signature, which
  remains $(0,+)$. To reduce clutter we refrain from introducing Wick
  rotated variables here. } points $p\in \mathcal{M}$ along the action
\begin{equation}\label{eq:identification}
    p\sim e^{\beta \xi }\cdot p \qquad \qquad \beta \in \mathbb{R}^+ 
\end{equation}
where $e^{\beta \xi }\, \cdot $ means flowing the point $p$ along the
integral curve of $\xi$ by a parameter difference $\beta$.

Having in mind a Carroll analogue of Euclidean dilaton gravity, we
introduce a (cut-off) boundary at some large positive value of the
dilaton, $X(r_c)=X_c$. The two topologies of interest to us are
cylinder and disk. C-thermal manifolds, by definition, require the
latter.

The natural topology for a 2d Carroll manifold with compactified
time direction is a cylinder, i.e., a direct product manifold of the
spatial line and the temporal cycle. Assuming the temporal cycle to be
finite and non-zero globally, only cylinders are possible. Thus, if we
insisted on the absence of Carroll structure singularities
C-thermal manifolds, which have disk topology, would be impossible.

A smooth disk is obtained by demanding the temporal cycle to shrink to
a point at the locus $\XH\to 0$ such that the manifold is smooth
there. Before tackling this issue, we address general aspects of
Carroll manifolds with a Carroll structure singularity induced by
$\XH\to 0$ in the centre of the disk.

The frame we define on such a manifold is given by
\begin{equation}
  \label{eq:thermal_carroll_frame}
    v=\frac{1}{e^Q \XH}\,\partial _t \qquad \qquad e=\partial _r
\end{equation}
where the coordinate $t$ is compactified, $t\sim t+\beta $. Similarly
to a frame adapted to polar coordinates on a Euclidean disk, it
becomes singular at the origin. This is not only because of the
divergence in $v$ but also because the very notion of tangent and
radial directions cannot be defined at the origin. The way to still
obtain a global orthonormal frame field in the Euclidean case is to
perform an $SO(2)$-rotation to a Cartesian frame that can be extended
to the origin. In the general case, this could be done locally such
that, e.g., asymptotically one still has a polar frame while one
switches to a Cartesian one in a neighbourhood of the centre with
corresponding transition functions on the overlap.

In the Carroll case, however, this is not possible: The
transformations acting on the frame belong to the homogeneous Carroll
group $\text{Carr}(2,\mathbb{R})=ISO(1)$, which leaves $v$ invariant.
So, starting with \eqref{eq:thermal_carroll_frame} asymptotically,
one necessarily arrives at a singular description of the origin. This
is just another way of stating the presence of a Carroll structure
singularity at this locus.

One can quantify this singularity by picking a loop around the origin
$\gamma:[0,1]\to \mathcal{M}$ parametrized by
$\sigma \mapsto x^\mu (\sigma )$ and computing its associated holonomy
$H_\gamma (\omega )$ for the connection \eqref{eq:car117}. The
parallel transport equation
\begin{align}
    \frac{\dd }{\dd \sigma }V^A+\frac{\dd x^\mu }{\dd \sigma }\omega ^A{}_{\mu B}V^B =0
\end{align}
is solved by
\begin{align}
    V^A\big (\gamma (\sigma =1)\big)=\exp \Big[-\int _0^1 \dd \sigma \, \Dot{x}^\mu \omega _\mu  \Big]^A{}_B\, V^B\big(\gamma (\sigma =0)\big)=\big(H_\gamma \big)^A{}_B V^B\big(\gamma (\sigma =0)\big)
\end{align}
where $\Dot{x}^\mu $ denotes $\frac{\dd }{\dd \sigma }x^\mu $ and $\Dot{x}^\mu \omega _\mu $ is a matrix with components $\Dot{x}^\mu \omega ^A{}_{\mu B}$ describing the representation of the homogeneous Carroll algebra on tangent space (see also subsection \ref{sec:gauging_procedure}). Using the solution \eqref{eq:car117} and choosing the loop $x^\mu (\sigma )=\big(\sigma \beta ,r_0\big)$ with $r_0=\text{const.}$ we obtain 
\begin{equation}
    H_\gamma (\omega )=\begin{pmatrix}
        1 && -\beta e^Q\mathcal{V}(X,\XH )\big \vert _{r=r_0}\\
        0 && 1
    \end{pmatrix} ~.
\end{equation}
Then, contracting $\gamma $ to a point at the origin yields
\begin{align}\label{eq:contr_holonomy}
    \lim _{r_0 \to 0} H_\gamma (\omega )=\begin{pmatrix}
        1 & -\beta w'(X) \\
        0 & 1
    \end{pmatrix} \Big \vert _{r= 0}
\end{align}
where, without loss of generality, the integration constant in
\eqref{eq:car114} was chosen such that $\XH (r=0)=0$. We stress that
while the Carroll spin connection is ambiguous and only defined up to
the addition of a term $\rho \,e$ (see section \ref{sec:2.1.2}), this
ambiguity does not enter here because the loop is chosen such that
$\Dot{x}^\mu e_\mu =0$.

Let us return to the smoothness condition at the origin of the
disk. In a Euclidean theory of gravity, one requires that the closer
$\gamma $ approaches the origin, the more the holonomy approaches the
one of a flat disk. The geometry of the latter is given by
\begin{align}
    v^{\text{\tiny E,Disk}}=\frac{1}{r}\,\partial _\varphi \qquad \quad e^{\text{\tiny E,Disk}}=\partial _r \qquad \quad \big(\omega ^{\text{\tiny E,Disk}}\big)^a{}_b=\begin{pmatrix}
        0 & \dd \varphi \\
        -\dd \varphi & 0
    \end{pmatrix} 
\end{align}
with the torsion-free spin connection  $\omega ^{\text{\tiny E,Disk}}$ and $\varphi \sim \varphi +2\pi $. Explicitly, the condition reads
\begin{align}
    \lim _{r_0\to 0}H_\gamma (\omega ^{\text{\tiny E}})\overset{!}{=}H_\gamma (\omega ^{\text{\tiny E,Disk}})=\exp \begin{pmatrix}
        0 & -2\pi \\
        2\pi & 0
    \end{pmatrix} 
\end{align}
where $\omega ^E$ is some curved connection. This condition fixes the Hawking
temperature. We use the same definition for the Carroll
case: A flat Carroll disk is given by
\begin{align}
   v^{\text{\tiny C,Disk}}=\frac{1}{r}\,\partial _\varphi \qquad \quad e^{\text{\tiny C,Disk}}=\partial _r \qquad \quad \big(\omega ^{\text{\tiny C,Disk}}\big)^a{}_b=\begin{pmatrix}
        0 & \dd \varphi \\
        0 & 0
    \end{pmatrix} ~,
\end{align}
where the ambiguity in the spin connection has again been neglected because it does not contribute to the holonomy integral. The condition we arrive at is
\eq{
\boxed{
\phantom{\Bigg(}
 \lim _{r_0\to 0}H_\gamma (\omega )\overset{!}{=}H_\gamma (\omega ^{\text{\tiny C,Disk}})=\exp \begin{pmatrix}
        0 & -2\pi \\
        0 & 0
    \end{pmatrix} 
\phantom{\Bigg)}}
}{eq:hol_cond}
Therefore, in Carroll theories of gravity, the holonomy is never equal to the identity for contractible loops around the origin, which is precisely because of the Carroll structure singularity. 
 
As another equivalent way to ensure a smooth disk, we use the
Gauss--Bonnet formula rewritten in first-order variables,
\begin{equation}\label{eq:gen_GB}
    2\pi \chi =\int _{\mathcal{M}}\dd \omega -\int _{\partial \mathcal{M}}\omega 
\end{equation}
where implicitly, we assume the bulk term $\dd\omega$ does not yield
$\delta$-like contributions (corresponding to deficit
angles).\footnote{ In general, there is an additional subtlety.
  Namely, depending on the frame, one may need to subtract an auxiliary
  connection $\omega_0$ to ensure gauge invariance. The boundary
  integral is equivalent to an integral of the second fundamental
  form. However, in our chosen frame, this turns out to be
  unnecessary.} This assumption, in general, is incorrect unless the
periodicity $\beta$ takes certain values.

In other words, while often a formula like $\eqref{eq:gen_GB}$ is used
to compute $\chi$ for given geometrical data on a manifold, we reverse
the logic: Taking $\chi=1$ and demanding
\eq{
\boxed{
\phantom{\Bigg(}
  2\pi \overset{!}{=}\int _{\mathcal{M}}\dd \omega -\int _{\partial \mathcal{M}}\omega   
\phantom{\Bigg)}}
}{eq:temp_fix}
ensures that no conical defects appear while $\mathcal{M}$ is
topologically a disk provided the periodicity $\beta$ is chosen
appropriately, which we shall do in the next section. Here, the
boundary integral is understood along the surface $X=X_c$ with
outward-pointing unit normal form $n=-\XH ^{-1} \dd X$ and with a
volume form $\text{vol}_{\partial \mathcal{M}}$ induced by
$\tau \wedge e=n\wedge \text{vol}_{\partial \mathcal{M}}$ such that
Stokes' theorem holds. In particular, this implies
$\int _{\partial \mathcal{M}}\omega =-\int _0^\beta e^Q\mathcal{V}\dd
t$.

While the spin connection is undefined at the origin of the disk,
the integrand of \eqref{eq:temp_fix} is defined up to this isolated
point, and we can continuously extend
$\dd \omega =2e^\mu v^\nu \partial _{[\mu }\omega _{\nu ]}\tau \wedge
e$ to the origin. On-shell the limit
\begin{align}
   \lim _{r\to 0}2e^\mu v^\nu \partial _{[\mu } \omega _{\nu ]}\big|_{\textrm{\tiny EOM}} =- w'' (X)\big \vert _{r=0}
\end{align}
exists whenever $w''(X)\vert_{r=0}$ is finite. Thus, we can continue
the integrand to the origin in such cases. As the resulting
contribution to the integral has measure zero, we find that the formula
\eqref{eq:temp_fix} is not even sensitive to the Carroll structure
singularity and therefore provides another suitable device to probe
disk topology. We shall see in the next section how
\eqref{eq:temp_fix} can be used to fix the Carroll temperature in
terms of the dilaton potential.

\section{Carroll thermal properties}
\label{sec:3}

In this section, we discuss the C-thermal properties of the linear
dilaton solutions derived in section \ref{sec:5.1.2}. In subsection
\ref{sec:3.1}, we derive the energy from the usual boundary charges.
In subsection \ref{sec:3.2}, we define two different notions of
temperature and show that they coincide with each other. In subsection
\ref{sec:3.3}, we address entropy and the first law of Carroll
thermodynamics. In subsection \ref{sec:dimensions}, we perform a
dimensional analysis to ensure dimensionless entropy. Finally,
in subsection \ref{sec:3.5}, we calculate the specific heat.

\subsection{Energy}\label{sec:3.1}
The canonical codimension-2 charge variations for a generic PSM
\eqref{eq:PSM} are (see for instance~Eq.~(6.1) in \cite{Grumiller:2017qao}) 
\begin{align}\label{eq:car200}
    \slashed \delta \mathcal{Q}_\lambda =\frac{k}{2\pi}\, \lambda_I\, \delta X^I \Big \vert_{\partial \mathcal{M}}
\end{align}
with boundary condition-preserving gauge parameters\footnote{%
These
  parameters, in general, depend on the fields, which can make this
  expression non-integrable in field space. To account for this
  possibility, we denote the charge variation by $\slashed \delta$.}
$\lambda_I$. We assume that the boundary conditions imposed on $X$ and
$\XH$ are such that they allow arbitrary variations of the mass
parameter $M$. Since we do not want to be too specific about these
boundary conditions at this stage, we just impose that diffeomorphisms
generated by the Killing vector $\xi=\partial_t$ are part of the
asymptotic symmetries that preserve the boundary conditions. The
associated gauge parameters are given by
$\lambda _X=\omega_t=e^Q\mathcal{V}$, $\lambda_H=\tau_t=-e^Q\XH$,
$\lambda_P=e_t=0$.

The charge variation \eqref{eq:car200} associated with unit time translations is given by
\eq{
\delta \mathcal{Q}_{\partial_t} = \frac{k}{2\pi}\,\big(e^Q\mathcal{V}(X,\XH )\,\delta X -e^Q\XH\,\delta\XH\big)\big \vert_{\partial \mathcal{M}} = \frac{k}{2\pi}\,\delta M
}{eq:car201}
where in the last equality, we used the (variation of the) Casimir relation \eqref{eq:car112}. Since $M$ is totally conserved, $\dd M=0$, it does not matter where this quantity is evaluated, which is why we dropped the indicator $\vert_{\partial \mathcal{M}}$. 

The charge \eqref{eq:car201} is integrable in field space and gives a simple expression for the energy, $E=\mathcal{Q}_{\partial_t}$, in terms of the mass parameter:
\eq{
\boxed{
\phantom{\Bigg(}
E=\frac{k}{2\pi}\,M
\phantom{\Bigg)}}
}{eq:car202}

\subsection{Temperature}
\label{sec:3.2}

Following the discussion in section \ref{sec:thermal_sols} we impose
equation \eqref{eq:temp_fix} to ensure having a Carroll disk
without any defects. Inserting the solutions of section
\ref{sec:5.1.2} and choosing an orientation such that
$\tau \wedge e =:e^Q\dd t \dd X$ we find
\begin{align}
    \int _{\mathcal{M}}\dd \omega &=-\int _{\mathcal{M}}\partial _X\mathcal{V}\, \tau \wedge e =\int _0^\beta \int _{X_{\textrm{\tiny min}}}^{X_c }\dd t\dd X \partial _X\Big( U (w-M)-\partial _X w \Big) \\
    &=\beta \Big(U (w-M)-\partial _X w \Big)\Big \vert ^{X_c }_{X_{\textrm{\tiny min}}}\\
    \int _{\partial \mathcal{M}}\omega &=-\int _0^\beta e^Q\mathcal{V}\dd t\Big \vert ^{X_c}=-\beta\Big(\partial _X w-U(w-M)\Big)\Big\vert ^{X_c} 
\end{align}
such that \eqref{eq:temp_fix} reads
\begin{align}
    2\pi  \overset{!}{=}\beta \, \partial _X w \big \vert _{X_{\textrm{\tiny min}}} ~.
\end{align}
Here $X_{\textrm{\tiny min}}$ is the value of the dilaton at the locus $\XH =0$, taking the positive branch in \eqref{eq:car9}. Interpreting $\beta=T^{-1}$ as inverse Carroll temperature establishes
\eq{
\boxed{
\phantom{\Bigg(}
T = \frac{\pot'(X_{\textrm{\tiny min}})}{2\pi}\,.
\phantom{\Bigg)}}
}{eq:temp5}
The result for the Carroll temperature \eqref{eq:temp5} is equivalent to the corresponding Lorentz\-ian result for the Hawking temperature of 2d dilaton gravity with the same potentials \eqref{eq:UV}. 

In addition to this topological derivation of Carroll temperature, there is also a definition in terms of Carroll surface gravity.
\eq{
\nabla_\mu\big(e^Q e^\nu \partial _\nu X\big)\big|_{\XH=0}=:\hat{\kappa }\, e_\mu\big|_{\XH=0} 
}{eq:kappa}
The quantity in parentheses is proportional to $\XH$ on-shell and thus vanishes at $\XH=0$. In this sense, the definition of $\kappa$ in \eqref{eq:kappa} is analogous to the definition of surface gravity in a Lorentzian theory. Taking the solutions \eqref{eq:sol} yields $\hat{\kappa } =w'(X_{\textrm{\tiny min}})$. Therefore, we recover the anticipated relation
\eq{
T = \frac{\hat{\kappa }}{2\pi}
}{eq:ladeeda}
between Carroll temperature $T$ and Carroll surface gravity $\hat{\kappa }$.

\subsection{Entropy and first law}\label{sec:3.3}

As the last missing piece for the first law, let us inspect the
definition of the entropy along the lines of Wald \cite{Wald:1993nt}.
Working in the covariant phase space formalism of first-order Carroll
dilaton gravity (see, e.g., \cite{Ruzziconi:2020wrb}) the symplectic
form is given by
\eq{
     \varpi (\delta _1\phi ,\delta _2\phi )=\frac{k}{2\pi }\Big (\delta _2X^I\delta _1A_I-\delta _1X^I\delta _2A_I\Big )
}{eq:boring1}
where we used PSM variables \eqref{eq:PSMmap} for convenience and denote the collection of fields by $\phi $. Contracting in a diffeomorphism generated by some vector field $\xi $ on the worldsheet and evaluating on-shell yields the fundamental theorem of covariant phase space
\eq{
    \varpi (\delta \phi ,\delta _\xi \phi )\approx \dd \Big (\delta Q_\xi -Q_{\delta \xi }-i_\xi \Theta (\delta \phi )\Big ) =:\dd \slashed \delta \mathcal{Q}_\xi 
}{eq:boring2}
where $Q_\xi $ is the Noether--Wald charge. The variation of the codimension-2 charges is given by
\eq{
    \slashed \delta \mathcal{Q}_\xi = \frac{k}{2\pi}\xi^\mu A_{I\,\mu}\,\delta X^I ~,
}{eq:boring3}
which just reproduces the special case $\lambda _I=A_{I\mu }\xi ^\mu $ of the more general result \eqref{eq:car200}. 
We choose $\xi$ to be the Carroll Killing vector associated with unit time translations,
\eq{
    \xi = \partial_t 
}{eq:boring4}
which implies $\varpi (\delta \phi ,\delta _\xi \phi )=0$. Additionally, we pick a constant time hypersurface $\Sigma $ extending from the point $\mathcal{E}=\{p\in \mathcal{M}:\XH =0\}$ in the interior to a point $\mathcal{B}\in \partial \mathcal{M}$ on the asymptotic boundary such that $\partial \Sigma =\mathcal{E}\cup \mathcal{B}$. Integrating \eqref{eq:boring2} over $\Sigma $ leads to the on-shell identity
\eq{
    \int _\Sigma \varpi (\delta \phi ,\delta _\xi \phi )\approx \int _\Sigma \dd  \slashed \delta \mathcal{Q}_\xi = \slashed \delta \mathcal{Q}_\xi \Big \vert _{\mathcal{B}}-\slashed \delta \mathcal{Q}_\xi \Big \vert _{\mathcal{E}}=0 ~.
}{eq:first_law} 
Explicitly, $\slashed \delta \mathcal{Q}_\xi$ reads on-shell
\begin{align}
    \slashed \delta \mathcal{Q}_\xi =\frac{k}{2\pi }\Big(e^Q\Big(V-\frac{U}{2}X_{\textrm{\tiny H}}^2\Big)\delta X-e^QX_{\textrm{\tiny H}}\delta X_{\textrm{\tiny H}} \Big) 
\end{align}
and evaluates at the two points of $\partial \Sigma$ to
\begin{align}
     \slashed \delta \mathcal{Q}_\xi \Big \vert _\mathcal{B}&=\frac{k}{2\pi}\,\delta M &
      \slashed \delta \mathcal{Q}_\xi \Big \vert _{\mathcal{E}}&=\frac{k}{2\pi }e^Q V\,\delta X\Big\vert_{\mathcal{E}} ~.
\end{align}
From \eqref{eq:first_law} we therefore find 
\begin{align}
    \frac{k}{2\pi}\,\delta M =\Big(\frac{e^{Q(X)}V(X)}{2\pi}\,k\,\delta X \Big)\Big\vert_{\mathcal{E}}
\end{align}
which together with the results \eqref{eq:car202} and \eqref{eq:temp5} takes the form of a first law,
\eq{
    \delta E=T\,\delta S ~.
}{eq:wtfwhynolabel}
The new thermodynamic quantity defined here is the entropy 
\eq{
\boxed{
\phantom{\Bigg(}
S := k\,X_{\textrm{\tiny min}}\,.
\phantom{\Bigg)}}
}{eq:ent0}
and arises as a Noether charge, just as in the relativistic case. Its functional form in terms of the dilaton also matches precisely with the one of relativistic dilaton gravity \cite{Gegenberg:1994pv}.  In words, entropy is given by the value of the dilaton at the Carroll extremal surface (defined in the next section), times the coupling constant.

\subsection{A word about dimensions}
\label{sec:dimensions}

If we assign standard units to all variables in the action, then
entropy turns out to have a non-standard dimension of velocity. The
quickest way to see this is first to verify that the connection
$\omega$ necessarily has the dimension of inverse velocity, assuming
that $\tau$ has time dimension and $e$ length dimension. This
statement follows from the Carroll torsion equation \eqref{eq:2}. The
first term in the action \eqref{eq:car24} has a dimension of
$[kX\omega]$ and, in units where $\kappa =1$, this combination must be
dimensionless. This follows from $\kappa ^{-1}$ being the prefactor of the exponent in a putative tantum gravity partition function (see also section 
\ref{sec:tant_partitionfunction}). Thus, we find that the dimension of entropy
\eqref{eq:ent0},
$[S]=[kX] = \frac{\textrm{length}}{\textrm{time}} =
\textrm{velocity}$, is unusual.

If one wants to recover a dimensionless entropy (e.g.~measured in
$e$-bits), one has to introduce a velocity as a conversion factor. In
Carroll theories, there is no natural choice for such a conversion
factor, but if viewed as an expansion of a Lorentzian theory, we have
a velocity available, namely the speed of light. Even for intrinsic
Carroll theories, we shall assume the presence of some quantity
with the dimension of velocity and convert time into length. This
assumption permits $\tau$ to have length dimension and hence $\omega$
to be dimensionless. In the following, we denote the length dimensions
by integers $[\bullet]=n$, meaning that the corresponding quantity
$\bullet$ has length dimension $n$ (if $n$ is negative, the quantity
$\bullet$ has corresponding inverse length dimension). For instance,
our choice above means $[e]=[\tau]=[r]=[t]=1$ and $[\omega]=0$.

The remaining freedom is which length dimension to assign to the
dilaton field. From an intrinsic 2d perspective, the only natural
choice is to assume dimensionless dilaton, $[X]=0$, implying also
$[k]=0$. The dimensions of all other quantities follow from this
assignment and compatibility with the equations of motion
\eqref{eq:eom}: $[{\mathcal{V}}]=-2$, $[\XH]=[\XP]=[M]=[w]=[v]=-1$,
$[\Omega]=0$, $[e^Q]=+1$. The only subtlety (known already from the
Lorentzian counterpart) is the last assignment and can be attributed
to a length dimension carried by the (otherwise irrelevant)
multiplicative integration constant inherent to the definition of
$e^Q$, see \eqref{eq:car111}.\footnote{%
  Since $[X]=0$ but $[{\mathcal{V}}]=-2$, the potential generically
  contains some relevant coupling constant. We can always use an
  appropriate power of that constant for the multiplicative
  integration constant in $e^Q$. }

We deduce the dimensions of our thermodynamical quantities as
$[E]=[T]=-1$ and $[S]=0$. In particular, the entropy is dimensionless,
and inverse temperature $\beta$ has length dimension consistently with
our starting point of assigning time a length dimension.

\subsection{Specific heat}
\label{sec:3.5}

With the quantities obtained so far, we define a Carroll specific heat as
\eq{
    C:=\frac{\dd E}{\dd T} =T\,\frac{\dd S}{\dd T}
}{eq:Cdef}
yielding
\eq{
    C=k\,\frac{w'(X_{\textrm{\tiny min.}})}{w''(X_{\textrm{\tiny min.}})}~.
}{eq:Cres}
The equivalence of this expression to the Lorentzian case \cite{Grumiller:2007ju} is worth highlighting. Assuming positive temperature, $w'(X_{\textrm{\tiny min.}})>0$, specific heat is positive if and only if $w''(X_{\textrm{\tiny min.}})>0$.

Having investigated the C-thermal properties of linear dilaton solutions with Carroll structure singularities, we define Carroll extremal surfaces and Carroll black holes in the next section.

\section{Carroll extremal surfaces}
\label{sec:4}

In this section, we introduce the geometric notion of Carroll extremal
surfaces, guided by corresponding Lorentzian results. We start in
subsection \ref{sec:4.1} with a translation of standard (relativistic)
extremal surfaces into the PSM formulation. We copy this definition in
subsection \ref{sec:4.2} and apply it to define Carroll extremal
surfaces. We translate back this definition into first- and
second-order formulations of Carroll gravity in subsection
\ref{sec:4.3}. Finally, we collect our results to define Carroll black
holes in subsection \ref{sec:4.4}.

\subsection{Standard extremal surfaces in PSM formulation}
\label{sec:4.1}

Our first task is to translate the notion of an extremal surface (both
null expansions vanish, see, e.g.~\cite{Grumiller:2022qhx}) into the
PSM formulation. For relativistic 2d dilaton gravity in the PSM
formulation the Poisson tensor is given by \cite{Grumiller:2021cwg}
\eq{
P^{IJ} = \begin{pmatrix}
          0 & -X^+ & X^- \\
          P^{+\text{\tiny X}}=-P^{\text{\tiny X}+} & 0 & {\mathcal{V}}(X,\,X^+X^-) \\
          P^{-\text{\tiny X}}=-P^{\text{\tiny X}-} & P^{-+}=-P^{+-} & 0
         \end{pmatrix}
}{eq:rel1}
and the worldsheet metric is written in terms of the zweibein as 
\begin{equation}
    \dd s^2=2e^+e^- ~.
\end{equation}
On-shell the latter is given by
\eq{
\dd s^2 = 2e^Q\dd v\,\dd X + 2e^{2Q}X^+X^-\,\dd v^2
}{eq:rel2}
where $Q$ is a known function of the dilaton $X$ and of an integration
constant, the mass $M$. The Lorentz invariant combination $X^+X^-$
also can be expressed as such a function, using the conserved Casimir
inherent to the Poisson tensor \eqref{eq:rel1}. The metric
\eqref{eq:rel2} allows expressing worldsheet features in terms of
conditions on the target space coordinates $X^I$. In the table below
we summarize the relativistic interpretation of various loci on the
worldsheet in terms of the signs of $X^\pm$.

\smallskip

\begin{center}
{\footnotesize
\noindent \begin{tabular}{c|ccc}
signs & $X^+>0$ & $X^+<0$ & $\boldsymbol{X^+=0}$ \\ \hline
$X^->0$ & anti-trapped & anti-normal & marginally anti-trapped \\
$X^-<0$ & normal & trapped & marginally trapped  \\
$\boldsymbol{X^-=0}$ & marginally anti-trapped & marginally trapped & {\textbf{extremal}}
\end{tabular}
}
\end{center}

\smallskip

\begin{figure}
\def\L{2.3}
\begin{center}
\begin{tikzpicture}
 \draw[black] (-\L,-\L) coordinate (lole) -- (\L,\L) coordinate (upre);
 \draw[black] (\L,-\L) coordinate (lore) -- (-\L,\L) coordinate (uple);
 \draw[black] (upre) node[right] (scrip) {marginally trapped};
 \draw[black] (lore) node[right] (scrip) {marginally anti-trapped};
 \draw[black] (1/2*\L,0) node[right] (scrip) {normal};
 \draw[black] (-1/2*\L,0.2*\L) node[left] (scrip) {anti-normal};
 \draw[black] (0,-1/2*\L) node[below] (scrip) {anti-trapped};
 \draw[black] (0,1/2*\L) node[above] (scrip) {trapped};
 \draw[red,thick] (0,0) coordinate (center) node[left] (scrip) {extremal\;};
 \draw[red,thick] (center) circle(0.02*\L);
\end{tikzpicture}
\end{center}
\caption{Kruskal diagram of Lorentzian eternal black hole}
\label{fig:1.1}
\end{figure}
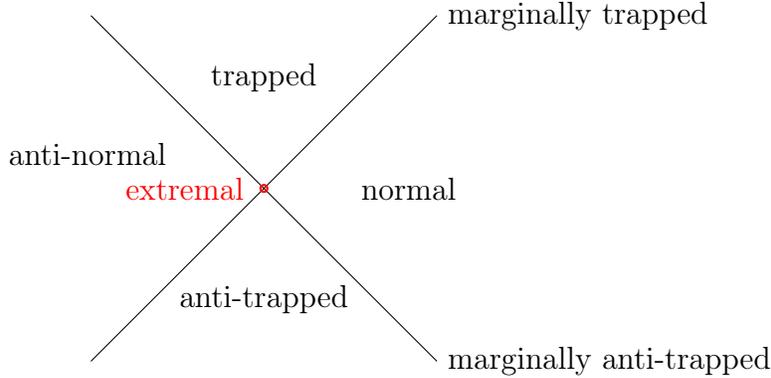

For Kruskal-type of spacetimes ``normal'' refers to the outside
region, ``anti-normal'' to the second outside region, ``trapped'' to
the black hole region, ``anti-trapped'' to the white hole region,
``marginally trapped'' and ``marginally anti-trapped'' to the
bifurcate Killing horizon and ``extremal'' to the bifurcation point.
See Fig.~\ref{fig:1.1} for a reminder.

A nice way of expressing what is special about extremal surfaces from a PSM perspective is to consider the action of relativistic boosts on the target space coordinates, 
\eq{
\delta_\lambda X=0\qquad\qquad\delta_\lambda X^\pm=\mp X^\pm\,\lambda\,.
}{eq:rel3}
Comparing with the table above, marginally trapped or anti-trapped surfaces are fixed lines (though not fixed-point lines) under boosts, since e.g.~every locus $X^+=0$ is mapped to another locus where $X^+=0$. This provides a target space notion of marginally \mbox{(anti-)}trapped surfaces as fixed lines under boosts.

Similarly, by inserting the definition of extremal loci from the table above, we see that extremal surfaces are fixed points with respect to boosts: all the target space coordinates are invariant under boosts on the extremal locus.

\paragraph{PSM definition of relativistic extremal surfaces.} Relativistic extremal surfaces are loci in the PSM target space that are fixed points under relativistic boosts.
\paragraph{}

This is the kind of property we were after. We have a definition of
extremal surfaces as loci in the target space where the Poisson tensor
is invariant under the gauge symmetries associated with boosts. Since
we do not need the boosts to be relativistic for this definition to
apply, it readily generalizes to Carroll boosts and thus allows
defining Carroll extremal surfaces, which we shall do in the next
subsection. Before doing so, we translate back the definition above
into a more familiar language.

In the second-order formulation, the target space coordinates $X^\pm$ do not exist but are replaced on-shell by directional derivatives of the dilaton field projected along the vielbein components.
\eq{
X^\pm \approx \pm e^\mu_\pm\partial_\mu X
}{eq:car50}
An extremal surface in the second-order formulation is thus a locus where the dilaton has a saddle point or an extremum. 
\eq{
\textrm{relativistic\;extremal\;surface:}\qquad e^\mu_\pm\partial_\mu X = 0\qquad\qquad X>0
}{eq:car52}
This is compatible with higher-dimensional intuition if 2d dilaton gravity is viewed as a dimensional reduction from higher-dimensional gravity. The condition of positive $X$ was added to eliminate ludicrous cases of fake extremal surfaces, which from a higher-dimensional perspective are spherical coordinate singularities, and from a 2d perspective, are singular loci where the effective gravitational coupling diverges.\footnote{%
A simple example of such a fake extremal surface is the centre of 4d Minkowski space in spherical coordinates, with $X=r^2$. The quantity $e^\mu_\pm\partial_\mu X \propto\pm\partial_r X=2r$ vanishes at the origin $r=0$.
}  

Similarly, we define (non-extremal) eternal black holes in terms of thermal and target space properties:

\paragraph{PSM definition of relativistic eternal black holes.} Relativistic eternal black holes are thermal states with finite entropy that have a relativistic extremal surface.
\paragraph{}

Thermality is needed to exclude extremal black holes, finite entropy
is needed to exclude constant dilaton solutions, and the presence of a
relativistic extremal surface is needed to ensure there is a special
locus in target space that lies on the bifurcate Killing horizon of
the worldsheet geometry. We are ready for the Carroll
generalization of extremal surfaces and black holes.

\subsection{Carroll extremal surfaces in PSM formulation}
\label{sec:4.2}

Trying to mimic the relativistic classification of loci is less rich
in the Carroll case, since the target space coordinate $\XP$ does not
appear on the right-hand side of the Carroll boost transformation laws
\eqref{eq:car25}, which we re-display here for convenience.
\eq{
\delta_\lambda X = \delta_\lambda \XH = 0 \qquad\qquad \delta_\lambda \XP = \XH\,\lambda
}{eq:carboost}
As expected, there is no natural notion of a Carroll horizon (since ``everything moves with the speed of light''). However, there still is the notion of an extremal surface, as evident from \eqref{eq:carboost}, namely when $\XH=0$. The three different cases are summarized in the table below. (We label negative $\XH$ as ``normal'' since, in most applications, $\XH$ is negative between the asymptotic region and the extremal locus.) 

\bigskip

\begin{center}
\begin{tabular}{c|ccc}
signs & $\XH>0$ & $\XH<0$ & $\boldsymbol{\XH=0}$ \\ \hline
 & anti-normal & normal & {\textbf{extremal}} \\
\end{tabular}
\end{center}

\bigskip

Thus, we have a similar definition of Carroll extremal surfaces as in
the relativistic case.\footnote{%
  Perhaps also $|\XH|=\infty$ has a similar interpretation, but since
  such loci typically are not part of the physical Carroll spacetime,
  we disregard this possibility here. These loci could appear at
  asymptotic boundaries, for instance, separating future and past null
  infinity. }

\paragraph{PSM definition of Carroll extremal surfaces.} Carroll extremal surfaces are loci in the PSM target space that are fixed points under Carroll boosts.
\paragraph{}

\smallskip

Note that every line of constant $X$ and $\XH$ (but varying $\XP$) is
a fixed-line under Carroll boosts, so in that sense, every such line
is ``marginally (anti-)trapped'' and the whole Carroll geometry could
be viewed as a ``horizon''. On this ``horizon'' there can still be an
exceptional point, the Carroll extremal surface, reminiscent of the
bifurcation surface of relativistic black holes.

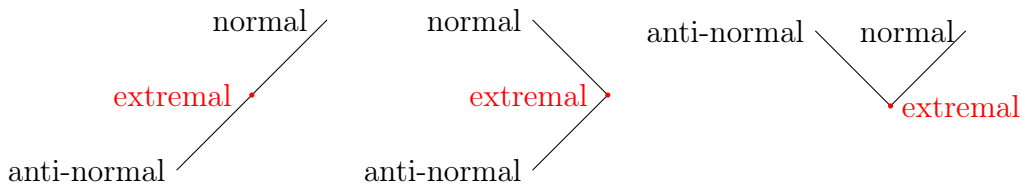
\begin{figure}[ht!]
\def\L{1.0}
\begin{center}
\begin{tikzpicture}
 \draw[black] (-\L,-\L) coordinate (lole) -- (\L,\L) coordinate (upre);
 \draw[black] (lole) node[left] (scrip) {anti-normal};
 \draw[black] (upre) node[left] (scrip) {normal\;};
 \draw[red,thick] (0,0) coordinate (center) node[left] (scrip) {extremal\;};
 \draw[red,thick] (center) circle(0.02*\L);
\end{tikzpicture}
\,
\begin{tikzpicture}
 \draw[black] (-\L,-\L) coordinate (lole) -- (0,0) coordinate (center);
 \draw[black] (center) -- (-\L,\L) coordinate (uple);
 \draw[black] (lole) node[left] (scrip) {anti-normal};
 \draw[black] (uple) node[left] (scrip) {normal};
 \draw[red,thick] (center) node[left] (scrip) {extremal\;};
 \draw[red,thick] (center) circle(0.02*\L);
\end{tikzpicture}
\,
\begin{tikzpicture}
 \draw[black] (\L,\L) coordinate (upri) -- (0,0) coordinate (center);
 \draw[black] (center) -- (-\L,\L) coordinate (uple);
 \draw[black] (upri) node[left] (scrip) {normal};
 \draw[black] (uple) node[left] (scrip) {anti-normal};
 \draw[red,thick] (center) node[right] (scrip) {extremal};
 \draw[red,thick] (center) circle(0.02*\L);
 \draw[draw=none] (0,-\L) coordinate (lole) -- (center); 
 \draw[draw=none] (lole) node[left] (scrip) {};
\end{tikzpicture}
\end{center}
\caption{Three diagrams to visualize Carroll black holes}
\label{fig:1.2}
\end{figure}

There is no Carroll analogue of Carter--Penrose diagrams, but we
can still draw diagrams similar to Fig.~\ref{fig:1.1} to highlight
different regions in the Carroll manifold with different signs of
$\XH$. Naturally, the diagrams in Fig.~\ref{fig:1.2} are less rich in
structure since there are fewer possibilities in the table above
compared to the Lorentzian case. While we chose to draw the lines at
45$^{\circ}$ (as a reminder that null hypersurfaces have Carroll
structures), at this stage there is no significance to this angle.
From a limiting perspective, one can understand these diagrams as
emerging from infinite boosts of $t=\mathrm{const.}$ hypersurfaces
(``wormholes'') of Lorentzian black hole Carter--Penrose diagrams.

The three diagrams in Fig.~\ref{fig:1.2} represent the same entity and
emphasise different ways of boosting the constant time slice of the
parent Carter--Penrose diagram. For instance, in the right diagram,
the $t=\mathrm{const.}$ hypersurface in the parent Carter--Penrose diagram
is boosted all the way to the future event horizon to both sides of
the extremal surface. We shall elaborate on the relation to the
wormhole picture in a higher-dimensional context in section
\ref{sec:CS_4d}.

\subsection{Carroll extremal surfaces in first- and second-order formulations}
\label{sec:4.3}

In the second-order version, the quantity $\XH$ does not exist, but
through the equations of motion \eqref{eq:eom} it is related on-shell
to the directional derivative of the dilaton field projected onto the
spatial inverse vielbein,
\eq{
\XH \approx -e^\mu\,\partial_\mu X \,.
}{eq:car49}
Thus, in the second-order formulation, the criterion for an extremal surface is that the directional derivative of the dilaton field projected onto the spatial inverse vielbein vanishes.
\begin{equation}
  \label{eq:car51}
\boxed{
\phantom{\Bigg(}
\textrm{Carroll\;extremal\;surface:}\qquad e^\mu\,\partial_\mu X = 0\qquad\qquad X>0
\phantom{\Bigg)}}
\end{equation}
This definition is analogous to the relativistic one \eqref{eq:car52} and is on-shell Carroll boost invariant, since $\delta_\lambda e^\mu\,\partial_\mu X=-\lambda v^\mu\,\partial_\mu X\approx \XH\lambda \,v^\mu e_\mu =0$. Note that one could add to \eqref{eq:car51} for free the condition $v^\mu\,\partial_\mu X=0$ since $X$ does not depend on $t$ on-shell.

\subsection{Carroll black holes}
\label{sec:4.4}

Equipped with our definition of extremal surfaces, we define Carroll black holes.

\medskip

\fbox{
\parbox{0.9\textwidth}{\vspace{0.5cm}
\noindent\textbf{Definition of Carroll black holes.} Carroll black holes are C-thermal states with finite entropy that have a Carroll extremal surface.
\vspace{0.5cm}}}

\medskip

In particular, we need the condition of finite entropy to exclude
Carroll constant dilaton solutions, which definitely should not be
referred to as Carroll black holes. In the next section we turn to the evaluation of the tantum gravity partition function for these solutions. 

\section{Evaluating the tantum gravity partition function}
\label{sec:tant_partitionfunction}
Now that we have explored the classical solution space to the action \eqref{eq:car24} and given physical interpretation to the massive states, we would like to investigate how the tantum gravity partition function for these models can be evaluated \cite{Ecker:2024czh}. We shall see that in a saddle-point approximation it indeed reproduces the correct thermodynamics.

The used argument is based on the Euclidean path integral approach to Einstein gravity, as formulated by Gibbons and Hawking in \cite{Gibbons:1976ue}. Their pivotal insight involves expanding the complete gravitational action $\Gamma[g]$, including relevant boundary terms, around a classical saddle-point $g_{\textrm{\tiny cl}}$. Which boundary terms are relevant is determined by demanding a well-defined variational principle, $\delta\Gamma[g_{\textrm{\tiny cl}};\,\delta g]=0$, as well as a finite on-shell action $\Gamma[g_{\textrm{\tiny cl}}]$. Under these hypotheses, the Euclidean partition function of general relativity
\begin{align}
    Z_{\textrm{GR}}=\int \mathcal{D}g\, \exp (-\frac{1}{\hbar }\,\Gamma_{\textrm{GR}} [g]) \approx \exp (-\frac{1}{\hbar }\,\Gamma_{\textrm{GR}}[g_{\textrm{\tiny cl}}])
\end{align}
is classically well-approximated by the exponential of the on-shell action provided $\hbar$ is sufficiently small. We need to specify boundary conditions before reducing the partition function to its classical approximation. In the simplest case (on which we shall focus below), this amounts to fixing the periodicity of Euclidean time, which in turn can be interpreted as the inverse temperature \cite{Gibbons:1976ue} together with some fall-offs for the dynamical fields as they approach the boundary. 

We shall do the same for the tantum gravity partition function using the example of two-dimensional Carroll dilaton gravity. According to the scaling limit \eqref{eq:tg1} we now use $\kappa $ as a semi-classical expansion parameter and write 
\begin{align}\label{eq:tg_part_fun}
    Z_{\textrm{TG}}=\int \mathcal{D}A_I \mathcal{D}X^I \exp \Big(-\frac{1}{\kappa }\Gamma _{\textrm{TG}}[A_I,X^I]\Big)
\end{align}
where 
\begin{align}
    \Gamma _{\textrm{TG}}[A_I,X^I]=I_{1^{\mathrm{st}}}+I_{\partial \mathcal{M}}\end{align}
is the first order action \eqref{eq:car24} augmented by some boundary term that makes the variational principle well-defined, a notion that will be made explicit below. While it is indeed possible to show that the exponent in \eqref{eq:tg_part_fun} arises from a limiting procedure \cite{Ecker:2024czh} we shall work directly in the limiting theory and show its internal consistency. Notably, the prefactor of the action in \eqref{eq:tg_part_fun} depends on the product of the two TG coupling constants \eqref{eq:tg1} (since $k\sim G_M^{-1}$). Thus, even though we formally sent $\hbar\to\infty$ in the Bronstein cube, there is still a semi-classical limit corresponding to sufficiently small $\kappa$\footnote{The semi-classical limit of the action \eqref{eq:whatever} exists whenever the product $\kappa G_M$ is small, which leads to two possible limits, either $\kappa\to 0$ while keeping $G_M\sim k^{-1}$ finite, or $G_M\to 0$ while keeping $\kappa$ finite. However, the latter possibility has to be discarded since it is at odds with our hypotheses to keep the Schwarzschild radius and the energy finite.}. In other words, as long as $\kappa$ is sufficiently small as compared to the action, we can expect a well-defined saddle-point approximation to the Euclidean path integral. While in the other sections we set $\kappa =1$ we shall keep it here for the sake of explicitness.

The choice of boundary conditions is usually motivated by the physical situation one wants to study. In the present case, we would like to choose them such that the solutions found in section \ref{sec:5.1.2} are allowed but not much more than that. Also, as often done for dilaton gravity models \cite{Grumiller:2007ju}, we choose the boundary $\partial \mathcal{M}$ to be a dilaton isosurface at some large $X=X_c >>1$. It is therefore practical to take the dilaton itself as a radial coordinate. The solutions from section \ref{sec:5.1.2} for generic $U(X)$ and $V(X)$ then read
\begin{equation}
    \begin{gathered}
        \XH = -\sqrt{2e^{-Q}(w-M)} \\
        v = e^{-Q}\,\frac{1}{\XH}\,\partial_t
    \end{gathered}
    \qquad\qquad
    \begin{gathered}
        \XP = 0 \\
        \tau = -\XH\,e^{Q}\,\dd t
    \end{gathered}
    \qquad\qquad
    \begin{gathered}
        \omega = e^Q\,\mathcal V\,\dd t \\
        e = -\frac{1}{\XH}\,\dd X
    \end{gathered}
    \label{eq:solutions 1st order}
\end{equation}
where $M$ is again a real constant parameter and the functions $Q(X)$ and $w(X)$ are defined by \eqref{eq:car111}. The fall-off conditions are chosen such that the following variations are allowed 
\begin{equation}
    \begin{gathered}
        \delta \XH = -\frac{\delta M}{\XH e^Q} \\
        \delta X = 0
    \end{gathered}
    \qquad\qquad
    \begin{gathered}
        \delta \XP = 0 \\
        \delta\tau = \frac{\delta M}{\XH}\,\dd t
    \end{gathered}
    \qquad\qquad
    \begin{gathered}
        \delta\omega = U\,\delta M\,\dd t \\
        \delta e = -\delta M\,\frac{\dd X}{\XH^3 e^Q}
    \end{gathered}
\end{equation}
where all the components have to be thought of as functions of $X$. Following the discussion in subsection \ref{sec:dimensions}, we assume that the coordinate $t$ has dimensions of length and is periodically identified $t\sim t+\beta \kappa $ where 
\begin{align}
    \beta \kappa := \lim _{X_c \to \infty }\oint \tau \Big \vert _{X=X_c}
\end{align}
is the proper length of the temporal cycle at infinity and $\beta $ is the inverse temperature measured in units of inverse energy ($k_B=1$). Keeping $\beta $ fixed completes our set of boundary conditions and defines a canonical ensemble. 

Since the dominant saddle-points typically correspond to smooth classical solutions \cite{Grumiller:2012rt}, we can use the tools of subsection \ref{sec:thermal_sols} to single out a unique $M$ for a given $\beta $ by the implicit relation 
\begin{equation}\label{eq:carr_temp}
     \beta ^{-1}=T= \frac{\kappa }{2\pi} w'(X_{\textrm{\tiny ext.}}(M)) ~,
\end{equation}
where $X_{\textrm{\tiny ext.}}(M)$ is the solution to the equation $w(X)=M$. To determine the boundary term in the action $\Gamma _{\textrm{TG}}$ we need to make sure that the present class of solutions indeed extremizes the action on the given space of allowed field configurations and that the on-shell value of the action is finite, i.e.,
\begin{align}
    \delta \Gamma _{\textrm{TG}}\,\big \vert _{\scriptscriptstyle\text{on-shell}} = 0 && \Gamma _{\textrm{TG}}\,\big \vert _{\scriptscriptstyle\text{on-shell}} < \infty ~.
\end{align}
The way these conditions are evaluated in practice is to keep the cutoff $X_c$ finite at first and only send $X_c$ to infinity at the very end. We start by an ansatz for the boundary action 
\begin{align}
    I_{\partial \mathcal{M}}=\frac{k}{2\pi }\int _{\partial \mathcal{M}} \Big(\alpha _1\, X\,\omega +\alpha _2\, X_H\, \tau +\alpha _3\, \tau\,\sqrt{2\,w\,e^{-Q}} \Big ) && \alpha _i \in \mathbb{R}
\end{align}
with three undetermined constants $\alpha _i \in \mathbb{R}$, $i=1,2,3$. Using that a large class\footnote{%
This class includes all Schwarzschild--Tangherlini--Reissner--Nordstr{\"o}m--AdS black holes as well as Witten and JT black holes and deformations thereof. See \cite{Grumiller:2007ju} for details.} of dilaton gravity models obeys
\begin{equation}
    \lim\limits_{X\to \infty }w(X)=\infty 
\end{equation} 
we can write the variation of the full action on-shell as a boundary term
\begin{align}
    \delta \Gamma_{\textrm{\tiny TG}} \Big \vert _{\scriptscriptstyle\text{on-shell}} = \frac{k}{2\pi }\int _{\partial \mathcal{M}}\!\!\! \dd t\,\Big((\alpha _1+1)XU+2\alpha _2-\alpha _3+1+\mathcal{O}(X_c^{-1})\Big)\,\delta M \overset{!}{=}0
\end{align}
which fixes the value of the first coefficient, $\alpha_1=-1$, and further gives the algebraic constraint $2\alpha_2 = \alpha_3-1$. The on-shell action then simplifies to
\begin{equation}
    \begin{split}
        \Gamma_{\textrm{\tiny TG}} &\Big \vert _{\scriptscriptstyle\text{on-shell}} =\\
        &\frac{k}{2\pi }\int _0^{\beta \kappa }\!\!\! \dd t \,\int _{X_{\scriptscriptstyle\mathrm{ext.}}}^{X_c}\!\!\! \dd X \, \partial _X\Big(X\partial _Xw(X)-XU(X)(w-M)-2w(X)\Big)\\
        &\quad +\frac{k}{2\pi }\int _0^{\beta \kappa } \!\!\! \dd t \,\Big(-X\partial _Xw+XU(w-M)+(\alpha _3+1)w-M+\mathcal{O}(X_c^{-1})\Big) \\ 
        &=\frac{\beta \kappa k }{2\pi }\Big(2M-\frac{2\pi T}{\kappa } X_{\scriptscriptstyle\mathrm{ext.}}\Big)+\frac{k}{2\pi }\int _0^{\beta\kappa }\!\!\! \dd t \,\Big((\alpha _3-1)w-M+\mathcal{O}(X_c^{-1})\Big)
    \end{split}
\end{equation}
where we used \eqref{eq:carr_temp} in the last equality. One can see that the on-shell action is finite in the limit $X_c\to \infty $ if we set $\alpha_3=1$. In summary, we have 
\begin{equation}
    \alpha _1 =-1 \qquad\qquad \alpha _2 =0 \qquad\qquad \alpha _3=1 \,.
\end{equation}
The holographically renormalized first-order action therefore reads
\begin{equation}
  \addtolength{\fboxsep}{7pt}
   \boxed{
\begin{aligned}
   \Gamma_{\textrm{\tiny TG}}\,[A_I,X^I] =\frac{k}{2\pi }\int _{\mathcal{M}}\Big(X\,\dd \omega &+ \XH \,(\dd \tau +\omega \wedge e)+ \XP \,\dd e+\mathcal{V}\,\tau \wedge e \Big) \\
   &- \frac{k}{2\pi }\int _{\partial \mathcal{M}}\Big(X\,\omega -\tau\, \sqrt{2e^{-Q}w}\Big)
\end{aligned}
   }
\end{equation} 
and in the limit $X_c\to \infty $ evaluates on-shell to
\begin{equation}
    \Gamma_{\textrm{\tiny TG}} \Big \vert _{\scriptscriptstyle\text{on-shell}}=\frac{\beta k\kappa }{2\pi } \Big(M-\frac{2\pi T}{\kappa } X_{\scriptscriptstyle\text{ext.}}\Big)\,.
\end{equation}
The free energy of the solutions \eqref{eq:solutions 1st order} is then
\begin{align}
    F=\frac{1}{\kappa \beta} \Gamma_{\textrm{\tiny TG}} \Big \vert _{\scriptscriptstyle\text{on-shell}} = \frac{k}{2\pi }\Big(M-\frac{2\pi T}{\kappa }X_{\scriptscriptstyle\text{ext.}}\Big) = F (T)
    \label{eq:F 2}
\end{align}
and upon using the standard thermodynamic relations
\begin{align}
    S=-\frac{\partial F}{\partial T} && E=F+TS
\end{align}
one precisely recovers the thermodynamic quantities of section \ref{sec:3} (with $\kappa $ reinstalled),
\begin{equation}
    E = \frac{k}{2\pi }M \qquad\qquad S = \frac{k}{\kappa }X_{\scriptscriptstyle\text{ext.}} ~.
    \label{eq:E and S 2}
\end{equation}
This makes it evident that these relations hold for generic dilaton gravity models and shows the consistency of the partition function with the first law. 

The analysis in this section was done for the two-dimensional setting but to some extent these results also apply to the spherically symmetric sector of higher-dimensional magnetic Carroll gravity. After all, we saw in subsection \ref{sec:spher-reduct-magn} that the latter admits a classically equivalent formulation in terms of a two-dimensional Carroll dilaton gravity theory. However, since in the Lorentzian setting dimensional reduction and quantization do not commute \cite{Frolov:1999an} (``dimensional reduction anomaly'') we in general have to expect a similar behaviour in a Carroll setting. Therefore, one should think of the spherically reduced theory as a separate quantum gravity model, the classical limit of which coincides with the classical limit of spherically symmetric magnetic Carroll gravity. In the saddle-point approximation, on the other hand, the dimensional reduction anomaly plays no role since the latter is a 1-loop effect and a consequence of the UV behavior of the theory when coupled to matter.

Our conclusions of this section are in line with previous results in
the literature: taking the standard Carroll limit in thermal partition functions leads to infinities which are ultimately related to a degeneracy in the energy spectrum \cite{Figueroa-OFarrill:2023qty,
deBoer:2023fnj,Cotler:2024xhb}. For instance, for the theory of a free electric scalar field \eqref{eq:elec_scalar} in a finite volume a na\"ive evaluation of an expression like $\text{Tr}(e^{-\beta H})$ involves a sum over momenta. But, since the Hamiltonian does not contain any spatial gradients, any choice of momentum leads to the same energy making the sum diverge. This is in contrast to the Euclidean case where the contributions of large momenta are exponentially damped.  
Proposals to circumvent this problem are introducing an imaginary chemical potential \cite{Poulias:2025eck} to regulate this sum or keeping a UV-regulator in the form of a finite lattice spacing \cite{Cotler:2024xhb}. However, in all of these treatments $\hbar $ was left untouched. Following the discussion in section \ref{sec:bronstein}, our notion of thermodynamics for Carroll black holes precisely uses this handle of scaling $\hbar $ to obtain a sensible notion of thermodynamics for Carroll theories.  

We now turn to several examples of Carroll black hole geometries and investigate their thermodynamical properties. 

\section{Examples for 2d Carroll black holes}
\label{sec:5}

In this section, we apply our general analysis to specific models, the
Carroll JT model, the Carroll--Schwarzschild model, the Carroll CGHS
model, and the Carroll Witten black hole. In each case, we ignore the
constant dilaton vacua and focus exclusively on the linear dilaton
sector. Here, we work again in units where $\kappa =1$.

The Carroll thermodynamic quantities of solutions in the black hole
sector of these models are listed in Table \ref{table:1}. The table
also includes some other, more generic cases like the
Carroll--Schwarzschild--Tangherlini solution and the Carroll
$ab$-family.

\begin{table}[ht!]
\centering
\begin{tabular}{||c || c c c | c c c c ||} 
 \hline
 Model & $U(X)$ & $V(X)$ & $w(X)$ & $E$ & $T$ & $S$ & $C$ \\ [0.5ex] 
 \hline\hline
 \text{CJT} & $0$ & $\frac{X}{\ell ^2}$ & $\frac{X^2}{2\ell ^2}$ & $\frac{k}{2\pi }M$ & $\frac{\sqrt{2M}}{2\pi \ell }$ & $k\ell \sqrt{2M}$ & $2\pi k \ell ^2 T$ \\[.5em]
 \text{CS} & $-\frac{1}{2X}$ & $\frac{\lambda ^2}{4}$ & $\frac{\lambda ^2}{2}\sqrt{X}$ & $\frac{k}{2\pi }M$ & $\propto \frac{1}{M}$ & $\propto M^2$ &  $\propto -T^{-2}$ \\[.5em]
 \text{CST} & \eqref{eq:choice_UV} & \eqref{eq:choice_UV} & $\frac{\lambda ^2}{2}X^{\frac{d-2}{d-1}}$ & $\frac{k}{2\pi }M$ & $\propto M^{\scriptscriptstyle \frac{1}{2-d}}$ & $\propto M^{\scriptscriptstyle \frac{d-1}{d-2}}$ &  $\propto -T^{1-d}$ \\[.5em]
 \text{CCGHS} & $0$ & $\Lambda >0$ & $\Lambda X$ &$\frac{k}{2\pi}M$ & $\frac{\Lambda}{2\pi}$ & $\frac{kM}{\Lambda}$ & $\infty $ \\[.5em]
 \text{CWBH} & $-\frac{1}{X}$ & $\frac{\lambda^2}{2}X$ & $\frac{\lambda^2}{2}X$ & $\frac{k}{2\pi}M$ & $\frac{\lambda^2}{4\pi}$ & $\frac{2kM}{\lambda^2}$ & $\infty $ \\[.5em]
 \text{Cab} & $-\frac{a}{X}$ & $\frac{B}{2}X^{a+b}$ & $\frac{B}{2(b+1)}X^{b+1}$ & $\frac{k}{2\pi }M$ & $\propto M^{\frac{b}{b+1}}$ & $\propto M^{\frac{1}{b+1}}$ &  $\frac{k}{b}\Big(\frac{4\pi T}{B}\Big)^{\frac{1}{b}}$ \\[.5em]
 \hline
\end{tabular}
\caption{Carroll thermodynamic quantities for the Carroll JT model (CJT), Carroll--Schwarzschild (CS), Carroll--Schwarzschild--Tangherlini (CST) in $d+1$ spacetime dimensions, Carroll CGHS (CCGHS), Carroll Witten black hole (CWBH), and the Carroll $ab$-family (Cab). As $w''=0$ for CCGHS and CWBH the specific heat diverges for these models. In some expressions, we left out state-independent prefactors for brevity which is denoted by $\propto $.}
\label{table:1}
\end{table}

In the first two subsections \ref{sec:carjt}-\ref{sec:5.3}, we show the spectrum, thermodynamics, and an example for boundary conditions of the CJT model. In subsection \ref{sec:carr_schwarzsch}, we present the 2d perspective of the Carroll--Schwarzschild black hole. In subsection \ref{sec:5.4}, we investigate the CCGHS model. In subsection \ref{sec:5.5}, we present the Carroll--Witten black hole.

\subsection{Carroll JT model}
\label{sec:carjt}

The Jackiw--Teitelboim (JT) model \cite{Jackiw:1984,Teitelboim:1983ux} was the first 2d model of gravity. It is
particularly elegant, as it allows a reformulation as non-abelian BF
theory \cite{Isler:1989hq,Chamseddine:1989yz}, in contrast to nearly
all other 2d dilaton gravity models. All its solutions are locally
(A)dS$_2$ so that JT gravity is tailor-made for a holographic
description \cite{Strominger:1998yg,Maldacena:1998uz,Brigante:2002rv,Gupta:2008ki,Alishahiha:2008tv,Hartman:2008dq,Castro:2008ms}. Especially the SYK/JT correspondence
\cite{Kitaev:15ur,Sachdev:1992fk,Sachdev:2010um,Maldacena:2016hyu}
has reinvigorated the interest in JT gravity and its holographic
description. Due to its simple BF formulation (see also subsection \ref{sec:BF_theories}), the JT model was the
starting point for Carroll limits of 2d dilaton gravity
\cite{Grumiller:2020elf}. Here, we summarize the key results for the
Carroll JT (CJT) model, and in subsection \ref{sec:5.3}, we discuss
boundary conditions for CJT.

The CJT model is given by the Lagrangian \eqref{eq:car1} with the potentials
\eq{U_{\textrm{\tiny CJT}}(X) =0 \qquad \quad
V_{\textrm{\tiny CJT}}(X) = \frac{1}{\ell^2}\,X\,.
}{eq:JT1}
The function $\pot$ defined in \eqref{eq:car6} for CJT is given by 
\eq{
\pot_{\textrm{\tiny CJT}}(X) = \frac{X^2}{2\ell^2}\,.
}{eq:JT3}

Applying our general analysis of section \ref{sec:3} to the choice \eqref{eq:JT1} yields the linear dilaton solutions (we fix the integration constant coming from integrating \eqref{eq:car9} without loss of generality by a shift of the origin of the spatial coordinate $r$, and we take the positive branch of the square-root function)
\begin{subequations}\label{eq:sols_newgauge}
 \begin{align}
  X &= \frac{1}{2}\,e^{r/\ell} + M\ell^2\,e^{-r/\ell} & \omega &= \frac{X}{\ell^2}\,\dd t \\
  \XH &= -\frac{1}{2\ell}\,e^{r/\ell} + M\ell\,e^{-r/\ell} & 
  \tau &= -\XH\,\dd t\\
  \XP &= 0  &
  e &= \dd r \,.
 \end{align}
\end{subequations}
Translating the 1-forms into second-order notation, the solution above reads
\eq{
\dd s^2 = \dd r^2\qquad\qquad v=\frac{2\ell\,e^{-r/\ell}}{ 2M\ell^2\,e^{-2r/\ell}-1}\,\partial_t \qquad\qquad X = \frac{1}{2}\,e^{r/\ell} + M\ell^2\,e^{-r/\ell}\,.
}{eq:car47}

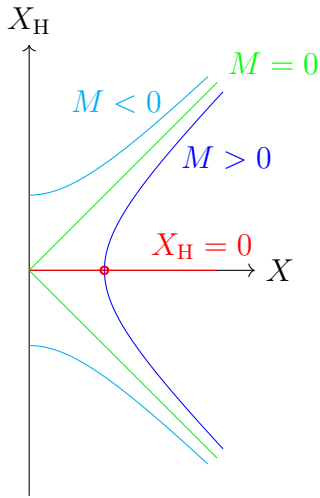
\begin{figure}
\def\L{2.5}
\begin{center}
\begin{tikzpicture}
 \draw[black,thin,->] (0,0) coordinate (le) -- (1.2*\L,0) coordinate (re);
 \draw[black,thin,->] (0,-1.2*\L,0) coordinate (dn) -- (0,1.2*\L) coordinate (up);
 \draw[green,thin] (0,0) coordinate (lole) -- (\L,\L) coordinate (upre);
 \draw[green,thin] (0,0) coordinate (lore) -- (\L,-\L) coordinate (uple);
 \draw[red,thin] (0,0) -- (\L,0) coordinate (rex);
 \draw[blue] plot[domain=0:0.64*\L] ({cosh(\x)},{sinh(\x)});
 \draw[blue] plot[domain=0:0.64*\L] ({cosh(\x)},{-sinh(\x)});
 \draw[cyan] plot[domain=0:0.64*\L] ({sinh(\x)},{cosh(\x)});
 \draw[cyan] plot[domain=0:0.64*\L] ({sinh(\x)},{-cosh(\x)});
 \draw[red] (rex) node[above] {\!\!\!\!\!\!$\XH=0$};
 \draw[black] (re) node[right] (scrip) {$X$};
 \draw[black] (up) node[above] (scrip) {$\XH$};
 \draw[green] (\L,1.1*\L) node[right] (scrip) {$M=0$};
 \draw[blue] (0.75*\L,0.6*\L) node[above,right] (scrip) {$M>0$};
 \draw[cyan] (0,0.9*\L) node[above,right] (scrip) {$\quad M<0$};
 \draw[red,thick] (0.4*\L,0) circle(0.02*\L);
\end{tikzpicture}
\end{center}
 \caption{Target space picture of Carroll JT by plotting the level sets of \eqref{eq:car112}, restricted to the region $X\geq 0$. Extremal points are red circles and exist only for $M> 0$. The solutions given in \eqref{eq:sols_newgauge} cover the lower half of this diagram. }
 \label{fig:3}
\end{figure}

In Fig.~\ref{fig:3} we depict a constant $\XP$ slice of the PSM target space associated with the CJT model.\footnote{
For positive mass, at the Carroll extremal surface $X=\ell\sqrt{2M}$ the solution can be joined to one where $\XH\to-\XH$ and hence $v\to -v$.} The spectrum of CJT falls into three classes, depending on the sign of the mass parameter $M$: 
\begin{itemize}
    \item $M<0$: no Carroll black hole, since $\XH<0$ everywhere, reminiscent of the global AdS$_2$ solution of JT
    \item $M=0$: limiting case, where $\XH\to 0$ as $r\to-\infty$, reminiscent of the Poincar\'e horizon of the massless JT solution
    \item $M>0$: Carroll black holes, since $\XH=0$ has the solution $X=\ell\sqrt{2M}$ or, equivalently, $r=\frac\ell2\,\ln(2M\ell^2)$, reminiscent of black hole solutions of JT 
\end{itemize}
We focus on the positive mass sector since it features Carroll extremal surfaces. Furthermore, to have positive entropy we restrict to the branches with $X> 0$.

Energy, entropy, temperature, and specific heat of these solutions are given in Table \ref{table:1}, and the first law is satisfied, as shown in section \ref{sec:3.3}. Expressing the entropy as a function of the energy shows a relation similar to the Cardy formula for a chiral half of a 2d conformal field theory,
\begin{align}
    S=\frac{\pi^2c\,T}{3} = 2\pi\sqrt{\frac{c\,E}{6}}
\end{align}
provided the central charge is chosen as
\begin{align}
    c = \frac{6k\ell ^2}{\pi} ~.
\end{align}
This is again reminiscent of the relativistic case \cite{Grumiller:2017qao}. 

\subsection{Example of boundary conditions for CJT}
\label{sec:5.3}

One can interpret \eqref{eq:car47} as radial Gaussian coordinates and provide a ``Feffer\-man--Graham'' expansion for the vector field and the dilaton
\eq{
v = 2\ell e^{-r/\ell}\,\big(-1+{\mathcal{O}}(e^{-2r/\ell})\big)\,\partial_t\qquad\qquad X = \frac12\,e^{r/\ell}\,\big(1+{\mathcal{O}}(e^{-2r/\ell})\big)
}{eq:car48}
where the leading terms are fixed, and the subleading terms contain state-dependent information. Similarly to JT gravity, there are numerous inequivalent choices for boundary conditions \cite{Grumiller:2017qao}. It is not our intention to exhaustively discuss the possibilities for CJT gravity. Instead, we provide just one example for boundary conditions and leave a more comprehensive study for future work.

The Brown--Henneaux-like boundary conditions
\begin{subequations}
    \label{eq:JT9}
\begin{align}    
X &= \frac12\,e^{r/\ell} + M(t)\,\ell^2 e^{-r/\ell}  & \omega &= \frac{1}{2\ell^2}\,e^{r/\ell}\,\dd t + {\mathcal{O}}(e^{-r/\ell}) \\
\XH &= -\frac{1}{2\ell}\,e^{r/\ell} + M(t)\,\ell\,e^{-r/\ell} & \tau &= \frac{1}{2\ell}\,e^{r/\ell}\,\dd t + {\mathcal{O}}(e^{-r/\ell})\\
\XP &= 0  & e &= \dd r 
\end{align}
\end{subequations}
with $\delta M\neq 0$ are preserved by the gauge transformations \eqref{eq:car25}-\eqref{eq:car26a} with gauge parameters $\lambda_{\textrm{\tiny P}} = 0$ and
\eq{
\lambda = e^{r/\ell}\,\frac{\eta}{\ell} + 2\ell M(t)\eta\,e^{-r/\ell} \qquad\qquad\lambda_{\textrm{\tiny H}} = e^{r/\ell}\,\eta - 2\ell^2 M(t)\eta\,e^{-r/\ell} 
}{eq:JT42}
where $\eta$ is the transformation parameter. The equations of motion \eqref{eq:eom} are solved by the field configuration \eqref{eq:JT9}, up to subleading terms (which can be determined in closed form, if desired). On-shell the mass function $M(t)$ is given by the Casimir $M=\frac{X^2}{2\ell^2}-\frac{\XH^2}{2}$.

The variation of the boundary charges
\eq{
\delta \mathcal{Q}[\lambda_I] = \frac{k}{2\pi}\,\big(\lambda\,\delta X+\lambda_{\textrm{\tiny H}}\,\delta\XH+\lambda_{{\textrm{\tiny P}}}\,\delta\XP\big)
}{eq:JT38}
in the present case can be integrated (in field space) to a single boundary charge
\eq{
\mathcal{Q}[\eta] =\frac{k\ell}{\pi}\,\eta\,M(t)
}{eq:JT40}
which is finite as the radial coordinate approaches the asymptotic boundary, $r\to\infty$. On-shell it is also conserved, $\partial_t\mathcal{Q}[\eta]\approx 0$. We assumed here a slicing of the phase space where $\eta$ is state-independent (see, e.g., \cite{Adami:2020ugu, Grumiller:2021cwg} for a discussion of different phase space slicings in Lorentzian 2d dilaton gravity).

The asymptotic symmetry algebra trivially is abelian in the present case since we have only one boundary charge, namely the Casimir $M$.
\eq{
\{\mathcal{Q}[\eta_1],\,\mathcal{Q}[\eta_2]\} \approx \delta_{\eta_2} \mathcal{Q}[\eta_1] \approx \frac{k\ell}{\pi}\,\eta_1\,\delta_{\eta_2}M = 0
}{eq:JT39}

\subsection{Carroll--Schwarzschild black hole, 2d perspective}
\label{sec:carr_schwarzsch}

As reviewed in section \ref{sec:2.2.1}, spherical reduction of
Einstein gravity leads to a specific 2d dilaton gravity model, the
solutions of which reproduce the Schwarzschild black hole. There is an
expansive history of spherically reduced gravity \cite{Berger:1972pg,Unruh:1976db,Benguria:1977in,Thomi:1984na} that predates the
developments of 2d dilaton gravity. Here, we consider the Carroll
limit of the Schwarzschild black hole from a 2d perspective. See section \ref{sec:CS_4d} for a 4d perspective.

\begin{figure}
\def\L{2.5}
\begin{center}
\begin{tikzpicture}
 \draw[black,thin,->] (0,0) coordinate (le) -- (1.2*\L,0) coordinate (re);
 \draw[black,thin,->] (0,-1.2*\L,0) coordinate (dn) -- (0,1.2*\L) coordinate (up);
 \draw[red,thin] (0,0) -- (\L,0) coordinate (rex);
 \draw[scale=0.5,samples=500, domain=2:\L*2, smooth, variable=\x, blue] plot ({\x}, {1.5*sqrt(\x-sqrt(2)*sqrt(\x))});
 \draw[scale=0.5,samples=500, domain=2:\L*2, smooth, variable=\x, blue] plot ({\x}, {-1.5*sqrt(\x-sqrt(2)*sqrt(\x))});
  \draw[scale=0.5,samples=500, domain=0:\L*2, smooth, variable=\x, green] plot ({\x},{1.5*sqrt(\x)});
 \draw[scale=0.5, domain=0:\L*2,samples=500, smooth, variable=\x, green] plot ({\x}, {-1.5*sqrt(\x)});
 \draw[cyan] plot[ samples=500,domain=0:\L,smooth] ({\x},{1.2*sqrt(\x+sqrt(2)*sqrt(\x))});
 \draw[cyan] plot[samples=500,smooth,domain=0:
 \L] ({\x},{-1.2*sqrt(\x+sqrt(2)*sqrt(\x))});
 \draw[red] (rex) node[above] {\!\!\!\!\!\!$\XH=0$};
 \draw[black] (re) node[right] (scrip) {$X$};
 \draw[black] (up) node[above] (scrip) {$\XH$};
 \node [blue] at (3.2,1.0) {$M>0$};
  \node [green] at (3.2,1.7) {$M=0$};
   \node [cyan] at (3.2,2.6) {$M<0$};
 \draw[red,thick] (0.4*\L,0) circle(0.02*\L);
\end{tikzpicture}
\end{center}
 \caption{Target space picture of spherically reduced Carroll--Schwarzschild black hole. Extremal points are red circles and exist only for $M> 0$. The other symplectic leaves do not exhibit such points as for $M=0$ the point would be at $X=0$ and for $M<0$ the leaves do not contain points with $\XH =0$ at all (they are not simply connected). The black hole sector is thus given by $M>0$. The solutions \eqref{eq:CSS_1}-\eqref{eq:CSS_3} describe the lower half of the diagram.}
 \label{fig:5}
\end{figure}
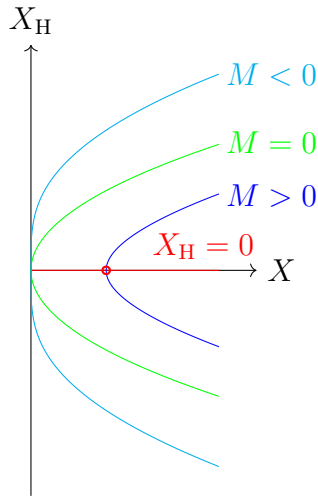

The spherically reduced Carroll--Schwarzschild (CS) model is given by the potentials
\eqref{eq:choice_UV}, which for $d=3$ are
\begin{align}
    U_{\text{\tiny CS}}(X)=-\frac{1}{2X} && V_{\text{\tiny CS}}(X)=\frac{\lambda ^2}{4} ~.
\end{align}
The functions $w_{\text{\tiny CS}}$ and $e^{Q_{\text{\tiny CS}}}$ are
\begin{align}
    e^{Q_{\text{\tiny CS}}}=\frac{1}{2\sqrt{X}} && w_{\text{\tiny CS}}(X)=\frac{\lambda ^2}{4}\sqrt{X} 
\end{align}
where we chose the integration constant of the second
integral in \eqref{eq:car111} accordingly. This model is described by a target space diagram
given in Fig.~\ref{fig:5}. The solutions never take negative values of
the dilaton. Moreover, the black hole sector of the model is given by
$M>0$ as all other solutions do not lead to states with finite entropy
$S\sim X_{\text{\tiny ext.}}$.
Let us choose $\lambda =2$ for convenience. This
implies that the dilaton measures the surface radius as seen from the
higher-dimensional setting, i.e., the spherical part of the 4d metric
reads $X\dd \Omega ^2_{S^2}$ (see also \eqref{eq:carr_spher_ansatz},
\eqref{eq:redef_dilaton}). Applying our analysis from section
\ref{sec:3} yields
\begin{align}
\XH &=-\sqrt{4X-4M\sqrt{X}} & \omega &=\frac{2\sqrt{X}-M}{2X}\dd t \label{eq:CSS_1}\\
 \XP &=0 & \tau &=-\frac{\XH }{2\sqrt{X}}\dd t\label{eq:CSS_2}\\
& &  e&=\dd r \label{eq:CSS_3}
\end{align}
and a 2d Carroll curvature scalar
\begin{align}
    R=4e^\mu v^\nu \partial _{[\mu }\hat{\omega }_{\nu ]}=\frac{2M}{X^{\frac{3}{2}}} && \hat{\omega } =\omega -U(X)\XH \tau 
\end{align}
where $\hat{\omega }$ is defined as the torsion-free part of the Carroll connection [see \eqref{eq:spin_conn_split}]. The Carroll second-order variables read
\begin{equation}
    v=-\frac{1}{\sqrt{1-\frac{M}{\sqrt{X}}}}\partial _t \qquad \qquad h=\frac{\dd X^2}{4X-4M\sqrt{X}}
\end{equation}
where the vector field is normalized asymptotically, $\lim _{X\to \infty }v=-\partial _t$. For simplicity, in these solutions, the ambiguity in the torsion-free spin connection was fixed to $\Xphi=0$. To bring this into a more familiar form, we can define the radial coordinate
\begin{align}
  \label{eq:raddef}
  \rad ^2=X  
\end{align}
which together with the Schwarzschild mass $m=\frac{M}{2}$ leads to
\begin{equation}\label{eq:2dCS2ovars}
    v=-\frac{1}{\sqrt{1-\frac{2m}{\rad}}}\partial _t \qquad \qquad h=\frac{\dd \rad^2}{1-\frac{2m}{\rad}}~. 
\end{equation}

The Carroll thermodynamic quantities for the Carroll--Schwarzschild black hole
\begin{align}
    E=\frac{k}{\pi}\,m && T=\frac{1}{8\pi m} && S=4km^2
\label{eq:lala}
\end{align}
satisfy the first law,\footnote{%
From a 4d perspective, the coupling constant $k$ is given by $\pi/G_M$ in units where $\lambda =2$, see Eq.~\eqref{eq:choice_UV}.}
\begin{equation}
    \delta E=T\,\delta S ~.
\end{equation}
Generalizing Carroll--Schwarzschild to
Carroll--Schwarzschild--Tangherlini is straightforward, and the main
results are summarized in Table \ref{table:1}.

In section~\ref{sec:CS_4d}, we provide the 4d perspective on
these solutions.

\subsection{Carroll CGHS}
\label{sec:5.4}

The Callan--Giddings--Harvey--Strominger (CGHS) model
\cite{Callan:1992rs} is a 2d toy model for black hole evaporation. It
consists of a Lorentzian 2d dilaton gravity action with potentials
$U=0$, $V=\Lambda=\textrm{const.}$ plus some minimally coupled scalar
fields as carriers of the Hawking quanta. In the present discussion, we always
neglect interactions with matter, so when we refer to the CGHS model
or its Carroll counterpart, we solely mean the geometric part of the
model without matter. Besides the JT model, the CGHS model is arguably
the simplest 2d dilaton gravity model. A more precise version of this
statement is that only the JT and the CGHS model permit a
reinterpretation of the corresponding PSM as non-abelian BF theory.
This is the main reason why the CGHS model was the first one to
receive a holographic interpretation \cite{Afshar:2019tvp} after the
JT model.

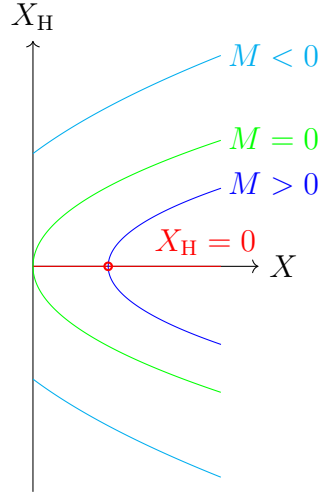
\begin{figure}
\def\L{2.5}
\begin{center}
\begin{tikzpicture}
 \draw[black,thin,->] (0,0) coordinate (le) -- (1.2*\L,0) coordinate (re);
 \draw[black,thin,->] (0,-1.2*\L,0) coordinate (dn) -- (0,1.2*\L) coordinate (up);
 \draw[red,thin] (0,0) -- (\L,0) coordinate (rex);
 \draw[scale=0.5, domain=2:\L*2, smooth,samples=500, variable=\x, blue] plot ({\x}, {1.2*sqrt(\x-2)});
 \draw[scale=0.5, domain=2:\L*2, smooth,samples=500, variable=\x, blue] plot ({\x}, {-1.2*sqrt(\x-2)});
  \draw[scale=0.5, domain=0:\L*2, smooth,samples=500, variable=\x, green] plot ({\x}, {1.5*sqrt(\x)});
 \draw[scale=0.5, domain=0:\L*2, smooth,samples=500, variable=\x, green] plot ({\x}, {-1.5*sqrt(\x)});
 \draw[cyan] plot[domain=0:\L,samples=500,smooth] ({\x},{1.5*sqrt(\x+1});
 \draw[cyan] plot[domain=0:
 \L,samples=500,smooth] ({\x},{-1.5*sqrt(\x+1});
 \draw[red] (rex) node[above] {\!\!\!\!\!\!$\XH=0$};
 \draw[black] (re) node[right] (scrip) {$X$};
 \draw[black] (up) node[above] (scrip) {$\XH$};
 \node [blue] at (3.2,1.1) {$M>0$};
  \node [green] at (3.2,1.7) {$M=0$};
   \node [cyan] at (3.2,2.8) {$M<0$};
 \draw[red,thick] (0.4*\L,0) circle(0.02*\L);
\end{tikzpicture}
\end{center}
 \caption{Target space picture of Carroll CGHS with $\XP =0$, restricted to the region $X\geq 0$. Extremal points are red circles and exist only for $M> 0$. The lower half is described by the solutions \eqref{eq:Carr_cghs_1}-\eqref{eq:Carr_cghs_3}.}
 \label{fig:4}
\end{figure}

The Carroll limit of the CGHS model (CCGHS) has the same potentials
\begin{equation}
    U_{\text{\tiny CCGHS}}=0 \qquad \qquad V_{\text{\tiny CCGHS}}=\Lambda =\text{const.} >0\,.
\end{equation}
The solutions of the linear dilaton sector are 
\begin{align}\label{eq:Carr_cghs_1}
    X&=\frac{\Lambda }{2}r^2+\frac{M}{\Lambda } & \omega &=\Lambda \dd t \\
    \XH &=-\Lambda r & \tau &=-\XH \dd t\\
    \XP &=0 & e &=\dd r \label{eq:Carr_cghs_3}
\end{align}
leading to a flat Carroll spacetime, i.e., $R=0$. Here, the radial
coordinate was fixed such that $r=0$ corresponds to $\XH =0$.
Investigating the spectrum of this model shows that Carroll extremal
surfaces exist only for positive values of $M$ (see Fig.~\ref{fig:4}). 
The various Carroll thermodynamic quantities of these solutions are given in Table \ref{table:1}. As the temperature is fixed to a single specific value in terms of the model-dependent constant $\Lambda $, the specific heat diverges. 

\subsection{Carroll Witten black hole}
\label{sec:5.5}

The Witten black hole \cite{Mandal:1991tz, Elitzur:1991cb,
  Witten:1991yr} emerges from 2d string theory. From the worldsheet
perspective, it is described by an SL$(2,\mathbb{R})/U(1)$ gauged
WZW-model. Interpreting the vanishing of the $\beta$-functions of this
conformal field theory as target space equations of motion yields as
target space action a Lorentzian 2d dilaton gravity model with
potentials $U=-\frac1X$ and $V=\frac{\lambda^2}{2}\,X$, where
$\lambda^2\propto 1/\alpha^\prime$ with the inverse string tension
$\alpha^\prime$. As common in the literature, we use the phrase
``Witten black hole'' as a label for the conformal field theory, the target space
theory, and the positive mass spectrum of solutions to the latter.
The Euclidean continuation of the Witten black hole is the famous
cigar geometry, $\dd s^2=\dd r^2+\tanh^2r\,\dd\tau^2$. For more
details on the Witten black hole, see section~2.1.2 in
\cite{Grumiller:2002nm} and Refs.~therein.

The Carroll limit of the Witten black hole features the same potentials
\begin{equation}
  U_{\text{\tiny CWBH}}=-\frac{1}{X} \qquad \qquad V_{\text{\tiny CWBH}}=\frac{\lambda^2}{2}\,X >0
\end{equation}
and is referred to as Carroll Witten black hole. Analogously to its
Lorentzian avatar (see e.g.~\cite{Emparan:2013xia}), it emerges as
$D\to\infty$ limit of the CST black hole and is conformally related to
the CCGHS model by a dilaton-dependent Weyl rescaling.

In the linear dilaton sector, the CWBH solutions
\begin{align}\label{eq:Carr_CW_1}
    X&=\frac{2M}{\lambda^2}\,\cosh^2\frac{\lambda r}{2} & \omega &= \Big(\lambda^2-\frac{M}{X}\Big)\,\dd t \\
    \XH &=-\sqrt{\lambda^2X^2-2MX} & \tau &=-\frac{\XH }{X} \dd t\\
    \XP &=0 & e &=\dd r \label{eq:Carr_CW_3}
\end{align}
lead to the same thermodynamical behaviour as the CCGHS model. In fact,
all thermodynamic formulas are equivalent for CCGHS and CWBH upon
replacing $\Lambda\to\tfrac{\lambda^2}{2}$.
\section{Carroll--Schwarzschild black hole, 4d perspective}
\label{sec:CS_4d}
In this section, we elaborate on the 4d perspective of the Carroll–-Schwarzschild black hole and the associated wormhole picture. While the following discussion easily generalizes to higher dimensions (section~\ref{sec:5}), we focus, for clarity, on $3+1$ dimensions. 

The Schwarzschild line element is given by
\begin{equation}
  \dd s^2=-\left(1-\frac{\rad_s}{\rad}\right)c^2\dd t^2+\frac{\dd\rad^2}{1-\frac{\rad_s}{\rad}}+\rad ^2\dd\Omega^2_{S^{2}}
\end{equation}
where $\rad _s=\frac{2mG_N}{c^2}$ and $\dd\Omega^2_{S^{2}}$ is the metric on the round 2-sphere. 


In the tantum gravity limit, where $\rad _s $ stays finite by construction we arrive at \cite{Perez:2021abf,Hansen:2021fxi}
\begin{align}
  \label{eq:4DCSChwarz}
    v & =  -\frac{1}{\sqrt{1-\frac{\rad_s}{\rad}}}\,\partial_t  &    h & = \frac{\dd\rad^2}{1-\frac{\rad_s}{\rad}}+\rad^2\,\dd\Omega^2_{S^2}
\end{align}
This is the lifted version of \eqref{eq:2dCS2ovars} and forms a solution of magnetic Carroll gravity \cite{Guerrieri:2021cdz,Perez:2021abf}. The
extension with a non-vanishing cosmological constant was described
in~\cite{Perez:2022jpr}.

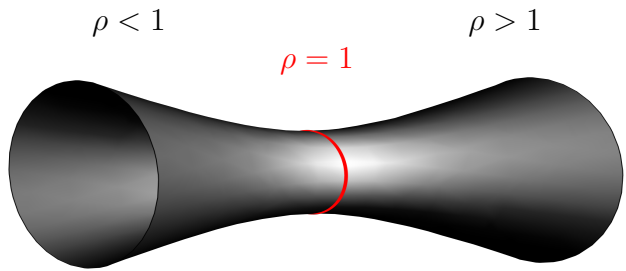
\begin{figure}
\centering
\begin{tikzpicture}
  \node at (0,0) {\includegraphics[width=0.5 \textwidth,height=3cm]{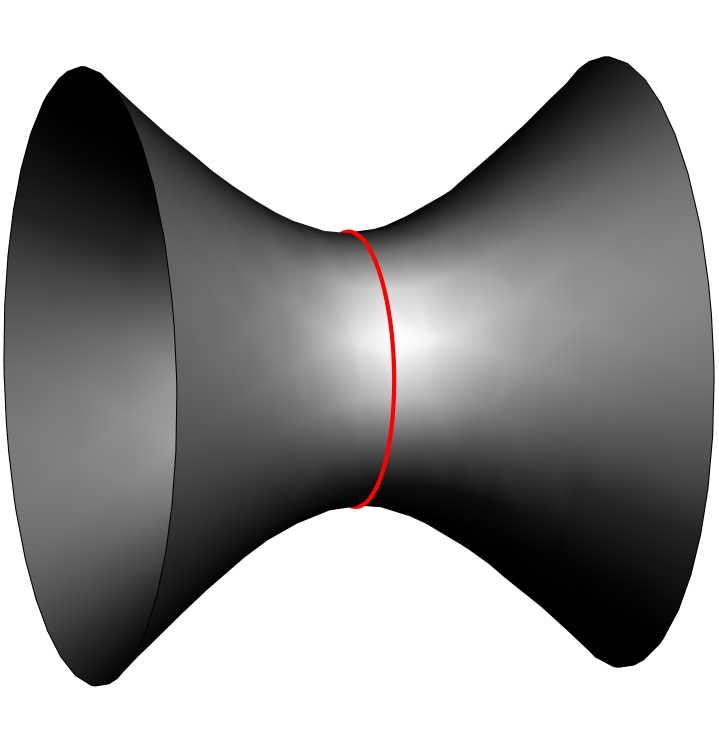}};
  \node at (0,1.5) {\color{red} $\rho=1$};
  \node at (-2.5,2) {$\rho<1$};
  \node at (2.5,2) {$\rho>1$};
\end{tikzpicture}
\caption{Sketch of spatial Carroll wormhole
  geometry~\eqref{eq:Carrollwormhole}. It corresponds to the spatial
  wormhole geometry of the maximally extended Schwarzschild black hole
  that cuts through the bifurcation sphere. In red, where $\rho=1$
  ($\rad=\rad_s$), we have encircled the Carroll extremal surface at the wormhole's throat.}
    \label{fig:wormhole}
\end{figure}

It is instructive to rewrite this configuration in terms of isotropic coordinates
obtained by the (double cover) coordinate transformation $\rad\mapsto\rho=\rho(\rad)$
given by
\begin{equation}
  \label{eq:coordtrafo}
    \rad=\frac{\rad_s}{4}\,\Big(\rho+\frac{1}{\rho}+2\Big)
\end{equation}
resulting in the Carroll wormhole geometry
\begin{align}
  \label{eq:Carrollwormhole}
  v  &=  -\frac{\rho+1}{\rho-1}\,\partial_t 
  &   h &=  \rad_s^2\left(\frac{(\rho+1)^2}{4\rho^2}\right)^2\left({\dd}\rho^2+\rho^2\, \dd\Omega^2_{S^2}\right)\,.
\end{align}
The spatial Carroll metric $h$ is $\rho\rightarrow 1/\rho$ symmetric, and $v$ changes sign under this map, corresponding to the well-known fact that the Killing time runs opposite in the universe on the other side of the wormhole (see Fig.~\ref{fig:wormhole}).

Let us scan for Carroll extremal surfaces, cf.\ section~\ref{sec:4}.
By definition, they satisfy $e^M_a \partial_MX =0$, which is the
condition that these surfaces are invariant under any linear
deviations, regardless of the direction. The dilaton $X$ is the surface
area of the $2$-spheres that foliate our spherically symmetric
spacetime. For~\eqref{eq:4DCSChwarz} 
it is given by $X=\rad^2$ [with $\lambda=2$ in
\eqref{eq:redef_dilaton}]. This means that, using the conventions of
section \ref{sec:spher-reduct-magn}, only the radial part of the
inverse vielbein $e^M_1\partial _M=:e^M\partial_M$ leads to a nontrivial
condition
\begin{align}
  \label{eq:4dextr}
  e^M \partial_MX =
  \left(
  1-\frac{\rad_{s}}{\rad} 
  \right)\partial_{\rad}X \stackrel{!}{=}0 \, .
\end{align}
The Carroll extremal surface is at $\rad=\rad_s$ or, equivalently, at
$\rho=1$ (see Fig.~\ref{fig:wormhole}).

From the 4d perspective, it is natural to assign the dilaton the length dimension $[X]=2$, which implies $[\rad]=1$ and $[k]=-2$ (cf.~the discussion in section \ref{sec:dimensions}). The relativistic entropy, temperature, and energy of the Schwarzschild black hole are given by
\begin{equation}
    S_{\textrm{\tiny rel}}=\frac{\pi c^3\,\rad^2_s}{\hbar G_N} \qquad\qquad T_{\textrm{\tiny rel}}=\frac{\hbar c}{4\pi\,\rad_s} \qquad \qquad E_{\textrm{\tiny rel}}=\frac{c^4}{2G_N}\,\rad _s 
\end{equation}
where we restored all conversion factors except for Boltzmann's constant, which we fix to one. The tantum gravity limit \eqref{eq:tg1} leads to the leading-order Carroll quantities 
\eq{
    S=\frac{\pi\,\rad^2_s}{G_M} \qquad\qquad T=\frac{1}{4\pi\,\rad_s} \qquad \qquad E=\frac{1}{2G_M}\,\rad_s \,.
}{eq:thermo} 
where we set $\kappa =\hbar c=1$ again. The results \eqref{eq:thermo} coincide with the general results in 2d derived in section \ref{sec:3}, using the 2d-4d dictionary (note that our choices imply $e^Q=1/(2\sqrt{X})$)
\eq{
k=\frac{\pi}{ G_M}\qquad\qquad X=\rad^2\qquad\qquad  X_{\textrm{\tiny min}}=\rad_s^2\qquad\qquad w(X)=\sqrt{X}\,.
}{eq:2d4d}
These dimensions work as required since in units where $\kappa =1$ both $G_M$ and $X$ have dimensions of square metres. So entropy is dimensionless, while temperature and energy are measured in Joule.

We show next that the expressions \eqref{eq:thermo} can be computed using 4d geometric arguments that are similar to what one does in general relativity. As shown above, the Carroll entropy is proportional to the area of the wormhole's throat. At this locus we have $h\vert_{\rho=1}=\rad_s^2\dd\Omega^2_{S^{2}}$. As discussed above and in section \ref{sec:3}, we ensured that $S$ is dimensionless.

We next turn to temperature. The 4d Carroll boost and rotation connections $\omega^a$ and $\omega^{ab}=-\omega^{ba}$ are solutions to the Cartan zero torsion-equations (cf. sections \ref{sec:gauging_procedure} and \ref{sec:mag_carr})
\eq{
\dd\tau +\omega^a\wedge e^a =0=  \dd e^a+\omega^{ab}\wedge e^b \,.
}{eq:Cartantorsioneq1}
For the Carroll wormhole solution \eqref{eq:4DCSChwarz}, we choose the vielbeins
\eq{
    \tau =  f(\rad)\,\dd t \qquad\qquad
    e^1 =  f^{-1}(\rad)\,\dd\rad \qquad\qquad
    e^l =  \rad\,\Bar{e}^l
}{eq:viel}
where $l=2,3$ correspond to the 2-sphere tangent space directions, $\Bar{e}^l$ are round unit 2-sphere vielbeins, and we defined $f(\rad)=(1-\rad_s/\rad)^{1/2}$. The most general solution to these equations is
\eq{
    \omega^ 1 = f'\tau+C^{11} \,e^1+C^{1l} e^l \qquad\qquad
    \omega^l = C^{l1} e^1+C^{lm}e^m \qquad\qquad
    \omega^{1l} =  f\Bar{e}^l 
}{eq:omega1}
and $\omega^{lm}=\Bar{\omega}^{lm}$ is the connection on the unit 2-sphere. 

We next consider the pullback of \eqref{eq:Cartantorsioneq1} onto the 2d submanifold obtained by fixing
a point on the 2-sphere. In order to avoid clutter, we denote the
pullbacks of the vielbeins by the same symbols. If the 2-sphere has
coordinates $\theta, \phi$ we are considering the manifold
$\theta=\theta_0$ and $\phi=\phi_0$ where $\theta_0$ and $\phi_0$ are
constants. The equations \eqref{eq:Cartantorsioneq1} on this submanifold become
\eq{
  \dd\tau +\omega^1\wedge e^1 = \dd e^1 = 0
}{eq:Cartantorsioneq}
with $\tau=f(\rad)\dd t$, $e^1=f^{-1}\dd\rad $ and $\omega^1 = f'\tau+C^{11}\, e^1$.

To define the Carroll temperature $T$, we first Wick rotate by the prescription
\begin{equation}
    t= it_{\textrm{\tiny W}} \qquad \qquad \tau = i\tau _{\textrm{\tiny W}} \qquad \qquad \omega ^1 = i\omega ^1_{\textrm{\tiny W}} \qquad \qquad v= iv_{\textrm{\tiny W}} 
\end{equation}
where it is understood that the conversion factor has already been
absorbed such that the length dimensions are
$[\tau _{\textrm{\tiny W}}]=[t_{\textrm{\tiny W}}]=1$ and
$[\omega _{\textrm{\tiny W}}]=0$ (again, see section
\ref{sec:dimensions}). The Wick rotation does not change the signature
of the geometry and the holonomy of the manifold (unlike in general
relativity) but it allows us to consider \eqref{eq:Cartantorsioneq}
for a periodic $t_{\textrm{\tiny W}}$. Now, $t_{\textrm{\tiny W}}$ has
the dimension of a length such that it can be compactified with period
$1/T$. In the Wick rotated setting, 
\eqref{eq:Cartantorsioneq} becomes
    $\dd\tau_{\textrm{\tiny W}} +\omega^1_{\textrm{\tiny W}}\wedge e^1=0$,
where $\tau_{\textrm{\tiny W}}=f(\rad)\dd t_{\textrm{\tiny W}}$ has
dimensions of length and
$\omega^1_{\textrm{\tiny W}}=f'\tau_{\textrm{\tiny W}}+C^{11}
e^1$ is dimensionless. The 2d geometry described by
$t_{\textrm{\tiny W}}\sim t_{\textrm{\tiny W}}+1/T$ and
$\rad_s<\rad<\rad_c$ where $\rad_c$ is some arbitrary number is a
cigar-like geometry that we will denote by $\Sigma$. The boundary of
$\Sigma$ is given by the circle at $\rad=\rad_c$. The Carroll boost
connection $\omega^1_{\textrm{\tiny W}}$ contains the undetermined
function $C^{11}$. We assume that $C^{11}$ is globally
well-defined on the cigar geometry so that it is periodic in
$t_{\textrm{\tiny W}}$. Then, a direct calculation tells us
\begin{equation}
    \int_\Sigma \dd\omega^1_{\textrm{\tiny W}}-\int_{\partial\Sigma}\omega_{\textrm{\tiny W}}^1=\frac{1}{2\rad_s}\frac{1}{T}
\end{equation}
where $\partial\Sigma$ is the circle at $\rad=\rad_c$. The bulk
orientation is chosen such that
$\dd t_{\textrm{\tiny W}}\wedge \dd r=:\dd t_{\textrm{\tiny W}}\dd r$
and the boundary orientation is induced similarly to section
\ref{sec:thermal_sols}. The left-hand side is $2\pi$ times the Euler
character $\chi$ of $\Sigma$ which is topologically a disk so that
$\chi=1$. This procedure recovers the result for temperature announced in \eqref{eq:thermo}.

In \cite{Perez:2021abf} the energy $E$ of the Carroll solution
discussed here was computed, where the author additionally set $16\pi G_M=1$. Restoring this constant the result is the same as for the
Schwarzschild black hole, namely
\begin{equation}
    E=\frac{\rad_s}{2G_M}=2T S
\end{equation}
in agreement with \eqref{eq:thermo}. Of course, the first law
\begin{equation}
    \delta E=T\,\delta S
\end{equation}
is obeyed.

All the thermodynamical relations above follow immediately from the general analysis of section \ref{sec:3}, using the 2d-4d dictionary \eqref{eq:2d4d}.

\section{Charged and rotating Carroll black holes}
\label{sec:8}

All examples so far can be understood entirely in terms of 2d Carroll
dilaton gravity or one of its dimensional uplifts to higher
dimensions. In this section, we go beyond this case by considering
charged or rotating Carroll black holes (mostly from a dimensionally
reduced, 2d, perspective), where a generalization to 2d
Carroll--Maxwell dilaton gravity is required.

\subsection{General remarks on charged Carroll black holes in 2d}
\label{sec:8.1}

In the PSM formulation, adding a Maxwell field amounts to adding
another target space coordinate $Y$ and adding to the Poisson tensor
an extra row and column of zero entries. The potential in
\eqref{eq:car1} can depend on this additional target space coordinate
as well, and the Lagrange 2-form acquires an additional term
$Y\,\dd A$, where $A=A_\mu\,\dd x^\mu$ is the Maxwell gauge field
1-form. \eq{ {\mathcal{L}} = Y\,\dd A +
  X\,\dd\omega+\XH\,\big(\dd\tau+\omega\wedge e\big)+\XP\,\dd e
  + {\mathcal{V}}(X,\,\XH,\,Y)\,\tau\wedge e }{eq:charged1} The additional
$U(1)$ gauge symmetry generated by some transformation parameter
$\Lambda$ acts trivially on all fields except on the Maxwell gauge
field, $\delta_\Lambda A = \dd\Lambda$. The equations of motion
\eqref{eq:eom} are essentially unchanged, with the replacement
${\mathcal{V}}(X,\,\XH)\to{\mathcal{V}}(X,\,\XH,\,Y)$. There are two additional
equations of motion from varying with respect to $Y$ and $A$:
\begin{align}
    & \delta Y: & \dd A &= -\frac{\partial{\mathcal{V}}(X,\,\XH,\,Y)}{\partial Y}\,\tau\wedge e \label{eq:charged13}\\
    & \delta A: & \dd Y &= 0 \label{eq:charged14}
\end{align}
The quantity $Y$ on-shell is a second conserved Casimir, 
\eq{
Y=q=\textrm{const.} 
}{eq:charged3}
and physically corresponds to a conserved $U(1)$ charge (though in
some applications, it might have a different interpretation, e.g., as
angular momentum in higher dimensions).

If the Lagrange 2-form \eqref{eq:charged1} emerges from some
Carroll limit, it could happen that the first and last terms are
multiplied by some powers of the speed of light $c$. In that case, we
can always first rescale $Y$ with an appropriate factor of $c$ to
eliminate $c$ from the last term and then rescale $A$ with an
appropriate factor of $c$ to eliminate $c$ from the first term. Thus,
without loss of generality, we assume there are no explicit
factors of $c$ in the Lagrange 2-form \eqref{eq:charged1}.

Solving the equations of motion can be done exactly as in section
\ref{sec:5.1}. The solutions for the spatial metric and the temporal
vector field will, in general, depend not only on the mass parameter $M$
but also on the $U(1)$ charge $q$. As a consequence of this
dependence, there can be BPS-like bounds and extremality conditions.
(Since we already use the word ``extremal'' to denote Carroll extremal
surfaces, we call the confluent case ``degenerate'' instead.)
Relatedly, new constant dilaton vacua can emerge and typically have
an interpretation as ``near horizon extremal geometries''. (To clarify
also here the vocabulary, we refer to such geometries as
``near-Carroll-extremal-surface degenerate geometries'' in a
Carroll context.) Other than these marginal changes, our general
discussion of section \ref{sec:5.1} applies.

A prototypical form of the potential is given by
\eq{
{\mathcal{V}}(X,\,\XH,\,Y) = \hat{{\mathcal{V}}}(X,\,\XH) - \frac{Y^2}{4F(X)}\,.
}{eq:charged2}
In this case, the charge $q$ on-shell is related to the electric field, $E=\ast\dd A$, and the dilaton,
\eq{
q=Y=2F(X)\,\ast\dd A=2F(X)\,E
}{eq:charged4}
where we used the Carroll--Hodge-$\ast$ relation $\ast(\tau\wedge e):=1$. For a definition of this operator, we refer to \cite{Fecko:2022shq}. Charge conservation implies
\eq{
\dd\big(F(X)\,\ast\dd A\big) = 0\,.
}{eq:charged5}
Integrating out the scalar field $Y$ by its own equation of motion yields a (non-minimally coupled) Maxwell term in the Lagrange 2-form (with the usual expression for the field strength, $F_{\mu\nu}=\partial_\mu A_\nu-\partial_\nu A_\mu$).
\begin{equation}
  \label{eq:charged6}
  {\mathcal{L}} = X\,\dd\omega+\XH\,\big(\dd\tau+\omega\wedge e\big)+\XP\,\dd e + \hat{\mathcal{V}}(X,\,\XH)\,\tau\wedge e + F(X) \underbrace{(\ast\dd A) \dd A}_{\sim F_{\mu\nu}F^{\mu\nu}\,\textrm{vol}}
\end{equation}
Consistently, varying \eqref{eq:charged6} with respect to the Maxwell
connection $A$ yields the equation of motion \eqref{eq:charged5}. For
models that come from a dimensional reduction of higher-dimensional
Einstein--Maxwell theories the coupling function $F(X)$ typically is
linear in the dilaton since the Maxwell term gets the same volume
factor as the curvature term $X\,\dd\omega$.

Finally, we note that often it is sufficient to consider the charge
$q$ as a parameter in the action rather than as a constant of motion.
In that case, one can use the potential \eqref{eq:charged2} with $Y$
replaced by its on-shell value $q$.

\subsection{Carroll--Reissner--Nordstr\"om}
\label{sec:8.2}

There are different paths to obtaining CRN black holes. We found it
simplest to first reduce spherically Schwarzschild to 2d, then take
the Carroll limit, and finally, add a Maxwell field. This means we take
the Schwarzschild results for the potential $\hat{\mathcal{V}}$ and set
$F(X)=X$,
\eq{
{\mathcal{V}}_{\textrm{\tiny CRN}}(X,\,\XH,\,Y) = \frac{\lambda^2}{4} + \frac{\XH^2}{4X} - \frac{Y^2}{4X} \,.
}{eq:charged7}
Without loss of generality, we set $\lambda=2$.

In the coordinates introduced in section \ref{sec:5.1}, the CRN solution is given by
\eq{
\dd s^2=\dd r^2 = \frac{\dd X^2}{\XH^2}\qquad\qquad v=\frac{2\sqrt{X}}{\XH}\,\partial_t
}{eq:charged8}
and
\eq{
\XH = \pm \sqrt{4X+4\,q_e^2-8m\sqrt{X}}\,,
}{eq:charged9}
where $q_e=\frac{q}{2}$ is the electric charge and $m=\frac{M}{2}$ is the mass. We again chose the integration constant in \eqref{eq:car111} such that $e^{-Q}=2\sqrt{X}$ to achieve an asymptotic normalization of the vector field, $\lim _{X\to \infty }v=-\partial _t$.
The solution for the gauge field follows from \eqref{eq:charged13},
\eq{
\dd A = \frac{q_e}{2X}\,\tau\wedge e
}{eq:charged15}
which in Coulomb gauge leads to the usual Coulomb-potential
\eq{
A = \frac{q_e}{\sqrt{X}}\,\dd t\,.
}{eq:charged16}
The Carroll limit on the gravity side is a magnetic limit, while the
Carroll limit on the Maxwell side is an electric limit, so this is an
example of electric Carroll Maxwell theory coupled to magnetic Carroll
gravity. 
Carroll extremal surfaces in the CRN geometry arise at two loci,
\eq{
\sqrt{X}_\pm = m \pm \sqrt{m^2-q_e^2} 
}{eq:charged10}
provided the mass is positive and the charge obeys the BPS-bound
\eq{
|q_e|\leq m\,.
}{eq:charged11}
When saturated, $q_e^2=m^2$, the two loci coalesce to a single degenerate Carroll extremal surface with vanishing Carroll temperature, similar to extremal Reissner--Nordstr\"om black holes.

While the CS model (see table \ref{table:1}) does not have any constant dilaton vacuum solution, the CRN model has such a solution for the value of the dilaton
\eq{
X = q_e^2\,.
}{eq:charged12}
Since this is the same value the dilaton takes at the degenerate extremal Carroll surface in the confluent case, one can interpret this constant dilaton vacuum analogously to the Robinson--Bertotti solution, i.e., as near-Carroll-extremal-surface degenerate geometry. Generalizations of CRN to arbitrary dimension is straightforward; one
just has to replace the first two terms in the potential
\eqref{eq:charged7} by the corresponding CST potentials corresponding
to the desired dimension, see table \ref{table:1}. We end our investigations on charged Carroll black holes here and refer to the outlook chapter \ref{ch:conclusion} for more discussions and generalizations.
In the next subsection, we
address a different pertinent example, namely Carroll BTZ.

\subsection{Carroll BTZ}\label{sec:8.3}

It is not obvious how to take a Carroll limit of the BTZ
black hole \cite{Banados:1992wn, Banados:1992gq}.  We take the following
route. First, we Kaluza--Klein reduce along the azimuthal angle
$\varphi$ to obtain the 2d Ach\'ucarro--Ortiz model
\cite{Achucarro:1993fd} and only then we take the Carroll limit.

The first step leads to a charged 2d dilaton gravity model, where the
Maxwell field is the one appearing in the Kaluza--Klein ansatz
($\alpha,\beta,\gamma\in\{0,1\}$)
\eq{
\dd s^2 = g_{\alpha\beta}(x^\gamma)\,\dd x^\alpha\dd x^\beta + X^2(x^\gamma)\,\big(\dd\varphi + A_\alpha(x^\gamma)\,\dd x^\alpha\big)^2
}{eq:BTZ1}
and the associated $U(1)$ charge is the BTZ angular momentum $J$. The dimensionally reduced 2d model has the Ach\'ucarro--Ortiz potential (see section 6.3 in \cite{Grumiller:2007ju})
\eq{
V_{\textrm{\tiny AO}}(X,\,Y) = \frac{X}{\ell^2} - \frac{Y^2}{X^3}
}{eq:BTZ2}
where $\ell$ is the 3d AdS radius, which we set to one, $\ell=1$. On-shell $Y=J$.

The second step consists of importing the potential
\eqref{eq:BTZ2} into generic 2d Carroll dilaton
gravity~\eqref{eq:car1}. This yields Carroll black
hole solutions we refer to as CBTZ. They are given by
\begin{equation}
  \label{eq:BTZ3}
  \dd s^2 = \dd r^2 = \frac{\dd X^2}{\XH^2}\qquad\qquad v=\frac{1}{\XH}\,\partial_t
\end{equation}
with
\eq{
\XH=\pm\sqrt{X^2+\frac{J^2}{X^2}-2M}\,.
}{eq:BTZ4}
The solution for the gauge field follows from \eqref{eq:charged13},
\begin{equation}
  \label{eq:BTZ7}    
  \dd A = \frac{2J}{X^3}\,\tau\wedge e
\end{equation}
which, in Coulomb gauge, leads to
\eq{
A = \frac{J}{X^2}\,\dd t\,.
}{eq:BTZ8}

As expected, there are two loci with Carroll extremal surfaces,
\eq{
X^2_\pm = M \pm \sqrt{M^2-J^2}
}{eq:BTZ5}
provided the mass is positive and the angular momentum obeys the BPS bound
\eq{
|J|\leq M\,.
}{eq:BTZ6}
When saturated, $J^2=M^2$, the Carroll extremal surface degenerates and has vanishing Carroll temperature, similar to extremal BTZ.

In summary, the Carroll BTZ black hole~\eqref{eq:BTZ3}-\eqref{eq:BTZ8} defined here is the Carroll limit of the Ach\'ucarro--Ortiz model, which in turn is a Kaluza--Klein reduction of the BTZ black hole. The Carroll BTZ black hole is a positive mass solution of 2d Carroll--Maxwell dilaton gravity \eqref{eq:charged1} with the potential function \eqref{eq:BTZ2}, subject to the BPS bound \eqref{eq:BTZ6}.
\chapter{Quantum effects: The Carroll Hawking effect}\label{ch:Hawking}
\setlength{\epigraphwidth}{0.8\textwidth}
\epigraphfontsize{\small\itshape}
\epigraph{``I have noticed that even those who assert that everything is predestined and that we can change nothing about it still look both ways before they cross the street.''}{--- \textup{Stephen Hawking}, Black holes and baby universes, and other essays}
In the previous chapter, we introduced Carroll black holes as certain exact solutions of 2d Carroll dilaton gravity that admit Carroll thermal properties. In particular, all these solutions have a finite Carroll temperature which is defined by entirely geometric ingredients. However, at this point it is not clear how far the analogy with a true, physical temperature goes. We shall further investigate this here by coupling a Carroll matter field theory to such backgrounds and observing how the vacuum states of its quantized version behave. As we shall show, given certain regularity conditions, the asymptotic energy density acquires a non-zero value which is compatible with the Stefan--Boltzmann law. Therefore, in this semi-classical regime the Carroll black hole temperature may indeed be associated with the temperature of a thermal state of a matter field coupled to it. This chapter is based on the original work \cite{Aggarwal:2024yxy}.

Before we delve into the Carroll version, we shall start with a brief summary of the Lorentzian/Euclidean argument to familiarize ourselves with the tools. One needs to keep in mind that at all stages the background geometry remains fixed and classical, the Hawking effect and its Carroll avatar are one-loop effects of the matter theory.

\section{Quantum states on Lorentzian static backgrounds}
\label{eq:rev_Hawking}
While Hawking's original derivation \cite{Hawking:1975vcx} proceeded by quantizing the modes of a scalar field directly, we shall follow the approach by Christensen and Fulling \cite{Christensen:1977jc} which uses the Weyl anomaly of a conformally coupled field theory to infer properties of its vacuum states. 

Consider the two-dimensional description of a 4d Schwarzschild black hole. It is given by a 2d metric together with a dilaton that is a function of the areal radius $\rad $, 
\begin{align}\label{eq:ahso}
    \dd s^2=-\xi (\rad)\dd t^2+\frac{\dd \rad^2}{\xi (\rad)} && X=\rad^2 
\end{align}
where $\xi (\rad )=1-\frac{\rad _s}{\rad }$. The uplift to the 4d Schwarzschild metric is given by a similar expression as in the Carroll case (see sections \ref{sec:spher-reduct-magn} and \ref{sec:CS_4d}),
\begin{align}
    ds^2_{(4d)}=\dd s^2+X(\rad)\dd \Omega ^2_{S^{(2)}} ~.
\end{align}
Next, denoting the two-dimensional metric components by $G_{\mu \nu }$, consider a massless scalar field that minimally couples to $G_{\mu \nu }$. It is given by the action 
\begin{align}\label{eq:scalar.action}
    I_m[\phi ]=-\frac12 \int \dd ^2x \sqrt{G} \, G^{\mu \nu }\partial _\mu  \phi \partial _\nu \phi 
\end{align}
which classically enjoys Weyl invariance under $\delta _\rho G_{\mu \nu }=2\rho G_{\mu \mu }$ and $\delta _\rho \phi =0$, i.e. $\delta _\rho I_m=0$. Since we are interested in a corresponding quantum theory, we shall now define the path integral for this theory. The Weyl symmetry will acquire an anomaly in that case.

\subsection{Path integral and conformal anomaly}
For the derivation of the Weyl anomaly we switch to Euclidean signature. The Euclidean partition function for the massless scalar is given by
\begin{align}
    Z[G]=e^{-W}=\int \mathcal{D}\phi \, e^{-I^E_m[\phi ,G]}
\end{align}
where $I^E_m[\phi ]$ is the Euclidean action corresponding to \eqref{eq:scalar.action} by performing a Wick rotation $t\to it$ and setting $I_m=iI^E_m$. The vacuum expectation value of the stress-energy tensor $\mathfrak{T}^{\mu \nu }$ is obtained by taking a functional variation with respect to the background metric, 
\begin{align}\label{eq:stress_deff}
    \langle \mathfrak T^{\mu \nu }\rangle =\frac{2}{\sqrt{G}}\frac{\delta W}{\delta G_{\mu \nu }} ~.
\end{align}
For defining the path integral above it is important to choose the measure $\mathcal{D}\phi $. This may be done in several inequivalent ways and must be handled with care if one requires some covariance aspects of the background to be preserved. Since we are in a setting where ultimately dynamical gravity could be coupled to the matter fields, it is natural to require diffeomorphism invariance to be preserved. The path integral measure is then defined by 
\begin{align}\label{eq:meas_def}
    1=\int \mathcal{D}\phi \, e^{-\frac12 \int \dd ^2x\, \sqrt{G}\phi ^2} 
\end{align}
which violates Weyl invariance. We stress that this is a choice motivated by the specific background. There are indeed situations where this choice does not lead to a sensible definition \cite{Grumiller:2023ahv}, since the fall-off of the configurations $\phi $ might not be strong enough to render the exponent in \eqref{eq:meas_def} finite. In such cases, one may instead define a Weyl invariant measure which, however, is not invariant under general diffeomorphisms \cite{Karakhanian:1994gs,Jackiw:1995qh,Amelino-Camelia:1995nk}. In general, there is no way to define a measure that preserves both symmetries. Sticking to the former choice \eqref{eq:meas_def} we formally evaluate the partition function by 
\begin{align}
    W[G]=\frac12 \log \det (L) ~.
\end{align}
Since this expression is divergent, it needs to be evaluated in a regularized way before extracting correlation functions which can be done elegantly using heat-kernel methods \cite{Vassilevich:2003xt,Grumiller:2002nm}. We are specifically interested into an infinitesimal Weyl variation which was computed in the above references to be 
\begin{align}
    \delta _\rho W=\frac{1}{24\pi }\int \dd ^2x\, \sqrt{G}\, R\,\rho 
\end{align}
where $R$ is the Ricci scalar associated to $G_{\mu \nu }$. Together with \eqref{eq:stress_deff} this implies 
\begin{align}\label{eq:eucl_weyl}
    \langle \mathfrak T^\mu {}_\mu \rangle =\frac{1}{24\pi }R 
\end{align}
which is the well-known expression for the two-dimensional Weyl anomaly on a Riemannian background. Since diffeomorphism invariance is preserved the conservation of the stress-energy tensor continues to hold at the quantum level, 
\begin{align}\label{eq:eucl_diff}
    \nabla _\mu \langle \mathfrak T^{\mu \nu }\rangle =0 ~.
\end{align}
As a next step we solve the two Ward identities \eqref{eq:eucl_weyl}, \eqref{eq:eucl_diff} for the vacuum expectation values $\langle \mathfrak T^{\mu \nu }\rangle $ on static backgrounds.

\subsection{Hawking effect}
\label{sec_Lore_hawking}
Switching back to Lorentzian signature it is convenient to change the coordinates $(t,\rad)$ to $x^\pm $ defined by
\begin{align}
x^\pm =\frac{t\pm z}{\sqrt{2}} && \frac{\dd z}{\dd \rad }=\frac{1}{\xi (\rad)}    
\end{align}
such that the metric in \eqref{eq:ahso} takes the form
\begin{align}
    \dd s^2=-2e^{2\Omega }\dd x^+\dd x^- && \Omega =\frac12 \log \xi(\rad ) ~.
\end{align}
The only non-vanishing Christoffel symbols are $\Gamma ^\pm {}_{\pm \pm }=2\partial _\pm \Omega $, and we may write the two equations \eqref{eq:eucl_diff} as 
\begin{align}
    \partial _- \langle \mathfrak T_{++}\rangle +\partial _+\langle \mathfrak T_{+-}\rangle -2\partial _+ \Omega \langle \mathfrak T_{+-}\rangle &=0 \\
    \partial _+ \langle \mathfrak T_{--}\rangle +\partial _-\langle \mathfrak T_{+-}\rangle -2\partial _- \Omega \langle \mathfrak T_{+-}\rangle &=0 ~.
\end{align}
Since we consider static backgrounds for which $\partial _+ = -\partial _-=\frac{1}{\sqrt{2}}\partial _z$ they may be simplified to 
\begin{align}\label{eq:yoyo2}
    \partial _z \langle \mathfrak T_{\pm \pm }\rangle - \partial _z \langle \mathfrak T_{+-}\rangle + 2\partial _z\Omega \langle \mathfrak T_{+-}\rangle =0 ~.
\end{align}
On the other hand, the trace Ward identity \eqref{eq:eucl_weyl} implies 
\begin{align}
    \langle \mathfrak T_{+-}\rangle = -\frac{1}{48\pi }e^{2\Omega }R=\frac{\partial _z^2\Omega }{24\pi }
\end{align}
such that insertion into \eqref{eq:yoyo2} leads to two ordinary differential equations for $\langle \mathfrak T_{\pm \pm }\rangle$. The general solution is 
\begin{align}
    \langle \mathfrak T_{\pm \pm }\rangle =\frac{1}{24\pi }\Big(\partial _z^2\Omega -(\partial _z\Omega )^2\Big)+t_\pm  && t_\pm \in \mathbb{R} 
\end{align}
where $t_\pm $ are two integration constants. They parametrize a family of static quantum states of the model. Expressing the solution in terms of the metric function, $\xi (\rad )$ this is 
\begin{align}\label{eq:tinSScords}
    \langle \mathfrak T_{\pm \pm }\rangle =\frac{1}{96\pi }\Big(2\xi \partial _\rad ^2\xi  -(\partial _\rad \xi )^2\Big)+t_\pm  && t_\pm \in \mathbb{R} ~.
\end{align}
In this two-parameter family of states there are three special cases which differ in their regularity properties at infinity and at the horizon. First, note that $\langle \mathfrak T_{++}\rangle $ describes the matter flux through surfaces of constant $x^-$ which are ingoing null rays. Similarly, $\langle \mathfrak T_{--}\rangle $ may be understood as describing outgoing matter flux. Expanding \eqref{eq:tinSScords} for $r\to \infty $ we find that the asymptotic values of these fluxes are given by 
\begin{align}
    \mathfrak T _{--} \Big \vert _{\mathcal{I}_+} = t_- && \mathfrak T _{++} \Big \vert _{\mathcal{I}_-} = t_+
\end{align}
where $\mathcal{I}_\pm $ denote future and past null infinity. On the other hand, one may expand the fluxes around the horizon at $\rad =\rad _s$. For this it is necessary to switch to regular coordinates there which is achieved by the Kruskal patch (see, e.g., \cite{Grumiller:2022qhx} for a review), 
\begin{align}
    U=-e^{-x^-/\sqrt{2}\rad _s} && V=e^{x^+/\sqrt{2}\rad _s} ~.
\end{align}
The future horizon $\mathcal{H}_+$ is then given by the locus $U=0$ while the past horizon $\mathcal{H}_-$ is at $V=0$. Taking as an example the outgoing flux component we find 
\begin{align}
    \langle \mathfrak{T}_{UU}\rangle &= \Big ( \frac{\dd U}{\dd x^-}\Big) ^{-2}\langle \mathfrak{T}_{--}\rangle  ~.
\end{align}
This component is finite at $\mathcal{H}_-$ but expands around $\mathcal{H}_+$ as 
\begin{align}
    \langle \mathfrak{T}_{UU}\rangle \sim \frac{2\rad _s^4}{e^2(\rad -\rad _s)^2}\Big(t_--\frac{1}{96\pi \rad _s^2}\Big)-\frac{4\rad _s^3}{e^2(\rad -\rad _s)}\Big(t_--\frac{1}{96\pi \rad _s^2}\Big)+\mathcal{O}(\rad -\rad _s)^0
\end{align}
which diverges unless we choose the integration constant as $t_-=1/(96\pi \rad _s^2)$. A similar expansion for $\langle \mathfrak{T}_{++} \rangle $ around the past horizon also gives a divergence unless we choose $t_+=1/(96\pi \rad _s^2)$. Therefore, demanding regularity at $\mathcal{H}_\pm$ leads to a non-zero flux at $\mathcal{I}_\pm $. The three prominent choices for the constants $t_\pm$ emerging from this behaviour are given in table \ref{tab:quantum_states}.
\begin{table}
\centering
\renewcommand{\arraystretch}{1.4} 
\setlength{\tabcolsep}{12pt}      
\begin{tabular}{||c | c c c ||} 
\hline
 & $t_+$ & $t_-$ & Properties \\ 
\hline\hline
Boulware \cite{Boulware:1975dm} & 0 & 0 & no flux at $\mathcal{I}_\pm$, sing. at $\mathcal{H}_\pm $ \\ 
\hline
Unruh \cite{Unruh:1976db} & 0 & $\displaystyle 1/(96\pi r_s^2)$ & only outgoing flux, reg. at $\mathcal{H}_+$ \\ 
\hline
Hartle--Hawking \cite{Hartle:1976tp}&  $\displaystyle 1/(96\pi r_s^2)$ &  $\displaystyle 1/(96\pi r_s^2)$ & flux at $\mathcal{I}_\pm$, reg. at $\mathcal{H}_\pm $ \\ 
\hline
\end{tabular}
\label{tab:quantum_states}
\caption{Three prominent quantum states for a scalar field on a Schwarzschild background.}
\end{table}
They are describing different physical situations. 

The Boulware state forbids flux at infinity but pays the price of being singular along the entire horizon. Its physical significance lies in the description of quantum states around a static star. The analog of the Unruh state was in fact used by Hawking in his original derivation for particle creation around black hole horizons \cite{Hawking:1975vcx}. It describes the late-time behaviour of a collapsing star, there is no flux coming in from $\mathcal{I}_-$. This state is singular at the past horizon which is, however, not relevant in Hawking's original scenario of a collapsing star. The white hole region is just regarded an unphysical artefact in that case. The Hartle--Hawking state describes the configuration of an eternal black hole being coupled to a thermal bath at infinity. This is the only state that is regular across the entire horizon of the maximally extended spacetime, in particular at the bifurcation sphere. The temperature of the bath is fixed in terms of the horizon radius of the black hole as the Hawking temperature
\begin{align}
    T=\frac{1}{4\pi \rad _s}
\end{align}
such that the asymptotic flux follows the two-dimensional Stefan--Boltzmann law 
\begin{align}
    \langle \mathfrak{T}_{++}\rangle \Big \vert _{\mathcal{I}_-}=\frac{\pi }{6}T^2 = \langle \mathfrak{T}_{--}\rangle \Big \vert _{\mathcal{I}_+} ~.
\end{align}
In other words, the asymptotic matter flux corresponds to true black-body radiation at a specific temperature fixed by the black hole geometry. Similarly, the matter energy density measured by a static observer at infinity is 
\begin{align}
    \lim _{\rad \to \infty }\langle \mathcal{E} \rangle = \lim _{\rad \to \infty }\langle \mathfrak{T}^t{}_t \rangle = \frac{\pi }{6}T^2 ~.
\end{align}
We shall find a similar conclusion for the Carroll case to which we turn now.

\section{Carroll--Hawking effect from a limit}
\label{sec:Hawking_limit}
As a first approach to investigate whether some analogue of the Hawking effect exists for Carroll black holes, we perform a limit from the Lorentzian case. The background we are interested in is the Carroll Schwarzschild black hole as described in section \ref{sec:carr_schwarzsch}. In its two-dimensional formulation it is given by 
\begin{align}\label{eq:HE_background}
    h_{\mu \nu }\dd x^\mu \dd x^\nu =\frac{\dd \rad ^2}{\xi (\rad )} && \tau =\sqrt{\xi (\rad )}\dd t && v=-\frac{1}{\sqrt{\xi (\rad )}}\partial _t && X=\rad ^2 
\end{align}
where $\xi (\rad )=1-\frac{\rad_s}{\rad }$ with $\rad _s=2m$. 

Consider the vacuum expectation values of the stress-energy energy tensor associated to a massless scalar field as in the previous section. For taking the limit $c\to 0$ it is important to reinsert $c$ into these expressions, i.e., we define the light-cone coordinates by 
\begin{align}
    x^\pm =\frac{ct\pm z}{\sqrt{2}} ~.
\end{align}
With this modification one arrives again at the components of a Lorentzian stress-energy tensor 
\begin{align}
    \langle \mathfrak T_{\pm \pm }\rangle &=\frac{1}{96\pi }\Big(2\xi \partial _\rad ^2\xi  -(\partial _\rad \xi )^2\Big)+t_\pm  && t_\pm \in \mathbb{R} \label{eq:jaja}\\
    \langle \mathfrak T_{+-}\rangle &=\frac{1}{48\pi }\xi \partial _\rad ^2\xi  
\end{align}
which solve the diffeomorphism and anomalous Weyl Ward identities. Note, that no explicit factors of $c$ appear in these expressions. We need to transform these components back to a static coordinate system to make them compatible with the chosen background. As a final step, we adapt the index structure to obtain a proper Carroll stress-energy tensor with one upper and one lower index (see eq. \eqref{eq:Carroll_EMT}). This leads to 
\begin{subequations}
\begin{align}
     \langle \mathfrak T^{t} {}_{t} \rangle &=\frac{1}{24\pi }\Big(\frac{1}{4\xi }(\partial _\rad \xi )^2-\partial _\rad ^2\xi -12\pi \frac{t_++t_-}{\xi }\Big)\\
    \langle \mathfrak T^{t} {}_\rad \rangle &=\frac{t_--t_+}{2\,c\,\xi ^2}\\
    \langle \mathfrak T^\rad {}_{t} \rangle &=\frac{c}{2}(t_--t_+) \\
    \langle \mathfrak T^\rad {}_\rad \rangle &=\frac{1}{24\pi }\Big(-\frac{(\partial _\rad \xi )^2}{4\xi }+12\pi \frac{t_++t_-}{\xi }\Big)
\end{align}
\end{subequations}
In this limit, $\langle\mathfrak T^{t}{}_r\rangle\rightarrow\infty$ unless we assume 
\begin{align}
    t_+-t_-=ct_0
\end{align}
with some fixed constant $t_0$. With this assumption, we obtain
\begin{subequations}
\begin{align} \label{eq:T limit sch}
    \langle \mathfrak T^{t} {}_{t} \rangle \to \langle \mathbf T^{t} {}_{t} \rangle &=-\frac{1}{24\pi }\Big(\partial _\rad^2\xi -\frac{(\partial _\rad\xi )^2}{4\xi }+ \frac{24\pi t_+}{\xi }\Big)\\
    \langle \mathfrak T^{t} {}_\rad \rangle \to \langle \mathbf T^{t} {}_\rad\rangle &=-\frac{1}{2}\frac{t_0 }{\xi ^2}\\
    \langle \mathfrak T^\rad{}_{t} \rangle \to \langle \mathbf T^\rad{}_{t} \rangle &=0\\
    \langle \mathfrak T^\rad{}_\rad\rangle \to \langle \mathbf T^\rad{}_\rad\rangle &=\frac{1}{24\pi }\Big(-\frac{(\partial _\rad\xi )^2}{4\xi }+ \frac{24\pi t_+}{\xi }\Big)~.
\end{align}
\end{subequations}
This result satisfies the Carroll boost and diffeomorphism Ward identities presented in section \ref{eq:sec311}, 
\begin{align}
    h_{\mu \alpha }\langle \mathbf T ^\mu {}_\nu \rangle v^\nu =0 && \nabla _\mu \langle \mathbf{T}^\mu {}_\nu \rangle =0
\end{align}
where we already inserted $K_{\mu \nu }=0$ for the Carroll Schwarzschild background. The trace Ward identity stays anomalous after the limit. Taking the limit of the flux components \eqref{eq:jaja} leads to
\begin{align}\label{eq:flux_limit}
    \langle \mathbf T_{\pm \pm }\rangle  := \lim _{c\to 0 } \langle \mathfrak T_{\pm \pm }\rangle =\frac{1 }{96\pi }\Big(2\xi \partial _r^2\xi -(\partial _r\xi )^2\Big)+t_+ 
\end{align}
which shows that both fluxes have to agree, $\langle \mathbf T_{++}\rangle =\langle \mathbf T_{--}\rangle $. This is unlike the situation we saw previously in the Lorentzian theory (see section \ref{sec_Lore_hawking}), where $\langle\mathfrak T_{\pm\pm}\rangle$ would be associated with in- and outgoing matter fluxes. They could be chosen independently of each other, according to the physical situation at hand. The fact that both fluxes have to agree in the present case is just another manifestation of no (net) energy flux being possible in a Carroll theory \cite{deBoer:2021jej}. It furthermore implies that not all vacuum choices of the analogous Lorentzian theory are possible anymore. In particular, local Carroll boost-invariance is inconsistent with the Unruh vacuum.
Thus, there are only two of the three special cases in table \ref{tab:1} remaining, Boulware and Hartle--Hawking, which may be distinguished by the energy density at infinity and at the Carroll extremal surface. Its general expression reads
\begin{align}
    \langle \mathcal{E} \rangle =-\tau_\mu \langle T^\mu {}_\nu \rangle v^\nu =\frac{1}{24\pi }\Big(\partial _r^2\xi -\frac{(\partial _r\xi )^2}{4\xi }\Big)+ \frac{t_+}{\xi } 
\end{align}
demanding a vanishing $\langle \mathcal E \rangle$ at infinity fixes $t_+=0$. However, analogous to the Lorentzian case the energy density blows up at the Carroll extremal surface which corresponds to the Boulware vacuum. Making instead the unique choice $t_+=\frac{1}{96\pi{\rad }_s^2}$ renders $\langle \mathcal{E}\rangle $ regular at $\rad _s$ and defines the Carroll analog of the Hartle--Hawking vacuum with asymptotic energy density
\eq{
    \lim_{\rad \to\infty} \langle \mathcal{E}\rangle =\frac{1}{96\pi \rad _s^2}=\frac{\pi}{6}\,T^2 ~.
}{eq:angelinajolie2}
In the second equality we used the result for the Carroll temperature $T$ of the Carroll--Schwarzschild background, $T^{-1}=4\pi \rad_s$ as derived in section \ref{sec:carr_schwarzsch}. This equality is our main result \cite{Aggarwal:2024gfb} and shows that the asymptotic energy density \eqref{eq:angelinajolie2} is compatible with the 2d Stefan-Boltzmann law. 

\section{The conformal anomaly in Carroll theories}
\label{sec:Conf_anomaly_Carr}
Instead of taking Carrollian limits, we consider in this section the magnetic scalar action \eqref{eq:mag_scalar} from the start and solve the Carroll Ward identities, analogous to the procedure in the Lorentzian case (see section \ref{sec_Lore_hawking}). We set the potential $V(\phi )=0$ for simplicity. Plugging it into the path integral yields
\begin{align}\label{Zmag1}
    Z=\int \mathcal{D}\chi \mathcal{D}\phi \exp \Big(-I_{\textrm{M}}[\chi ,\pi ]\Big) =e^{-W}
\end{align}
where 
\begin{align}
    I_M[\chi ,\phi ]=-\int \dd ^{d+1}x\,\mathfrak{e}\,\Big(\chi v^\mu \partial _\mu \phi -\frac12 h^{\mu \nu }\partial _\mu \phi \partial _\nu \phi \Big ) ~.
\end{align}
Integrating out $\chi $ produces a functional $\delta$-function $\delta(v^\mu\partial_\mu\phi)$ so that we remain with a path integral over a 1d time-independent scalar field, but with a Jacobian factor $\mathcal{J}=(\det(v^\mu\partial_\mu))^{-1}$. The operator $v^\mu\partial_\mu$ contains a derivative along the time direction but no derivative along the spatial direction. This means that the operator is not elliptic, so there is no regular method known to us to define its determinant\footnote{This situation is similar to one with a Faddeev--Popov determinant in axial gauge on a curved background \cite{Vassilevich:1995cz}.}. Since a direct method fails, we try a less direct one. 

We assume that the path integral \eqref{Zmag1} exists (see also the more recent treatment \cite{Vassilevich:2024vei}) and write a general variation with respect to the background fields as
\begin{align}\label{eq:gen:var}
    \delta W=\int _{\mathcal{M}}\tau \wedge e\,\big(\langle T_{(\tau )}^\mu \rangle \delta  \tau _\mu  +\langle T_{(e)}^\mu \rangle  \delta  e_\mu +\langle T_{(\omega )}^\mu \rangle  \delta  \omega_\mu \big) 
\end{align}
where the functions $\langle T_{(\tau )}^\mu \rangle$, $\langle T_{(e)}^\mu \rangle$ and $\langle T_{(\omega )}^\mu \rangle$ are the response functions to varying $\tau $, $e$ and $\omega$, respectively. In particular, a Weyl variation of the effective action is
\begin{align}
    \delta _\rho W
    =\int _{\mathcal{M}}\tau \wedge e\,\big(\langle T_{(\tau )}^\mu \rangle \delta _\rho \tau _\mu  +\langle T_{(e)}^\mu \rangle  \delta _\rho e_\mu +\langle T_{(\omega )}^\mu \rangle  \delta _\rho \omega_\mu \big) ~.
\end{align}
Ultimately, the background is given by some solution of magnetic Carroll dilaton gravity, which we shall again take to be the 2d Carroll--Schwarzschild solution. In a first-order formulation the background field content is therefore given by $(\tau ,e,\omega )$ and $(X,\XH ,\XP )$ (cf. section \ref{sec:2.1.1}) and we also assumed in \eqref{eq:gen:var} that the matter theory does not couple to the latter background data\footnote{A common generalization of this is to allow a coupling to the dilaton $X$, which typically arises from spherically reducing higher-dimensional matter models \cite{Grumiller:2002nm}.}. Demanding that the conformal anomaly is local and Carroll boost invariant, there are only two choices with the correct mass dimension,
\begin{align}\label{eq:WA_proposal}
    \delta _\rho W=\int _{\mathcal{M}}\; \rho \big( \alpha _1 \,\dd \omega +\alpha _2 \dd e\big) && \alpha _1,\alpha _2 \in \mathbb{R} 
\end{align}
where we used that under a Carroll boost $\delta _\lambda \omega =\dd \lambda $. Moreover, since the term $\dd e$ vanishes for any solution to magnetic Carroll dilaton gravity it does not lead to anomalous terms in the Weyl Ward identity. We therefore drop this term from now on. The dimensionless coefficient $\alpha _1$ stays undetermined at this point and will be fixed later by demanding consistency with the asymptotic energy density derived in the previous section. 

For computing the Weyl Ward identity we first need to prescribe the transformation behaviour of the Carroll boost connection $\omega $. While, in general, there is some freedom in choosing this transformation if the background theory is unspecified, we may use the Weyl-transformation properties of 2d Carroll dilaton gravity introduced in section \ref{sec:2.1.1} as a guidance. Demanding on-shell consistency with \eqref{eq:weyl7} we propose 
\begin{align}
    \delta _\rho \omega =\big(-\tau e^\mu +ev^\mu \big)\partial _\mu \rho ~.
\end{align}
Indeed, one may check that for a dilaton dependent $\rho (X)$ this reduces to \eqref{eq:weyl7} once the equations of motion and the gauge condition $\XP =0$ are used. On static backgrounds infinitesimally close to flat Carroll spacetime this also makes the Weyl anomaly match with the result of \cite{Bagchi:2021gai}. Finally, the expression \eqref{eq:WA_proposal} with $\alpha _2=0$ can be shown to be consistent in the sense of Bardeen and Zumino \cite{Bardeen:1984pm}, i.e., 
\begin{align}
    [\delta _{\rho _1},\delta _{\rho _2}]W=0 ~.
\end{align}
This is another reason for setting $\alpha _2=0$, since this term would not lead to a consistent anomaly. Using that the frame variables transform as in section \ref{sec:Weyl} with $z=1$, i.e. $\delta _\rho \tau =\rho \tau $ and $\delta _\rho e=\rho e$, we may determine the Weyl Ward identity as
\begin{align}
    \mathfrak{e}\big(\langle T^\mu _{(\tau )}\rangle \tau _\mu +\langle T^\mu _{(e)}\rangle e_\mu \big)-\partial _\alpha \Big(\mathfrak{e}\langle T^\mu _{(\omega )}\rangle (-\tau _\mu e^\alpha +e_\mu v^\alpha ) \Big)=\alpha _1 2\mathfrak{e} e^\mu v^\nu \partial _{[\mu }\omega _{\nu ]} \label{ah4}~.
\end{align}
Similarly, the Carroll boost and diffeomorphism Ward identities are given by
\begin{gather}
    \mathfrak{e}\langle T_{(\tau )}^\mu \rangle e_\mu +\partial _\mu (\mathfrak{e}\langle T^\mu _{(\omega )}\rangle )=0 \label{ah5}\\
    \partial _\mu \Big(\mathfrak{e}\langle T^\mu _{(\tau )}\rangle \tau _\alpha +\mathfrak{e}\langle T^\mu _{(e )}\rangle e _\alpha +\mathfrak{e}\langle T^\mu _{(\omega )}\rangle \omega _\alpha\Big)=\mathfrak{e}\langle T^\mu _{(\tau )}\rangle \partial _\alpha \tau _\mu +\mathfrak{e}\langle T^\mu _{(e )}\rangle \partial _\alpha e _\mu +\mathfrak{e}\langle T^\mu _{(\omega )}\rangle \partial _\alpha \omega _\mu ~. \label{ah6}
\end{gather}
We proceed with solving these identities for the 2d Carroll Schwarzschild black hole of section \ref{sec:carr_schwarzsch}. The background fields are given by \eqref{eq:HE_background} together with 
\begin{align}
    \omega =\frac{2\rad  -\rad _s}{\rad ^2}\dd t ~.
\end{align}
Note, that in these coordinates we have $\mathfrak{e}=1$. We may first use the Carroll boost- and Weyl Ward identities to algebraically solve for $T^\mu _{(\tau )}$, 
\begin{align}
    T_{(\tau )}=\Big[-\frac{\rad ^2\langle T_{(e)}^\rad \rangle + \xi ^3\alpha _1}{\rad ^2\xi ^2}+\frac{\rad (\xi ^2-1)\langle T_{(\omega )}^t\rangle +\xi ^2\rad ^2\partial _\rad \langle T_{(\omega )}^t\rangle}{\rad ^2\xi }\Big]\partial _t - \xi \partial _\rad \langle T_{(\omega )}^\rad \rangle \partial _\rad ~. \label{eq:tsolut}
\end{align}
Inserting into the diffeomorphism Ward identities then yields two equations in four unknowns,
\begin{align}
    2\rad ^2\xi ^2 \partial _\rad ^2 \langle T_{(\omega )}^\rad \rangle+(\rad -3\rad \xi ^2)\partial _\rad \langle T_{(\omega )}^\rad \rangle +2\xi ^2&\langle T_{(\omega )}^\rad \rangle =0 \label{eq:firsteq}\\
    \partial _\rad \langle T_{(e)}^\rad \rangle + \frac{1-\xi ^2}{2\rad \xi ^2}\langle T_{(e)}^\rad \rangle + \alpha _1 \frac{\xi ^3-\xi }{2\rad ^3}&=\frac{2\xi ^2-3\xi ^4-1}{2\rad ^2\xi }\langle T_{(\omega )}^t \rangle +\frac{\xi ^3-\xi }{2\rad }\partial _\rad \langle T_{(\omega )}^t \rangle \label{eq:secondeq}
\end{align}
which may be solved for $\langle T_{(\omega )}^\rad \rangle$ and $\langle T_{(e)}^\rad \rangle$ once $\langle T_{(\omega )}^t \rangle$ is given. The component $\langle T_{(e)}^t \rangle $ does not feature in these equations and may be chosen arbitrarily. Thus, in contrast to the Lorentzian case, we do not have enough conditions to define all components of the Carroll stress-energy tensor. We leave a more general analysis of the space of solutions to future work (see also chapter \ref{ch:conclusion}) and for now just focus on the simple choices
\begin{align}
    \langle T_{(\omega )} ^t\rangle = 0 && \langle T_{(e)}^t \rangle = q(\rad ) 
\end{align}
where $q(\rad )$ is some function of $\rad $. Equations \eqref{eq:firsteq} and \eqref{eq:secondeq} may then be solved by 
\begin{align}
    \langle T_{(\omega )} ^\rad\rangle = 0 && \langle T_{(e)} \rangle = q(\rad )\partial _t + \frac{1}{\sqrt{\xi }}\Big(c_0+\alpha _1 \frac{3\rad _s^2-4\rad _s\rad }{24 \rad ^4 }\Big)\partial _\rad 
\end{align}
with an integration constant $c_0$. In summary, this provides us with a class of static solutions to the Ward identities \eqref{ah4}, \eqref{ah5} and \eqref{ah6} given by 
\begin{align}
    \langle T_{(\omega )} \rangle = 0 && \langle T_{(e)} \rangle = q(\rad )\partial _t + \frac{1}{\sqrt{\xi }}\Big(c_0+\alpha _1 \frac{3\rad _s^2-4\rad _s\rad }{24 \rad ^4 }\Big)\partial _\rad  && T_{(\tau )}=-\frac{\rad ^2\langle T_{(e)}^\rad \rangle + \xi ^3\alpha _1}{\rad ^2\xi ^2}\partial _t \label{eq:state_sol}
\end{align} 
parametrized by $c_0$ and $q(\rad )$. Let us look at the physical properties of the corresponding states. We may construct a conserved current for each background symmetry analogous to what was done in section \ref{sec:coupl_conn}. A symmetry is defined by a vector field $\zeta $ that leaves the background invariant up to local Carroll transformations (see also the discussions in sections \ref{sec:symmetries} and \ref{sec:5.1.3}),
\begin{align}
    \mathcal{L}_\zeta \tau + \delta _{\lambda (\zeta )}\tau=0 && \mathcal{L}_\zeta e=0 && \mathcal{L}_\zeta \omega +\delta _{\lambda (\zeta )}\omega =0 \\  \mathcal{L}_\zeta X=0 &&  \mathcal{L}_\zeta \XH =0 &&  \mathcal{L}_\zeta \XP + \delta _{\lambda (\zeta )}\XP=0 ~. 
\end{align}
One can then show by using the generalized covariant conservation equation \eqref{ah6} that the current density defined by 
\begin{align}
    \langle j^\mu \rangle =\mathfrak{e}\langle T_{(\tau )}^\mu \rangle \tau _\nu \zeta ^\nu +\mathfrak{e}\langle T_{(e )}^\mu \rangle e_\nu \zeta ^\nu +\mathfrak{e}\langle T_{(\omega )}^\mu \rangle \omega _\nu \zeta ^\nu +\mathfrak{e}\langle T_{(\omega )}^\mu \rangle \lambda (\zeta )
\end{align}
is indeed conserved, $\partial _\mu \langle j^\mu \rangle =0$. Since for our background time translations $\zeta =\partial _t$ are such a symmetry with $\lambda (\zeta )=0$ we obtain an associated energy density 
\begin{align}
    \langle \mathcal{E} \rangle = \langle j^t \rangle = \langle T_{(\tau )}^t \rangle \sqrt{\xi }+\langle T_{(\omega )}^t \rangle \omega _t 
\end{align}
which for states with \eqref{eq:state_sol} is regular at $\rad \to \rad _s$ if the integration constant is fixed as
\begin{align}
    c_0 = \frac{\alpha _1}{24 \pi \rad _s^2} ~.
\end{align}
This in turn produces an asymptotic energy density
\eq{
    \lim _{r\to \infty }\langle \mathcal{E}\rangle  =-\frac{\alpha _1}{24r_s^2} 
}{eq:result}
which coincides precisely with the Carroll--Hartle--Hawking energy density \eqref{eq:angelinajolie2} for $\alpha_1=-\frac{1}{4\pi}$. As a consequence of the Carroll boost Ward identity, there is again no energy flux allowed for the chosen state, i.e., $\partial _t \langle \mathcal{E} \rangle =0$. 

While this derivation is not entirely independent of the previous one, since it required choosing the anomaly coefficient $\alpha _1$ appropriately it still suggests a similar behaviour of Carroll field theories coupled to two-dimensional Carroll black holes: The states of such field theories acquire temperature dependent expectation values of the asymptotic energy density once regularity at the Carroll extremal surface is enforced. However, in neither case can the Carroll black hole evaporate since the energy flux is identically vanishing. We shall conclude our analysis here and refer to the outlook chapter \ref{ch:conclusion} for further discussions.

\chapter{Conclusion and Outlook}
\label{ch:conclusion}
We shall conclude our explorations in the world of Carroll symmetries by pointing out a few open ends and suggesting several directions that may be worthwhile to investigate in future research.

\paragraph{Rotating Carroll black holes in 3d}
While we gave a possible definition for a rotating Carroll black hole in 3d in subsection \ref{sec:8.3}, the Carroll BTZ black hole, this approach was somewhat indirect: It used an intrinsically two-dimensional description in terms of Carroll dilaton gravity obtained after performing a Kaluza--Klein reduction of three-dimensional Einstein gravity with $\Lambda <0$. For convenience, we present this solution here again, 
\begin{align}
    h_{\mu \nu }\dd x^\mu \dd x^\nu =\frac{\dd X^2}{\XH ^2} && v^\mu \partial _\mu = -\frac{1}{\XH }\partial _t && \tau _\mu \dd x^\mu =\XH \,\dd t&& A=\frac{J}{X}\dd t
\end{align}
where $\XH =-\sqrt{X^4-2MX^2+J^2}/X$. From a 2d point of view, the angular momentum ends up being encoded into an additional constant of motion $J$ that is just the $U(1)$ charge of a 2d electric Carroll Maxwell field. One may ask whether there still exists an uplift of these two-dimensional variables that is a solution to some higher-dimensional theory of Carroll gravity. A difficulty arising here is a certain asymmetry in the Carroll limit: The 2d metric part is described by a magnetic limit, while the Maxwell part is described by an electric limit. It is not clear how taking such an ``electromagnetic limit'' should work directly in the 3d setting, where there is really just a metric and no Maxwell field.

Nevertheless, recent investigations of three-dimensional magnetic AdS-Carroll gravity \cite{Aviles:2025ygw} show that there do exist solutions with non-zero angular momentum. The latter is thereby defined as an asymptotic boundary charge, and the corresponding bulk geometries exhibit thermal properties that indeed satisfy a first law, albeit with an entropy term that has a sign more reminiscent of cosmological solutions. In this regard, those geometries behave differently from the ones discussed here suggesting that the three-dimensional theory we are looking for is not magnetic Carroll gravity.

\paragraph{Rotating Carroll black holes in higher dimensions}
So far, we did not touch a higher-dimensional version of rotating Carroll black holes such as a Carroll limit of the Kerr spacetime. A naive limit of the latter in, e.g., Boyer--Lindquist coordinates may be attempted by first writing the Lorentzian metric in terms of the energy $E$ and the coupling constant $G_M$, which we both want to hold fixed (see section \ref{sec:bronstein}),
\begin{align}
    \dd s^2 = -\Big(1-\frac{2EG_Mr}{\Sigma }\Big)c^2\dd t^2 -\frac{4EG_Mar\sin ^2\theta }{\Sigma }&c\,\dd t \dd \phi +\frac{\Sigma }{\Delta }\dd r^2+\Sigma \dd \theta ^2\\
    &+\Big(r^2+a^2+\frac{2EG_Ma^2\sin ^2\theta }{\Sigma }\Big)\sin ^2\theta \dd \phi ^2 \nonumber
\end{align}
where $\Delta = r^2-2G_MEr+a^2$ and $\Sigma =r^2+a^2\cos ^2\theta $. The angular momentum parameter is given by $a=Jc/E$, and one observes two possible ways of taking the limit on the level of this constant: Either we fix $J$, which implies $a\to 0$ in the limit and thus reduces the geometry to the Carroll--Schwarzschild spacetime. Alternatively, we might fix $a$, which yields a diagonal Carroll metric 
\begin{align}
    h_{\mu \nu }\dd x^\mu \dd x^\nu =\frac{\Sigma }{\Delta }\dd r^2+\Sigma \dd \theta ^2+\Big(r^2+a^2+\frac{2EG_Ma^2\sin ^2\theta }{\Sigma }\Big)\sin ^2\theta \dd \phi ^2 ~.
\end{align}
However, such a metric does not solve the constraints of magnetic Carroll gravity as introduced in section \ref{sec:mag_carr}. In particular, one can show that the Ricci scalar of the associated ADM-metric is non-zero, $R^{(3)}[h_{ij}]\neq 0$. Thus, it seems that there is no Carroll limit of the Kerr spacetime that is a solution of magnetic Carroll gravity and at the same time carries non-zero angular momentum. This was made more precise in a recent no-go theorem \cite{Kolar:2025ebv}.

There is one catch in this story that we would like to point out, and that constitutes a possibly interesting research avenue: It addresses Carroll geometries as limits of Lorentzian geometries. However, there is evidence that not all features of Carroll theories can be obtained from a $c\to 0$ limit, e.g., on a level of representation theory \cite{Figueroa-OFarrill:2022mcy} or on a level of quantum field theories \cite{Cotler:2024xhb}. In other words, there might still be a possibility of defining a truly rotating higher-dimensional black hole in magnetic Carroll gravity - it just does not arise from a limit. Further evidence in support of this assertion comes from an asymptotic symmetry analysis \cite{Perez:2021abf}, which shows that there do exist well-defined boundary charges associated to angular momentum. It could be interesting to start from there and try to construct geometries that exhibit non-vanishing values of these charges. 

\paragraph{Charged Carroll black holes}
Similarly to the Carroll BTZ black hole, we defined charged Carroll black holes in a spherically reduced theory after taking the Carroll limit. It would be pleasing to see that first taking such a limit
in higher dimensions and then performing a spherical reduction
leads to the same 2d solutions. Additionally, in \cite{deBoer:2023fnj} the combined magnetic gravity and magnetic Maxwell limits were investigated for a four-dimensional Reissner--Nordstr\"om black hole carrying both electric and magnetic charge. It would be interesting to examine how these features manifest after performing a spherical reduction and to determine what kind of effective two-dimensional model emerges in this case. 

Generalizations to completely different types of charged Carroll black
holes are possible as well and can be done on a case-by-case basis.
Examples that come to mind are 2d type 0A string theory with equal
number $q_e$ of electric and magnetic D0 branes \cite{Douglas:2003up,
  Gukov:2003yp} and the dimensionally reduced Chern--Simons term
\cite{Guralnik:2003we, Grumiller:2003ad}.

\paragraph{Carroll fields coupled to the connection}
We have seen in section \ref{sec:gauging_procedure} that, once requiring a Carroll geometry to carry a torsion-free and metric-compatible connection, there are still independent degrees of freedom remaining in a part of the connection. This is in contrast to the Riemannian case. On a (magnetic) Carroll gravity level, this part is quite important, since it plays the role of the canonical momenta \eqref{eq:renaming_C} in an ADM formulation or the one of Lagrange multipliers in a second order formulation (see eq. \eqref{eq:second_order_mag_action}), ensuring local Carroll boost invariance of the action. 

When analyzing field theories on such backgrounds, it is natural to also consider couplings to this part of the connection. We presented an example of a scalar field with this property in \eqref{eq:nonminmag}, but otherwise focused on the simplified case in which this coupling is absent. It would be interesting to explore this possibility in more detail and to study explicit solutions of the resulting theories. Since the Carroll boost Ward identity acquires additional terms in this setting (see section \ref{sec:coupl_conn}), statements such as the absence of energy flux will have to be reconsidered.

A specific application that comes to mind in this case is holography in asymptotically flat spacetimes. While there is additional conformal symmetry present in that case, the undetermined part of the connection still has a similar structure and encodes the gravitational shear. Thus, any holographically dual theory should also couple to this field if it is to carry information about the radiative structure of spacetime. 

\paragraph{Asymptotic symmetries of 4d asymptotically flat spacetimes} Related to the previous point, we have seen in section \ref{sec:Quantum} how an intrinsically Carroll derivation of stress-energy tensor brackets is able to reproduce the asymptotic symmetries of three-dimensional asymptotically flat spacetimes for Einstein gravity. The only input from holography was the presence of a Carroll boost anomaly which could be related to a certain central extension being switched on. It would be interesting to perform a similar computation for the four-dimensional case. An important difference in that regard is the additional coupling to the gravitational shear, which introduces further response functions in the variation of the partition function \cite{Hartong:2025jpp}.
A Carroll derivation of the BMS charge brackets—originally obtained from bulk considerations in \cite{Barnich:2011mi}—could provide a deeper understanding of their geometric origin and illuminate how these structures emerge naturally within the Carroll framework. This may also shed further light on the holographic principle applied to asymptotically flat spacetimes.

\paragraph{One-loop effects on Carroll Schwarzschild spacetime}
We have seen in section \ref{sec:Conf_anomaly_Carr} that the Ward identities of a Carroll quantum field theory on a 2d Carroll Schwarzschild background involve additional response functions $\langle T_{(\omega )}^\mu \rangle$ associated to varying the Carroll boost connection. We only considered a special family of states where these responses identically vanish - this allowed us to connect with Carroll versions of the Hartle--Hawking or Boulware states obtained as a limit from the Lorentzian setting. It could be worthwile to study more general scenarios where $\langle T_{(\omega )}^\mu \rangle\neq 0$ and see which possibilities arise from demanding regularity at the Carroll extremal surface or at infinity. Given the discussions in the previous paragraphs about rotating Carroll geometries, it is plausible that there exist cases which do not follow from any limit of a Lorentzian theory.  

\paragraph{Tantum gravity limit on the field theory side}
We mostly looked at the tantum gravity limit from a bulk perspective, since we were interested in the solutions of its magnetic Carroll gravity approximation. Following the typical procedure in holography, one might also ask about possible asymptotic falloff conditions and the associated asymptotic symmetries in order to find out more about a putative holographically dual theory. There are already some investigations in these directions in two \cite{Grumiller:2020elf}, three \cite{Aviles:2025ygw} and four spacetime dimensions \cite{Perez:2022jpr}, mostly on the level of asymptotic symmetries. It would also be interesting to investigate whether, for example in the case of the Carroll JT model, the Cardy-like formula derived in subsection \ref{sec:carjt} admits an interpretation from a dual viewpoint based on microscopic state counting. Understanding whether such a statistical perspective exists could shed further light on the thermodynamic structure of Carroll black holes.

\paragraph{Carroll de Sitter}
Finally, we mention that one may also look at Carroll limits of cosmological solutions. As a simple scenario consider again a description in terms of two-dimensional Carroll dilaton gravity \eqref{eq:car24} where, e.g., the choice of potentials 
\begin{align}
    U(X)=-\frac{1}{2X} && V(X)=1-\Lambda X
\end{align} 
together with $X=\rad ^2$ leads to the Carroll metric data
\begin{align}
    \dd s^2=\frac{\dd \rad ^2}{1-\frac{M}{\rad }-\frac{\Lambda }{3}\rad ^2} && v=-\frac{1}{\sqrt{{1-\frac{M}{\rad }-\frac{\Lambda }{3}\rad ^2}}}\partial_t ~.
\end{align}
For $\Lambda >0$, this describes a Carroll limit of a Schwarzschild--de Sitter black hole in the static patch. It would be interesting to investigate the thermodynamics and Carroll extremal surfaces of this spacetime and explore whether it is again possible to get a smooth solution for some $M$ and choice of periodicity $\beta$, i.e., try to apply the methods we used in subsection \ref{sec:thermal_sols}.

\backmatter

\bibliographystyle{fullsort}
\bibliography{review}

\end{document}